\documentclass[11pt]{article}

\usepackage{customcommands}
\usepackage{nicefrac}
\usepackage{etoolbox}
\usepackage[multiple]{footmisc}
\usepackage{afterpage}
\usepackage{xr}
\usepackage{algpseudocode}
\usepackage{rotating}

\newcommand{\supplementtitle}[1]{
    \clearpage
    \thispagestyle{empty}
    \begin{center}
        \LARGE  #1
    \end{center}
    \vspace{0.5cm}
}

\algnewcommand\algorithmicforeach{\textbf{for each}}
\algdef{S}[FOR]{ForEach}[1]{\algorithmicforeach\ #1\ \algorithmicdo}

\AtBeginEnvironment{thebibliography}{\linespread{1.2}\selectfont}

\title{\vspace{-2cm} Identification of Dynamic Panel Logit Models with Fixed Effects\\~\\ 
	\footnotetext{We thank Victor Aguirregabiria, Roger Koenker, Ismael Mourifi\'e and Stanislav Volgushev for useful discussion. We are grateful to numerous seminar participants for their feedback, and are especially grateful to Francesca Molinari and three anonymous referees for their helpful comments. All errors are our own.}	
	\vspace{-.4in}
}
\date{\small First Version: April 15, 2021, This Version: \today}

\author{
    \begin{tabular}{cccc}
        Christopher Dobronyi\footnote{Christopher Dobronyi, University of Toronto. Email: dobronyi@google.com.
 	} &&&  Jiaying Gu\footnote{Jiaying Gu, Department of Economics, University of Toronto, 150 St. George Street, Toronto, Ontario, M5S3G7, Canada. Email: jiaying.gu@utoronto.ca.} \\  
 	\textit{\small University of Toronto} &&& \textit{\small University of Toronto}\\ \\
        Kyoo il Kim\footnote{Kyoo il Kim, Department of Economics, Michigan State University, 486 W. Circle Dr, East Lansing, MI 48864, USA. Email: kyookim@msu.edu. 
 	} &&& Thomas M. Russell\footnote{Thomas M. Russell, Department of Economics, Carleton University, 1125 Colonel By Drive, Ottawa, Ontario, K1S 5B6, Canada. Email: thomas.russell3@carleton.ca.} \\
 	\textit{\small Michigan State University} &&& \textit{\small Carleton University}\\ \\
    \end{tabular}
}

\begin{document}

\maketitle
\thispagestyle{empty}
\vspace{-.4in}

\begin{abstract}
\noindent We show that identification in a general class of dynamic panel logit models with fixed effects is related to the \textit{truncated moment problem} from the mathematics literature. We use this connection to show that the identified set for structural parameters and functionals of the distribution of latent individual effects can be characterized by a finite set of conditional moment equalities subject to a certain set of shape constraints on the model parameters. In addition to providing a general approach to identification, the new characterization can deliver informative bounds in cases where competing methods deliver no identifying restrictions. We then present an estimation and inference procedure that uses semidefinite programming methods, is applicable with continuous or discrete covariates, and can be used for models that are either point- or partially-identified. Finally, we illustrate our identification results with a number of examples, and we provide an empirical application to employment dynamics using data from the National Longitudinal Survey of Youth. 
\end{abstract}


\clearpage
\pagenumbering{arabic}

\section{Introduction} \label{sec1}

Dynamic panel logit models are valuable empirical tools for modeling repeated choices made by households, firms and individual consumers. These models are favored in part because they can account for \textit{permanent unobserved heterogeneity}, allowing the researcher to distinguish between \textit{true dynamics}  induced by lagged choice dependence, and \textit{spurious dynamics}, which are a result of persistent individual heterogeneity (see \cite{heckman1981heterogeneity}). Work on identification in these models has a long history, and can be roughly divided into two main areas: the sufficient statistics approach (e.g. \cite{rasch1960probabilistic}, \cite{rasch1961general}, \cite{andersen1970asymptotic}, \cite{chamberlain1980analysis}, \cite{chamberlain1985heterogeneity}, \cite{honore2000panel}, \cite{magnac2000subsidised}, \cite{hahn2001information}, \cite{gu2023information}) and the functional differencing approach (e.g. \cite{johnson2004identification},  \cite{bonhomme2012functional}, \cite{honore2024moment}). 

In this paper we study a new and general approach to identification in a class of dynamic panel logit models with latent individual effects, providing an alternative to the sufficient statistics and functional differencing approaches. Using the structure of the logistic distribution, we show that the likelihood for these models can be written as a polynomial in certain \textit{generalized moments} of the distribution of latent individual effects, revealing a connection to the \textit{truncated moment problem} dating back to \cite{chebyshev1874valeurs} and \cite{stieltjes1894recherches}. Through this connection, we show that the identified set of structural parameters can be characterized by a set of conditional moment equalities subject to a certain set of shape restrictions on the model parameters. Estimation and inference is then based on repeatedly solving semidefinite programs, a special kind of convex program which can be solved quickly and reliably. We then show how to adapt the inference procedure in \cite{chernozhukov2023constrained} to construct confidence sets for the model parameters. A key advantage of our method is its ability to handle partially identified models, and we show that the new characterization can deliver sharp bounds in cases where existing methods deliver no identifying restrictions. In addition, unlike many existing approaches, our approach can be used to construct the identified set of certain functionals of the distribution of latent individual effects, including average marginal effects and the average structural function. 

There are two main challenges when studying dynamic panel logit models: the initial conditions problem, and the incidental parameters problem. The initial conditions problem arises because the joint distribution of the initial choices and the individual fixed effects is not nonparametrically point-identified (e.g. see \cite{heckman1981incidental} and \cite{wooldridge2005fixed}). The incidental parameters problem refers to the fact that, when the number of time periods is fixed, it is generally not possible to consistently estimate individual fixed effects, and attempting to do so can bias the estimates of the structural parameters (e.g. \cite{neyman1948consistent}).  This paper focuses on the incidental parameters problem, for which there are two common approaches: the random effects approach, and the fixed effects approach.\footnote{For a more complete survey of the literature, we refer the readers to \cite{arellano2001panel}.} The (correlated) random effects approach places restrictions on the joint distribution of the initial conditions and the individual effects using a parametric distribution or a finite mixture (e.g. \cite{chamberlain1980CRC}, \cite{wooldridge2005initial}). When these assumptions are satisfied, the structural parameters and various functionals of the latent variable distribution are point-identified and can be consistently estimated. In contrast, the fixed effects approach treats the latent individual effects as random, but is entirely agnostic about their distribution and their dependence on the initial conditions.\footnote{Consistent with the existing literature, if the distribution of the time-invariant individual effects is not parametrically specified and is allowed to depend arbitrarily on covariates and initial conditions, then we refer to this as the ``fixed effects'' approach. See for instance \cite{honore2006bounds} p. 612 for similar terminology. Throughout the paper we used ``fixed effects'' and ``latent individual effects'' interchangeably.} As a result, the fixed effects approach is more flexible, but presents a number of interesting identification and estimation challenges.

Under the fixed effects approach, in some cases the structural parameters are identified and can be consistently estimated using \textit{conditional maximum likelihood}, pioneered by \cite{rasch1960probabilistic}, \cite{rasch1961general}, \cite{andersen1970asymptotic}, \cite{chamberlain1980analysis}, and \cite{chamberlain1985heterogeneity}. This method involves finding a minimally sufficient statistic for the fixed effects, and constructing a partial likelihood that conditions on this statistic. By the definition of sufficiency, this partial likelihood no longer depends on the fixed effects. If this partial likelihood also depends on the structural parameters, then the first-order conditions to maximize the partial likelihood provide moment conditions that can be used for identification and estimation. \cite{honore2000panel} extend this approach to dynamic logit models with time-varying covariates. Unfortunately, few models admit nontrivial sufficient statistics, and so the method does not always result in useful identifying restrictions. 
Even when it does, it can fail to exhaust all of the model's identifying content, and so can it can deliver nonidentification in cases when the structural parameters are point- or partially-identified.\footnote{There is one exception: if the likelihood of the sufficient statistics no longer depends on the structural parameters, then the conditional maximum likelihood method utilizes all relevant identifying information for the structural parameters. In many cases, including in the dynamic panel logit model, this condition is generally not satisfied.} In contrast, our approach always delivers the sharp set of model restrictions, and can always be used to construct the sharp identified set for the structural parameters. Later, we show specific examples where we are able to construct the sharp identified set for the structural parameters when conditional maximum likelihood delivers no identifying restrictions.

Our analysis also sheds light on the functional differencing approach proposed by \cite{johnson2004identification} and \cite{bonhomme2012functional} and used for a similar class of models by \cite{honore2024moment}. At a high level, functional differencing searches for a collection of moment functions that do not depend on the latent variables, but that deliver some identifying information about the structural parameters. Functional differencing proceeds on a case-by-case basis, searching for moment conditions specific to each model. Finding the relevant collection of moment functions has historically been challenging. It is also difficult to determine whether all relevant moment functions have been found, and whether a collection of moment functions exhaust all the identifying restrictions of the model. Even given all relevant moment functions, it can be difficult to prove point identification of the structural parameters, a precondition for using standard estimation and inference methods. Despite these challenges, recent progress was made by \cite{honore2024moment}, who found new moment conditions for the structural parameters in the AR(1) dynamic panel logit model with covariates, and proved point identification under certain conditions. They also found moment conditions in models for which the conditional maximum likelihood approach provides no identifying restrictions, such as the AR(2) dynamic panel logit model. 

Relative to functional differencing, the advantage of our approach is its generality. In particular, it relies on a general structure of the logistic likelihood function that makes it relatively straightforward to apply to different models. It is also able to handle both point- and partially-identified models---such as short panels or models with limited covariate variation---and can be used to study functionals of the distribution of fixed effects. However, our approach also has an interesting connection to functional differencing: as a by-product of our analysis we show how the moment conditions from functional differencing can be constructed from the basis of the left null space of a certain matrix that arises in our approach. This allows us to provide a simple geometric explanation for why our approach sometimes provides more identifying restrictions than approaches based on functional differencing, and we provide a number of examples to illustrate when this is the case. This connection also suggests a new method for constructing moment conditions for functional differencing, which may be a promising avenue of future research.



As mentioned repeatedly, an important feature of our approach is its ability to study functionals of the distribution of fixed effects, including certain counterfactual parameters. This is done by linking the functional of interest to the generalized moments of the distribution of the latent individual effects. In contrast, both the conditional maximum likelihood approach and the functional differencing approach aim at removing the individual effects to derive moment conditions for the structural parameters. As a result, they cannot be used to study functionals of the distribution of latent individual effects. 
Our results on functionals relate to \cite{aguirregabiria2024identification}, who were the first to show that the average marginal effect of the lagged choice in the AR(1) dynamic logit model is point-identified. While \cite{aguirregabiria2024identification} restrict attention to models in which the structural parameters and the functional of interest are both point-identified, we generalize their setting to allow for partially-identified models, and cover a broader class of functionals. We also provide easily-checked sufficient conditions under which functionals are  point-identified even when the latent variable distribution is not point-identified.

Outside of the sufficient statistics and functional differencing approaches, other approaches have been proposed that are based on discretizing the distribution of latent individual effects. This includes the linear programming approach in \cite{honore2006bounds} and the quadratic programming approach in \cite{chernozhukov2013average}. These approaches have similar advantages to our method, including the ability to handle point- or partially-identified models and the ability to study functionals of the latent variable distribution. However, both of these approaches require choosing a finite grid for the support of the latent distribution of individual effects. Furthermore, neither paper studies the impact of the approximation on rates of convergence or inference, and both papers focus on discrete covariates. Rather than construct a finite approximation to the infinite-dimensional latent variable distribution, we instead show that the latent variable distribution can be completely summarized by a finite vector of moments.  Furthermore, our method maintains a similar computational cost.\footnote{For instance, in a simulation exercise with an AR(1) model with $T=3$ and $n=10^3$, across $100$ replications on average the quadratic program in \cite{chernozhukov2013average} took about $0.0042$ seconds to solve with a grid size of $61$ points for the $\alpha$ distribution, and about $0.2374$ seconds to solve for a grid size of $601$ points. For the same model, our proposed semidefinite program took an average of $0.0082$ seconds to solve. 
}

The rest of the paper is organized as follows. Section \ref{section_methodology} introduces the identification problem and our main assumptions, and works through an example to illustrate our approach. General identification results and connections to the existing literature are presented in Section \ref{section_general_results}. Estimation and inference using semidefinite programming is presented in Section \ref{section_consistency_and_inference}. An empirical application is presented in Section \ref{section_application}, and Section \ref{section_conclusion} concludes. The proofs of the main results, and additional material including a brief Monte Carlo study, can be found in the Online Supplementary Material. 

\section{Methodology}\label{section_methodology}
\subsection{Main Assumptions and Examples}
We begin with some examples of models that fit into our framework.
\begin{example}[$AR(1)$ dynamic logit binary choice] \label{example_AR1} Consider a model of panel binary choice:
\begin{align*}
Y_{it} = 1\{ \alpha_i + \beta Y_{it-1} + \bm X_{it}^\top \gamma \geq \epsilon_{it}\},
\end{align*}
where the researcher observes $(Y_{i0}, \bm Y_i) = (Y_{i0}, Y_{i1}, \dots, Y_{iT}) \in \{0,1\}^{T+1}$ and covariates $\bm X_i = (\bm X_{i1}, \dots, \bm X_{iT})\in \mathcal{X}^T$ for individuals $i = 1, \dots, n$, and the i.i.d.\ utility shocks $(\epsilon_{it})_{t=1}^{T}$ follow a standard logistic distribution. Here, the latent variable $\alpha_i$ characterizes persistent unobserved heterogeneity, and is allowed to depend on the initial choice $Y_{i0}$ and the covariates $\bm X_i$. The utility shocks $(\epsilon_{it})_{t=1}^{T}$ are assumed to be independent of $(Y_{i0}, \bm X_{i},\alpha_{i})$. When $\beta = 0$, the static version of the model is known as the Rasch model (see \cite{rasch1960probabilistic}). The dynamic version of the model is analyzed in \cite{chamberlain1985heterogeneity} using conditional maximum likelihood. This model is widely used to study unemployment and labor force participation (e.g. \cite{card1987measuring}). 
\end{example}


\begin{example}[$AR(p)$ dynamic logit binary choice]\label{example_ARp} Consider the following model for a panel of binary choices:
\begin{align*}
	Y_{it} = 1\left\{\alpha_i + \sum_{k=1}^p \beta_k Y_{it-k} + \bm X_{it}^\top \gamma \geq \epsilon_{it}\right\},
\end{align*}
where the researcher observes $(Y_{i,1-p}, \dots, Y_{i0}, \bm Y_i) = (Y_{i,1-p}, \dots, Y_{i0}, Y_{i1}, \dots Y_{iT})$ and covariates $\bm X_i = (X_{i1}, \dots, X_{iT})$. Again, assume $\alpha_{i}$ is a latent individual-specific variable with an unrestricted distribution, and assume the per-period utility shocks $(\epsilon_{it})_{t=1}^{T}$ are i.i.d.\ with a standard logistic distribution, independent of $(Y_{i0}, \bm X_{i},\alpha_{i})$. \cite{chamberlain1985heterogeneity}, \cite{honore2019identification} and \cite{honore2024moment} analyzed the AR(p) model for the special case with $p = 2$. The approach in this paper applies for any finite $p$. 
\end{example}

\begin{example}[Dynamic AR(1) ordered logit model]\label{example_ODC} Consider the following panel ordered choice model with $M$ choice options: 
	\[
	Y_{it} = \begin{cases} 1 &\text{ if } \alpha_i + \sum_{m=1}^M \beta_m 1\{Y_{it-1} = m\} + \bm X_{it}^\top \eta + \epsilon_{it} \in (-\infty, \gamma_1],\\
		2 &\text{ if } \alpha_i + \sum_{m=1}^M \beta_m 1\{Y_{it-1} = m\} + \bm X_{it}^\top \eta + \epsilon_{it} \in (\gamma_1, \gamma_2],\\
		\vdots & \qquad\qquad\qquad\qquad\qquad\qquad\vdots \\
		M &\text{ if } \alpha_i + \sum_{m=1}^M \beta_m 1\{Y_{it-1} = m\} + \bm X_{it}^\top \eta + \epsilon_{it} \in (\gamma_{M-1},+\infty),
		\end{cases} 
	\]
	where we observe $(Y_{i0}, \bm Y_i) = (Y_{i0}, Y_{i1}, \dots, Y_{iT})$ and covariates $\bm X_i = (X_{i1}, \dots, X_{iT})$. The utility shocks $(\epsilon_{it})_{t=1}^{T}$ are i.i.d.\ with a standard logistic distribution, and are independent of $(Y_{i0},\bm X_{i},\alpha_{i})$. If the coefficients $\{\beta_m\}_{m=1}^{M}$ are all zero, then we have the static version of the panel ordered logit model. The latent variable $\alpha_i$ is allowed to have any distribution, which can also depend on $(Y_{i0}, \bm X_i)$. The static version of the model is analyzed in \cite{muris2017estimation} and the dynamic model was analyzed in \cite{muris2023dynamic} and applied to study patterns of self-reported health status. Further identification results for the dynamic model are provided in \cite{honore2025dynamic}. 
\end{example}

\begin{example} [Dynamic AR(1) binary choice logit-type and mixed logit errors] \label{example_logittype} Consider the dynamic panel discrete choice model: 
	\[
	Y_{it} = 1\{ \alpha_i + \beta Y_{it-1} + \bm X_{it}^\top \gamma \geq \epsilon_{it}\}, 
	\]
where the per-period utility shocks $(\epsilon_{it})_{t=1}^{T}$ are i.i.d.\ and independent of $(Y_{i0}, \bm X_{i},\alpha_{i})$. Now consider two extensions beyond the logistic distribution: logit-type errors and mixed logit errors. For the first extension, assume that the distribution $F_{1}$ of $\epsilon_{it}$ takes the form: 
\[
\frac{F_1(u)}{1-F_1(u)} = \sum_{k=1}^K q_k \exp(\lambda_k u),
\]
where $K$ and $1= \lambda_1 < \dots < \lambda_K$ are known integers, and the weights $q_k>0$ are unknown parameters. For the second extension, assume that the distribution $F_{2}$ of $\epsilon_{it}$ takes the form: 
\[
F_2(u) = \sum_{k=1}^K q_k \frac{\exp(\lambda_k u)}{1+\exp(\lambda_k u)},
\]
where we assume $K$ and $1= \lambda_1 < \dots < \lambda_K$ are known integers and the unknown mixture weights $q_k$ belong to the unit simplex. The first extension is considered in \cite{davezies2023fixed} for the static model ($\beta = 0$), and the second extension considers the distribution as a scale mixture of logistic distributions. 
\end{example} 
We now present a general assumption that nests these examples as a special case. In the following, we let $\bm Y = (Y_{1}, \dots, Y_{T}) \in \mathcal{Y}^T$ denote a vector of observed choices, and we let $\bm X = (\bm X_{1}, \dots, \bm X_{T}) \in \mathcal{X}^T$ denote a vector of observed covariates. Throughout, we use $\bm W = (\bm W_{1}, \dots, \bm W_{T}) \in \mathcal{W}$ to denote a generic vector of conditioning variables, which includes any covariates $\bm X$ and may also include the initial conditions $(Y_{1-p}, Y_{2-p}, \ldots, Y_{0}) \in \mathcal{Y}^p$, depending on the model. Finally, the model also includes a latent individual effect $\alpha \in \mathbb{R}$ and a vector of structural parameters $\theta \in \Theta \subset \mathbb{R}^{d_{\theta}}$. 
\begin{assumption}\label{assumption_main}
	There exists a complete and nonatomic probability space $(\Omega,\mathfrak{F},P)$, random vectors $\bm Y: \Omega \to \mathcal{Y}^T \subset \mathbb{R}^T$ and $\bm W:\Omega \to \mathcal{W}\subseteq \mathbb{R}^{d_w}$, a random variable $\alpha: \Omega \to \mathbb{R}$, and a vector  $\theta_{0}\in \Theta \subset \mathbb{R}^{d_{\theta}}$ such that:
	\begin{align}
	P(\bm Y = \bm y \mid \bm W = \bm w, \alpha) = f(\bm y \mid \bm w, \alpha; \theta_{0}),\label{eq_prob}
	\end{align}
	almost surely (a.s.) for some known (likelihood) function $f(\,\cdot\, \mid \bm w, \alpha; \theta)$ of the form:
	\begin{equation}\label{modelf} 
	f(\bm y \mid \bm w, \alpha; \theta) = \kappa(\bm w, \alpha, \theta) \cdot \sum_{s=0}^S \exp(\alpha)^{s}\cdot g_{s}(\bm y,\bm w,\theta), 
	\end{equation}
	where $S$ is finite, $\{g_{s}(\bm y,\bm w,\theta)\}_{s=0}^{S}$ are finite, nonnegative, not all zero, and continuously differentiable in $(\bm w, \theta)$, and $0<\kappa(\bm w, \alpha, \theta)<1$ is measurable in $(\bm w, \alpha)$ and is such that $\kappa(\bm w, \alpha, \theta)^{-1}$ is a polynomial of degree $S$ in $\exp(\alpha)$ for all $(\bm w,\alpha,\theta) \in \mathcal{W} \times \mathbb{R} \times \Theta$.  Furthermore, $T$ is finite, the support $\mathcal{Y}$ is finite, and $\alpha \mid \bm W \sim Q_{\alpha \mid \bm W}$.
\end{assumption}

Assumption \ref{assumption_main} covers discrete choice models with idiosyncratic errors independent from the covariates and the fixed effects.\footnote{See \cite{aristodemou2021semiparametric} and \cite{khan2023identification} for results when this independence assumption is relaxed.} The assumption restricts attention to models whose conditional likelihood $f(\,\cdot\, \mid \bm w, \alpha; \theta)$ can be written as a polynomial in $\exp(\alpha)$, up to a common factor of $\kappa(\bm w, \alpha, \theta)$. Here $\kappa(\bm w, \alpha, \theta)^{-1}$ itself is a strictly positive polynomial of degree $S$ in $\exp(\alpha)$, which ensures that the function $f(\,\cdot\, \mid \bm w, \alpha; \theta)$ is bounded in $\alpha \in \mathbb{R}$. This will be important for our theoretical results.\footnote{Note this is actually implied by \eqref{modelf} and the other positivity assumptions from Assumption \ref{assumption_main}: summing over $\bm y \in \mathcal{Y}^T$, we have $1 = \kappa(\bm w, \alpha, \theta) \cdot \sum_{s=0}^S \exp(\alpha)^{s}\cdot \sum_{\bm y \in \mathcal{Y}^T} g_{s}(\bm y,\bm w,\theta)$, and rearranging for $\kappa(\bm w, \alpha, \theta)^{-1}$ shows it must be a strictly positive polynomial of degree $S$ in $\exp(\alpha)$.} The term $\kappa(\bm w, \alpha, \theta)$ changes depending on the model, but often its choice is obvious (e.g.\ see Example \ref{example_AR1} below). Assumption \ref{assumption_main} also fixes attention to the case where the support $\mathcal{Y}^T$ is finite, and emphasizes that $\alpha$ will be treated as a random variable with an unknown conditional distribution. Importantly, Assumption \ref{assumption_main} imposes no assumptions on the moments of $\alpha$, and no assumptions on the dependence between $\alpha$ and $\bm W$. Assumption \ref{assumption_main} also allows for interactions between the lagged outcomes and covariates, and known (up to a finite vector of parameters) nonlinear functions of the covariates, lagged outcomes, and model parameters to enter the index functions. The structure of the likelihood in \eqref{modelf} in Assumption \ref{assumption_main} is essential to our approach, but is satisfied by a general class of logit models, including Examples \ref{example_AR1} - \ref{example_logittype}. Throughout, let $\Lambda(u) := \frac{\exp(u)}{1+\exp(u)}$, and let $\bm g(\bm y, \bm w,\theta) = (g_{s}(\bm y, \bm w,\theta))_{s=0}^{S}$ denote an $(S+1)\times 1$ vector. We now show how Assumption \ref{assumption_main} applies to the examples introduced above.
\setcounter{example}{0}
\begin{example}[$AR(1)$ dynamic logit binary choice, continued]  Recall the $AR(1)$ dynamic logit binary choice model from Example \ref{example_AR1}. Consider the case with $T = 2$ and let $\bm w = (y_0,  x_1, x_2) \in \{0,1\}\times \mathcal{X}^2$, $\theta = (\beta, \gamma)$, and:\label{example1_Gmat}
	\[
	f(\bm y \mid \bm w, \alpha; \theta) = \prod_{t=1}^T \Lambda(\alpha + \beta y_{t-1} + \bm x_{t}^\top \gamma)^{y_t} (1-\Lambda(\alpha + \beta y_{t-1}+\bm x_{t}^\top \gamma))^{1-y_t}.
	\]
Now set $\kappa(\bm w, \alpha, \theta) = (1-\Lambda(\alpha + \beta y_0 + \bm x_1^\top\gamma))(1-\Lambda(\alpha + \bm x_2^\top\gamma))(1-\Lambda(\alpha + \beta + \bm x_2^\top\gamma))$. Then $S = 3$, and a simple calculation shows that we can set: 
	\begin{align*}
\bm g((0,0), \bm w, \theta)^\top &= \begin{bmatrix} 1 & \exp(\beta + \bm x_2^\top\gamma) & 0 & 0 \end{bmatrix}, \\
\bm g((1,0), \bm w, \theta)^\top & = \begin{bmatrix} 0 & \exp(\beta y_0+\bm x_1^\top\gamma) & \exp(\beta y_0 + \bm x_1^\top\gamma + \bm x_2^\top\gamma) & 0\end{bmatrix}, \\
\bm g((0,1), \bm w, \theta)^\top & = \begin{bmatrix} 0 & \exp(\bm x_2^\top\gamma) & \exp(\beta + 2 \bm x_2^\top\gamma) & 0 \end{bmatrix}, \\
\bm g((1,1),\bm w, \theta)^\top & = \begin{bmatrix} 0 & 0 & \exp(\beta (y_0 + 1) + \bm x_1^\top\gamma + \bm x_2^\top\gamma) & \exp(\beta(y_0+1) + \bm x_1^\top\gamma + 2\bm x_2^\top\gamma) \end{bmatrix}.
		\end{align*}
Note that the choice of $\kappa(\bm w, \alpha, \theta)$ in this example (and in all other examples) is just the common denominator of all likelihood terms $f(\bm y \mid \bm w, \alpha; \theta)$ across $\bm y \in \mathcal{Y}^T$. The procedure for generating the function $\kappa(\bm w, \alpha,\theta)$ and the vectors $\bm g(\bm y, \bm w,\theta)$ can also be generalized to any finite $T$.\footnote{For details, see the Additional Online Supplementary Material, which can be accessed \href{https://arxiv.org/abs/2104.04590}{here}. Note: in the general case, $S=2T-1$.}
\end{example}

\begin{example}[$AR(p)$ dynamic logit binary choice, continued] 
Recall the $AR(p)$ dynamic logit binary choice model from Example \ref{example_ARp}. Consider the case when $T = 2$ and $p = 2$ and let $\bm w = (y_{-1}, y_0, x_1,x_2) \in \{0,1\}^2 \times \mathcal{X}^2$, $\theta = (\beta_1, \beta_2, \gamma)$, and: 
\[
f(\bm y \mid \bm w, \alpha; \theta) = \prod_{t=1}^T\Lambda(\alpha + \beta_1 y_{t-1}+\beta_2 y_{t-2} + \bm x_{t}^\top\gamma)^{y_t} (1-\Lambda(\alpha + \beta_1 y_{t-1}+\beta_2 y_{t-2} + \bm x_{t}^\top\gamma))^{1-y_t}.
\]
Now set $\kappa( \bm w, \alpha, \theta) = (1-\Lambda(\alpha + \beta_1 y_0 + \beta_2 y_{-1} + \bm x_1^\top\gamma))(1-\Lambda(\alpha + \beta_2 y_0 + \bm x_2^\top\gamma))(1-\Lambda(\alpha + \beta_1 + \beta_2 y_0 + \bm x_2^\top\gamma))$. Then $S = 3$, and a simple calculation shows that we can set: 
\begin{align*}
	\bm g((0,0), \bm w, \theta)^\top & = \begin{bmatrix} 1 & \exp(\beta_1 + \beta_2 y_0 + \bm x_2^\top\gamma) & 0 & 0 \end{bmatrix}, \\[5pt]
	\bm g((1,0), \bm w, \theta)^\top & = \begin{bmatrix} 0 & \exp(\beta_1 y_0 + \beta_2 y_{-1} + \bm x_1^\top\gamma) & \exp\left(
\begin{aligned}
&\beta_1y_0 + \beta_2(y_{-1}+y_0)\\
&\quad + (\bm x_1 + \bm x_2)^\top\gamma
\end{aligned}\right) & 0 \end{bmatrix}, \\[5pt]
	\bm g((0,1),\bm w, \theta)^\top & = \begin{bmatrix} 0 & \exp(\beta_2 y_0 + \bm x_2^\top\gamma) & \exp(\beta_1 + 2\beta_2 y_0 + 2\bm x_2^\top\gamma) & 0\end{bmatrix},\\[10pt]
	\bm g((1,1), \bm w, \theta)^\top & = \begin{bmatrix}
0 & 0 & \exp\left(
\begin{aligned}
    &\beta_1(y_0+1) + \beta_2 (y_{-1}+y_0) \\
    &\qquad+ (\bm x_1 + \bm x_2)^\top\gamma
\end{aligned}\right) &
\exp\left(
\begin{aligned}
    &\beta_1(y_0+1) + \beta_2(y_{-1}+2y_0) \\
    &\qquad+ (\bm x_1+2\bm x_2)^\top\gamma
\end{aligned}\right)
\end{bmatrix}.
	\end{align*}
The procedure for generating the function $\kappa(\bm w, \alpha,\theta)$ and the vectors $\bm g(\bm y, \bm w,\theta)$ can also be generalized to any finite $p$ and $T$.\footnote{See the Additional Online Supplementary Material, which can be accessed \href{https://arxiv.org/abs/2104.04590}{here}. Note: in the general case, we have $S=2^T-1$ if $T\leq p+1$, and $S= (2^p-1) + 2^p(T-p)$ if $T > p+1$.} 
\end{example}

\begin{example}[Dynamic AR(1) ordered logit, continued]  
	Recall the panel ordered choice model from Example \ref{example_ODC}. Consider $T = 1$ and $M = 3$ and let $\bm w = (y_0, x_1) \in \{1,2,3\} \times \mathcal{X}$, $\theta = (\beta_1, \beta_2, \beta_3, \eta, \gamma_1,\gamma_2)$. Define $\beta_{y_0} := \sum_{m=1}^M \beta_m 1\{y_{0} = m\}$, and note:
	\begin{align*}
	f(1\mid \bm w, \alpha; \theta) &= 1-\Lambda(\alpha + \beta_{y_0}+ \bm x_1^\top\eta-\gamma_1),\\
	f(2\mid \bm w,\alpha;\theta) & = \Lambda(\alpha + \beta_{y_0} + \bm x_1^\top\eta-\gamma_1)-\Lambda(\alpha + \beta_{y_{0}} + \bm x_1^\top\eta-\gamma_2),\\
	f(3\mid \bm w, \alpha; \theta) &= \Lambda(\alpha + \beta_{y_0} + \bm x_1^\top\eta - \gamma_2).
	\end{align*} 
Now set $\kappa(\bm w,\alpha,\theta) = (1-\Lambda(\alpha + \beta_{y_0} + \bm x_1^\top\eta - \gamma_1))(1-\Lambda(\alpha + \beta_{y_0} + \bm x_1^\top\eta - \gamma_2))$. Then $S = 2$, and a simple calculation shows that: 
	\begin{align*}
		\bm g(1, \bm w, \theta)^\top & = \begin{bmatrix} 1 & \exp(\beta_{y_0} + \bm x_1^\top\eta - \gamma_2) & 0\end{bmatrix}, \\
		\bm g(2,\bm w,\theta)^\top & = \begin{bmatrix} 0 & \exp(\beta_{y_0}+\bm x_1^\top\eta) (\exp(-\gamma_1)-\exp(-\gamma_2)) & 0 \end{bmatrix}, \\
		\bm g(3,\bm w, \theta)^\top & = \begin{bmatrix} 0 & \exp(\beta_{y_0} + \bm x_1^\top\eta - \gamma_2) & \exp(2\beta_{y_0} + 2\bm x_1^\top\eta - \gamma_1 - \gamma_2)\end{bmatrix}.
		\end{align*} 
The procedure for generating the function $\kappa(\bm w, \alpha,\theta)$ and the vectors $\bm g(\bm y, \bm w,\theta)$ can also be generalized to any finite $M$ and $T$.\footnote{For details, see the Additional Online Supplementary Material, which can be accessed \href{https://arxiv.org/abs/2104.04590}{here}. Note: in the general case we have $S=M-1+(T-1)M(M-1)$.}
\end{example}

\begin{example} [Dynamic AR(1) with logit-type or mixed logit errors, continued]  
Recall the dynamic panel discrete choice model with logit-type or mixed logit errors from Example \ref{example_logittype}. For both distributions, consider $T = 1$, $K = 2$, $\lambda_2 = 2$, and $\bm w = (y_0, x_1) \in \{0,1\}\times\mathcal{X}$. With logit-type errors we have $f(y_1 \mid \bm w, \alpha, \theta) =F_1(\alpha + \beta y_{0}+\bm x_1^\top\gamma)^{y_1}(1-F_1(\alpha + \beta y_{0} + \bm x_1^\top\gamma))^{1-y_1}$. Set $\kappa( \bm w, \alpha,\theta) = 1-F_1(\alpha + \beta y_0 + \bm x_1^\top\gamma)$. Then $S = 2$, and we can set: 
\begin{align*}
	\bm g(0, \bm w, \theta)^\top = \begin{bmatrix} 1 & 0 & 0 \end{bmatrix}, &&	\bm g(1, \bm w, \theta)^\top = \begin{bmatrix} 0 & q_1 \exp(\beta y_0 + \bm x_1^\top\gamma) & q_2\exp(2\beta y_0 + 2\bm x_1^\top\gamma)\end{bmatrix}.
	\end{align*}
For the case of a logit mixture, we have $f(y_1 \mid \bm w, \alpha, \theta) =F_2(\alpha + \beta y_{0}+\bm x_1^\top\gamma)^{y_1}(1-F_2(\alpha + \beta y_{0} + \bm x_1^\top\gamma))^{1-y_1}$. Now set $\kappa(\bm w,  \alpha, \theta) =(1-\Lambda(\alpha + \beta y_0 + \bm x_1^\top\gamma))(1-\Lambda(2(\alpha + \beta y_0 + \bm x_1^\top\gamma)))$, and denote $\Gamma(\bm w, \theta) = \exp(\beta y_0 + \bm x_1^\top\gamma)$. Then $S = 3$, and we can set:
\begin{align*}
	\bm g(0, \bm w, \theta)^\top & = \begin{bmatrix} 1 & (1-q_1) \Gamma(\bm w, \theta) & q_1 \Gamma^2(\bm w, \theta) & 0 \end{bmatrix}, \\
	\bm g(1, \bm w, \theta)^\top & = \begin{bmatrix} 0 & q_1 \Gamma(\bm w, \theta) &  q_2 \Gamma^2(\bm w, \theta) & \Gamma^3(\bm w, \theta) \end{bmatrix}.
	\end{align*} 
\end{example} 

With these examples in hand, we now describe the general identification problem for models governed by Assumption \ref{assumption_main}. Define $p(\bm y \mid \bm w):=P(\bm Y = \bm y \mid \bm W = \bm w)$, and fix a pair $(\theta, \bm w) \in \Theta\times\mathcal{W}$. Let $\mathcal{Q}$ denote the set of all Borel probability measures on $\mathbb{R}$, and consider a candidate conditional distribution $Q_{\alpha \mid \bm W}^\dagger \in \mathcal{Q}$ for the latent individual effect $\alpha$. We say that the conditional distribution $Q_{\alpha \mid \bm W}^\dagger$ can \textit{rationalize} the observed conditional choice probabilities at $\theta\in \Theta$ if and only if:
\begin{align}
p(\bm y \mid \bm w)=  \int f(\bm y \mid \bm w, \alpha; \theta) dQ_{\alpha \mid \bm W}^\dagger(\alpha \mid \bm w),\label{eq_rationalize}
\end{align}
a.s.\ for all $\bm y \in \mathcal{Y}^T$. The collection of all conditional probability measures $Q_{\alpha \mid \bm W}^\dagger$ that can rationalize the observed conditional choice probabilities for a fixed pair $(\theta, P)$ is given by:
\begin{equation} \label{Qset_y0obs}
\mathcal{Q}(\theta,P) = \left\{Q_{\alpha \mid \bm W}^\dagger \in \mathcal{Q} : Q_{\alpha \mid \bm W}^\dagger \text{ satisfies \eqref{eq_rationalize} $P_{\bm W}-$a.s. for all $\bm y \in \mathcal{Y}^T$} \right\}. 
\end{equation}
Note that, depending on the value of $\theta \in \Theta$, this set may be empty. The set of all $\theta \in \Theta$ for which this set is nonempty is precisely the identified set of structural parameters. 
\begin{definition}[Identified Set]\label{definition_identified_set} Under Assumption \ref{assumption_main}, the identified set for the structural parameter $\theta \in \Theta$ is $\Theta_I(P) := \{\theta \in \Theta : \mathcal{Q}(\theta,P) \neq \emptyset \}$. 
\end{definition}



To construct the identified set in practice, for each $\theta \in \Theta$ we must ask whether there exists a probability measure $Q_{\alpha \mid \bm W}^\dagger \in \mathcal{Q}$ that rationalizes the observed vector of conditional choice probabilities through \eqref{eq_rationalize}, $P_{\bm W}-$a.s. Since a probability measure is an infinite-dimensional object, verifying the existence of such a conditional probability measure is an \textit{infinite-dimensional existence problem}.\footnote{This is a common feature of partially identified models. As far as we know, this terminology was first used by \cite{torgovitsky2019partial}.} We now illustrate that the structure of the likelihood function $f(\,\cdot\mid \bm w, \alpha; \theta)$ in Assumption \ref{assumption_main} allows us to convert the \textit{infinite-dimensional existence problem} to a tractable finite-dimensional problem. 


\subsection{An Example of the Methodology: The AR(1) Model with $T=2$}\label{section_AR1_T2}
Consider Example \ref{example_AR1} with $T = 2$ and $\gamma=0$ (i.e. without covariates).\footnote{The literature on the AR(1) model with $T=2$ is quite sparse. \cite{halliday2007testing} studies testing for state dependence in an AR(1) model in which covariates are allowed to have time-varying coefficients, but focused on hypothesis testing. \cite{chamberlain2023identification} proved the impossibility of point identification in the AR(1) model with bounded covariates when $T = 2$. Finally, in \cite{dobronyi2021identification}, a previous working paper version of the current paper, we derived analytical bounds for the dynamic coefficient.} This simple example helps to illustrate a fundamental connection between identification in models governed by Assumption \ref{assumption_main} and the \textit{truncated moment problem} in mathematics.\footnote{See \cite{schmudgen2017moment} for a recent textbook treatment.} We use this simple example to provide the intuition for our approach before presenting our general identification results. This simple case is also interesting in itself: using functional differencing, \cite{honore2024moment} show that there are no identifying restrictions for the parameter $\beta$. In contrast, we will show that the model still provides information about the structural parameters through a finite set of moment equalities and shape constraints.  In particular, conditional on observing $Y_{0} = y_{0}$, the logistic distribution for $\epsilon_{t}$ implies that for any $\bm y \in \{0,1\}^2$:
\[
f(\bm y \mid y_0 , \alpha; \theta) =\prod_{t=1}^2 \Lambda(\alpha + \beta y_{t-1})^{y_t} (1-\Lambda(\alpha + \beta y_{t-1}))^{1-y_t}.
\]
Now let $A := \exp(\alpha)$ and $B := \exp(\beta)$ and choose $\kappa(y_0,\alpha,\beta) =(1-\Lambda(\alpha + \beta y_0)) (1-\Lambda(\alpha)) (1-\Lambda(\alpha + \beta))$. 
Then we can write the likelihood as: 
\begin{equation}\label{likelihood} 
\begin{bmatrix} f((0,0)\mid y_0 , \alpha; \theta) \\
	f((1,0)\mid y_0, \alpha; \theta) \\
	f((0,1)\mid y_0 , \alpha; \theta) \\
	f((1,1)\mid y_0 , \alpha; \theta)
	\end{bmatrix} = \kappa(y_0,\alpha,\beta) \underbrace{\begin{bmatrix} 1 & B & 0 & 0\\
	0 & B^{y_0} & B^{y_0} & 0\\
	0 & 1 & B & 0\\
	0 & 0 & B^{y_0+1} & B^{y_0+1}\end{bmatrix}}_{=:\bm G(y_0,\beta)}\begin{bmatrix} 1 \\ A \\ A^2 \\ A^3\end{bmatrix}.
\end{equation}
Relating to \eqref{modelf} in Assumption \ref{assumption_main}, in this example we have $S = 3$, and the entries in the rows of the matrix $\bm G(y_0,\beta)$ represent the coefficients $g_{s}(\bm y, y_0, \beta)$ of the polynomials of $A$ for the history $\bm y \in \mathcal{Y}^2$. Integrating the likelihood from \eqref{likelihood} with respect to any conditional distribution $Q_{\alpha\mid y_0}^\dagger(\alpha \mid y_0)$ for the individual effect yields:
\[
\bm G(y_0,\beta) \begin{bmatrix} \int_{\mathbb{R}}\kappa(y_0,\alpha,\beta)   \, dQ_{\alpha \mid y_0}^\dagger(\alpha\mid y_0)  \\\int_{\mathbb{R}} \kappa(y_0,\alpha,\beta)    \exp(\alpha)\, dQ_{\alpha \mid y_0}^\dagger(\alpha\mid y_0)  \\ \int_{\mathbb{R}}\kappa(y_0,\alpha,\beta)    \exp(2\alpha)\, dQ_{\alpha \mid y_0}^\dagger(\alpha\mid y_0)  \\ \int_{\mathbb{R}} \kappa(y_0,\alpha,\beta)   \exp(3\alpha)\, dQ_{\alpha \mid y_0}^\dagger(\alpha\mid y_0)  \end{bmatrix}= \bm G(y_0,\beta)  \begin{bmatrix} \int_{[0,\infty)}  1\,  d\bar Q_{A \mid y_0}^\dagger(A\mid y_0)\\ \int_{[0,\infty)}  A\,  d\bar Q_{A \mid y_0}^\dagger(A\mid y_0)\\ \int_{[0,\infty)}  A^2\, d\bar Q_{A \mid y_0}^\dagger(A\mid y_0)\\ \int_{[0,\infty)}  A^3\,d\bar Q_{A \mid y_0}^\dagger(A\mid y_0)\end{bmatrix}. 
\]
To arrive at the second equality, we perform the change of measure:
\begin{align*}
\bar Q_{\alpha\mid y_0}^\dagger(C\mid y_0) := \int_{C} \kappa(y_0,\alpha,\beta) \,dQ_{\alpha \mid y_0}^\dagger(\alpha \mid y_0),
\end{align*}
and then let $\bar Q_{A\mid y_0}^\dagger(\,\cdot\,\mid y_0)$ denote the push-forward measure of $\bar{Q}_{\alpha\mid y_0}^\dagger(\,\cdot\,\mid y_0)$ under the map $\alpha\mapsto \exp(\alpha)$.\footnote{While this measure depends on the unknown parameters, the proof of our main identification results show that this fact has no identifying power for the structural parameters.} Since $\kappa(y_0,\alpha,\beta)$ is bounded and positive for all $\alpha \in \mathbb{R}$ by Assumption \ref{assumption_main}, the measure $\bar Q_{A\mid y_0}^\dagger(\,\cdot\,\mid y_0)$ is a finite nonnegative Borel measure on $(\mathbb{R}_{+},\mathcal{B}(\mathbb{R}_{+}))$.\footnote{The fact that $\bar Q_{A\mid y_0}^\dagger(\emptyset \mid y_0)=0$ is obvious. Countable additivity follows by dominated convergence. } Now define the vector:
\begin{align*}
	\bm{r}(y_0) := 
	\begin{bmatrix}
	r_{0}(y_0)\\
	r_{1}(y_0)\\
	r_{2}(y_0)\\
	r_{3}(y_0)\\
	\end{bmatrix}= \begin{bmatrix} \int_{[0,\infty)}  1\,  d\bar Q_{A \mid y_0}^\dagger(A\mid y_0)\\ \int_{[0,\infty)}  A\,  d\bar Q_{A \mid y_0}^\dagger(A\mid y_0)\\ \int_{[0,\infty)}  A^2\, d\bar Q_{A \mid y_0}^\dagger(A\mid y_0)\\ \int_{[0,\infty)}  A^3\,d\bar Q_{A \mid y_0}^\dagger(A\mid y_0)\end{bmatrix}.
	\end{align*}
Then $\bm{r}(y_0)$ is a vector of moments of the variable $A$ up to order $3$ with respect to the measure $\bar Q_{A\mid y_0}^\dagger(A\mid y_0)$. We refer to $\bm{r}(y_0)$ as the vector of \textit{generalized moments} of $\alpha$ throughout. Now let $\bm p(y_0)$  denote the vector of conditional probabilities $p(\bm y \mid y_{0})$ stacked across $\bm y \in \mathcal{Y}^2$.\footnote{The ordering of the choice sequence should match the order in \eqref{likelihood}. We maintain a consistent ordering of choice sequences throughout: when the time period increases by one, we always append $0$ to all existing choice sequences, and then append $1$. } Then the question of whether a particular $\beta$ belongs to the identified set is equivalent to the question of whether, for each $y_0 \in \{0,1\}$, there exists a measure---specifically, a nonnegative Radon measure---whose moment vector $\bm{r}(y_0)$ satisfies $\bm p(y_0) = \bm G(y_0,\beta)\bm{r}(y_0)$.\footnote{When specialized to Euclidean space, a Radon measure is a nonnegative Borel measure that is finite on all compact sets. On Euclidean space, all finite nonnegative Borel measures are Radon, although not all Radon measures are finite measures; for example, the Lebesgue measure is a Radon measure.  } 
This result reveals a fundamental connection between the identification of structural parameters in dynamic logit models and the \textit{moment problem} from the mathematics literature.\footnote{See \cite{karlin1966tchebycheff}, \cite{krein1977markov}, and \cite{schmudgen2017moment} for comprehensive treatments of this subject.} One of the main questions studied in the literature on the moment problem is whether there exists a Radon measure that rationalizes a  sequence of real numbers as its moments. Given an infinite sequence of real numbers, this problem is referred to as the \textit{full moment problem}. Given a finite sequence of real numbers, this problem is referred to as the \textit{truncated moment problem}. When the Radon measure is restricted to have support on $\mathbb{R}_+$, as in our context, the truncated moment problem is known as the \textit{truncated Stieltjes moment problem}, as it was first raised and analyzed by \cite{stieltjes1894recherches}.

Let $\mathcal{P}_{+}$ denote the set of all nonnegative Radon measures on $(\mathbb{R}_{+},\mathcal{B}(\mathbb{R}_{+}))$, and define the moment space:
\begin{align}
\mathcal{M}_{S} :=\left\{\bm c \in \mathbb{R}^{S+1}: \exists \mu \in \mathcal{P}_{+} \text{ s.t. } c_s = \int_0^{+\infty} A^s d\mu(A) \text{ for } s =0,1,\dots S\right\}.\label{eq_moment_space}
\end{align}
Referring back to Definition \ref{definition_identified_set}, for the AR(1) model with $T=2$ we can write the identified set as:
\begin{align*}
\Theta_I(P) = \left\{\beta \in \Theta: \exists \bm{r}(y_0) \in \mathcal{M}_3 \text{ s.t. $\bm p(y_0) = \bm G(y_0,\beta)\bm{r}(y_0)$ $\forall y_0 \in \{0,1\}$}\right\}. 
\end{align*}
This characterization of the identified set is not useful without a tractable means of verifying whether a vector $\bm{r}(y_0)$ belongs to the moment space $\mathcal{M}_S$ from \eqref{eq_moment_space}. However, the geometric structure of the moment space $\mathcal{M}_S$ has been studied extensively, and results from the literature on the moment problem lead to the following theorem. 
\begin{theorem} \label{theorem_AR1_T2}
	Suppose Assumption \ref{assumption_main} holds, and consider the dynamic logit model in Example \ref{example_AR1} with $T = 2$ and $\gamma = 0$. Then $\beta \in \Theta_{I}(P)$ if and only if there exists vectors $\bm{r}(0),\bm{r}(1) \in \mathbb{R}^4$ satisfying:
	\begin{enumerate}[label=(\roman*)]
	 	\item $\bm p(0) = \bm G(0,\beta)\bm{r}(0)$ and $\bm p(1) = \bm G(1,\beta)\bm{r}(1)$;
	 	\item $\sum_{j = 0}^3 \eta_{0,j} r_{j}(0) \geq 0$ and $\sum_{j = 0}^3 \eta_{1,j} r_{j}(1) \geq 0$ for every real-valued sequence of coefficients $\{\eta_{0,j}\}_{j=0}^3$ and $\{\eta_{1,j}\}_{j=0}^3$ satisfying $\sum_{j=0}^3 \eta_{0,j} A^j \geq 0$ and $\sum_{j=0}^3 \eta_{1,j} A^j \geq 0$ for every $A \in [0, \infty)$;
	 	\item For some real coefficients $a_{0,1}$, $a_{0,2}$, $a_{1,1}$, and $a_{1,2}$:
	 	\begin{align*}
	 	r_{2}(0) =  a_{0,1} r_{0}(0) + a_{0,2} r_{1}(0), &&r_{2}(1) =  a_{1,1} r_{0}(1) + a_{1,2} r_{1}(1),\\
	 	r_{3}(0) =  a_{0,1} r_{1}(0) + a_{0,2} r_{2}(0), &&r_{3}(1) =  a_{1,1} r_{1}(1) + a_{1,2} r_{2}(1).
	 	\end{align*}
	 	
	 \end{enumerate}

\end{theorem}

Theorem \ref{theorem_AR1_T2} uses Theorem 5.1 in \cite{curto1991recursiveness}, with parts $(ii)$ and $(iii)$ providing a means of verifying whether the vectors $\bm{r}(0)$ and $\bm{r}(1)$ belong to the moment space $\mathcal{M}_{S}$ when $S=3$.\footnote{See also Theorems 9.35 and 9.36 in \cite{schmudgen2017moment}.} To understand condition $(ii)$, the key insight is that the moment space $\mathcal{M}_{S}$ is a convex cone. As such, it has an associated dual cone given by:
\begin{align*}
\mathcal{M}_S^{*} := \{\bm \eta\in \mathbb{R}^{S+1} : \bm \eta^\top \bm{c} \geq 0 \text{ for all  } \bm{c} \in \mathcal{M}_S\}. 
\end{align*}
Theorem II 9.1 in \cite{karlin1966tchebycheff} derives the specific form of the dual cone, and shows that $\mathcal{M}_S^* = \mathcal{P}_{S}$, where:
\begin{align*}
\mathcal{P}_S = \left\{\bm \eta \in \R^{S+1} : \sum_{i=0}^{S} \eta_i A^i \geq 0 \text{ for all } A \geq 0\right\}.
\end{align*}
In particular, $\mathcal{P}_{S}$ is the set of coefficients that produce a nonnegative polynomial on $\mathbb{R}_{+}$. By standard results in convex analysis, taking the dual of the dual cone $\mathcal{M}_{S}^*$ again recovers the closure of the moment space $\mathcal{M}_S$; that is, $(\mathcal{M}_{S}^*)^* = \text{cl}(\mathcal{M}_{S})$. Since $\mathcal{M}_{S}^* = \mathcal{P}_{S}$, the dual of the dual cone is:
\begin{align*}
\text{cl}(\mathcal{M}_{S}) = (\mathcal{M}_{S}^*)^* = (\mathcal{P}_{S})^* = \left\{\bm c \in \R^{S+1} : \bm{\eta}^\top \bm c \geq 0 \text{ for all } \bm \eta \in \mathcal{P}_S\right\}. 
\end{align*}
Thus, $\bm{r}(0)$ and $\bm{r}(1)$ belong to $\text{cl}(\mathcal{M}_{S})$ if and only if they satisfy condition $(ii)$ in Theorem \ref{theorem_AR1_T2}. 

To see how to check condition $(ii)$ from Theorem \ref{theorem_AR1_T2} in practice, consider the case when $y_0 = 0$. Every nonnegative polynomial of $A$ with an odd degree $2m+1$ for some $m\in \mathbb{N}$ has a representation of the form:\footnote{For the even case, $\sum_{j = 0}^{2m} \eta_j A^j = f^2(A) + A q^2(A)$ where $f(A)$ are polynomials of A of at most order $m$ and $q(A)$ is a polynomial of $A$ of at most order $m-1$. See Corollary 8.1 in Chapter V of \cite{karlin1966tchebycheff} and the further discussion in Section 10 of Chapter V. Also see Corollary 3.5 of \cite{schmudgen2017moment}.} 
\[
\sum_{j=0}^{2m+1} \eta_{0,j} A^j= A f^2(A) + q^2(A)\geq 0,
\]
for all $A\in [0, \infty)$, where $f(A)$ and $q(A)$ are polynomials up to order $m$. In our AR(1) example with $T=2$, $S = 2m+1 = 3$, and thus $f(A)$ and $q(A)$ are polynomials of at most degree 1. Therefore, nonnegativity implies that we can write $f(A) = \xi_0 + \xi_1 A$ and $q(A) = \lambda_0 + \lambda_1 A$ for any coefficients $(\xi_0, \xi_1)$ and $(\lambda_0, \lambda_1)$ satisfying:
\[
\sum_{j = 0}^{3} \eta_{0,j} A^j = A(\xi_0 + \xi_1A)^2 + (\lambda_0 + \lambda_1A)^2 \geq 0. 
\]
Retrieving the corresponding coefficients $\eta_{0,j}$, the condition $\sum_{j=0}^3 \eta_{0,j} r_{j}(0) \geq 0$ requires:
\begin{align*}
	\lambda_0^2 r_{0}(0) + 2\lambda_0 \lambda_1 r_{1}(0) + \lambda_1^2 r_{2}(0) + \xi_0^2 r_{1}(0) + 2\xi_0 \xi_1 r_{2}(0) + \xi_1^2 r_{3}(0) \geq 0, 
\end{align*} 
which can be equivalently stated as:
\begin{align}
\begin{bmatrix} \lambda_0 & \lambda_1\end{bmatrix}  \begin{bmatrix} r_{0}(0)  & r_{1} (0)\\
	r_{1} (0) & r_{2}(0)  \end{bmatrix} \begin{bmatrix} \lambda_0 \\ \lambda_1 \end{bmatrix} + \begin{bmatrix} \xi_0 & \xi_1 \end{bmatrix}  \begin{bmatrix} r_{1}(0) & r_{2}(0) \\ 
	r_{2}(0)  & r_{3}(0)
\end{bmatrix} \begin{bmatrix} \xi_0 \\
	\xi_1
\end{bmatrix} \geq 0,\label{eq_semidefinite1}
\end{align}
for \textit{all} coefficients $(\lambda_0, \lambda_1)$ and $(\xi_0, \xi_1)$. This condition is equivalent to checking that the two square matrices in \eqref{eq_semidefinite1}, defined using the elements of $\bm r(0)$, are positive semidefinite. These matrices are known as \textit{Hankel matrices} in the truncated moment problem literature.\footnote{See Section 3.2 in \cite{schmudgen2017moment}.} 

Note that condition $(ii)$ ensures only that $\bm{r}(0)$ and $\bm{r}(1)$ belong to $\text{cl}(\mathcal{M}_{S})$, and not necessarily to $\mathcal{M}_{S}$.\footnote{To see what can go wrong, consider the vector $\bm{r}(0)^\top = [0,0,0,1]$. Then the matrices in \eqref{eq_semidefinite1} are positive semidefinite, but clearly $\bm{r}(0)$ cannot be rationalized as a moment vector of a nonnegative Radon measure with support on $\mathbb{R}_{+}$, so that $\bm{r}(0)\notin\mathcal{M}_{S}$. This example is ruled out by condition $(iii)$: there are no coefficients satisfying $r_{3}(0) = a_{0,1} r_{1}(0) + a_{0,2} r_{2}(0)$, showing that $\bm{r}(0)$ cannot be rationalized as a moment vector.} Here condition $(iii)$ plays a role. When conditions $(ii)$ and $(iii)$ are combined, simple linear algebra combined with the discussion above shows that they are equivalent to checking if:
\begin{align}
\bm H_{1}^*(\bm r, \varsigma):=
 \begin{bmatrix} 
 r_{0}(0)  & r_{1}(0)& r_{2}(0)\\
r_{1} (0) & r_{2}(0) & r_{3}(0) \\
r_{2} (0) & r_{3}(0) & \varsigma 
\end{bmatrix}, 
&&\bm B_{1} (\bm r) :=\begin{bmatrix} 
r_{1}(0) & r_{2}(0) \\ 
r_{2}(0)  & r_{3}(0)
\end{bmatrix}, \label{eq_hankel_extension}
\end{align}
are positive semidefinite for some $\varsigma\geq 0$ (see Lemma 2.3 in \cite{curto1991recursiveness}). The matrix $\bm H_{1}^*(\bm r, \varsigma)$ is called the \textit{Hankel extension} of the corresponding Hankel matrix in \eqref{eq_semidefinite1}. Following a similar logic as above, positive semidefiniteness of these matrices is equivalent to:
\begin{align}
(r_{0}(0),r_{1}(0),r_{2}(0),r_{3}(0),\varsigma) \in \text{cl}(\mathcal{M}_{S+1}). \label{eq_extended_moment}
\end{align}
Theorem V 3.1 in \cite{karlin1966tchebycheff} then shows that $\text{cl}(\mathcal{M}_{S+1})$ can be expressed as:
\begin{align}
\text{cl}(\mathcal{M}_{S+1}) = \mathcal{M}_{S+1} + \{(0,\ldots,0,\lambda) : \lambda \geq 0 \}, \label{eq_extended_moment2}
\end{align}
so that the closure of the moment space is equal to the original moment space $\mathcal{M}_{S+1}$, but also includes a ray from the origin. Combining \eqref{eq_extended_moment} with \eqref{eq_extended_moment2}, we see that condition $(ii)$ and $(iii)$ are equivalent to checking if $(r_{0}(0),r_{1}(0),r_{2}(0),r_{3}(0)) \in \mathcal{M}_{S}$.\footnote{In particular, if $(r_{0}(0),r_{1}(0),r_{2}(0),r_{3}(0),\varsigma) \in \text{cl}(\mathcal{M}_{S+1})$ then there exists a $\lambda\geq 0$ such that $(r_{0}(0),r_{1}(0),r_{2}(0),r_{3}(0),\varsigma-\lambda) \in \mathcal{M}_{S+1}$. But then there exists a measure that supports this vector as its four moments, and this same measure must also support  $(r_{0}(0),r_{1}(0),r_{2}(0),r_{3}(0))$ as its first three moments. This implies $(r_{0}(0),r_{1}(0),r_{2}(0),r_{3}(0)) \in \mathcal{M}_{S}$. }  

Using Theorem \ref{theorem_AR1_T2} we see that, in the specific case of the AR(1) model with $T=2$, the identified set can be constructed by checking two conditional moment equalities, and by checking if there exists a constant $\varsigma\in \mathbb{R}$ such that the matrices in \eqref{eq_hankel_extension} are positive semidefinite.\footnote{For this specific model, it is possible to further derive analytical bounds on the parameter $\beta$ by converting matrix nonnegativity to inequalities on the determinants of all of its principal minors. See \cite{dobronyi2021identification}.} By making a connection to the moment problem, our approach is able to obtain sharp restrictions on the structural parameters in examples like the AR(1) model with $T=2$ where competing approaches fail to deliver any nontrivial identifying restrictions.\footnote{See the discussion of this model in \cite{honore2024moment}.} Intuitively, this is because our approach exploits two new facts that have not been considered by other methods: $(i)$ in a large class of models, the fixed effect distribution can be completely summarized by a finite vector of generalized moments, and $(ii)$ this vector of generalized moments must satisfy certain constraints which have nontrivial identifying content for the structural parameters. 

While this section was meant to introduce the main assumptions and ideas through a simple example, in the next section we expand on the connection to the truncated moment problem and apply it to a larger class of models.

\section{General Results}\label{section_general_results}

\subsection{Identification}\label{section_general_identification}

With the results from the dynamic panel logit model for $T = 2$ and $\gamma = 0$ in hand, we now generalize the identification analysis to all models governed by Assumption \ref{assumption_main}. For the following, let $J:=|\mathcal{Y}|^T$, and define:
	\begin{align}
	\bm G(\bm w,\theta) := \begin{bmatrix} 
	g_{0}(\bm y_1,\bm w,\theta) & g_{1}(\bm y_1,\bm w,\theta) & \ldots & g_{S}(\bm y_1,\bm w,\theta)\\
	g_{0}(\bm y_2,\bm w,\theta) & g_{1}(\bm y_2,\bm w,\theta) & \ldots & g_{S}(\bm y_2,\bm w,\theta)\\
	\vdots & \vdots & \ddots & \vdots\\
	g_{0}(\bm y_J,\bm w,\theta) & g_{1}(\bm y_J,\bm w,\theta) & \ldots & g_{S}(\bm y_J,\bm w,\theta)
	\end{bmatrix},\label{eq_general_G}
	\end{align}
	where $\bm y_1,\ldots,\bm y_{J},$ denotes an enumeration of the support $\mathcal{Y}^T$, and where $g_{s}(\bm y_j,\bm w,\theta)$ are the coefficients from Assumption \ref{assumption_main}. The following is the main identification result of the paper. 

\begin{theorem}\label{theorem_main}
	Suppose Assumption \ref{assumption_main} holds. Then $\theta \in \Theta_I(P)$ if and only if $\bm{p}(\bm w) = \bm G(\bm w, \theta) \bm r(\bm w)$ for some $\bm r(\bm w) \in \mathcal{M}_S$, $P_{\bm W}-$a.s.
\end{theorem}

Theorem \ref{theorem_main} shows that the identified set for the structural parameters $\theta \in \Theta$ for the class of models satisfying Assumption \ref{assumption_main} can be characterized by a set of moment equality conditions imposed on the conditional probabilities, as well as additional semidefinite shape restrictions on the parameter $\bm r(\bm w)$ coming from the moment space restrictions. 

The following theorem shows the necessary and sufficient conditions to have $\bm r(\bm w) \in \mathcal{M}_S$, generalizing conditions $(ii)$ and $(iii)$ in Theorem \ref{theorem_AR1_T2}. The result follows from classic results in the moment literature (e.g. \cite{curto1991recursiveness}). Here we use the notation $\bm A \succeq 0$ to represent the fact that the square matrix $\bm A$ is positive semidefinite.

\begin{theorem}\label{theorem_hankel}

	Let $\bm{r} := \{r_0, r_1, \dots, r_S\} \in\mathbb{R}^{S+1}$.
	\begin{enumerate}[label=(\roman*)]

	\item Suppose $S = 2m+1$ for some $m \in \mathbb{N}$ (i.e. $S$ is odd, or $S+1$ is even), and consider the matrices: 
\begin{align*}
	\bm{H}_{m}^*(\bm{r},\varsigma)  := 	\left( \begin{array}{ccccc}
		r_0 & r_1 & \cdots & r_m & r_{m+1}\\
		r_1 & r_2 & \cdots & r_{m+1}& r_{m+2}\\
		\vdots & \vdots & \ddots & \vdots& \vdots\\
		r_m & r_{m+1} & \cdots & r_{2m}& r_{2m+1}\\
		r_{m+1}& r_{m+2} & \cdots & r_{2m+1} & \varsigma
	\end{array} \right), &&
	\bm{B}_{m}(\bm{r}) := 	\left( \begin{array}{cccc}
		r_1 & r_2 & \cdots & r_{m+1}\\
		r_2 & r_3 & \cdots & r_{k+2}\\
		\vdots & \vdots & \ddots & \vdots \\
		r_{m+1}& r_{m+2} & \cdots & r_{2m+1}
	\end{array} \right).
\end{align*}
	Then $\bm{r} \in \mathcal{M}_{S}$ if and only if there exists $\varsigma \geq 0$ such that $\bm H_{m}^*(\bm{r},\varsigma) \succeq 0$ and $\bm B_m(\bm{r}) \succeq 0$.

	\item Suppose $S = 2m$ for some $m \in \mathbb{N}$ (i.e. $S$ is even, or $S+1$ is odd), and consider the matrices: 
\begin{align*}
	\bm{H}_{m}(\bm{r})  := 	\left( \begin{array}{cccc}
		r_0 & r_1 & \cdots & r_m\\
		r_1 & r_2 & \cdots & r_{m+1}\\
		\vdots & \vdots & \ddots & \vdots \\
		r_m & r_{m+1} & \cdots & r_{2m}
	\end{array} \right),&&
	\bm{B}_{m}^*(\bm{r},\varsigma) := 	\left( \begin{array}{ccccc}
		r_1 & r_2 & \cdots & r_{m} & r_{m+1}\\
		r_2 & r_3 & \cdots & r_{m+1}& r_{m+2}\\
		\vdots & \vdots & \ddots & \vdots& \vdots \\
		r_{m}& r_{m+1} & \cdots & r_{2m-1}& r_{2m}\\
		r_{m+1} & r_{m+2} & \cdots & r_{2m} & \varsigma
	\end{array} \right).
\end{align*}
	Then $\bm{r} \in \mathcal{M}_{S}$ if and only if there exists $\varsigma \geq 0$ such that $\bm H_m(\bm{r}) \succeq 0$ and $\bm B_{m}^*(\bm{r},\varsigma) \succeq 0$. 
\end{enumerate}
\end{theorem}

Theorem \ref{theorem_hankel} shows that, to check that a vector $\bm r \in \mathbb{R}^{S+1}$ belongs to the moment space $\mathcal{M}_{S}$, it is both necessary and sufficient to check that two matrices are positive semidefinite. Checking if a matrix is positive semidefinite is equivalent to checking that all principle minors of the matrix are nonnegative.\footnote{See \cite{meyer2000matrix} p.566. Recall that an $r\times r$ principle submatrix of an $n\times n$ matrix $\bm A$ is obtained by deleting the same set of $n-r$ rows and columns from the matrix $\bm A$. The principle minors of a matrix $\bm A$ are the determinants of the principle submatrices of $\bm A$. See \cite{meyer2000matrix} p.494.  } In this sense, the semidefinite restrictions on the matrices from Theorem \ref{theorem_hankel} can be viewed as nonlinear shape restrictions on the unknown vector of moments $\bm r(\bm w) \in \mathbb{R}^{S+1}$. Combining this idea with Theorem \ref{theorem_main}, verifying whether a vector $\theta \in \Theta$ belongs to the identified set amounts to checking whether a certain set of conditional moment equalities hold subject to a set of shape restrictions on $\bm r(\bm w) \in \mathbb{R}^{S+1}$, $P_{\bm W}-$a.s. To formalize this, let $\mathcal{S}_{+}^d$ denote the space of symmetric $d \times d$ positive semidefinite matrices, and define the moment function:
\begin{align}
m_{j}(\bm y,\bm w,\theta,\bm r) = 1\{\bm y = \bm y_{j} \} - \bm g(\bm y_{j}, \bm w,\theta)^\top \bm r(\bm w). \label{eq_mom_func}
\end{align}
Finally, let $\bm m(\bm y,\bm w,\theta,\bm r)$ denote the $J\times 1$ vector of moment functions of the form \eqref{eq_mom_func} stacked across $j=1,\ldots,J$, and let $L^0(\mathcal{E}_1,\mathcal{E}_2)$ denote the set of all measurable functions from $\mathcal{E}_1$ to $\mathcal{E}_2$, where $\mathcal{E}_1$ and $\mathcal{E}_2$ are (subsets of) Euclidean space equipped with the Borel $\sigma-$algebra. The following is a simple corollary of Theorems \ref{theorem_main} and \ref{theorem_hankel}.  
\begin{corollary}\label{corollary_main}
Suppose Assumption \ref{assumption_main} holds. 
\begin{enumerate}[label=(\roman*)]
	\item If $S = 2m+1$ for some $m \in \mathbb{N}$ (i.e. $S$ is odd, or $S+1$ is even), then:
\begin{align*}
\Theta_{I}(P) = \left\{\theta \in \Theta:  \exists \bm r \in L^0(\mathcal{W},\mathbb{R}^{S+1}), \varsigma\in  L^0(\mathcal{W},\mathbb{R}) \text{ s.t.}\,\, 
\begin{array}{l} 
E_{P}[\bm m(\bm Y,\bm W,\theta,\bm r) \mid \bm W] = \bm 0 \text{ a.s.,}\\ 
\bm B_{m}(\bm r(\bm w)) \in \mathcal{S}_+^{m+1}\text{ a.s.,}\\
\bm H_{m}^*(\bm r(\bm w),\varsigma(\bm w)) \in \mathcal{S}_+^{m+2}\text{ a.s.}
\end{array}\right\}.
\end{align*}
	\item If $S = 2m$ for some $m \in \mathbb{N}$ (i.e. $S$ is even, or $S+1$ is odd), then:
\begin{align*}
\Theta_{I}(P) = \left\{\theta \in \Theta:  \exists \bm r \in L^0(\mathcal{W},\mathbb{R}^{S+1}), \varsigma\in  L^0(\mathcal{W},\mathbb{R}) \text{ s.t.}\,\, 
\begin{array}{l} 
E_{P}[\bm m(\bm Y,\bm W,\theta,\bm r) \mid \bm W] = \bm 0 \text{ a.s.,}\\ 
\bm H_{m}(\bm r(\bm w)) \in \mathcal{S}_+^{m+1} \text{ a.s.,}\\
\bm B_{m}^*(\bm r(\bm w),\varsigma(\bm w)) \in \mathcal{S}_+^{m+1}\text{ a.s.}
\end{array}\right\}. 
\end{align*}
\end{enumerate}
\end{corollary}
In certain special cases, one can say more about the structure of the identified set. For instance, in the AR(1) model with $T=2$ and no covariates, the identified set for the dynamic coefficient is convex. For the AR(1) model with $T=3$ where the sole covariate is a time trend, the identified set for the dynamic coefficient can consist of at most two isolated points.\footnote{See \cite{dobronyi2021identification} for these results.} In practice, we can check whether a given $\theta \in \Theta$ belongs to the identified set by solving a semidefinite program. To see this, for now consider the case when $\mathcal{W} = \{\bm w_1,\ldots, \bm w_{L} \}$ is finite, and $S=2m+1$ for some $m \in \mathbb{N}$ (i.e. $S$ is odd). Now consider the following optimization problem:
\begin{align}
&\min_{\xi_0, \xi_{11}, \xi_{12},\ldots,\xi_{JL},\bm r(\bm w_1),\ldots, \bm r(\bm w_L), \varsigma(\bm w_1),\ldots, \varsigma(\bm w_L)} \xi_{0},\tag*{SDP($\theta$)}\label{eq_sdp_program}\\[15pt]
\text{ subject to: } \qquad&(i)\quad \xi_0 \geq \left(\sum_{j=1}^{J} \sum_{\ell=1}^{L} \xi_{j\ell}^2\right)^{1/2},\nonumber \\[10pt]
 \qquad&(ii)\quad \xi_{j\ell} = E_{P}[m_{j}(\bm Y, \bm W,\theta, \bm r(\bm w_{\ell}))1\{\bm W = \bm w_{\ell} \}], \quad\text{$j=1,\ldots,J$,\,\, $\ell=1,\ldots,L$,}\nonumber\\[10pt] 
\qquad&(iii)\quad\bm B_{m}(\bm r(\bm w_\ell)) \in \mathcal{S}_+^{m+1}\text{ and } \bm H_{m}^*(\bm r(\bm w_\ell),\varsigma(\bm w_\ell)) \in \mathcal{S}_+^{m+2}, \quad\text{$\ell=1,\ldots,L$}.\nonumber 
\end{align}
Both constraints $(i)$ and $(iii)$ in \ref{eq_sdp_program} can be written as semidefinite constraints, which enforce the positive semidefiniteness of a matrix.\footnote{If $\bm \xi = (\xi_{11},\ldots,\xi_{JL})^\top$, then:
\begin{align*}
\xi_0 \geq \left(\sum_{j=1}^{J} \sum_{\ell=1}^{L} \xi_{j\ell}^2\right)^{1/2} \iff 
\begin{bmatrix}
\bm I \cdot \xi_{0}  &\bm \xi \\
\bm \xi^\top  & \xi_{0}\\
\end{bmatrix}\succeq 0. 
\end{align*}
}  The constraints in $(ii)$ are linear constraints. This makes the program \ref{eq_sdp_program} a \textit{semidefinite program}.\footnote{In general, semidefinite programs are programs that involve optimizing a linear objective function subject to linear constraints and semidefinite constraints. For an introduction see Section 4.6 in \cite{boyd2004convex}, or Chapter 3 in \cite{ben2001lectures}. } Semidefinite programs are convex optimization problems, are a special case of conic programs, and can be solved quickly and reliably with most commercially available solvers.\footnote{All computational results presented in this paper were obtained using the MOSEK interface in R. } For instance, for the AR(1) model with $T=2$ studied in the previous section, the average time to solve the corresponding semidefinite program is approximately $0.0029$ seconds. Average computational times for other models can be found in Section \ref{appendix_simulation} of the Online Supplementary Material. It is straightforward to see that, in the case when $\mathcal{W} = \{\bm w_1,\ldots, \bm w_{L} \}$, by Corollary \ref{corollary_main} we have $\theta \in \Theta_{I}(P)$ if and only if $\text{val}$(\ref{eq_sdp_program})$=0$. In Section \ref{section_consistency_and_inference} we propose an estimator that replaces the population moment conditions in constraint $(ii)$ of the program \ref{eq_sdp_program} with their sample analogs, and we study consistency and propose a method of inference. We also show how to extend the semidefinite programming approach introduced above to cases where $\bm W$ may be continuous or discrete.

\subsection{Identification of Functionals of Unobserved Heterogeneity}\label{section_functionals}

In addition to providing a tractable representation of the identified set of structural parameters, our approach can be used when the researcher's parameter of interest is a functional of the distribution of latent individual effects. In particular, let $\psi: \mathcal{W} \times \mathbb{R} \times \Theta  \to \mathbb{R}$ be a function of the form:
\begin{align}
\psi(\bm w,\alpha, \theta) := \kappa(\bm w,\alpha,\theta) \cdot \sum_{s=0}^S \exp(\alpha)^s \cdot \eta_{s}(\bm w, \theta),\label{eq_psi}
\end{align}
for some known sequence of coefficients $\bm{\eta}(\bm w, \theta) := (\eta_{0}(\bm w, \theta), \ldots, \eta_{S}(\bm w, \theta))^\top$, where $\kappa(\bm w,\alpha,\theta)$ is as in \eqref{modelf}. Then the function $\psi(\bm w,\alpha, \theta)$ is sum of polynomials with the same order and the same factor $\kappa(\bm w,\alpha,\theta)$ as the likelihood in \eqref{modelf} from Assumption \ref{assumption_main}. Now suppose the researcher's parameter of interest is:
\begin{align*}
\Psi(\bm w,\theta_{0}) = E_{Q_{\alpha \mid \bm W}}[\psi(\bm w,\alpha, \theta_0)\mid \bm W = \bm w] = \int \psi(\bm w, \alpha, \theta_0) \,dQ_{\alpha \mid \bm W}(\alpha\mid \bm w),
\end{align*}
for $\bm w \in \mathcal{W}$. Given this representation for $\Psi(\bm w,\theta_0)$, and given the form of $\psi(\bm w,\alpha, \theta_0)$ from \eqref{eq_psi}, for any given $\bm w \in \mathcal{W}$ we have: 
\begin{align}
\Psi(\bm w, \theta_0) = \bm{\eta}(\bm w, \theta_0)^\top \bm r(\bm w),\label{eq_functionals_form}
\end{align}
for a known (up to $\theta_0$) vector $\bm{\eta}(\bm w, \theta_0)$. As we will show, a number of interesting functionals, including the average marginal effect of the lagged choice, can be written in this form. For the AR(1) model in Example \ref{example_AR1}, the point identification of the average marginal effect of lagged choice was first discovered by \cite{aguirregabiria2024identification}. Our results generalize to other functionals of the form \eqref{eq_functionals_form} for models satisfying Assumption \ref{assumption_main}, and also allow for partial identification. 
The ability to bound functionals is also an advantage of our method over existing approaches like conditional maximum likelihood and functional differencing. 

Note that if both $\theta\in \Theta$ and $\bm r(\bm w)$ are point-identified, then $\Psi( \bm w,\theta)$ is point-identified. Furthermore, point-identification of $\Psi( \bm w,\theta)$ can often be easily established using our framework.
\begin{proposition}\label{proposition_functional_point_identification}
Suppose Assumption \ref{assumption_main} holds, suppose $\theta\in\Theta$ is point-identified, and suppose that the matrix $\bm G(\bm w, \theta)$ from \eqref{eq_general_G} has full column rank at $\bm w \in \mathcal{W}$. Then $\Psi(\bm w, \theta)$ is point-identified.
\end{proposition}
Proposition \ref{proposition_functional_point_identification} provides a simple sufficient condition for point identification of the functional $\Psi(\bm w, \theta)$ that can be used even when the conditional distribution $Q_{\alpha \mid \bm W}$ is not point-identified. In particular, if $\theta\in\Theta$ is point-identified and $\bm G(\bm w, \theta_{0})$ has full column rank, then the generalized moments $\bm r(\bm w)$ are point-identified from the equation $\bm p(\bm w) = \bm G(\bm w, \theta_{0})\bm r(\bm w)$. Point identification of $\Psi(\bm w,\theta)$ then follows from \eqref{eq_functionals_form}. We illustrate how to use this result in the examples ahead, which include functionals like the average marginal effect and the average structural function in the AR(1) model.

\setcounter{example}{0}

\begin{example}[$AR(1)$ dynamic logit binary choice, continued] Recall the $AR(1)$ dynamic logit binary choice model from Example \ref{example_AR1}, and suppose that $T = 3$ and $\gamma = 0$. Here we show that the average marginal effect of a lagged choice is point-identified, confirming the results of \cite{aguirregabiria2024identification}. Suppose the researcher's quantity of interest is the average marginal effect of a lagged choice, which is the average difference in the counterfactual choice probabilities when the lagged choice $Y_{it-1}$ takes the value 1 versus 0: 
\[
\Psi(y_0,\beta_0) = E_{Q_{\alpha \mid y_0}}\Big [\Lambda(\alpha_{i} + \beta_0) -\Lambda(\alpha_{i}) \,\big|\, Y_{i0} = y_0\Big ].
\]
For simplicity, fix $y_0 = 0$. Set $\kappa(0,\alpha,\beta) = (1-\Lambda(\alpha))^3(1-\Lambda(\alpha + \beta))^2$, and note that $S=5$. Straightforward calculation shows that we can represent $f(\bm y\mid  y_0; \beta)$ as in \eqref{modelf} using this choice of $\kappa(0,\alpha,\beta)$.\footnote{For details, see the Additional Online Supplementary Material, which can be accessed \href{https://arxiv.org/abs/2104.04590}{here}.} Furthermore:
\begin{align*}
	\Lambda(\alpha + \beta) -\Lambda(\alpha) &= \kappa(0,\alpha,\beta) \left(e^{\alpha+\beta} (1+e^{\alpha})^3(1+e^{\alpha + \beta}) - e^{\alpha}(1+e^{\alpha})^2(1+e^{\alpha + \beta})^2\right), 
\end{align*}
which implies $\Psi(0, \beta) = \bm \eta(0, \beta)^\top \bm r(0)$ with: 
\begin{align*}
	\bm{\eta}(0,\beta) = (e^{\beta}-1) \begin{bmatrix} 0 & 1 & 2+e^{\beta} & 1+2e^{\beta} & e^{\beta} & 0\end{bmatrix}^\top. 
\end{align*}
\cite{chamberlain1985heterogeneity} establishes that $\beta$ is point identified in the AR(1) dynamic logit model without covariates whenever $T \geq 3$. Point identification of $\Psi(y_0,\beta)$ then follows from Proposition \ref{proposition_functional_point_identification} after observing that $\bm G(0, \beta)$ is full column rank in this example. A similar result holds when $y_0=1$.
\end{example} 



\setcounter{example}{0}
\begin{example}[$AR(1)$ dynamic logit binary choice, continued] 
Recall the $AR(1)$ dynamic logit binary choice model with $T=2$ from Example \ref{example_AR1}, and consider the case with $\beta = 0$ and with a scalar covariate. This leads to the static binary choice model considered in \cite{rasch1960probabilistic} and more recently by \cite{davezies2021identification}. Here we consider identification of the average marginal effect of a covariate, as in \cite{davezies2021identification}. To be concrete, consider the case when $T = 2$, suppose $X_{it}$ is a scalar, and consider the marginal effect of a change in $X_{it}$ at some fixed period $t^\star \in \{1,2\}$. The conditional average marginal effect is:
\begin{align*}
	\Psi(\bm x,\gamma_0) &=  E_{Q_{\alpha\mid \bm X}} \left[ \nabla_{x_{t^\star}}P(Y_{it} = 1\mid \bm X_{i} = \bm x, \alpha) \mid \bm X_{i} = \bm x \right] \\
	&= \gamma_0 E_{Q_{\alpha\mid \bm X}} \left[ \Lambda(\alpha_{i} + \gamma_0 x_{t^\star}) (1-\Lambda(\alpha_{i} + \gamma_0 x_{t^\star}))\mid \bm X_{i} = \bm x  \right]. 
\end{align*}
Set $\kappa(\bm x, \alpha, \gamma) = (1-\Lambda(\alpha + \gamma x_1))(1-\Lambda(\alpha + \gamma x_2))(1-\Lambda(\alpha + \gamma x_{t^\star}))$. Furthermore, let $\bm f(\bm x, \alpha; \gamma)$ denote the vector with typical element $f(\bm y \mid \bm x, \alpha; \gamma)$. Then we have:
\begin{align}
\begin{bmatrix} \bm f(\bm x, \alpha; \gamma) \\ \gamma \Lambda(\alpha + \gamma x_{t^\star})(1-\Lambda(\alpha  + \gamma x_{t^\star}))\end{bmatrix} = \kappa(\bm x, \alpha,\gamma) \begin{bmatrix} \bm G(\bm x, \gamma) \\ \bm{\eta}(\bm x,\gamma)^\top \end{bmatrix}  \begin{bmatrix} 1 \\ A \\ A^2 \\ A^3 \end{bmatrix}, \label{eq_desired_format}
\end{align}
where:
\begin{align*}
	\bm G(\bm x, \gamma) = \begin{bmatrix} 1 & e^{\gamma x_{t^\star}} & 0 & 0 \\ 0 & e^{\gamma x_1} & e^{\gamma(x_1+x_{t^\star})} & 0 \\ 0 & e^{\gamma x_2} & e^{\gamma (x_2+x_{t^\star})}& 0 \\ 0 & 0 & e^{\gamma(x_1+x_2)} & e^{\gamma(x_1+x_2+x_{t^\star})} \end{bmatrix}, &&
	\bm{\eta}(\bm x,\gamma) =  \gamma \begin{bmatrix} 0 \\ e^{\gamma x_{t^\star}} \\ e^{\gamma(x_1+x_2)} \\ 0 \end{bmatrix}.
\end{align*}
\cite{rasch1960probabilistic} shows that $\gamma$ is point-identified in this model. However, there are multiple vectors $\bm r(\bm x) \in \R^4$ satisfying $\bm p(\bm x) = \bm G(\bm x,\gamma) \bm r(\bm x)$, since $\bm G(\bm x,\gamma)$ does not have full column rank.\footnote{The third row of $\bm G(\bm x, \gamma)$ is a exactly $e^{\gamma(x_2 - x_1)}$ times the second row.} This means that Proposition \ref{proposition_functional_point_identification} does not apply, and the functional $\Psi(\bm x,\gamma_0)$ will generally be partially identified.\footnote{Although not immediately obvious from our analysis, in the special case when $x_1 =x_2$ the functional $\Psi (\bm x,\gamma_0)$ is point-identified. This is consistent with the finding of \cite{hoderlein2012nonparametric}. It also shows that the converse of Proposition \ref{proposition_functional_point_identification} (as currently stated) is generally false.} Our approach delivers the same sharp identified set as in \cite{davezies2021identification}: see Section \ref{app: DDLcompare} of the Online Supplementary Material for more discussion. However, note that we have modified both the function $\kappa(\bm x, \alpha, \gamma)$ and the matrix $\bm G(\bm x, \gamma)$ from our initial choices in Example \hyperref[example1_Gmat]{1} in order to accommodate the functional of interest, making this example a nontrivial extension of the method introduced above. 
\end{example}

\setcounter{example}{0}

\begin{example}[$AR(1)$ dynamic logit binary choice, continued]
Recall the $AR(1)$ dynamic logit binary choice model from Example \ref{example_AR1}, and consider the case with a scalar covariate. For a fixed value of $(y^\star, x^\star) \in \{0,1\} \times \mathcal{X}$, the conditional average structural function is:
\begin{align}
\Psi(\bm w,\theta_0) = E_{Q_{\alpha\mid \bm W}}\left[ \Lambda(\alpha_{i} + \beta_0 y^\star + \gamma_0 x^{\star} ) \mid \bm W_{i} = \bm w\right].  \label{casf}
\end{align}
Suppose that $T = 3$, $y_0 = 0$, and $\bm w = (y^\star, x^\star,x^\star,x^\star)$. Set $\kappa(\bm w, \alpha,\theta) = \prod_{t=1}^T (1-\Lambda(\alpha + \gamma x_t)) \prod_{t=2}^T (1-\Lambda(\alpha  +\beta + \gamma x_t))$ and $S = 5$. Then we can represent $\Psi(\bm w, \theta)$ as in \eqref{eq_functionals_form} where $\bm{\eta}(\bm w, \theta)$ is the vector of coefficients of $\exp(\alpha)$ in the polynomial function $e^{\alpha + \beta + \gamma x_3}(1+e^{\alpha + \beta + \gamma x_2})(1+e^{\alpha  + \gamma x_1})(1+e^{\alpha + \gamma x_2})(1+e^{\alpha + \gamma x_3})$. Since $\theta$ is point-identified (e.g. see \cite{chamberlain1985heterogeneity}), and since the matrix $\bm G(\bm w, \theta)$ has full column rank for our chosen $\kappa(\bm w, \alpha,\theta)$, Proposition \ref{proposition_functional_point_identification} implies the conditional average structural function is point-identified.\footnote{For details, see the Additional Online Supplementary Material, which can be accessed \href{https://arxiv.org/abs/2104.04590}{here}.} However, this case is somewhat special: when the pair $(y^\star,x^\star)$ does not enter the conditioning argument in \eqref{casf}, it is no longer possible to choose $\kappa(\bm w, \alpha,\theta)$ to ensure that the matrix $\bm G(\bm w, \theta)$ has full column rank. As a result, the conditional average structural function is generally only partially identified. Since the average structural function can be recovered by integrating the conditional average structural function with respect to the distribution of $\bm W_{i}$, this implies that the average structural function is also partially identified in general.
\end{example}

In the general case, the identified set for $\Psi(\bm w, \theta)$ can also be constructed using semidefinite programming.  To see this, consider again the simplified case when $\mathcal{W} = \{\bm w_1,\ldots, \bm w_{L} \}$ is finite and $S=2m+1$ for some $m \in \mathbb{N}$ (i.e. $S$ is odd). Let $\bm w \in \mathcal{W}$ be some value, and consider the following optimization problem:
\begin{align}
&\min_{\xi_0, \xi_{11}, \xi_{12},\ldots,\xi_{JL},\xi_{\Psi},\bm r(\bm w_1),\ldots, \bm r(\bm w_L), \varsigma(\bm w_1),\ldots, \varsigma(\bm w_L)} \xi_{0},\tag*{SDP($\theta,\Psi$)}\label{eq_sdp_program2}\\[15pt]
\text{ subject to: } \qquad&(i)\quad \xi_0 \geq \left(\xi_{\Psi}^2+\sum_{j=1}^{J} \sum_{\ell=1}^{L} \xi_{j\ell}^2\right)^{1/2},\nonumber \\[10pt]
 \qquad&(ii)\quad \xi_{j\ell} = E_{P}[m_{j}(\bm Y, \bm W,\theta, \bm r(\bm w_{\ell}))1\{\bm W = \bm w_{\ell} \}], \quad\text{$j=1,\ldots,J$,\,\, $\ell=1,\ldots,L$,}\nonumber\\[10pt] 
\qquad&(iii)\quad\bm B_{m}(\bm r(\bm w_\ell)) \in \mathcal{S}_+^{m+1}\text{ and } \bm H_{m}^*(\bm r(\bm w_\ell),\varsigma(\bm w_\ell)) \in \mathcal{S}_+^{m+2}, \quad\text{$\ell=1,\ldots,L$,}\nonumber\\
\qquad&(iv)\quad\xi_{\Psi}=\Psi - \bm{\eta}(\bm w, \theta)^\top \bm r(\bm w).\nonumber
\end{align}
Compared to program \ref{eq_sdp_program} introduced earlier, the program \ref{eq_sdp_program2} includes the additional constraint $(iv)$, and also adds an additional parameter $\xi_{\Psi}$ to constraint $(i)$. Since constraint $(iv)$ is linear in $\bm r(\bm w)$, the program \ref{eq_sdp_program2} remains a semidefinite program. It is straightforward to see that, in the case when $\mathcal{W} = \{\bm w_1,\ldots, \bm w_{L} \}$, the pair $(\theta,\Psi)$ belongs to the identified set if and only if $\text{val}$(\ref{eq_sdp_program2})$=0$. The approach introduced in Section \ref{section_consistency_and_inference} can also be used to extend the semidefinite program introduced here to cases where $\bm W$ may be continuous or discrete.

\subsection{Connections with Functional Differencing}\label{section_functional_differencing} 
Functional differencing was proposed by \cite{johnson2004identification} and \cite{bonhomme2012functional} and recently used by \cite{honore2024moment}, \cite{honore2025dynamic} and \cite{davezies2023fixed}. This method aims to find a vector of nonzero moment functions $\bm h(\,\cdot\,,\theta): \mathcal{Y}^{T} \times \mathcal{W} \to \mathbb{R}^{d_h}$ that satisfy:
\begin{equation}
E_{P}[\bm h(\bm Y,\bm W, \theta_0) \mid \bm W, \alpha] = \bm 0, \label{FD}
\end{equation}
$P_{\bm W}-$a.s.\ for all $\alpha$.\footnote{Note the number of moment functions $d_{h}$ is typically not known ahead of time. } Appealing to the discrete nature of $\bm Y$ under Assumption \ref{assumption_main}, we can rewrite the moment conditions in \eqref{FD} as:
\begin{equation}
\sum_{\bm y \in \mathcal{Y}} \bm h(\bm y, \bm W, \theta_{0}) f(\bm y \mid \bm W, \alpha; \theta_{0}) = \bm 0.\label{FD2}
\end{equation}
If \eqref{FD2} holds for all $\alpha \in \R$, then it holds regardless of the true distribution of fixed effects. Provided the functions $\bm h(\,\cdot\,,\theta)$ are known, they can be used to obtain valid moment conditions to identify $\theta \in \Theta$. In particular, let $\bm f(\bm w, \alpha; \theta)$ denote the $J\times 1$ vector that stacks the likelihood function $f(\bm y\mid \bm w, \alpha; \theta)$ across all $\bm y\in \mathcal{Y}^T$. Then the set of moment functions that satisfy \eqref{FD2} are given by:\footnote{Without loss of generality, we focus on finding moment functions that satisfy \eqref{FD2} for all $(\bm w, \alpha)$, rather than $P_{\bm W}-$a.s.\ for all $\alpha$.}
\begin{align*}
\bm D(\theta):= \{\bm h(\,\cdot\,, \theta) \in L^0(\mathcal{W},\mathbb{R}^{J}) : \bm h(\bm w, \theta)^\top \bm f(\bm w, \alpha, \theta) = \bm 0, \,\,\,\forall (\bm w, \alpha) \in \mathcal{W}\times\mathbb{R}\}.
\end{align*}
Connecting with Assumption \ref{assumption_main}, it is also clear that the collection of conditional moment functions satisfy $\bm h(\bm W, \theta)^\top \bm p(\bm W) = 0$ a.s. 
The challenge of using functional differencing lies in finding the functions $\bm h(\,\cdot\,, \theta)$. In some cases, these functions can be constructed numerically with the aid of a computer (see a detailed procedure in \cite{honore2024moment}). However, these functions need to be found model-by-model and for each specific $T$. 

In order to better compare our approach with functional differencing, we first provide a unified analytical method to find these functions for any model that has a likelihood function satisfying Assumption \ref{assumption_main}. 

\begin{theorem}\label{theorem_connectionFD}
	Suppose Assumption \ref{assumption_main} holds, and let $\bm G(\bm w,\theta)$ be the $J \times (S+1)$ matrix from \eqref{eq_general_G}. Furthermore, consider the set:
	\[
	\bm M(\theta) := \{\bm v(\,\cdot\,, \theta) \in L^0(\mathcal{W},\mathbb{R}^{J}) : \bm v(\bm w, \theta)^\top \bm G(\bm w, \theta)  = \bm 0, \,\,\,\forall \bm w \in \mathcal{W} \}.
	\]
	Then $\bm M(\theta)=\bm D(\theta)$ for every $\theta \in \Theta$.
\end{theorem}

Intuitively, Theorem \ref{theorem_connectionFD} suggests that the left null space of $\bm G(\bm w,\theta)$ provides a basis that spans the set $\bm D(\theta)$. Since $\bm G(\bm w,\theta)$ is a known matrix for fixed $\theta \in \Theta$ and $\bm w \in \mathcal{W}$, constructing a basis for the left null space can be done analytically, or by using symbolic computation with the aid of a computer.\footnote{Note that, as with the procedure of \cite{honore2024moment}, there is no guarantee that all moment conditions are functions of $\theta$, and so some may be uninformative.} Checking whether \eqref{FD} holds at $\theta\in\Theta$ is then equivalent to checking if $\bm v(\bm w, \theta)^\top \bm p(\bm w)=0$ for all basis vectors $\bm v(\bm w, \theta)$ in the left null space of $\bm G(\bm w,\theta)$. 

This connection provides additional insight into some results obtained earlier in the literature. For example, in the AR(1) model from Example \ref{example_AR1} with  $T = 2$ and $\gamma = 0$, the $4\times 4$ matrix $\bm G(\bm w,\theta)$ has full rank for each $(\bm w,\theta)$, so that its left null space consists only of the zero vector. This explains why there are no moment conditions for $\beta$ using the functional differencing approach, a result reported by \cite{honore2024moment}. Despite this, our approach still delivers identifying restrictions through the constraints $\bm p(\bm w) = \bm G(\bm w, \theta) \bm r(\bm w)$ and $\bm r(\bm w) \in\mathcal{M}_{S}$. 

As another example of how Theorem \ref{theorem_connectionFD} can be helpful, consider the AR(1) model from Example \ref{example_AR1} with general $T$ and $\gamma=0$. For this model, \cite{honore2024moment} find $2^T - 2T$ linearly independent moment conditions using a numerical search method, and they conjecture that these are all the moment conditions available. To use the approach suggested by Theorem \ref{theorem_connectionFD}, first note that the matrix $\bm G(\bm w, \theta)$ is of dimension $2^T \times 2T$ and has full column rank.\footnote{For details, see the Additional Online Supplementary Material, which can be accessed \href{https://arxiv.org/abs/2104.04590}{here}.} Therefore, the left null space of $\bm G(\bm w, \theta)$ provides a basis with exactly $2^T - 2T$ linearly independent moment conditions, verifying the conjecture of \cite{honore2024moment}. This result is also useful since ex ante it is not known how many linearly independent moment functions exist when using functional differencing. Our result suggests that \cite{honore2024moment} have indeed found all the relevant moment functions.

As a final example, consider the AR(1) dynamic ordered logit model from Example \ref{example_ODC} with $M$ choice options and $T$ periods.  The corresponding matrix $\bm G(\bm w, \theta)$ has dimension $M^T \times ((T-1)M^2 - (T-2)M)$ and is of full column rank.\footnote{For details, see the Additional Online Supplementary Material, which can be accessed \href{https://arxiv.org/abs/2104.04590}{here}.} Theorem \ref{theorem_connectionFD} thus confirms the conjecture made in \cite{honore2025dynamic} that there are $M^T - (T-1)M^2 + (T-2)M$ linearly independent moment conditions available in this model. 


Using Theorem \ref{theorem_connectionFD}, the difference between functional differencing and our approach can be explained geometrically. For a fixed $(\bm w,\theta) \in \mathcal{W}\times\Theta$, let $\bm p_{\bm G}(\bm w)$ denote the projection of the choice probability vector $\bm p(\bm w)$ onto the column space of $\bm G(\bm w,\theta)$, and let $\bm r^*(\bm w,\theta)\in \mathcal{M}_{S}$ denote the vector that minimizes $|| \bm p(\bm w) - \bm G(\bm w, \theta) \bm r||$ over all $\bm r \in \mathcal{M}_{S}$. Note by Theorem \ref{theorem_main} we have $\theta\in \Theta_{I}(P)$ if and only if $||\bm p(\bm w) - \bm G(\bm w, \theta) \bm r^*(\bm w,\theta)||=0$, $P_{\bm W}-$a.s.\ It is straightforward to show that the vectors $\bm p(\bm w) - \bm p_{\bm G}(\bm w)$ and $\bm p_{\bm G}(\bm w) - \bm G(\bm w, \theta) \bm r^*(\bm w,\theta)$ are orthogonal, so that by Pythagoras' Theorem:\footnote{In particular, $\bm p(\bm w) - \bm p_{\bm G}(\bm w)$ is the least-squares residual, which lies in the null space of $\bm G(\bm w, \theta)^\top$, and so is orthogonal to the column space of $\bm G(\bm w, \theta)$. Thus, it is orthogonal to $\bm p_{\bm G}(\bm w) - \bm G(\bm w, \theta) \bm r^*(\bm w,\theta)$, which lies in the column space of $\bm G(\bm w,\theta)$.}
\begin{figure}
\centering
\includegraphics[scale=0.65]{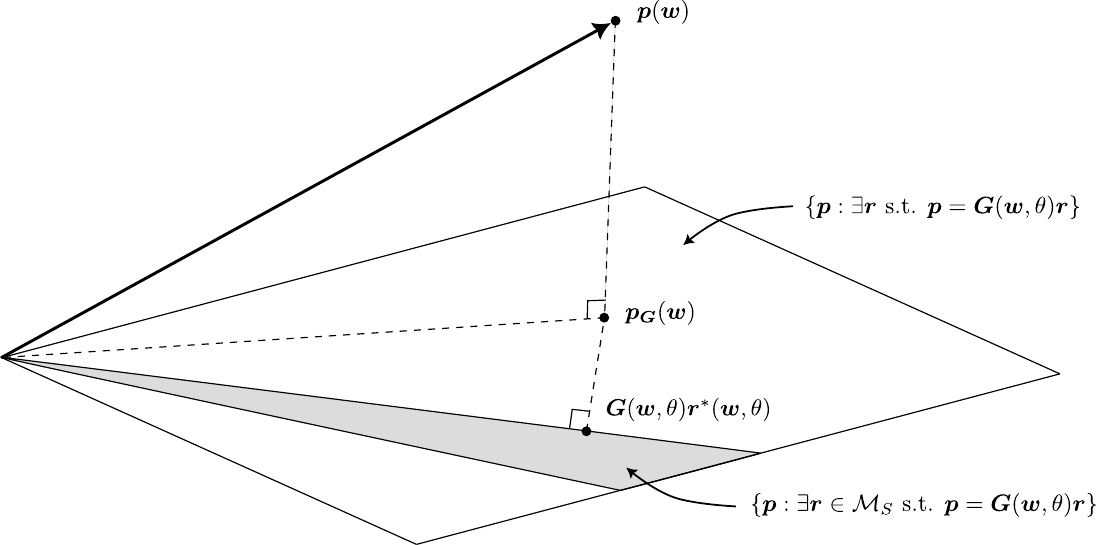}
\caption{The orthogonal decomposition of the vector $\bm p(\bm w) - \bm G(\bm w,\theta)\bm r^*(\bm w,\theta)$ into the vectors $\bm p(\bm w) - \bm p_{\bm G}(\bm w)$ and $\bm p_{\bm G}(\bm w) - \bm G(\bm w,\theta)\bm r^*(\bm w,\theta)$. Functional differencing checks if $||\bm p(\bm w) - \bm p_{\bm G}(\bm w)||=0$, but this is not sufficient to verify whether $||\bm p(\bm w) - \bm G(\bm w,\theta)\bm r^*(\bm w,\theta)||=0$. }\label{fig_orthogonal}
\end{figure}
\begin{align}
|| \bm p(\bm w) - \bm G(\bm w, \theta) \bm r^*(\bm w,\theta)||^2 = \underbrace{|| \bm p(\bm w) - \bm p_{\bm G}(\bm w)||^2}_{(i)} + \underbrace{||\bm p_{\bm G}(\bm w) - \bm G(\bm w, \theta) \bm r^*(\bm w,\theta)||^2}_{(ii)}.\label{eq_decomposition} 
\end{align}
See Figure \ref{fig_orthogonal} for an illustration. 
Now by Theorem \ref{theorem_connectionFD} and its following discussion, functional differencing searches for vectors $\bm v(\bm w, \theta)$ that form a basis for the left nullspace of $\bm G(\bm w,\theta)$, and that are orthogonal to $\bm p(\bm w)$. By the Fundamental Theorem of Linear Algebra, the condition $\bm v(\bm w, \theta)^\top \bm p(\bm w)=0$ holds for all basis vectors $\bm v(\bm w, \theta)$ in the left null space of $\bm G(\bm w,\theta)$  if and only if $\bm p(\bm w)$ lies in the column space of $\bm G(\bm w,\theta)$; that is, if and only if $\bm p(\bm w) = \bm p_{\bm G}(\bm w)$. By this reasoning, functional differencing is equivalent to checking whether term $(i)$ in \eqref{eq_decomposition} is equal to zero, which is a necessary but not sufficient condition to have $\bm p(\bm w) = \bm G(\bm w, \theta) \bm r^*(\bm w,\theta)$ under the constraint $\bm r^*(\bm w,\theta) \in \mathcal{M}_S$. In contrast, our approach requires that both terms $(i)$ and $(ii)$ in \eqref{eq_decomposition} are equal to zero. Seen in this way, functional differencing misses a piece of the orthogonal decomposition of $\bm p(\bm w) - \bm G(\bm w, \theta)\bm r^*(\bm w,\theta)$, and as a result it generally fails to pick up all relevant identifying restrictions.

In addition to providing a general approach to identification and allowing us to bound functionals of the distribution of the latent individual effects, our procedure delivers the sharp identified set even when there are no moment conditions available using functional differencing, it provides sharp bounds in cases where the functional differencing approach cannot, and it allows us to test for model misspecification.\footnote{Even when the structural parameters are point-identified from the functional differencing moment conditions, in some cases adding additional (binding) constraints on the model parameters can reduce asymptotic mean squared error. This was shown for the empirical likelihood estimator and the GMM estimator with an optimal weighting matrix by \cite{moon2009estimation} in the specific case when the model parameters are point-identified by a set of moment equalities and the researcher has access to a single additional (drifting-to-)binding moment inequality. } We now illustrate these points using examples.

\setcounter{example}{0}

\setcounter{example}{0}
\begin{example}[$AR(1)$ dynamic logit binary choice, continued] 
Recall the $AR(1)$ dynamic logit binary choice model from Example \ref{example_AR1}. Suppose that $T=3$, and consider the case when the only covariate is a time trend. For simplicity, fix $y_0 = 0$, let $B := \exp(\beta)$, and let $C := \exp(\gamma)$. The matrix $\bm G(y_0,\theta)$ is given by: 
	\begin{align*}
\bm G(0,\theta) = 
\begin{bmatrix} 
	1 & BC^2(1+C) & B^2C^5 & 0 & 0 & 0 \\
	0 & C & C^3(1+BC) & BC^6 & 0 & 0 \\
	0 & C^2 & C^4(B+C) & BC^7 & 0 & 0 \\
	0 & 0 & BC^3 & BC^5(1+C) & BC^8 & 0 \\
	0 & C^3 & BC^5(1+C) & B^2C^8 & 0 & 0 \\
	0 & 0 & C^4 & C^6(1+BC) & BC^9 & 0 \\
	0 & 0 & BC^5 & BC^7(B+C) & B^2C^{10} & 0 \\
	0 & 0 & 0 & B^2C^6 & B^2C^8(1+C) & B^2C^{11}
\end{bmatrix}.
\end{align*}
Symbolic computation shows that the left null space for $\bm G(0,\theta)$ is spanned by the following two vectors: 
	\begin{align*}
	\bm v_1(\theta) & = \begin{bmatrix} 0 & -1& C^{-1}  & (C-1)&0 & -1 &(BC^2)^{-1} & 0 \end{bmatrix}^\top,\\
	\bm v_2(\theta) & = \begin{bmatrix}0 & C^2 & -1  & 0& (C^{-1}-1) & BC & -1 & 0 \end{bmatrix}^\top.
	\end{align*}
	We now demonstrate a numerical example where our method provides point identification but functional differencing leads to partial identification with an identified set containing two points. Suppose $Q_{\alpha\mid y_0}$ is a discrete distribution with equal mass at $-2$ and $1$, and suppose $\theta_0 = (0.5, 0.8)$. Figure \ref{figure:momentequalities} shows that there are two values of $\theta$ ($\theta_0$ and $\tilde \theta = (1.15, 0.3)$) which satisfy the moment conditions $\bm v_1(\theta)^\top \bm p(0) = 0$ and $\bm v_2(\theta) ^\top \bm p(0) = 0$. 
	However, at $\tilde \theta$ the Hankel matrix $\bm H_2(\bm r(0))$ is:\footnote{Here $\bm G(0,\tilde{\theta})$ is of full column rank, so $\bm r(0)$ is uniquely determined by the equation $\bm p(0) = \bm G(0,\tilde{\theta})\bm r(0)$. }  
	\[
	\bm H_2(\bm r(0)) = \begin{bmatrix} -0.246 & 0.046 & -0.006\\ 0.046 & -0.006 & 0.002 \\ -0.006 & 0.002 & 0.00006\end{bmatrix}, 
	\]
	which is clearly not positive semidefinite. Therefore, the Hankel extension $\bm H_2^*(\bm r(0),\varsigma)$ of $\bm H_2(\bm r(0))$ cannot be positive semidefinite for any $\varsigma \geq 0$. Thus, although $\tilde \theta$ satisfies all the moment conditions found by functional differencing, these exists no $\bm r(0)\in \mathcal{M}_{5}$ that satisfies $\bm p(0)=\bm G(0,\tilde \theta) \bm r(0)$. In contrast, our approach eliminates $\tilde \theta$ from the identified set and correctly concludes that $\theta_{0}$ is point-identified.\footnote{Even though we show the time trend model is point-identified in this specific numerical example, it is not known whether the time trend model is generically point-identified when $T=3$. Regardless, functional differencing always yields two solutions for $\theta$ when $y_0 = 0$ and $T = 3$. }

\begin{figure}
	\centering
	\includegraphics[scale = 0.1]{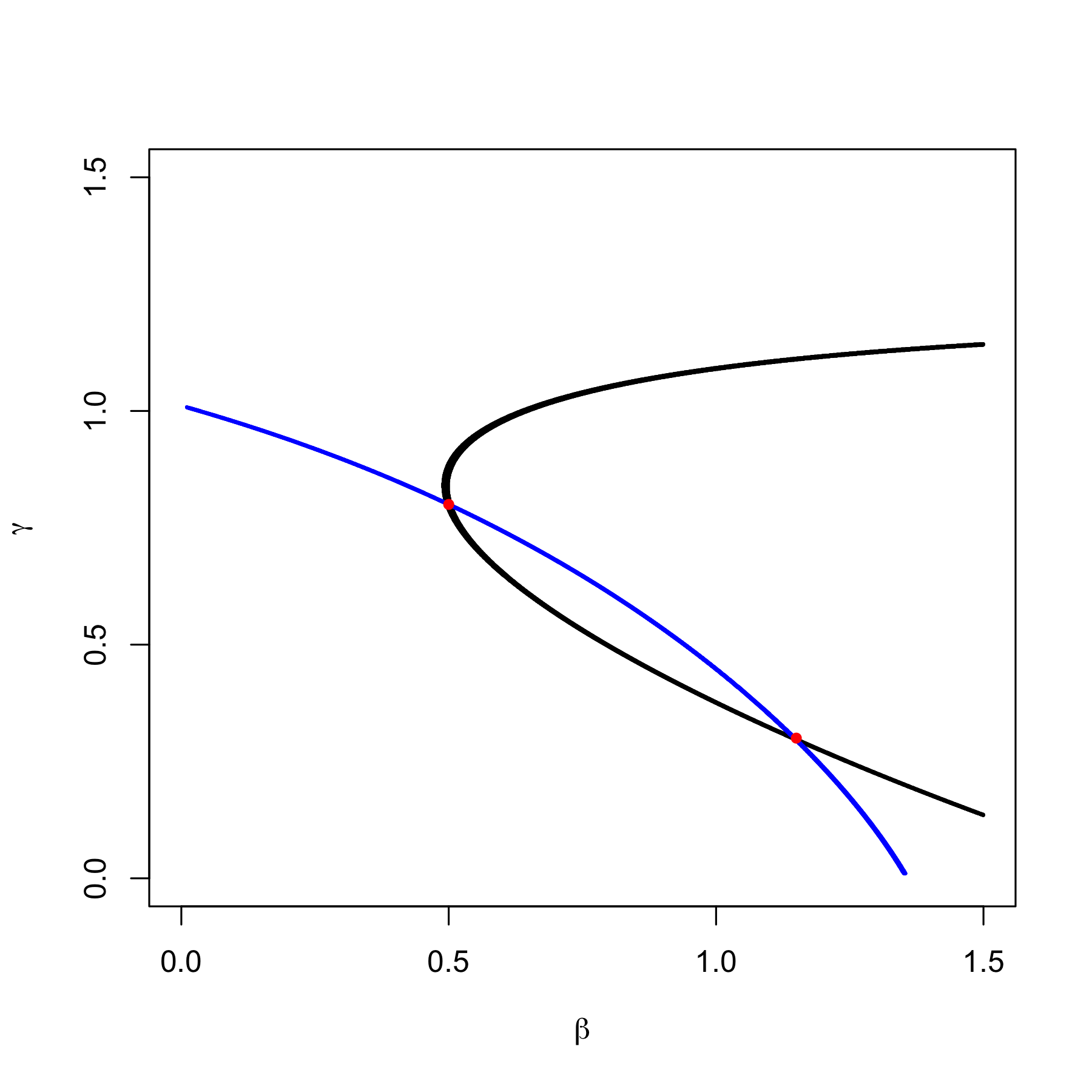}
	\caption{The black curve is the set of $\theta$ that satisfies $\bm v_1(\theta)^\top\bm p(0, \bm x) = 0$ and the blue curve is the set of $\theta$ that satisfies $\bm v_2(\theta)^\top\bm p(0, \bm x) = 0$. There are two values of $\theta$ that satisfy both moment restrictions. The underlying data generating process imposes $P(Y_{i0} = 0) = 1$, that the fixed effect distribution $Q_{\alpha}$ is discrete with equal mass at $-2$ and $1$, and that $(\beta_0,\gamma_0)=(0.50,0.80)$. } 
	\label{figure:momentequalities}
\end{figure}
\end{example}

\setcounter{example}{0}
\begin{example}[$AR(1)$ dynamic logit binary choice, continued] 
Recall the $AR(1)$ dynamic logit binary choice model from Example \ref{example_AR1}. Consider the $T=3$ case with fixed $y_0 = 0$, $\beta = 0.5$, and $\gamma = 0$. Suppose that $\alpha_{i}$ is uniformly distributed on $\{-2,2\}$. However, suppose now that the logit model is misspecified, and the data is instead generated by $\epsilon_{it} \sim N(0,1)$. If we incorrectly assume that $\epsilon_{it}$ is standard logistic, then the moment conditions developed in \cite{chamberlain1985heterogeneity} identify $\beta$ through the formula $\beta = \log (p_{011}/p_{101})=1.3$, where $p_{011}$ and $p_{101}$ are the conditional probabilities of the choice paths $(Y_{i1},Y_{i2},Y_{i3})=(0,1,1)$ and $(Y_{i1},Y_{i2},Y_{i3})=(1,0,1)$ given $Y_{i0}=0$, respectively. Using this value of $\beta$, the corresponding Hankel matrix is:\footnote{Again, here $\bm G(0,\beta)$ is of full column rank, so $\bm r(0)$ is uniquely determined by the equation $\bm p(0) = \bm G(0,\beta)\bm r(0)$. }  
	\[
	\bm H_{2}(\bm r(0)) = \begin{bmatrix} 0.406 & 0.027 & -0.011\\ 0.027 & -0.011 & 0.007 \\ -0.011 & 0.007 & -0.001 \end{bmatrix},
	\]
which is clearly not positive semidefinite. Therefore, the Hankel extension $\bm H_2^*(\bm r(0),\varsigma)$ of $\bm H_2(\bm r(0))$ cannot be positive semidefinite for any $c\in \mathbb{R}$. In fact, for this example, there exists no value of $\beta$ such that $\bm p(0) = \bm G(0,\beta) \bm r(0)$ for $\bm r(0)\in \mathcal{M}_5$. This shows our approach can produce testable implications for the logit model. The inference procedure introduced in Section \ref{section_consistency_and_inference} can also be used to conduct a formal test of model misspecification. See Remark \ref{remark_misspecification}.
	\end{example} 

\setcounter{example}{2}

\begin{example}[$AR(p)$ dynamic logit binary choice, continued] 
Recall the $AR(p)$ dynamic logit binary choice model from Example \ref{example_ARp}. Suppose that $p = 2$ and $T = 3$, let $\bm w = (y_{-1}, y_{0}, \bm x)$ and further fix $(y_{-1}, y_0) = (0,0)$. Now set: 
	\begin{align*}
	\kappa(\bm w, \alpha, \theta) &=(1-\Lambda(\alpha + \beta_2 + \gamma x_3)) (1- \Lambda(\alpha + \beta_1 + \beta_2 + \gamma x_3))\\
	&\qquad\qquad\times\Big( \prod_{t=1}^3 (1- \Lambda(\alpha + \gamma x_t)) \prod_{t=2}^3 (1-\Lambda(\alpha + \beta_1 + \gamma x_t)) \Big). 
	 \end{align*}
	The matrix $\bm G(\bm w,\theta)$ can be shown to be of dimension $8 \times 8$ with full rank for all $\bm w \in \mathcal{W}$, except when $x_2 = x_3$.\color{teal}\footnotemark[6] \color{black} When $x_2\neq x_3$, there are no moment conditions from functional differencing  since the left null space of $\bm G(\bm w, \theta)$ contains only the zero vector. However, even in this case our method can provide partial identification by leveraging the additional moment restrictions $\bm r(\bm w)\in \mathcal{M}_{7}$. 
\end{example}

\section{Estimation and Inference}\label{section_consistency_and_inference}

While our main results concern identification, in this section we propose a consistent estimator of the identified set that is applicable when the structural parameters are either point- or partially-identified, and we also propose an inference procedure. Our estimation and inference procedure allow for both discrete and continuous covariates, and our inference procedure is based on the procedure of \cite{chernozhukov2023constrained} (CNS hereafter). The CNS procedure is designed for inference on (possibly infinite-dimensional) shape-constrained parameters in models defined by conditional moments, and allows for both point and partial identification. CNS also allows for a general class of shape constraints defined by equality and inequality restrictions. In our setting, the relevant shape constraints are on the moment vectors, since by Theorem \ref{theorem_hankel} any valid moment vector must be such that the Hankel matrix and its Hankel extension are positive semidefinite. Note that positive semidefiniteness of a matrix is equivalent to nonnegativity of the determinants of all of its principal minors. Thus, positive semidefiniteness of a matrix can be enforced by imposing certain nonlinear inequality constraints on the entries of the matrix, connecting our setting to the shape constraints allowed by CNS.  However, in order to maintain the semidefinite programming structure discussed in the previous section, we use a conservative implementation of their procedure.\footnote{In the notation of CNS, we set $r_{n} = +\infty$, and take $\hat{V}_{n}(\theta,R \mid \ell_n)=\{\bm 0\}$. We also set the weighting matrix as the identity matrix. These are always feasible (but potentially conservative) choices. These choices greatly simplify computation by avoiding the need to optimize over the set $\hat{V}_{n}(\theta,R \mid \ell_n)$ in our bootstrap procedure, which would otherwise destroy the semidefinite programming structure of our bootstrap test statistic. Avoiding this minimization in the bootstrap test statistic leads to a larger-than-necessary critical value, but keeps the procedure tractable. } In addition to providing substantial computational gains, our simplified implementation also allows us to use a weaker set of assumptions than those provided in CNS. We outline this weaker set of assumptions in Appendix \ref{section_inference_appendix_assumptions}. 
To keep notation simple, we focus on providing results for the identified set of structural parameters, although our approach extends to the functionals from Section \ref{section_functionals} under minimal additional assumptions.

\subsection{Consistency}\label{section_consistency}

Recall from Corollary \ref{corollary_main} and equation \eqref{eq_mom_func} that the model constraints $\bm p(\bm w) = \bm G(\bm w, \theta) \bm r(\bm w)$ can be written as conditional moment equalities of the form:
\begin{align}
E_{P}[\bm m(\bm Y_{i},\bm W_{i},\theta,\bm r) \mid \bm W_{i}] = \bm 0 \text{ a.s.}, \label{eq_mom_func2}
\end{align}
where $\bm m(\bm Y_{i},\bm W_{i},\theta,\bm r)$ is a $J\times 1$ vector of moment functions with $j^{th}$ element:
\begin{align}
m_{j}(\bm Y_{i},\bm W_{i},\theta,\bm r) = 1\{\bm Y_{i} = \bm y_{j} \} - \bm g(\bm y_j,\bm W_{i},\theta)^\top \bm r(\bm W_{i}). \label{eq_mom_definition}
\end{align}
While $\bm g(\bm y_j, \bm W_{i},\theta)^\top$ is a known function of the covariates and structural parameters, $\bm r(\bm W_{i})$ is an unknown vector-valued function that must be estimated. Furthermore, from Corollary \ref{corollary_main}, we must also impose a number of shape constraints on these functions during estimation. Since the covariates may be continuous or discrete, it is desirable to allow for a flexible specification for the functions $\bm r:\mathcal{W}\to \mathbb{R}^{S+1}$, viewing the moments as a function of the covariates $\bm W_{i}$. Furthermore, the specification for these functions should be amenable to our implementation using semidefinite programming, even when the covariates are continuous. With these concerns in mind, we recommend a simple sieve approximation based on piecewise constant functions.\footnote{While this choice simplifies computation significantly, it is not necessary: researchers interested in other methods of approximation can consult Appendix \ref{section_inference_appendix_assumptions} for the minimal set of assumptions required by our procedure.}

Let $\mathcal{D}_{l_n}$ denote a growing partition of $\mathcal{W}$ into $l_n$ disjoint sets, and assume the sequence $\{\mathcal{D}_{l_n}\}_{n=1}^{\infty}$ is nested for all but finitely many $n\geq 1$. Now let $\mathcal{C}_{n}(\overline{\delta})$ denote the following set of functions:
\begin{align*}
\mathcal{C}_{n}(\overline{\delta}) := \left\{ f:\mathcal{W} \to \mathbb{R}_{+} : f(\bm w) = \sum_{D \in \mathcal{D}_{l_n}}  1\{\bm w \in D\} \delta_{D},\,\delta_D \in [0,\overline{\delta}] \right\}. 
\end{align*}
Note that $\mathcal{C}_{n}(\overline{\delta})$ is the class of piecewise constant functions with uniformly bounded coefficients. Using this collection, we define a sieve for the functions $\bm r(\,\cdot\,):\mathcal{W}\to \mathbb{R}^{S+1}$ using all vector-valued functions whose elements are piecewise constant functions on the partition $\mathcal{D}_{l_n}$:
\begin{align}
\mathcal{R}_{n} : = \left\{ \bm r : \mathcal{W} \to \mathbb{R}^{S+1} : \bm r(\bm w)^\top =  \left(f_{0}(\bm w), \ldots, f_{S}(\bm w)  \right), \text{ with $f_{s} \in \mathcal{C}_{n}(\overline{\delta})$ for $s=0,\ldots,S$} \right\}.\label{eq_pc}
\end{align}
Note that $\mathcal{R}_{n}$ is the set of all piecewise constant vector-valued functions of the form $\bm r_{n}(\bm w) = \sum_{D \in \mathcal{D}_{l_n}} \bm \delta_{D}\cdot 1\{\bm w \in  D\}$, where $\bm \delta_{D} \in [0,\overline{\delta}]^{S+1}$.\footnote{In many examples, $\overline{\delta}<\infty$ is guaranteed whenever $\mathcal{W}$ and $\Theta$ are compact.} Finally, let $\mathcal{R}$ denote the set of all functions that can be approximated as uniform limits of the sequences $\bm r_n \in \mathcal{R}_n$:\footnote{Other choices of the norm are possible.}
\begin{align*}
\mathcal{R}:= \left\{ \bm r : \mathcal{W} \to \mathbb{R}^{S+1} : \lim_{n\to \infty} ||\bm r-\bm r_{n}||_{\infty} =0 \text{ for some $\bm r_{n} \in \mathcal{R}_{n}$ $\forall n$} \right\}.
\end{align*}
Then $\mathcal{R}$ is a subset of a Banach space, although the precise properties of $\mathcal{R}$ will depend on the sequence of partitions $\{\mathcal{D}_{l_n}\}_{n=1}^{\infty}$ chosen by the researcher.

Now since the model is characterized in terms of conditional moment equalities, we first convert the conditional moments into unconditional moments using instrument functions. In particular, given a collection $\mathcal{D}_{k_n}$ of $k_n$ Borel subsets of $\mathcal{W}$, define the $k_{n}\times 1$ vector of instrument functions:
\begin{align}
\bm q^{k_n}(\bm w)
:=
\begin{bmatrix}
1\{\bm w \in D_1'\}&
1\{\bm w \in D_2'\}&
\ldots&
1\{\bm w \in D_{k_{n}}'\}
\end{bmatrix}^\top.\label{eq_instrument_functions}
\end{align}
For any such partition, the $J\times 1$ vector of conditional moment equalities of the form \eqref{eq_mom_func2} imply the following set of $J\cdot k_n \times 1$ vector of unconditional moment equalities:
\begin{align}
E_{P}[\bm m(\bm Y_{i},\bm W_{i},\theta,\bm r) \otimes \bm q^{k_n}(\bm W_{i})] = \bm 0. \label{eq_mom_func3}
\end{align}
With continuous covariates, we will generally require $k_n\uparrow \infty$ as $n\to\infty$. Now given an i.i.d.\ sample $\{(\bm Y_{i}, \bm W_{i})\}_{i=1}^{n}$, our estimator of the identified set is based on the minimizers of the following criterion function:
\begin{align}
Q_{n}(\theta,\bm r):= \left|\left| \frac{1}{n} \sum_{i=1}^{n} \bm m(\bm Y_{i},\bm W_i,\theta,\bm r) \otimes \bm q^{k_n}(\bm W_{i}) \right|\right|. \label{eq_sample_Qn}
\end{align}
In particular, define the following set of shape restrictions:
\begin{align*}
\mathcal{S} := 
\begin{cases}
	\left\{(\theta, \bm r) \in \Theta \times \mathcal{R} : \bm H_{m}^*(\bm r(\bm w),\varsigma^*(\bm w)) \succeq 0, \,\,\bm B_{m}(\bm r(\bm w)) \succeq 0, \forall \bm w \in \mathcal{W}  \right\}, &\text{ if $S=2m+1$,}\\
	\left\{(\theta, \bm r) \in \Theta \times \mathcal{R} : \bm H_{m}(\bm r(\bm w)) \succeq 0, \,\,\bm B_{m}^*(\bm r(\bm w),\varsigma^*(\bm w)) \succeq 0, \forall \bm w \in \mathcal{W}  \right\}, &\text{ if $S=2m$.}
\end{cases}
\end{align*}
Here $\varsigma^*(\bm w)$ is any choice that ensures either $\bm H_{m}^*(\bm r(\bm w),\varsigma^*(\bm w))\succeq 0$ (when $S$ is odd) or $\bm B_{m}^*(\bm r(\bm w),\varsigma^*(\bm w)) \succeq 0$ (when $S$ is even) whenever possible given a fixed $\bm r(\bm w)$.\footnote{Such a choice is always possible: see Lemma 2.3 in \cite{curto1991recursiveness}. For theoretical purposes, it is convenient to view $\varsigma^*(\bm w)$ as a deterministic function of $\bm r(\bm w)$.} If $\bm r \in \mathcal{R}$, then the joint identified set for $(\theta,\bm r)$ is given by: 
\begin{align}
\mathcal{I}^*(P) := \left\{ (\theta,\bm r) \in (\Theta \times \mathcal{R})\cap \mathcal{S} : E_{P}[m_{j}(\bm Y_{i},\bm W_{i},\theta,\bm r) \mid \bm W_{i}] = 0 \text{ a.s.}, \text{ for $j=1,\ldots, J$} \right\}.\label{eq_joint_identified_set}
\end{align}
Note that $\Theta_{I}(P)=\text{Proj}_{\Theta}(\mathcal{I}^*(P))$ is exactly the projection of $\mathcal{I}^*(P)$ onto $\Theta$. Our estimator for the joint identified set for $(\theta,\bm r)$ is given by:
\begin{align}
\hat{\mathcal{I}}_{n} := \left\{(\theta,\bm r) \in (\Theta \times \mathcal{R}_{n})\cap \mathcal{S} : Q_{n}(\theta,\bm r) \leq \inf_{(\theta,\bm r) \in (\Theta \times \mathcal{R}_{n}) \cap \mathcal{S}} Q_{n}(\theta,\bm r) + \tau_{n}\right\}, \label{eq_sample_identified_set_main2}
\end{align}
where $\tau_{n}\downarrow 0$ is a sequence of constants (see Remark \ref{eq_choice_taun}). Now let $\hat{\Theta}_{I,n}=\text{Proj}_{\Theta}(\hat{\mathcal{I}}_{n})$ denote the corresponding projection of $\hat{\mathcal{I}}_{n}$ on $\Theta$. This set can be written as:
\begin{align}
\hat{\Theta}_{I,n} := \left\{\theta \in \Theta :  \inf_{\bm r \in \Pi_{\mathcal{R}_n}(\mathcal{S})} Q_{n}(\theta,\bm r) \leq \inf_{(\theta,\bm r) \in (\Theta \times \mathcal{R}_{n}) \cap \mathcal{S}} Q_{n}(\theta,\bm r) + \tau_{n}\right\},\label{eq_sample_identified_set_main}
\end{align}
where $\Pi_{\mathcal{R}_n}(\mathcal{S}):=\{ \bm r \in \mathcal{R}_n : \exists \theta \in \Theta \text{ s.t. }(\theta,\bm r)\in \mathcal{S} \}$. The set $\hat{\Theta}_{I,n}$ represents our estimator for the identified set $\Theta_{I}(P)$. 

Our next result shows that the set estimator $\hat{\Theta}_{I,n}$ is consistent for the identified set $\Theta_{I}(P)$ in the Hausdorff metric, uniformly over a certain class of data generating processes (DGPs).\footnote{Recall the Hausdorff distance between two sets $A$ and $B$ is given by: 
\begin{align*}
d_{H}(A,B,||\,\cdot\,||) := \max\left\{\sup_{a \in A}\inf_{b \in B} ||a-b||, \sup_{b \in B}\inf_{a \in A} ||a-b|| \right\}.
\end{align*}
} Before introducing our result, we require two additional assumptions. In the following, let $\mathcal{P}$ denote a subset of the set of all distributions on $\mathcal{Y}^T\times \mathcal{W}$, and for each element $(\theta,\bm r)\in \Theta\times \mathcal{R}$ let $\Pi_{n}(\theta,\bm r)$ denote its approximation on $\Theta\times \mathcal{R}_{n}$ and define $\mathcal{I}_{n}^*(P) := \left\{ \Pi_{n}(\theta,\bm r) : (\theta,\bm r)\in\mathcal{I}^*(P) \right\}$.
\begin{assumption}\label{assumption_consistency}
$(i)$ $\{(\bm Y_{i}, \bm W_{i})\}_{i=1}^{n}$ is i.i.d.\ with $(\bm Y_{i},\bm W_{i}) \sim P \in \mathcal{P}$; $(ii)$ $\mathcal{W} \subset \mathbb{R}^{d_{w}}$ is compact; $(iii)$ $\Theta$ is compact; $(iv)$ there is a common sequence of partitions $\{\mathcal{D}_{m}\}_{m=1}^{\infty}$ of $\mathcal{W}$, nested for all $m$ sufficiently large, that determines both the piecewise constant functions $\bm r_n \in \mathcal{R}_n$ as in \eqref{eq_pc} and the vector of instrument functions $\bm q^{k_n}(\bm w)$ as in \eqref{eq_instrument_functions}; $(v)$ for every $P_n\in \mathcal{P}$ and $(\theta_n,\bm r_n) \in \mathcal{I}_n^*(P_n)$ there exists a corresponding $(\theta_{n}^*,\bm r_{n}^*) \in \mathcal{I}^*(P_n)$ such that $\sqrt{n}(E_{P_n}\left[||\bm r_{n}^*(\bm W_{i})- \bm r_{n}(\bm W_{i})||^2\right])^{1/2} = o((\log(n))^{-\beta_a})$ for some $\beta_a>0$.
\end{assumption}
Assumption \ref{assumption_consistency}$(i)-(iv)$ are straightforward. Assumption \ref{assumption_consistency}$(v)$ implies the ``asymptotic unbiasedness'' condition required in CNS.\footnote{$\beta_a = 1/2$ is assumed in all of CNS's examples: see CNS Assumption 4.1$(iv)$ (heterogeneity and demand analysis), Assumption A.2.8$(iv)$ (consumer demand), and Assumption A.2.14$(iii)$ (quantile treatment effects).} It can be seen as a condition on the quality of the sieve space, imposing the restriction that the true (but unknown) vector of moment functions $\bm r \in \mathcal{R}$ is well-approximated by piecewise constant functions. It holds trivially if regressors are discrete, but otherwise depends on the chosen sequence $\{\mathcal{D}_{l_n}\}_{n=1}^{\infty}$ and the properties of $\bm r \in \mathcal{R}$. 

For the next assumption, let $\vec{d}_{H}(A,B) = \sup_{a \in A} \inf_{b \in B}||a-b||$ denote the directed Hausdorff distance, and set:
\begin{align}
 Q_{n,P}(\theta,\bm r) := \left|\left| E_{P}[\bm m(\bm Y_{i},\bm W_i,\theta,\bm r) \otimes \bm q^{k_n}(\bm W_{i})] \right|\right|. \label{eq_QP}
\end{align}
That is, $Q_{n,P}(\theta,\bm r)$ is the analog of $Q_{n}(\theta,\bm r)$ when the sample moment conditions have been replaced by their population versions.
\begin{assumption}\label{assumption_consistency2}
For some constant $\delta>0$ and sequences $0<\nu_{n}^{-1} =O(1)$ and $b_{n}=o(\nu_n^{-1})$:
\begin{align*}
\nu_n^{-1} \min\{\delta, \vec{d}_{H}(\theta,\Theta_{I}(P))\} \leq \inf_{\bm r \in \Pi_{\mathcal{R}_n}(\mathcal{S})} Q_{n,P}(\theta,\bm r) - \inf_{(\theta',\bm r') \in (\Theta\times \mathcal{R}_n)\cap \mathcal{S}} Q_{n,P}(\theta',\bm r')+b_n,
\end{align*}
for every $\theta\in \Theta$ and $P \in \mathcal{P}$ for all $n$ sufficiently large.
\end{assumption}

Assumption \ref{assumption_consistency2} is similar to the standard polynomial minorant condition typically imposed in set estimation problems, going back to \cite{chernozhukov2007estimation} (see their Condition C.2).\footnote{See also \cite{kaido2022constraint} for an extensive discussion of this condition.} Intuitively, it requires that the criterion function \eqref{eq_QP} ``lifts off'' sufficiently fast in a neighborhood of the identified set.  However, Assumption \ref{assumption_consistency2} is stronger than the typical polynomial minorant condition, since it imposes constraints on both the quality of the sieve $\mathcal{R}_{n}$ and the strength of identification associated with the instrument functions. In general the condition depends on the interaction between the instrument functions and piecewise constant functions at the population level, and rules out weak identification. In certain cases simple sufficient conditions can be developed.\footnote{ For instance, in the AR(1) model with $T=3$ and no covariates, this condition is satisfied if all the choices probabilities are bounded away from zero. In the AR(1) model with $T=2$ from Section \ref{section_AR1_T2}, the condition is satisfied if $\theta=\beta$ is bounded away from zero, and if certain degenerate distributions are ruled out for $\alpha_i$. For details, see the Additional Online Supplementary Material, which can be accessed \href{https://arxiv.org/abs/2104.04590}{here}.} For added flexibility, an alternative assumption, which can be used to replace Assumption \ref{assumption_consistency2}, is presented in Section \ref{section_alternative} of the Online Supplementary Material.

Under these additional assumptions, we have the following consistency result. 
\begin{theorem}\label{theorem_consistency}
Suppose Assumptions \ref{assumption_main}, \ref{assumption_consistency} and \ref{assumption_consistency2}  hold, let $0<\frac{3}{2}\beta_k<\beta_\tau<\frac{1}{2}-\beta_k$, and let $\tau_{n}\downarrow 0$, $l_n\uparrow \infty$, and $k_n\uparrow \infty$ be sequences satisfying $\tau_n = O(n^{-\beta_\tau})$, $l_n\asymp k_n$, and $l_n\leq k_{n} =O(n^{\beta_k})$. Also, let $\nu_n \asymp k_n^{1/2}$. Then for any $\varepsilon>0$:
\begin{align*}
\limsup_{n\to \infty} \sup_{P \in \mathcal{P}} \text{Pr}_{P}\left( d_{H}(\hat{\Theta}_{I,n}, \Theta_{I}(P), ||\,\cdot\,||) >\varepsilon\right)=0.
\end{align*}
\end{theorem}
Theorem \ref{theorem_consistency} shows that our estimate of the identified set, given by \eqref{eq_sample_identified_set_main}, converges to the true identified set in the Hausdorff distance uniformly over the class of DGPs $\mathcal{P}$ implicitly defined by Assumptions \ref{assumption_main}, \ref{assumption_consistency} and \ref{assumption_consistency2}. Consistency requires that the sequence $\tau_{n}$ in \eqref{eq_sample_identified_set_main} tends to zero sufficiently slowly relative to the sample size and the number of instrument functions. We provide guidance on all tuning parameters at the end of this section.

\begin{remark}\label{eq_choice_taun}
The parameter $\tau_{n}$ is required for Hausdorff consistency of the identified set. While Theorem \ref{theorem_consistency} is theoretically applicable in models that are either point- or partially-identified, $\tau_{n}$ can be set to zero in models that are known to be point-identified. If it is not known a priori whether the model is point- or partially-identified, the researcher should choose $\tau_n$ to satisfy the conditions in Theorem \ref{theorem_consistency} (that is, as if the model is partially identified): this choice ensures consistency under both point- and partial identification, whereas setting $\tau_n=0$ ensures consistency only under point identification.  
\end{remark}
\begin{remark}\label{remark_not_necessary}
Neither the assumptions above, our choice of instrument functions, or our sieve approximation based on piecewise constant functions is necessary for consistency. In Section \ref{appendix_additional_inference} of the Online Supplementary Material we state the minimal set of assumptions---adapted from the assumptions in CNS---required for both our estimation and inference procedure. Lemma \ref{lemma_consistency} then proves consistency of our estimator under this weaker set of assumptions, allowing for a number of alternative modelling choices. Also note that it is not necessary to have $l_n\uparrow \infty$ and $k_n\uparrow \infty$ if all covariates are discrete. 
\end{remark}

As mentioned previously, our estimate of the identified set can be computed efficiently using semidefinite programming. In particular, let $\mathcal{D}_{l_n}:=\{D_1,\ldots,D_{l_n}\}$ and $\mathcal{D}_{k_n}:=\{D_1',\ldots,D_{k_n}'\}$. Since $\bm r \in \mathcal{R}_{n}$ implies that $\bm r(\bm w) = \sum_{\ell =1}^{l_n} \bm \delta_{\ell}\cdot 1\{\bm w \in  D_{\ell}\}$ for some vector of coefficients $\{\bm \delta_{\ell}\}_{\ell=1}^{l_n}$, for each $j=1,\ldots,J,$ we have:
\begin{align}
&\frac{1}{n} \sum_{i=1}^{n} m_{j}(\bm Y_{i},\bm W_i,\theta,\bm r) \otimes \bm q^{k_n}(\bm W_{i}) \nonumber\\
&=
\begin{bmatrix}
\frac{1}{n}\sum_{i=1}^{n} 1\{\bm W_{i}\in D_{1}'\} \left(1\{\bm Y_{i} = \bm y_j\} - g_{j}(\bm W_{i},\theta)^\top \left(\sum_{\ell =1}^{l_n} \bm \delta_{\ell}\cdot 1\{\bm W_{i}\in  D_\ell\} \right)\right)\\
\frac{1}{n}\sum_{i=1}^{n} 1\{\bm W_{i}\in D_{2}'\} \left(1\{\bm Y_{i} = \bm y_j\} - g_{j}(\bm W_{i},\theta)^\top \left(\sum_{\ell =1}^{l_n} \bm \delta_{\ell}\cdot 1\{\bm W_{i}\in  D_\ell\} \right) \right)\\
\vdots\\
\frac{1}{n}\sum_{i=1}^{n} 1\{\bm W_{i}\in D_{k_n}'\} \left(1\{\bm Y_{i} = \bm y_j\} - g_{j}(\bm W_{i},\theta)^\top \left(\sum_{\ell =1}^{l_n} \bm \delta_{\ell}\cdot 1\{\bm W_{i}\in  D_\ell\} \right) \right)\\
\end{bmatrix}\nonumber\\[10pt]
&=\begin{bmatrix}
\frac{1}{n}\sum_{i=1}^{n} 1\{\bm W_{i}\in D_{1}'\} \left(1\{\bm Y_{i} = \bm y_j\} - \sum_{\ell=1}^{l_n} g_{j}(\bm W_{i},\theta)^\top  \bm \delta_{1,\ell}1\{\bm W_{i} \in D_{1}' \cap D_{\ell} \}\right)\\
\frac{1}{n}\sum_{i=1}^{n} 1\{\bm W_{i}\in D_{2}'\} \left(1\{\bm Y_{i} = \bm y_j\} - \sum_{\ell=1}^{l_n} g_{j}(\bm W_{i},\theta)^\top  \bm \delta_{2,\ell}1\{\bm W_{i} \in D_{2}' \cap D_{\ell} \}\right)\\
\vdots\\
\frac{1}{n}\sum_{i=1}^{n} 1\{\bm W_{i}\in D_{k_n}'\} \left(1\{\bm Y_{i} = \bm y_j\} - \sum_{\ell=1}^{l_n} g_{j}(\bm W_{i},\theta)^\top  \bm \delta_{k_n,\ell}1\{\bm W_{i} \in D_{k_n}' \cap D_{\ell} \}\right)
\end{bmatrix},\label{eq_sample_moments}
\end{align}
where the $\bm \delta_{k,\ell}$'s are auxiliary parameter satisfying the constraints $\bm  \delta_{1,\ell} = \ldots = \bm  \delta_{k_n,\ell}$ for $\ell = 1, \ldots, l_n$. Now the semidefinite constraints $\bm B_{m}(\bm r(\bm W_{i})) \in \mathcal{S}_+^{m+1}$ and $\bm H_{m}^*(\bm r(\bm W_{i}),\varsigma(\bm W_{i})) \in \mathcal{S}_+^{m+2}$ are equivalent to $\bm B_{m}(\bm \delta_{k,\ell}) \in \mathcal{S}_+^{m+1}$ for $\ell=1,\ldots,l_n$ and $\bm H_{m}^*(\bm \delta_{k,\ell},\varsigma_{k,\ell}) \in \mathcal{S}_+^{m+2}$ for $\ell=1,\ldots,l_n$ for some sequence of coefficients $\varsigma_{k,1},\ldots, \varsigma_{k,l_n}$. Let $\bm \zeta_{k} = (\zeta_{j,k})_{j=1}^{J}$ denote a vector for $k=1,\ldots, k_n$. Then for each $\theta \in \Theta$, for both continuous and discrete covariates minimizing $Q_{n}(\theta,\bm r)$ over $\bm r \in \Pi_{\mathcal{R}_n}(\mathcal{S})$ can be accomplished by solving the optimization problem: 
\begin{align}
&\qquad\qquad\qquad\qquad\qquad\qquad\qquad\min_{\zeta_0,\, \bm \zeta_{1},\, \ldots,\,\bm \zeta_{k_n},\, \bm \delta_{1,1},\,\ldots, \,\bm \delta_{k_n,l_n},\, \varsigma_{1,1},\ldots,\, \varsigma_{k_n,\ell_n}} \zeta_{0},\tag*{$\text{SDP}_0(\theta$)}\label{eq_SDP}\\
&\text{ subject to: } \nonumber\\
&\qquad\qquad(1) \quad \zeta_{0} \geq \left(\sum_{k=1}^{k_n} \sum_{j=1}^{J} \zeta_{jk}^2\right)^{1/2},\nonumber\\[10pt]
&\qquad\qquad(2)\quad \zeta_{jk} = \frac{1}{n}\sum_{i=1}^{n} 1\{\bm W_{i}\in D_{k}'\} \left(1\{\bm Y_{i} = \bm y_j\} - \sum_{\ell=1}^{l_n} g_{j}(\bm W_{i},\theta)^\top  \bm \delta_{k,\ell}1\{\bm W_{i} \in D_{\ell} \cap D_{k}' \} \right), \text{ $\forall j, k$,}\nonumber\\[10pt]
&\qquad\qquad(3)\quad \bm B_{m}(\bm \delta_{k,\ell}) \in \mathcal{S}_+^{m+1}\text{ and } \bm H_{m}^*(\bm \delta_{k,\ell},\varsigma_{k,\ell}) \in \mathcal{S}_+^{m+2}, \text{ $\forall k, \ell$},\nonumber\\[10pt]
&\qquad\qquad(4)\quad \bm  \delta_{1,\ell} = \ldots = \bm  \delta_{k_n,\ell},\text{ $\forall\ell$}.\nonumber
\end{align}
The constraints in $(1)$ and $(3)$ are semidefinite constraints, and the constraints in $(2)$ and $(4)$ are linear constraints. This ensures that the program \ref{eq_SDP} is a semidefinite program, which can be computed efficiently for each fixed $\theta \in \Theta$. Minimizing $Q_{n}(\theta,\bm r)$ over all $(\theta,\bm r) \in (\Theta \times \mathcal{R}_{n})\cap \mathcal{S}$ can then be accomplished by establishing a fine grid of evaluation points $\Theta^\dagger\subset \Theta$, solving \ref{eq_SDP} at each $\theta\in \Theta^\dagger$, and then choosing the minimizing pair $(\theta,\bm r) \in (\Theta^\dagger \times \mathcal{R}_{n})\cap \mathcal{S}$. An estimate of the identified set can then be obtained by collecting all points $\Theta^\dagger$ satisfying the condition in \eqref{eq_sample_identified_set_main}. This procedure is summarized in Algorithm \ref{alg_estimation_and_inference} at the end of the next subsection.

\subsection{Inference}

Building on the results of the previous subsection, in this section we propose a method of confidence set construction using hypothesis test inversion. In particular, define the following slightly revised set $\mathcal{S}(\vartheta)$ representing the shape restrictions:
\begin{align*}
\mathcal{S}(\vartheta) := 
\begin{cases}
	\left\{(\theta, \bm r) \in \Theta \times \mathcal{R} : \theta = \vartheta, \, \bm H_{m}^*(\bm r(\bm w),\varsigma^*(\bm w)) \succeq 0, \,\,\bm B_{m}(\bm r(\bm w)) \succeq 0, \forall \bm w \in \mathcal{W}  \right\}, &\text{ if $S=2m+1$,}\\
	\left\{(\theta, \bm r) \in \Theta \times \mathcal{R} : \theta = \vartheta, \,  \bm H_{m}(\bm r(\bm w)) \succeq 0, \,\,\bm B_{m}^*(\bm r(\bm w),\varsigma^*(\bm w)) \succeq 0, \forall \bm w \in \mathcal{W}  \right\}, &\text{ if $S=2m$.}
\end{cases}
\end{align*}
Note that $\mathcal{S}(\vartheta)$ is the same as $\mathcal{S}$, but also has the additional restrictions that $\theta = \vartheta$ for some vector $\vartheta \in \Theta$. To construct a confidence set for $\theta$, we then invert the following hypothesis test:
\begin{align}
H_{0}: \mathcal{E}(P) \cap \mathcal{S}(\vartheta) \neq \emptyset \text{ v.s. } H_{1}: \mathcal{E}(P) \cap \mathcal{S}(\vartheta) = \emptyset,\label{eq_null}
\end{align}
where:
\begin{align*}
\mathcal{E}(P) := \left\{ (\theta, \bm r) \in \Theta \times \mathcal{R} : E_{P}[\bm m(\bm Y_{i},\bm W_i,\theta,\bm r) \mid \bm W_{i}] = \bm 0 \text{ a.s.} \right\}. 
\end{align*}
That is, the null hypothesis in \eqref{eq_null} tests whether there exists an $\bm r \in \mathcal{R}$ that satisfies all the moment conditions and semidefinite constraints when $\theta = \vartheta$. This will be the case if and only if $\vartheta \in \Theta_{I}(P)$, so that \eqref{eq_null} is equivalent to testing if $\vartheta \in \Theta_{I}(P)$. Due to the shape constraints on $\bm r \in \mathcal{R}$, we require an inference procedure that is valid under shape constraints, and we use a modified version of a procedure proposed by CNS. In particular, to test the null hypothesis from \eqref{eq_null}, we propose the following test statistic:
\begin{align}
T_{n}(\vartheta) := \inf_{(\theta,\bm r) \in (\Theta \times \mathcal{R}_{n})\cap \mathcal{S}(\vartheta) } \sqrt{n} Q_{n}(\theta,\bm r), \label{eq_test_statistic}
\end{align}
where $Q_{n}(\theta,\bm r)$ is as in \eqref{eq_sample_Qn}. Our rejection decision is then based on comparing $T_{n}(\vartheta)$ to a critical value constructed using a multiplier bootstrap procedure. For i.i.d.\ $\{\xi_{i}^{b}\}_{i=1}^{n}$ with $\xi_{i}^{b}\sim N(0,1)$ independent of $\{(\bm Y_{i},\bm W_{i})\}_{i=1}^{n}$, define the multiplier bootstrap process:
\begin{align}
\mathbb{G}_{n}^{b}(\theta,\bm r) := \frac{1}{\sqrt{n}} \sum_{i=1}^{n} \xi_{i}^b \left\{\bm m(\bm Y_{i},\bm W_i,\theta,\bm r) \otimes \bm q^{k_n}(\bm W_{i}) - \frac{1}{n} \sum_{i=1}^{n} \bm m(\bm Y_{i},\bm W_i,\theta,\bm r) \otimes \bm q^{k_n}(\bm W_{i}) \right\}.\label{eq_Gnb}
\end{align}
Then our bootstrap test statistic is given by:
\begin{align}
T_{n}^{b}(\vartheta) := \inf_{(\vartheta,\bm r) \in \hat{\mathcal{I}}_{n}(\vartheta)} \left|\left| \mathbb{G}_{n}^{b}(\vartheta,\bm r) \right|\right|,\label{eq_bootstrap_test_stat}
\end{align}
where:
\begin{align}
\hat{\mathcal{I}}_{n}(\vartheta) := \left\{(\theta,\bm r) \in (\Theta \times \mathcal{R}_{n})\cap \mathcal{S}(\vartheta) : Q_{n}(\theta,\bm r) \leq \inf_{(\theta,\bm r) \in (\Theta \times \mathcal{R}_{n}) \cap \mathcal{S}(\vartheta)} Q_{n}(\theta,\bm r) + \tilde{\tau}_{n}\right\}, \label{eq_sample_identified_set_lambda}
\end{align}
for some sequence $\tilde{\tau}_n = o(1)$ satisfying $\tilde{\tau}_n\leq \tau_n$. At level $\alpha$, our rejection decision is based on whether $T_{n}(\vartheta)$ exceeds the $1-\alpha+\delta$ quantile of the bootstrap distribution of $T_{n}^{b}(\vartheta)$, where $\delta$ is some infinitesimal constant.\footnote{The inclusion of $\delta$ allows us to avoid high-level assumptions on the continuity of the asymptotic distribution of $T_{n}(\vartheta)$ under the null. \cite{andrews2013inference} recommend $\delta=10^{-6}$.} Similar to estimation, the test statistic and bootstrap test statistic can be computed by solving a semidefinite program, which is demonstrated at the end of this section.


To introduce our next result, we require one final technical assumption to replace Assumption \ref{assumption_consistency2}. In the following, let $\{\tilde{D}_k\}_{k=1}^{\tilde{k}_n}$ denote an enumeration of all \textit{nonempty} sets $D_{l}\cap D_{k}'$ where $D_{l} \in \mathcal{D}_{l_n}$ is any set used in the construction of the piecewise constant functions, and $D_{k}'\in \mathcal{D}_{k_n}$ is any set used in the construction the instruments. Now define:
\begin{align}
\bm b_{n,j}(\bm y, \bm w,\theta)^\top = (\bm b_{n,j}^{(1)}(\bm y, \bm w,\theta)^\top,\bm b_{n,j}^{(2)}(\bm y, \bm w,\theta)^\top),\label{eq_bnj}
\end{align}
which is a $(k_n+\tilde{k}_n)\times 1$ vector with components: 
\begin{align*}
\bm b_{n,j}^{(1)}(\bm y, \bm w,\theta)&:=\begin{bmatrix}
1\{\bm y = \bm y_{j}, \bm w \in D_{1}\}& \ldots &1\{\bm y = \bm y_{j}, \bm w \in D_{k_{n}}\}
\end{bmatrix}^\top,\\
\bm b_{n,j}^{(2)}(\bm y, \bm w,\theta)&:=
\begin{bmatrix}
- c_{0}(\bm y, \bm w,\theta)1\{\bm w \in  \tilde{D}_{1}\}&
\ldots& 
-c_{S}(\bm y, \bm w,\theta)1\{\bm w \in  \tilde{D}_{\tilde{k}_{n}}\}
\end{bmatrix}^\top.
\end{align*}
As illustrated at the end of Section \ref{section_consistency}, each moment function $m_{j}(\bm y, \bm w, \theta, \bm r)$ can be written as a linear combination of the elements of the vector $\bm b_{n,j}(\bm y, \bm w,\theta)$ when $\bm r(\bm w)$ is a piecewise constant function. The properties of this vector, and the properties of the instrument vector $\bm q^{k_n}(\bm w)$, play an important role in determining the rate of the bootstrap coupling results in CNS which are crucial for our inference procedure. Define the matrices:
\begin{align*}
M_{n,P}^{(1)}(\theta):=E_{P}\left[ \bm q^{k_n}(\bm W_{i}) \otimes  \bm G(\bm W_{i}, \theta)\otimes \bm d^{l_n}(\bm W_{i})^\top\right], &&M_{n,P}^{(2)}:=E_{P}[\bm d^{l_n}(\bm W_{i}) \bm d^{l_n}(\bm W_{i})^\top],
\end{align*}
\begin{align*}
M_{n,P,j}^{(3)}(\theta):=\text{Var}_{P}(\bm q^{k_n}(\bm W_{i}) \otimes \bm b_{n,j}(\bm Y_{i}, \bm W_{i}, \theta)),
\end{align*}
and let $\sigma_{max}(A)$ and $\sigma_{min}(A)$ denote the smallest and largest singular values of a matrix $A$, respectively. 
\begin{assumption}\label{assumption_inference}
$(i)$ There exists a positive constant $c_1>0$ such that, for all $n$ sufficiently large, we have $\sigma_{max}(M_{n,P}^{(2)}) \leq c_1 k_{n} \sigma_{min}(M_{n,P}^{(1)}(\theta))$ uniformly in $P \in \mathcal{P}$ and $\theta \in \Theta_{I}(P)$; $(ii)$ there exists a positive constant $c_2>0$ such that $\sup_{P' \in \mathcal{P}} ||M_{n,P'}^{(1)}(\theta) \bm \delta||\leq c_2 ||M_{n,P}^{(1)}(\theta)\bm \delta||$ for every $\bm \delta\in \mathbb{R}^{(S+1)l_n}$, $P \in \mathcal{P}$, and $\theta\in \Theta_I(P)$; $(iii)$ there exists positive constants $c_1', c_2'>0$ such that, for all $n$ sufficiently large, $\sigma_{max}(M_{n,P,j}^{(3)}(\theta)) \leq c_1'$ and $\sigma_{min}(M_{n,P,j}^{(3)}(\theta)) \geq c_2' \cdot k_n^{-1}$, uniformly in $P \in \mathcal{P}$ and $\theta \in \Theta_{I}(P)$, for each $j$. 
\end{assumption}
Assumption \ref{assumption_inference} replaces Assumption \ref{assumption_consistency2} for our next result. Although it is not immediately obvious, Assumption \ref{assumption_inference}$(i)$ is conceptually related to sieve ill-posedness. To see why Assumption \ref{assumption_inference}$(i)$ is plausible, note that if $l_n$ and $k_n$ are of the same order and the singular values of $E_{P}[\bm G(\bm W_{i}, \theta) \mid \bm W_{i} \in D_{l}\cap D_{k}']$ are bounded away from zero uniformly in $P\in \mathcal{P}$, then the singular values of $M_{n,P}^{(1)}(\theta)$ and $M_{n,P}^{(2)}$ may be reasonably expected to decay at a rate of $O(k_n^{-1})$. Assumption \ref{assumption_inference}$(i)$ comfortably allows for this kind of behaviour.\footnote{These sufficient conditions appear to rule out the AR(1) model with $T=2$ and no covariates from Section \ref{section_AR1_T2} when $\beta=0$, since in this case the $4\times 4$ matrix $\bm G(\bm W_{i}, \theta)= \bm G(y_0, \theta)$ is rank deficient. However, when $\beta=0$ this model becomes the static panel logit model with fixed effects, and this model has a different $4 \times 3$ matrix $\bm G(y_0, \theta)$. Assumption \ref{assumption_inference}$(i)$ applies when the researcher uses this alternative matrix when $\beta=0$. } Assumption \ref{assumption_inference}$(ii)$ is a technical condition that places additional constraints on the class of DGPs $\mathcal{P}$. It requires that the nullspace of the matrix $M_{n,P}^{(1)}(\theta)$ does not change with $P$. It is trivially satisfied when $\mathcal{P}=\{P\}$ (``pointwise asymptotics''), and admits other possible classes, but it can fail, for instance, for classes $\mathcal{P}$ where the rank of the matrix $M_{n,P}^{(1)}(\theta)$ changes with $P$.  Finally, Assumption \ref{assumption_inference}$(ii)$ requires that the singular values of $M_{n,P}^{(3)}(\theta)$ are bounded away from infinity, and that they do not decay too fast. Again, if $l_n$ and $k_n$ are of the same order, the singular values of $M_{n,P}^{(3)}(\theta)$ may be reasonably expected to be of the order $O(k_{n}^{-1})$, which is allowed by Assumption \ref{assumption_inference}$(ii)$. Note Assumption \ref{assumption_inference}$(ii)$ is not required for our approach, but allows us to obtain a faster rate of convergence in the CNS bootstrap coupling result needed in the proofs of our main results, and allows us to maintain the same rate requirements on the sequences $a_{n}$ and $\tau_{n}$ as in Theorem \ref{theorem_consistency}.\footnote{Similar assumptions are used in the leading application in CNS: see CNS Assumption 4.1 and 4.2.}  With Assumption \ref{assumption_inference} in hand, the following theorem provides the uniform validity of the testing procedure described above.

\begin{theorem}\label{theorem_validity}
Suppose Assumptions \ref{assumption_main}, \ref{assumption_consistency} and \ref{assumption_inference} hold, let $\tilde{\tau}_n\leq \tau_n$, and suppose that the sequences $l_n$, $k_n$, $\tau_n$, and $\nu_n$ satisfy the conditions in Theorem \ref{theorem_consistency}. Furthermore, for any $\delta>0$, let $\hat{q}_{1-\alpha+\delta}(\vartheta)$ denote the $1-\alpha+\delta$ quantile of the bootstrap distribution of $T_{n}^{b}(\vartheta)$. Then:
\begin{align*}
\limsup_{n\to \infty} \sup_{P\in \mathcal{P}} \sup_{\vartheta \in \Theta_{I}(P)} \text{Pr}_{P}(T_{n}(\vartheta) > \hat{q}_{1-\alpha+\delta}(\vartheta)+\delta) \leq \alpha. 
\end{align*}
\end{theorem}
Theorem \ref{theorem_validity} shows the validity of our proposed testing procedure, uniformly over the class of DGPs $\mathcal{P}$ implicitly determined by Assumptions \ref{assumption_main}, \ref{assumption_consistency} and \ref{assumption_inference}. Using Theorem \ref{theorem_validity}, confidence sets for $\theta$ can be constructed via hypothesis test inversion by collecting the parameter vectors $\vartheta \in \Theta$ for which we fail to reject the null hypothesis in \eqref{eq_null}. In particular, define:
\begin{align}
C_{n,\alpha} : = \left\{ \theta \in \Theta :  T_{n}(\theta) \leq \hat{q}_{1-\alpha+\delta}(\theta)+\delta \right\},\label{eq_CI}
\end{align}
where $\hat{q}_{1-\alpha+\delta}(\theta)$ is as in Theorem \ref{theorem_validity}. The following is a straightforward immediate consequence of the previous result.
\begin{corollary}\label{corollary_validity}
Suppose the assumptions of Theorem \ref{theorem_validity} hold. Then:
\begin{align*}
\liminf_{n\to \infty} \inf_{P\in \mathcal{P}} \inf_{\theta \in \Theta_{I}(P)}  \text{Pr}_{P}(\theta \in C_{n,\alpha}) \geq 1- \alpha. 
\end{align*}
\end{corollary}
Theorem \ref{theorem_validity} and Corollary \ref{corollary_validity} justify the testing and inference procedure described above. Combining our approximation based on piecewise constant functions with semidefinite programming provides a computationally efficient means of constructing confidence sets for structural parameters in the models we consider. 
\begin{remark}\label{remark_tau_zero}
When computing our bootstrap test statistic, we can always set $\tilde{\tau}_{n}=0$ (for both point- and partially identified models) in the set \eqref{eq_sample_identified_set_lambda}, although this may make our procedure more conservative. To see why, note that, all else constant, setting $\tilde{\tau}_n=0$ instead of $\tilde{\tau}_n>0$ can only make the bootstrap test statistic in \eqref{eq_bootstrap_test_stat} (and thus also the critical value) larger. However, following Remark \ref{eq_choice_taun}, a strictly positive sequence $\tau_n$ is still required for the sets \eqref{eq_sample_identified_set_main2} and \eqref{eq_sample_identified_set_main} in order for our set estimator to be consistent when the model is partially-identified. 
\end{remark}
\begin{remark}
Similar to Remark \ref{remark_not_necessary}, not all the assumptions in Theorem \ref{theorem_validity} are necessary. In Section \ref{section_inference_appendix_assumptions} of the Online Supplementary Material we state a minimal set of required assumptions, which are adapted from the assumptions in CNS. Lemma \ref{lemma_validity} then provides a proof of the uniform validity of our testing procedure under these weaker assumptions.
\end{remark}
\begin{remark}\label{remark_misspecification}
As noted in Section \ref{section_functional_differencing}, our approach can be used to detect model misspecification. Our inference procedure can also be used to formally test model misspecification as a by-product. In particular, a formal test of model misspecification at the $\alpha$ significance level can be performed by checking whether the confidence set in \eqref{eq_CI} is empty. See the relevant discussion of the ``by-product'' test in \cite{bugni2015specification} and \cite{marcoux2024simple}. 
\end{remark}

To use our inference procedure in practice, we require an efficient method of computing the test statistic $T_{n}(\vartheta)$ and the bootstrap test statistic $T_{n}^{b}(\vartheta)$. Note that computing the test statistic $T_{n}(\vartheta)$ from \eqref{eq_test_statistic} is equivalent to solving \ref{eq_SDP} at $\theta=\vartheta$ (up to a rescaling by $\sqrt{n}$), so that our previous discussion of \ref{eq_SDP} applies to $T_{n}(\vartheta)$. Computing $T_{n}^{b}(\vartheta)$ from \eqref{eq_bootstrap_test_stat} requires only a few small modifications to this procedure. First, the objective function for $T_{n}^{b}(\vartheta)$ is different than $T_{n}(\vartheta)$. However, if $\bm r(\bm w) = \sum_{\ell =1}^{l_n} \bm \delta_{\ell}\cdot 1\{\bm w \in  D_{\ell}\}$, some thought shows that \eqref{eq_Gnb} is also linear in the coefficients $\{\bm \delta_{\ell}\}_{\ell=1}^{l_n}$. This makes the objective function for $T_{n}^{b}(\vartheta)$ the norm of a linear function, similar to the objective function for $T_{n}(\vartheta)$. Most of the constraints required to solve \eqref{eq_bootstrap_test_stat} are also identical to those required to compute $T_{n}(\vartheta)$, with the exception that we must also impose the constraint:
\begin{align}
Q_{n}(\vartheta,\bm r) \leq \inf_{(\vartheta,\bm r) \in (\Theta \times \mathcal{R}_{n}) \cap \mathcal{S}} Q_{n}(\vartheta,\bm r) + \tau_{n}. \label{eq_sdp_additional constraint}
\end{align}
The value of the infimum on the right is obtained as a by-product of estimating the identified set. As a result, this constraint can be added to the program as an additional semidefinite constraint. Summarizing, $T_{n}^{b}(\vartheta)$ can be computed by solving the following optimization problem at $\theta=\vartheta$: 
\begin{align}
&\qquad\qquad\qquad\qquad\qquad\qquad\qquad\min_{\gamma_0, \bm \gamma_{1}, \ldots,\bm \gamma_{k_n}, \zeta_0, \bm \zeta_{1}, \ldots,\bm \zeta_{k_n}, \bm \delta_{1},\ldots, \bm \delta_{l_n},\varsigma_{01},\ldots,\varsigma_{0l_n}} \gamma_{0},\tag*{$\text{SDP}_0^{b}(\theta$)}\label{eq_SDP_boot}\\
&\text{ subject to: } \nonumber\\
&\qquad(1) \quad \gamma_{0} \geq \left(\sum_{k=1}^{k_n} \sum_{j=1}^{J} \gamma_{jk}^2\right)^{1/2},\nonumber\\[10pt]
&\qquad(2)\quad \gamma_{jk} = \frac{1}{n} \sum_{i=1}^{n} \xi_{i}^{b} \left\{m_j(\bm Y_{i},\bm W_i,\theta,\bm \delta)1\{\bm W_{i} \in D_{k}'\} - \frac{1}{n} \sum_{i=1}^{n} m_j(\bm Y_{i},\bm W_i,\theta,\bm \delta) 1\{\bm W_{i} \in D_{k}'\} \right\} 
 \text{ $\forall j,k$,}\nonumber\\[10pt]
&\qquad(3)\quad \bm B_{m}(\bm \delta_{\ell}) \in \mathcal{S}_+^{m+1}\text{ and } \bm H_{m}^*(\bm \delta_{\ell},\varsigma_{0\ell}) \in \mathcal{S}_+^{m+2}, \text{ $\forall \ell$},\nonumber\\[10pt]
&\qquad(4) \quad \tau_{n} + \inf_{\theta \in \Theta} \text{\ref{eq_SDP}}\geq \left(\sum_{k=1}^{k_n} \sum_{j=1}^{J} \zeta_{jk}^2\right)^{1/2},\nonumber\\[10pt]
&\qquad(5)\quad \zeta_{jk} = \frac{1}{n}\sum_{i=1}^{n} 1\{\bm W_{i}\in D_{k}'\} \left(1\{\bm Y_{i} = \bm y_j\} - \sum_{\ell=1}^{l_n} g_{j}(\bm W_{i},\theta)^\top  \bm \delta_{k,\ell}1\{\bm W_{i} \in D_{\ell} \cap D_{k}' \} \right), \text{ $\forall j, k$,}\nonumber\\[10pt]
&\qquad(6)\quad \bm  \delta_{1,\ell} = \ldots = \bm  \delta_{k_n,\ell},\text{ $\forall\ell$}.\nonumber
\end{align}
Note that constraints $(4)$ and $(5)$ enforce the constraint \eqref{eq_sdp_additional constraint}.  Also note that the constraints in $(1)$, $(3)$ and $(4)$ are semidefinite constraints, and the constraints in $(2)$, $(5)$ and $(6)$ are linear constraints. This ensures that the program \ref{eq_SDP_boot} is a semidefinite program. 

Finally, we note that our proposed bootstrap procedure can be simplified dramatically at the cost of a conservative distortion. In particular, optimization in \eqref{eq_bootstrap_test_stat} can be avoided entirely by ``recycling'' the optimal vectors $\bm r_1,\ldots, \bm r_{l_n}$ obtained when computing the test statistic by substituting these optimal solutions into the bootstrap test statistic \eqref{eq_bootstrap_test_stat} rather than re-optimizing. Inspecting \eqref{eq_bootstrap_test_stat} and \eqref{eq_sample_identified_set_lambda}, this makes our test more conservative, but can also dramatically improves computation time, allowing the researcher to trade-off between these two concerns. See \cite{marcoux2024simple} for a similar procedure. The practical performance of our inference procedure is illustrated in a brief Monte Carlo exercise in Section \ref{appendix_simulation} of the Online Supplementary Material. In these simulation exercises, and in the application in the next section, we make use of the computational simplifications that come with ``recycling'' the optimal vectors $\bm r_1,\ldots, \bm r_{l_n}$ obtained when computing the test statistic in the bootstrap procedure.

\begin{algorithm}[t]\footnotesize
\setstretch{1.35}
\caption{Estimation and inference (for $S$ odd)}\label{alg_estimation_and_inference}\vspace{0.1cm}
\textbf{Input:} A sample $\{(\bm Y_{i},\bm W_{i})\}_{i=1}^{n}$, nested partitions $\mathcal{D}_{l_n}$ (for the moment vector) and $\mathcal{D}_{k_n}$ (for the instruments), a finite grid $\Theta^\dagger \subset \Theta$, an oracle to solve \ref{eq_SDP}, an oracle to solve \ref{eq_SDP_boot}, and scalars $\alpha$, $\delta$, and $\tau_{n}$.  \\
\textbf{Output:} $\hat{\Theta}_{I,n}$ (identified set) and $C_{n,\alpha}$ (confidence set).
\begin{algorithmic}[1]
\ForEach{$\theta \in \Theta^\dagger$}
\State Solve SDP: $Q^*(\theta) \leftarrow$\text{val}(\ref{eq_SDP}).
\For{$b=1,\ldots,B$}
\State Draw $\{\xi_{i}^b\}_{i=1}^{n}\overset{i.i.d.}{\sim}N(0,1)$.
\State Solve SDP: $Q_{b}^*(\theta) \leftarrow$\text{val}(\ref{eq_SDP_boot}).
\EndFor 
\State Set $\hat{q}_{1-\alpha+\delta}(\theta)\leftarrow$ $1-\alpha+\delta$ quantile of $\{\sqrt{n} Q_{b}^*(\theta) \}_{b=1}^{B}$. 
\EndFor 
\State Set $\hat{\Theta}_{I,n} \leftarrow \{\theta \in \Theta^\dagger : Q^*(\theta) \leq \min_{\theta \in \Theta^\dagger} Q^*(\theta) +\tau_n \}$. \Comment{Identified Set}
\State Set $C_{n,\alpha} \leftarrow \{\theta \in \Theta^\dagger : \sqrt{n} Q^*(\theta) \leq  \hat{q}_{1-\alpha+\delta}(\theta) + \delta \}$. \Comment{$1-\alpha$ Confidence Set}
\State \Return $\hat{\Theta}_{I,n}$ and $C_{n,\alpha}$.
\end{algorithmic}
\end{algorithm}

Our entire estimation and inference procedure for the odd case is provided in Algorithm \ref{alg_estimation_and_inference}. A similar algorithm works for the even case by replacing the semidefinite constraints $\bm B_{m}(\bm \delta_{\ell}) \in \mathcal{S}_+^{m+1}\text{ and } \bm H_{m}^*(\bm \delta_{\ell},\varsigma_{0\ell}) \in \mathcal{S}_+^{m+2}$ in \ref{eq_SDP} and \ref{eq_SDP_boot} with $\bm H_{m}(\bm \delta_{\ell}) \in \mathcal{S}_+^{m+1}\text{ and } \bm B_{m}^*(\bm \delta_{\ell},\varsigma_{0\ell}) \in \mathcal{S}_+^{m+2}$.  In terms of tuning parameters, for both estimation and inference the values $\tau_n=0.001 n^{-3/10}$, $\tilde{\tau}_n=0$, $k_n=1+\lceil n^{1/6} \rceil$, and $l_n = k_n-1$ meet all the theoretical requirements, and worked well in both the application in the next section and the Monte Carlo exercises in Section \ref{appendix_simulation} of the Online Supplementary Material. Setting $\tilde{\tau}_{n}=0$ and recycling the optimal $\bm r_1,\ldots, \bm r_{l_n}$ from the test statistic allows us to save substantial computational time by avoiding the need to re-optimize \eqref{eq_bootstrap_test_stat} during the bootstrap. Researchers who are willing to trade increased computation time for increased testing power can instead set $\tilde{\tau}_n=\tau_n$ and repeatedly solve \eqref{eq_bootstrap_test_stat} when computing the bootstrap test statistic. Finally, similar to the existing literature (e.g. \cite{andrews2013inference}), the value of $\delta$ in our testing procedure does not play an important role in practice, and can be set arbitrarily small (e.g. $\delta=10^{-6}$).
\section{Application}\label{section_application}

In this section, we illustrate the proposed identification, estimation and inference procedure by applying it to data from the National Longitudinal Survey of Youth 1997 (NLSY97). The longitudinal surveys are sponsored by the United States Bureau of Labor Statistics with the aim of documenting labor market outcomes over a prolonged period of time. The first round of surveys began in 1997. Here, we use data from the years 2008 to 2010, which we label as periods $t=1,2,3,$ respectively. The outcome variable $Y_{it}$ is a binary variable representing an individual's employment status in a given year, and is equal to $1$ if the respondent worked more than $1000$ hours in year $t$.\footnote{Here we use the same variable definition as \cite{honore2024moment}, who also use the NLSY97 data.} The value $Y_{i0}$ is defined similarly using data from the year 2007. Throughout we consider various cases of the following AR(1) model:
\begin{align}
Y_{it} = 1\{\alpha_{i} + Y_{it-1}\beta + t \gamma+ X_{it}\eta \geq \epsilon_{it} \}, \,\,t=1,2,3,\label{eq_application_model}
\end{align}
where $X_{it}$ is the respondent's spouse's income in hundreds of thousands of US dollars, $\epsilon_{it}$ is i.i.d.\ standard logistic, and $\alpha_{i}$ is the latent individual effect that can be arbitrarily dependent with all other random variables except $\epsilon_{it}$. In particular, in models of labor market outcomes it is especially important to distinguish between the true effect of state dependence, measured by $\beta$, and the effects of persistent unobserved heterogeneity, captured by the individual-specific effect $\alpha_{i}$ (see \cite{card1987measuring}). We consider four specifications, labelled (S1) - (S4), which are based on the general model in \eqref{eq_application_model}:
\begin{enumerate}
	\item [(S1)] \textbf{AR(1), $\mathbf{T=3}$}: a model with only a lagged effect:
\begin{align*}
Y_{it} = 1\{\alpha_{i} + Y_{it-1}\beta  \geq \epsilon_{it} \}, \,\,t=1,2,3.
\end{align*}
	This is a special case of model \eqref{eq_application_model} that arises by setting $\gamma=\eta=0$. 
	\item [(S2)] \textbf{AR(1), $\mathbf{T=3}$, with covariates}: a model with a lagged effect and covariates:
\begin{align*}
Y_{it} = 1\{\alpha_{i} + Y_{it-1}\beta +X_{it}\eta \geq \epsilon_{it} \}, \,\,t=1,2,3.
\end{align*}
	This is a special case of model \eqref{eq_application_model} that arises by setting $\gamma=0$. 
	\item [(S3)] \textbf{AR(1), $\mathbf{T=3}$, with a time trend}: a model with a lagged effect and time trend:
\begin{align*}
Y_{it} = 1\{\alpha_{i} + Y_{it-1}\beta +t\gamma \geq \epsilon_{it} \}, \,\,t=1,2,3.
\end{align*}
	This is a special case of model \eqref{eq_application_model} that arises by setting $\eta=0$. 
	\item [(S4)] \textbf{AR(1), $\mathbf{T=3}$, with a time trend and covariates}: this model is exactly  model \eqref{eq_application_model}. 
\end{enumerate}

We drop all observations with missing data either on hours worked or spouse's income over the period we consider, which leaves $5097$ individuals for estimation. Since our procedure requires compactness of the support of the covariates, we winsorize spouse's income $X_{it}$ at one hundred thousand. Since spouses income is in hundreds of thousands, this ensures $X_{it} \in [0,1]$ for $t=1,2,3$. For the instrument functions, we interact indicators $1\{Y_{i0}=0\}$ and $1\{Y_{i0}=1\}$ with indicators of the form $1\{\max\{X_{i1},X_{i2},X_{i3}\} \in D_k\}$ where $D_{k} = \left(\frac{k-1}{k_n}, \frac{k}{k_n}\right]$ for $k=1,\ldots,k_{n} = 1+\lceil n^{1/6} \rceil = 6$. For the piecewise constant approximation to the moment vector, we use a similar partition, but with only $l_n = k_n-1 = 5$ subsets. Furthermore, since it is not known if the time trend model is point- or partially-identified, as per Remark \ref{remark_tau_zero} we treat specifications (S3) and (S4) as if they were partially identified, and take $\tau_n = 0.001n^{-3/10}$.  We then compare the results to estimates from functional differencing in models $(S1)$ and $(S2)$ using the procedure described in \cite{honore2024moment}, which we refer to as ``HW'' in the results. Since it is not known whether the models $(S3)$ and $(S4)$ are point identified, we did not apply functional differencing. We also compare the results of our method to a model where $\alpha_{i}=\alpha$ for all $i=1,\ldots,n$, which is estimated using maximum likelihood. We refer to this comparison model as ``Logit ML'' in the results. Finally, we also include results from a model that estimates all the $\alpha_{i}$ as fixed effects using maximum likelihood, which we call ``Logit ML FE.'' Note that estimates from this model are inconsistent due to the incidental parameters problem (e.g. \cite{andersen1973conditional}).

\begin{table}[t!]
\centering
\caption{Estimated lagged effects and time trend effects for various specifications of the AR(1) model with $T=3$ using the NLSY97 data. The table displays (point and set) estimates of $\beta$ and $\gamma$, and also includes $95\%$ confidence intervals displayed below the estimates. The ``DGKR'' results use the proposed estimation and inference procedure in this paper with $999$ bootstrap replications. The ``HW'' results use the functional differencing procedure proposed in \cite{honore2024moment}. The ``Logit ML'' results set $\alpha_{i} = \alpha$ for all individuals, and uses maximum likelihood for estimation. The ``Logit ML FE'' includes a fixed effect (dummy) variable for all individuals, and uses maximum likelihood for estimation, producing inconsistent estimates.  }\label{table_application_results}
\begin{footnotesize}
\begin{tabular}{lcccc}
\toprule
            & $(S1)$              & $(S2)$              & $(S3)$              & $(S4)$              \\
\midrule
\vspace{-3mm} \\
\multicolumn{1}{l}{\textbf{DGKR}}        &                      &                      &                      &                      \\[5pt]
Lagged Effect ($\hat{\beta}$)  & $1.63$     	& $1.44$    		& $[1.41,1.57]$  & $[1.21,1.43]$  \\
                               & $(1.16, 2.33)$	& $(0.86, 2.43)$ 	& $(0.94, 2.37)$     & $(0.70,2.49)$     \\[10pt]
Time Trend ($\hat{\gamma}$)     &$-$ &$-$  & $[-0.09,-0.06]$   & $[-0.10,-0.05]$                \\
          &          &     & $(-0.25, 0.11)$    & $(-0.28, 0.13)$    \\
\vspace{-3mm} \\
\midrule
\vspace{-3mm} \\
\multicolumn{1}{l}{\textbf{HW}}        &                      &                      &                      &                      \\[5pt]
Lagged Effect ($\hat{\beta}$)  & $1.83$     & $1.89$    & $-$   & $-$ \\
                               & $(1.55, 2.19)$& $(1.51, 2.73)$ &     &     \\[10pt]
Time Trend ($\hat{\gamma}$)     &$-$ &$-$  & $-$    & $-$    \\
          &          &     &     &    \\
\vspace{-3mm} \\
\midrule
\vspace{-3mm} \\
\multicolumn{1}{l}{\textbf{Logit ML}}    &                      &                      &                      &                      \\[5pt]
Lagged Effect ($\hat{\beta}$)      & $3.11$                 & $3.11$                 & $3.11$                 & $3.11$                \\
                                & $(3.04, 3.18)$    & $(3.02, 3.20)$      & $(3.02, 3.20)$      & $(3.02, 3.20)$      \\[10pt]
Time Trend ($\hat{\gamma}$)                      &           $-$           &          $-$            & $-0.03$                & $-0.03$            \\
                                &                      &                      & $(-0.09, 0.02)$    & $(-0.09, 0.02)$    \\
\vspace{-3mm} \\
\midrule
\vspace{-3mm} \\
\multicolumn{1}{l}{\textbf{Logit FE ML}} &                      &                      &                      &                      \\[5pt]
Lagged Effect ($\hat{\beta}$)                & $-0.67$                & $-0.68$              & $-0.84$                & $-0.84$                \\
                                & $(-0.84, -0.5)$    & $(-0.85, -0.51)$   & $(-1.02, -0.66)$   & $(-1.02, -0.66)$   \\[10pt]
Time Trend ($\hat{\gamma}$)                       &          $-$            &          $-$            & $-0.39$                & $-0.38$                \\
           &                      &                      & $(-0.47, -0.3)$    & $(-0.47, -0.29)$\\\vspace{-3mm}\\\bottomrule  
\end{tabular}
\end{footnotesize}
\end{table}

The results are displayed in Table \ref{table_application_results}, which includes the (point and set) estimates of $\beta$ and $\gamma$, as well as $95\%$ confidence intervals displayed below the estimates. The results obtained using the methods developed in this paper are displayed under the heading ``DGKR.'' Across all specifications, we find that the effect of a lagged outcome is positive and significant at the $5\%$ level, indicating a strong and positive effect of the previous period's employment on future employment.  We find the effect of the time trend to be negative and insignificant. Our estimates in models $(S1)$ and $(S2)$ are also similar to those obtained using the method of \cite{honore2024moment}.  Interestingly, the qualitative conclusions from our approach agree with the conclusions of the benchmark ``Logit ML'' that constrains $\alpha_{i}=\alpha$ for $i=1,\ldots,n$. However, without properly accounting for the effects of individual-specific permanent unobserved heterogeneity, the results of this model suggest a state-dependence effect that is approximately twice as large. Consistent with our results, the Logit ML model suggests the time effect is small in magnitude and insignificant. Finally, the table also displays the ``Logit FE ML'' estimates which come from estimating all fixed effects using maximum likelihood. Due to the incidental parameters problem, all estimates in this model are inconsistent. Unlike the previous models, this model delivers estimates of state dependence of employment that are negative and significant, contrary to intuition. Furthermore, unlike the previous methods, this method produces estimates of the effect of the time trend that is negative and significant. These unintuitive but highly significant results serve as a warning against this model, and motivation for using estimation methods that are consistent in the presence of latent individual effects like the one developed in this paper.

\section{Conclusion}\label{section_conclusion}

This paper presents a new characterization of the identified set for structural parameters and functionals of the latent variables in a large class of dynamic panel
logit models. We do so by relating the problem of identification in these models to the truncated moment problem from the mathematics literature, which asks when a sequence of numbers can be rationalized as the moments of a Radon measure. In the case of structural parameters, we use this connection to show that the identified set can be characterized by a collection of conditional moment equalities subject to a certain set of shape restrictions on the model parameters.  In addition to providing a general approach to identification, our procedure delivers the sharp identified set even in cases where previous methods fail. Building on the results of \cite{chernozhukov2023constrained}, we present estimation and inference procedures that use semidefinite programming methods, are applicable with continuous or discrete covariates, and can be used if the model is point- or partially-identified. We also illustrate the usefulness of our results using a series of examples, and in an application to employment dynamics using data from the National Longitudinal Survey of Youth.

Although we did not pursue it here, our method might also be extended to accommodate environments where the initial outcome is unobserved, as in \cite{honore2006bounds}. The connection to the truncated moment problem also clearly extends beyond logit models (e.g. \cite{heckman1990testing}, \cite{d2017measuring}), and there also exists a class of models with multidimensional fixed effects which we believe can also be connected to the truncated moment problem. These include multinomial panel logit models, and bivariate models involving choices made by multiple interacting individuals (e.g. \cite{honore2019panel}, \cite{andaureo2021identification}, and \cite{AGM2021}). We therefore believe these tools will be useful in studying identification in a variety of other models.

\bibliographystyle{econometrica}
\bibliography{bibfile}

\clearpage
\appendix

\renewcommand{\thesection}{S.\arabic{section}}
\setcounter{section}{0}
\supplementtitle{Online Supplementary Material for ``Identification of Dynamic Panel Logit Models with Fixed Effects''}
\addcontentsline{toc}{section}{Online Supplementary Material}

\section{Proofs}\label{appendix_proofs}

Given two measures $\mu$ and $\nu$ on a measurable space $(X,\mathcal{A})$, we say that $\mu$ and $\nu$ are equivalent, denoted $\mu\sim \nu$, if $\mu \ll \nu$ and $\nu \ll \mu$. The following Lemmas will be useful in the proof of Theorem \ref{theorem_main}. We refer to \cite{bogachev_measure_theory_ii} p.179 for a discussion of both results.
\begin{lemma}\label{lemma_fact1}
 Let $\mu$ and $\nu$ be two finite and nonnegative measures on a measurable space $(X,\mathcal{A})$. Suppose $\nu \ll \mu$. Then $\nu \sim \mu$ if and only if  $d\nu/d\mu>0$ $\mu-$a.e.
\end{lemma}
\begin{lemma}\label{lemma_fact2}
Let $\mu_{1}$, $\mu_{2}$, and $\mu_{3}$ be three finite measures on a measurable space $(X,\mathcal{A})$ such that $\mu_1 \ll \mu_{2}$, and $\mu_{2}\ll \mu_3$. Then $\mu_{1}\ll \mu_{3}$ and $\frac{d\mu_{1}}{d\mu_{3}} = \frac{d\mu_{1}}{d\mu_{2}}\frac{d\mu_{2}}{d\mu_{3}}.$
In particular, if $\mu_{1}=\mu_3$ and $d\mu_{1}/d\mu_{2} >0$ $\mu_{2}-$a.e., then $\frac{d\mu_{2}}{d\mu_{1}} = \left(\frac{d\mu_{1}}{d\mu_{2}}\right)^{-1}$, $\mu_{1}$-a.e. (and also $\mu_{2}$-a.e.).
\end{lemma}

\begin{proof}[Proof of Theorem \ref{theorem_AR1_T2}]
See Corollary \ref{corollary_main} and Lemma 2.3 in \cite{curto1991recursiveness}. 
\end{proof}

\begin{proof}[Proof of Theorem \ref{theorem_main}]
Fix $\theta \in \Theta_{I}(P)$. Under Assumption \ref{assumption_main} we have: 
\begin{align*}
f(\bm y \mid \bm w, \alpha; \theta)  = \bm G(\bm w, \theta)\left(\begin{bmatrix} 1 & \exp(\alpha) & \ldots & \exp(\alpha)^S\end{bmatrix}\right)^\top \kappa(\bm w, \alpha,\theta).
\end{align*}
By Definition \ref{definition_identified_set} there exists a conditional distribution $Q_{\alpha \mid \bm W}$ for $\alpha$ given $\bm W$ satisfying: 
\begin{align}
\bm p (\bm w) = \bm G(\bm w, \theta) \int \begin{bmatrix} 1 & \exp(\alpha) & \dots & \exp(\alpha)^S\end{bmatrix}^\top \kappa(\bm w, \alpha,\theta) \,dQ_{\alpha \mid \bm W}(\alpha \mid \bm w),\label{eq_p_Gr}
\end{align}
almost surely, with the integral interpreted element-wise. Now define the $j^{th}$ entry of the vector $\bm r(\bm w)$ to be $r_{j}(\bm w) := \int \exp(\alpha)^{j-1} \kappa(\bm w, \alpha,\theta) \, dQ_{\alpha \mid \bm W}(\alpha \mid \bm w)$. By definition of a conditional distribution, $\bm w \mapsto Q_{\alpha \mid \bm W}(\,\cdot\, \mid \bm w)$ is measurable. Combined with Assumption \ref{assumption_main} and the integrability of $\exp(\alpha)^{j-1} \kappa(\bm w, \alpha,\theta)$ for each $\bm w \in \mathcal{W}$, the function $\bm w \mapsto r_{j}(\bm w)$ is measurable. Furthermore, by definition of a conditional distribution, for almost all $\bm w \in \mathcal{W}$ we have $B \mapsto Q_{\alpha \mid \bm W}(B \mid \bm w)$ is a probability measure. Since $\kappa(\alpha, \bm w,\theta)>0$ is bounded (and thus $Q_{\alpha \mid \bm W}-$integrable), we have that:
\begin{align*}
\bar{Q}_{\alpha \mid \bm W}(E \mid \bm w) := \int_{E}  \kappa(\alpha, \bm w,\theta) dQ_{\alpha \mid \bm W}(\alpha \mid \bm w),
\end{align*}
defines a finite nonnegative Borel (and thus, Radon) measure satisfying:
\begin{align*}
r_{j}(\bm w) = \int \exp(\alpha)^{j-1} \,d\bar{Q}_{\alpha \mid \bm W}(\alpha \mid \bm w),
\end{align*}
for $j=1,\ldots,S+1,$ so that $\bm r(\bm w) \in \mathcal{M}_{S}$ $P_{\bm W}-$almost surely. Thus, \eqref{eq_p_Gr} implies $\bm r(\bm w) \in \mathcal{M}_{S}$ and $\bm p(\bm w) = \bm G(\bm w, \theta) \bm r(\bm w)$, $P_{\bm W}-$almost surely.

For the opposite direction, fix $\theta \in \Theta$ and suppose $\bm r:\mathcal{W}\to \mathbb{R}^{S+1}$ is a measurable function satisfying $\bm r(\bm w) \in \mathcal{M}_S$ and $\bm{p}(\bm w) = \bm G(\bm w, \theta) \bm r(\bm w)$, $P_{\bm W}-$a.s. We will show that there exists a conditional distribution $Q_{\alpha \mid \bm W}$ satisfying:
\begin{align}
p(\bm y \mid \bm w)=  \int f(\bm y \mid \bm w, \alpha; \theta) dQ_{\alpha \mid \bm W}(\alpha \mid \bm w),\label{eq_sufficiency}
\end{align}
$P_{\bm W}-$a.s.\ for all $\bm y \in \mathcal{Y}^T$. Fix some $\bm w \in \mathcal{W}$ such that $\bm r(\bm w) \in \mathcal{M}_{S}$. By definition of $\mathcal{M}_{S}$, there exists a nonnegative Radon measure $B\mapsto \bar{Q}_{\alpha \mid \bm W}(B \mid \bm w)$ such that $r_j(\bm w) = \int \exp(\alpha)^{j-1} d\bar{Q}_{\alpha \mid \bm W}(\alpha \mid \bm w)$ for $j=1,\ldots,S+1$. Setting $A=\exp(\alpha)$, by definition of $\kappa(\alpha,\bm w,\theta)$ from Assumption \ref{assumption_main} we have:
\begin{align*}
1 =\kappa(\alpha,\bm w,\theta) \bm 1^\top \bm G(\bm w, \theta) \begin{bmatrix} 1 & A & \ldots & A^{S} \end{bmatrix}^\top \implies \frac{1}{\kappa(\alpha,\bm w,\theta)} = \bm 1^\top \bm G(\bm w, \theta) \begin{bmatrix} 1 & A & \ldots & A^{S} \end{bmatrix}^\top.
\end{align*}
Thus:
\begin{align}
\int \frac{1}{\kappa(\alpha,\bm w,\theta)} \,d\bar{Q}_{\alpha \mid \bm W} = \int \bm 1^\top \bm G(\bm w, \theta) \begin{bmatrix} 1 & A & \ldots & A^{S} \end{bmatrix}^\top \,d\bar{Q}_{\alpha \mid \bm W} = \bm 1^\top \bm G(\bm w, \theta) \bm r(\bm w) = 1, \label{eq_Q_prob_meas}
\end{align}
where we used the fact that $\bm{p}(\bm w) = \bm G(\bm w, \theta) \bm r(\bm w)$, $P_{\bm W}-$a.s. Now define:
\begin{align*}
Q_{\alpha \mid \bm W}(E \mid \bm w) := \int_{E} \left(\frac{1}{\kappa(\alpha,\bm w,\theta)}\right) \,d\bar{Q}_{\alpha \mid \bm W}(\alpha \mid \bm w).
\end{align*}
Then by \eqref{eq_Q_prob_meas}, $Q_{\alpha \mid \bm W}$ is a probability measure and $Q_{\alpha \mid \bm W}(\,\cdot\, \mid \bm w) \ll \bar{Q}_{\alpha \mid \bm W}(\,\cdot\, \mid \bm w)$. Thus, by the Radon-Nikodym Theorem we have $\left(dQ_{\alpha \mid \bm W}/d\bar{Q}_{\alpha \mid \bm W}\right)(\alpha, \bm w) = 1/\kappa(\alpha,\bm w,\theta)$ $\bar{Q}_{\alpha\mid \bm W}-$almost everywhere for almost every $\bm w \in \mathcal{W}$. Since $(dQ_{\alpha \mid \bm W}/d\bar{Q}_{\alpha \mid \bm W})(\,\cdot\,,\bm w) >0$ almost everywhere, we have $\bar{Q}_{\alpha \mid \bm W}(\,\cdot\, \mid \bm w) \sim Q_{\alpha \mid \bm W}(\,\cdot\, \mid \bm w)$ for almost every $\bm w \in \mathcal{W}$ by Lemma \ref{lemma_fact1}. Thus, Lemma \ref{lemma_fact2} implies:
\begin{align*}
&\left(\frac{d\bar{Q}_{\alpha \mid \bm W}}{dQ_{\alpha \mid \bm W}}\right)(\alpha, \bm w)=\left(\left(\frac{dQ_{\alpha \mid \bm W}}{d\bar{Q}_{\alpha \mid \bm W}}\right)(\alpha, \bm w)\right)^{-1} = \kappa(\alpha,\bm w, \theta),
\end{align*}
$\bar{Q}_{\alpha \mid \bm W}-$almost everywhere for almost every $\bm w \in \mathcal{W}$. Finally note that for $j=1,\ldots,S+1$:
\begin{align*}
p(\bm y_{j} \mid \bm w) =\bm g(\bm y_j, \bm w, \theta)^\top \bm r(\bm w)&= \sum_{s=0}^{S} g_{s}(\bm y_j, \bm w, \theta) \int\exp(\alpha)^s \,d \bar{Q}_{\alpha \mid \bm W}(\alpha \mid \bm w)\\
&= \sum_{s=0}^{S} g_{s}(\bm y_j, \bm w, \theta) \int \exp(\alpha)^s \left(\left(\frac{d\bar{Q}_{\alpha \mid \bm W}}{dQ_{\alpha \mid \bm W}}\right)(\alpha,\bm w) \right)\,d Q_{\alpha \mid \bm W}(\alpha \mid \bm w)\\
&=  \int \sum_{s=0}^{S} g_{s}(\bm y_j, \bm w, \theta) \exp(\alpha)^s \kappa(\alpha,\bm w, \theta) \,d Q_{\alpha \mid \bm W}(\alpha \mid \bm w)\\
&= \int f(\bm y_j \mid \bm w, \alpha ;\theta) \,d Q_{\alpha \mid \bm W}(\alpha \mid \bm w),
\end{align*}
for $P_{\bm W}-$almost all $\bm w\in \mathcal{W}$. Let $\bm f(\bm w, \alpha ;\theta)$ denote the likelihood stacked across $\bm y \in \mathcal{Y}^{T}$, and let $\mathcal{P}_{\alpha}$ denote the set of Borel probability measures on $\mathbb{R}$. Modifying $\bm p(\bm w)$ and $Q_{\alpha \mid \bm W}(\alpha \mid \bm w)$ on a null set if necessary, the derivation above shows that:
\begin{align*}
\mathcal{P}_{\bm f}(\bm w) := \left\{ \nu \in \mathcal{P}_{\alpha} : \bm p(\bm w) = \int \bm f(\bm w,\alpha; \theta)\,d\nu(\alpha) \right\},
\end{align*}
is nonempty for every $\bm w \in \mathcal{W}$, where here the integral is interpreted elementwise. Now define:
\begin{align*}
\bm \Upsilon (\bm w, \nu) := \bm p(\bm w) - \int \bm f(\bm w,\alpha; \theta)\,d\nu(\alpha).
\end{align*}
Under Assumption \ref{assumption_main}, each element of $\bm f(\bm w,\alpha; \theta)$ is a bounded and continuous function of $\alpha$. Thus $\nu\mapsto \bm \Upsilon (\bm w, \nu)$ is continuous and $\mathcal{P}_{\bm f}(\bm w)$ is closed in the weak$^*$ topology for every $\bm w\in \mathcal{W}$. Furthermore, since each element of $\bm f(\bm w,\alpha; \theta)$ is measurable in $\bm w$ and continuous in $\alpha$, it is jointly measurable in $(\bm w, \alpha)$ (see \cite{aliprantis2006infinite} Lemma 4.51). Since $\bm p(\bm w)$ is measurable by definition, we have $\bm w\mapsto \bm \Upsilon (\bm w, \nu)$ is measurable in $\bm w$ for every $\nu$ (see \cite{bogachev_measure_theory_i} Corollary 3.4.6). Conclude that $(\bm w, \nu)\mapsto \bm \Upsilon (\bm w, \nu)$ is jointly measurable (see \cite{aliprantis2006infinite} Lemma 4.51), and thus $\text{Graph}(\mathcal{P}_{\bm f} ) = \bm \Upsilon^{-1}(\{\bm 0\})$ is a Borel set. By the Jankov-von Neumann Selection Theorem, there exists a universally measurable selector $Q(\,\cdot \mid \bm w) \in \mathcal{P}_{\bm f}(\bm w)$ (see \cite{bogachev_measure_theory_ii} Theorem 6.9.2). Furthermore, there exists a Borel measurable $Q^*(\,\cdot\, \mid \bm w)$ such that $Q(\cdot \mid \bm w) = Q^*(\cdot\mid \bm w)$ $P_{\bm W}-$almost surely (see \cite{cohn2013measure} Proposition 2.2.5). Conclude that \eqref{eq_sufficiency} holds for $Q^*$, so that $Q^*\in \mathcal{Q}(\theta)$ and thus $\theta \in \Theta_{I}(P)$. 
\end{proof}

	\begin{proof}[Proof of Theorem \ref{theorem_hankel}]
		For any $l\times n$ matrix $A$, define $\text{Range}(A) = \{Au : u \in \R^n\}$. Let $\bm{H}_m(\bm r) = (r_{i+j})_{i,j=0}^m$, $\bm{B}_m(\bm r) = (r_{i+j+1})_{i,j=0}^m$ and $\bm r(m+1,m)= (r_{m+1}, \dots, r_{2m+1})$. Theorem 5.1 in \cite{curto1991recursiveness} shows that if $S = 2m+1$, then $\bm r \in \mathcal{M}_{2m+1}$ if and only if $\bm H_m(\bm r) \succeq 0$, $\bm B_m(\bm r) \succeq 0$, and $\bm r(m+1,m)$ is in $\text{Range}(\bm{H}_m(\bm{r}))$. Furthermore, Theorem 5.3 in \cite{curto1991recursiveness} shows that if $S = 2m$, then $\bm r \in \mathcal{M}_{2m}$ if and only if $\bm{H}_{m}(\bm r) \succeq 0$, $\bm{B}_{m-1}(\bm r) \succeq 0$, and $\bm r(m+1, m-1)$ is in $\text{Range}(\bm{B}_{m-1}(\bm{r}))$. Thus, it suffices to prove: (i) for $S = 2m+1$, $\bm{H}_{m}(\bm r) \succeq 0$ and $\bm r(m+1,m)$ is in $\text{Range}(\bm{H}_m(\bm{r}))$ if and only if there exists a value of $\varsigma \geq 0$ such that $\bm{H}^*_{m}(\bm r,\varsigma) \succeq 0$; and (ii) for $S = 2m$, $\bm{B}_{m-1}(\bm r)\succeq 0$ and $\bm r(m+1,m-1)$ is in $\text{Range}(\bm{B}_{m-1}(\bm{r}))$ if and only if there exists a value of $\varsigma\geq 0$ such that $\bm{B}_m^*(\bm r,\varsigma)\succeq 0$. We focus on proving (i) since the proof of (ii) is similar. First suppose there exists a $\varsigma\geq 0$ such that $\bm H_{m}^*(\bm r,\varsigma) \succeq 0$. By Lemma 2.3(i) in \cite{curto1991recursiveness}, this implies that $\bm H_{m}(\bm r) \succeq 0$ and $\bm r(m+1,m) \in \text{Range}(\bm H_m(\bm r))$. Now suppose $\bm H_{m}(\bm r)\succeq 0$ and $\bm r(m+1,m) \in \text{Range}(\bm H_m(\bm r))$. Then there exists a $\bm v \in \mathbb{R}^{m+1}$ such that $\bm r(m+1,m) = \bm H_m(\bm r) \bm v$. Now pick any $\varsigma \geq \bm v^\top \bm H_m(\bm r) \bm v$. Then by Lemma 2.3(ii) in \cite{curto1991recursiveness}, this implies that $\bm H_{m}^*(\bm r,\varsigma) \succeq 0$. 
	\end{proof}

	\begin{proof}[Proof of Corollary \ref{corollary_main}]
This follows from combining Theorem \ref{theorem_main} with Theorem \ref{theorem_hankel}.
\end{proof}

\begin{proof}[Proof of Proposition \ref{proposition_functional_point_identification}]
If $\bm G(\bm w,\theta_0)$ has full column rank, then:
\begin{align*}
\bm r(\bm w) = (\bm G(\bm w,\theta_0)^\top \bm G(\bm w,\theta_0))^{-1}\bm G(\bm w,\theta_0)^\top \bm p(\bm w), 
\end{align*}
so that $\bm r(\bm w)$ is point-identified. Since $\Psi(\bm w,\theta_0) = \bm \eta(\bm w,\theta_0)^\top \bm r(\bm w)$, conclude that $\Psi(\bm w,\theta_0)$ is point-identified.
\end{proof}

	\begin{proof}[Proof of Theorem \ref{theorem_connectionFD}]
Fix $\theta\in \Theta$ throughout. Let $\alpha_0,\ldots \alpha_S \in \mathbb{R}$ be any distinct set of points, and consider the $(S+1) \times (S+1)$ matrix: 
\[
V(\bm w,\theta) := \begin{bmatrix} \kappa(\bm w, \alpha_0, \theta) &  \exp(\alpha_0) \kappa(\bm w, \alpha_0, \theta) & \ldots &
\exp(\alpha_0)^S \kappa(\bm w, \alpha_0, \theta)\\
\kappa(\bm w, \alpha_1, \theta) &  \exp(\alpha_1) \kappa(\bm w, \alpha_1, \theta) & \ldots &
\exp(\alpha_1)^S \kappa(\bm w, \alpha_1, \theta)\\
\vdots & \vdots & \ddots & \vdots\\
\kappa(\bm w, \alpha_S, \theta) &  \exp(\alpha_S) \kappa(\bm w, \alpha_S, \theta) & \ldots &
\exp(\alpha_S)^S \kappa(\bm w, \alpha_S, \theta)
\end{bmatrix}. 
\]
Then the determinant of $V(\bm w,\theta)$ is:
\begin{align*}
\text{det}(V(\bm w,\theta)) =\left(\prod_{j=0}^S \kappa(\bm w, \alpha_j, \theta)\right) \prod_{0 \leq s < s' \leq S} (\exp(\alpha_{s'})-\exp(\alpha_s)).
\end{align*}
Since $\kappa(\bm w, \alpha, \theta)>0$, and since the set of points $\alpha_0,\ldots, \alpha_S$ are distinct, conclude that $\text{det}(V(\bm w,\theta)) \neq 0$. Furthermore, note that the same set of distinct points $\alpha_0,\ldots, \alpha_S$ can be used to obtain the same result for every pair $(\bm w,\theta)$. With this result in hand, define:
\begin{align*}
\bm q(\bm w, \alpha,\theta) := \begin{bmatrix} \kappa(\bm w, \alpha, \theta) & \exp(\alpha) \kappa(\bm w, \alpha; \theta) & \dots & \exp(\alpha)^S \kappa(\bm w, \alpha; \theta) \end{bmatrix}^\top.
\end{align*}
Under Assumption \ref{assumption_main}, we have $\bm{f}(\bm w, \alpha; \theta) = \bm G(\bm w, \theta) \bm q(\bm w,\alpha,\theta)$. Since every function $\bm h(\,\cdot\,,\theta) \in \bm D(\theta)$ satisfies $\bm h(\bm w,\theta)^\top  \bm{f}(\bm w, \alpha; \theta)=0$ $\forall (\bm w,\alpha)$, we have $\bm h(\bm w,\theta)^\top  \bm G(\bm w, \theta) \bm q(\bm w,\alpha,\theta)=0$. We also have:
\begin{align}
\bm h(\bm w,\theta)^\top  \bm G(\bm w, \theta) \bm q(\bm w,\alpha,\theta)=0 \quad \forall \alpha
&\implies \bm q(\bm w,\alpha_{s},\theta)^\top\bm G(\bm w, \theta)^\top \bm h(\bm w,\theta) =0, \,\,s=1,\ldots,S,\nonumber\\
&\implies V(\bm w,\theta) \bm G(\bm w, \theta)^\top \bm h(\bm w,\theta) = \bm 0,\label{eq_Vmat}
\end{align}
for every $\bm w \in \mathcal{W}$. Since the columns of $V(\bm w,\theta)$ are linearly independent for every $\bm w \in \mathcal{W}$ (since $\text{det}(V(\bm w,\theta)) \neq 0$), \eqref{eq_Vmat} can hold only if $\bm h(\bm w,\theta)^\top \bm G(\bm w, \theta) = \bm 0$ for every $\bm w \in \mathcal{W}$. Conclude that $\bm h(\,\cdot\,,\theta) \in \bm M(\theta)$. Now note that every $\bm v(\,\cdot\,,\theta) \in \bm M(\theta)$ satisfies $\bm v(\bm w,\theta)^\top \bm G(\bm w, \theta) = \bm 0$ for every $\bm w \in \mathcal{W}$, so that we must trivially also have $\bm v(\bm w,\theta)^\top \bm G(\bm w, \theta) \bm q(\bm w,\alpha,\theta)= \bm 0$ for every $(\bm w,\alpha) \in \mathcal{W} \times \mathbb{R}$. Since  $\bm{f}(\bm w, \alpha; \theta) = \bm G(\bm w, \theta) \bm q(\bm w,\alpha,\theta)$ under Assumption \ref{assumption_main}, this implies that $\bm v(\,\cdot\,,\theta) \in \bm D(\theta)$. 
\end{proof}

\begin{proof}[Proof of Theorem \ref{theorem_consistency}]
Theorem \ref{theorem_assumption_verification} shows that, under the assumed rate requirements on $\tau_{n}$, $l_{n}$ and $k_{n}$, Assumptions \ref{assumption_main} and \ref{assumption_consistency} imply Assumption \ref{assumption3.1_DGKR}, Assumption \ref{assumption3.2_DGKR} with $J_{n} = O(\sqrt{l_{n}})$ and $B_{n}=O(1)$, and Assumption \ref{assumption3.6_DGKR}$(ii)$ for any nonzero sequence $a_n=o((\log(n))^{-\beta_a})$ for $\beta_a>0$ from Assumption \ref{assumption_consistency}. Furthermore, the stated rate requirements on $\tau_n$, $l_n$, $k_n$, and $\nu_n$ satisfy \eqref{eq_taun}, and Assumption \ref{assumption_consistency2} implies condition \eqref{eq_weak_pm} in the statement of Lemma \ref{lemma_consistency}. The result follows from Lemma \ref{lemma_consistency}.
\end{proof}

\begin{proof}[Proof of Theorem \ref{theorem_validity}]
By Theorem \ref{theorem_assumption_verification}, under the assumed rate requirements on $l_n$, $k_{n}$, $\tau_{n}$, and $\nu_n$, Assumptions \ref{assumption_main}, \ref{assumption_consistency} and \ref{assumption_inference} imply Assumptions \ref{assumption3.1_DGKR} - \ref{assumption3.12_DGKR}. Thus, the result follows from Lemma \ref{lemma_validity}.  
\end{proof}
\begin{proof}[Proof of Corollary \ref{corollary_validity}]
Note that $\text{Pr}_{P}(\theta \in C_{n,\alpha}) = 1 - \text{Pr}_{P}(\theta \notin C_{n,\alpha}) =1 - \text{Pr}_{P}(T_{n}(\theta) > \hat{q}_{1-\alpha+\delta}(\theta)+\delta)$. The result then follows immediately from Theorem \ref{theorem_validity}.
\end{proof}

\section{Comparison to Davezies, D’Haultf{\oe}uille, and Laage (2021)} \label{app: DDLcompare}

In Section \ref{section_functionals} we discussed bounding average marginal effects in the panel static logit model $Y_{it} = 1\{\alpha_i + X_{it}\gamma \geq \epsilon_{it}\}$. This parameter was also considered in \cite{davezies2021identification}. We now provide a direct comparison in the case when $T = 2$ with a single covariate. The quantity of interest is the average marginal effect of a change in $X_{i2}$ evaluated at the value $X_{i2}=x_2$: 
\[
\Psi(\gamma_0)  = \gamma_0 E[\Lambda(\alpha + \gamma_0 x_2)(1-\Lambda(\alpha + \gamma_0 x_2))].
\]
This parameter is the average of the conditional marginal effect: 
\[
\Psi(\bm x,\gamma_0) = \gamma_0 E_{Q_{\alpha|\bm X}} [ \Lambda(\alpha + \gamma_0 x_2)(1-\Lambda(\alpha + \gamma_0 x_2))\mid \bm X = \bm x].
\]
Since $\gamma_0$ is point-identified (e.g. see \cite{rasch1960probabilistic}), we treat it as known. Now consider a fixed $\bm x \in \mathcal{X}$ and define $u := \Lambda(\alpha + \gamma_0 x_2)$. Then any conditional distribution $Q_{\alpha \mid \bm X}$ induces a corresponding distribution $Q_{u\mid \bm X}$ supported on $[0,1]$. By a change of variable, we have: 
\[
\Psi(\bm x,\gamma_0) = \gamma_0 E_{Q_{u \mid \bm X}}[u(1-u) \mid \bm X = \bm x]. 
\]
Expressed in terms of this new variable, the static logit model implies:
\begin{align*}
p((0,0) \mid \bm x) & = E_{Q_{u \mid \bm X}}\left[\left.\frac{(1-u)^2}{1-u + u\exp((x_1-x_2)\gamma_0)}\,\right|\,\bm X = \bm x\right] = E_{\bar Q_{u \mid \bm X}}[(1-u)^2 \mid \bm X = \bm x],\\
p((1,0) \mid \bm x) & = E_{Q_{u \mid \bm X}}\left[\left.\frac{u(1-u)\exp((x_1-x_2)\gamma_0)}{1-u + u\exp((x_1-x_2)\gamma_0)}\,\right|\,\bm X = \bm x\right] = E_{\bar Q_{u \mid \bm X}}[u(1-u)\exp((x_1-x_2)\gamma_0)\mid\bm X = \bm x],\\
p((0,1) \mid \bm x) & = E_{Q_{u \mid \bm X}}\left[\left.\frac{u(1-u)}{1-u + u\exp((x_1-x_2)\gamma_0)}\,\right|\,\bm X = \bm x\right] = E_{\bar Q_{u \mid \bm X}}[u(1-u) \mid \bm X = \bm x],\\
p((1,1) \mid \bm x) & = E_{Q_{u \mid \bm X}}\left[\left.\frac{u^2 \exp((x_1-x_2)\gamma_0)}{1-u + u\exp((x_1-x_2)\gamma_0)}\,\right|\, \bm X = \bm x\right] = E_{\bar Q_{u \mid \bm X}}[u^2 \exp((x_1-x_2)\gamma_0) \mid \bm X = \bm x],
\end{align*}
where we have defined $\bar Q_{u | \bm x} (E):= \int_{E}\frac{1}{1-u + u\exp((x_1-x_2)\gamma_0)} dQ_{u | \bm x}$ for any Borel set $E \subseteq [0,1]$. Although $\bar Q_{u \mid \bm X}$ is not-point identified, the choice probabilities point-identify moments of $\bar Q_{u \mid \bm X}$ up to order 2. In particular, let $\bm p(\bm x)$ denote the vector of choice probabilities and let:
\begin{align*}
\bm G(\bm x,\theta) = \begin{bmatrix} 1 &-2 &1 \\ 0 &\exp(\gamma_0(x_1-x_2)) & - \exp(\gamma_0(x_1-x_2)) \\ 0 & 1 & -1 \\ 0 & 0 & \exp(\gamma_0(x_1-x_2))\end{bmatrix}, &&\bm r(\bm x) = \begin{bmatrix} \int_0^1 d\bar Q_{u | \bm x} (u)\\ \int_0^1 u d\bar Q_{u|\bm x}(u) \\ \int_0^1 u^2 d\bar Q_{u|\bm x}(u) \end{bmatrix},
\end{align*}
where $\bm G(\bm x,\theta)$ is of full column rank. Then using the fact that $\bm p(\bm x) = \bm G(\bm x,\theta) \bm r(\bm x)$, we have $\bm r(\bm x)=(\bm G(\bm x,\theta)^\top \bm G(\bm x,\theta))^{-1} \bm G(\bm x,\theta)^\top \bm p(\bm x)$. Now the parameter of interest can be expressed as:
\[
\Psi(\bm x,\gamma_0) = \gamma_0 E_{\bar Q_{u \mid \bm X}}[u(1-u) (1-u + u \exp((x_1-x_2) \gamma_0)) \mid \bm X = \bm x],
\]
which involves the third order moment of $\bar Q_{u \mid \bm X}$. \cite{davezies2021identification} then make use of an \emph{extremal moment problem} result (see \cite{kreuin1977markov}) which provides closed-form bounds for the third-order moment of $\bar Q_{u \mid \bm X}$ once its moments up to order two are known. In contrast, our approach works with a change of variable from $\alpha$ to $\exp(\alpha)$. The choice probabilities are linked to moments of $\exp(\alpha)$ up to order 3 (see the details in the example in Section \ref{section_functionals}). We then show the quantity of interest $\Psi(\bm x,\gamma_0)$ is a linear combination of these moments. However, even though $\gamma_0$ is point identified, these moments of $\exp(\alpha)$ are not point-identified due to the rank deficiency of the corresponding matrix $\bm G(\bm x,\theta)$. Thus, $\Psi(\bm x,\gamma_0)$ remains partially-identified, and we provide a semidefinite programming approach to construct the identified set. Both approaches give the same sharp identified set for $\Psi(\bm x,\gamma_0)$, and hence also for the unconditional average marginal effect $\Psi(\gamma_0)$. 

\section{Additional Results for Consistency and Inference}\label{appendix_additional_inference}

In this section, we provide additional discussion and results to support the main consistency result (Theorem \ref{theorem_consistency}) and inference results (Theorem \ref{theorem_validity} and Corollary \ref{corollary_validity}). In Section \ref{section_inference_appendix_assumptions} we provide a set of weaker assumptions that are sufficient for our results which are adapted from the assumptions of \cite{chernozhukov2023constrained} (CNS hereafter). Using our reduced set of assumptions, in Sections \ref{section_additional_consistency} and Sections \ref{section_additional_inference} we show that versions of Lemma S.1.1, Theorem 3.1$(i)$, Theorem 3.2, and Corollary 3.1 in CNS continue to hold in our setting. These results are used in the proofs of Theorem \ref{theorem_consistency}, Theorem \ref{theorem_validity} and Corollary \ref{corollary_validity}. Finally, in Section \ref{section_verification} we verify the assumptions in Section \ref{section_inference_appendix_assumptions} using the assumptions stated in the main text (namely, Assumptions \ref{assumption_consistency} and \ref{assumption_inference}). 
In the proofs, we use the notation $\text{Pr}_{P}(\,\cdot\,)$ to denote the probability taken with respect to the sampling (or $n-$fold product) distribution. The coupling results in the proofs require changing the underlying probability space, but this is suppressed in the notation for simplicity.


\subsection{Assumptions}\label{section_inference_appendix_assumptions}

In this section we provide a set of weaker assumptions that are sufficient for our consistency result (Theorem \ref{theorem_consistency}) and our inference results (Theorem \ref{theorem_validity} and Corollary \ref{corollary_validity}). These assumptions are adapted from the assumptions of CNS. To begin, define the set:
\begin{align}
\mathcal{S}^* := \left\{(\theta, \bm r) \in \Theta \times \mathcal{R} : \Upsilon_{\bm F}(\theta,\bm r) = \bm 0 \text{ and }  \Upsilon_{\bm G}(\theta,\bm r) \leq \bm 0\right\}.\label{eq_S_CNS}
\end{align}
This set is similar to the set ``$R$'' defined in display $(13)$ in CNS, and can be used to impose shape restrictions on the parameters $(\theta, \bm r) \in \Theta \times \mathcal{R}$. In the main text, we set $\mathcal{S}^* = \mathcal{S}$ (for consistency) or $\mathcal{S}^* = \mathcal{S}(\vartheta)$ for some $\vartheta \in \Theta$ (for testing and inference). Our main shape constraint is positive semidefiniteness of the Hankel matrices (and their extensions). Since positive semidefiniteness of a matrix can be enforced by imposing nonnegativity of all its principal minors, the shape constraints in both $\mathcal{S}$ and $\mathcal{S}(\vartheta)$ can be written as in \eqref{eq_S_CNS}. The following is similar to CNS Assumption 3.1. 
\begin{assumption}\label{assumption3.1_DGKR}
$(i)$ $\{(\bm Y_{i},\bm W_{i}) \}_{i=1}^{n}$ is i.i.d. with $(\bm Y_{i},\bm W_{i})\sim P \in \mathcal{P}$; (ii) $\Theta\times \mathcal{R} \subseteq \bm B$, where $(\bm B, ||\,\cdot\,||_{\bm B})$ is a Banach space; (iii) in the case when $\mathcal{S}^*=\mathcal{S}$, $\Upsilon_{\bm F}:\bm B \to \bm F$ and $\Upsilon_{\bm G}: \bm B \to \bm G$, where $(\bm F, ||\,\cdot\,||_{\bm F})$ and $(\bm G, ||\,\cdot\,||_{\bm G})$ are Banach spaces. 
\end{assumption}
\begin{remark}
For computational reasons, our procedure does not make use of the linearization and local parameter space in CNS. As a result, we do not require that $(\bm G, ||\,\cdot\,||_{\bm G})$ is an AM space.\footnote{See Appendix A.1 in CNS for the definition of an AM space.}
\end{remark}
\noindent Recall the moment function $m_{j}(\bm Y_{i},\bm W_{i},\theta,\bm r)$ from \eqref{eq_mom_func}. Let $(\mathcal{R},||\,\cdot\,||_{\mathcal{R}})$ be a subset of a Banach space, and let $\mathcal{R}_{n} \subset \mathcal{R}$. Now define:
\begin{align}
\mathcal{F}_{n} := \left\{m_{j}(\,\cdot\,,\theta,\bm r) : (\theta, \bm r)\in (\Theta \times \mathcal{R}_{n}) \cap \mathcal{S}^* \text{ and }1\leq j \leq J  \right\}.\label{eq_Fn}
\end{align}
Note that $\mathcal{F}_{n}$ implicitly depends on the shape restrictions imposed by the set $\mathcal{S}^*$, although this dependence is suppressed for simplicity. Now define the bracketing integral:
\begin{align*}
J_{[\,\,]}(\delta,\mathcal{F}_{n},||\,\cdot\,||_{P,2}) := \int_{0}^\delta
\sqrt{1+ \log N_{[\,]}(\epsilon,\mathcal{F}_{n},||\,\cdot\,||_{P,2})}\,d\epsilon.
\end{align*}
The following assumption is similar to CNS Assumption 3.2.
\begin{assumption}\label{assumption3.2_DGKR}
(i) $\max_{1\leq k \leq k_{n}} ||q_{k}||_{\infty} \leq B_{n}$ with $B_{n}\geq 1$; (ii) in the case when $\mathcal{S}^* = \mathcal{S}$, the class $\mathcal{F}_{n}$ has an envelope $F_{n}$ satisfying $\sup_{P \in \mathcal{P}} ||F_{n}||_{P,2} <\infty$, and $\sup_{P \in \mathcal{P}} J_{[\,\,]}(||F_{n}||_{P,2},\mathcal{F}_{n},||\,\cdot\,||_{P,2}) \leq J_{n}$ with $J_{n}<\infty$. 
\end{assumption}
\begin{remark}
Since Assumption \ref{assumption3.2_DGKR}$(ii)$ holds for $\mathcal{S}^*=\mathcal{S}$, it also holds for $\mathcal{S}^* = \mathcal{S}(\vartheta_{n})$ for any sequence $\vartheta_{n} \in \Theta_{I}(P)$. This will be useful for the results ahead. CNS Assumption 3.2$(ii)$ is not required in our context. 
\end{remark}
Define the process:
\begin{align*}
\mathbb{G}_{n}(\theta, \bm r) := \frac{1}{\sqrt{n}} \sum_{i=1}^{n} \left\{\bm m(\bm Y_{i},\bm W_{i},\theta,\bm r) \otimes \bm q^{k_n}(\bm W_{i}) - E_{P}[\bm m(\bm Y_{i},\bm W_{i},\theta,\bm r)  \otimes \bm q^{k_n}(\bm W_{i})] \right\}. 
\end{align*}
The following assumption is similar to CNS Assumption 3.3, and is required to hold only for $\mathcal{S}^* = \mathcal{S}(\vartheta_{n})$ for any sequence $\vartheta_{n} \in \Theta_{I}(P)$. 
\begin{assumption}\label{assumption3.3_DGKR}
For any sequence $\vartheta_{n} \in \Theta_{I}(P)$: (i) $\sup_{(\theta,\bm r) \in (\Theta \times \mathcal{R}_{n})\cap \mathcal{S}(\vartheta_{n})} ||\mathbb{G}_{n}(\theta,\bm r) - \mathbb{G}_{P}(\theta,\bm r)|| = o_{P}(a_n)$ uniformly in $P \in \mathcal{P}$ for some $a_{n} = o(1)$ and Gaussian $\mathbb{G}_{P}$ (possibly depending on $n$) satisfying $E_{P}[\mathbb{G}_{P}(\theta,\bm r)]=0$ and $\text{Cov}_{P}(\mathbb{G}_{P}(\theta,\bm r),\mathbb{G}_{P}(\theta',\bm r')) = \text{Cov}_{P}(\mathbb{G}_{n}(\theta,\bm r),\mathbb{G}_{n}(\theta',\bm r'))$; (ii) there is a norm $||\,\cdot\,||_{\bm E}$, $\kappa_{m}>0$, and $K_{m}<\infty$ such that $E_{P}[||\bm m(\bm Y_{i},\bm W_{i},\theta,\bm r) - \bm m(\bm Y_{i},\bm W_{i},\theta',\bm r')||^{2}]\leq K_{m}^2 ||(\theta,\bm r) - (\theta',\bm r')||_{\bm E}^{2 \kappa_{m}}$, for all $(\theta,\bm r), (\theta',\bm r') \in (\Theta \times \mathcal{R}_{n})\cap \mathcal{S}(\vartheta_{n})$ and $P \in \mathcal{P}$. 
\end{assumption}

For the next assumption, let $\vec{d}_{H}(A,B,||\,\cdot\,||_{\bm E}) := \sup_{a\in A} \inf_{b \in B} ||a-b||_{\bm E}$ denote the directed Hausdorff distance, where $||\,\cdot\,||_{\bm E}$ is the norm from Assumption \ref{assumption3.3_DGKR}. Following CNS equation (15), for each element $(\theta,\bm r)\in \Theta\times \mathcal{R}$ let $\Pi_{n}(\theta,\bm r)$ denote its approximation on $\Theta\times \mathcal{R}_{n}$, and define:
\begin{align}
\mathcal{I}_{n}^*(P) &:= \left\{ \Pi_{n}(\theta,\bm r) : (\theta,\bm r)\in\mathcal{I}^*(P) \right\},&&\mathcal{I}_{n}^*(\vartheta,P) := \left\{ \Pi_{n}(\theta,\bm r) :(\theta,\bm r)\in\mathcal{I}^*(\vartheta,P)\right\},\label{eq_Isets}
\end{align}
for any $\vartheta \in \Theta_{I}(P)$. Here $\mathcal{I}^*(\vartheta,P)$ is the section of $\mathcal{I}^*(P)$ restricted to $\theta=\vartheta$. The following assumption is unchanged from CNS Assumption 3.4, but is required to hold only for $\mathcal{S}^* = \mathcal{S}(\vartheta_{n})$ for any sequence $\vartheta_{n} \in \Theta_{I}(P)$. 
\begin{assumption}\label{assumption3.4_DGKR}
For any sequence $\vartheta_{n} \in \Theta_{I}(P)$, there is a sequence of sets $\mathcal{V}_{n}(P)\subseteq (\Theta \times \mathcal{R}_{n})\cap \mathcal{S}(\vartheta_{n})$ and a sequence of constants $0<\nu_n^{-1}=O(1)$ such that: (i) for any $(\theta,\bm r) \in \mathcal{V}_{n}(P)$, it holds that:
\begin{align*}
\nu_n^{-1} \vec{d}_{H}\left((\theta,\bm r), \mathcal{I}_{n}^*(\vartheta_{n},P),||\,\cdot\,||_{\bm E}\right)\leq \sup_{(\tilde{\theta},\tilde{\bm r})\in \mathcal{I}_{n}^*(\vartheta_{n},P)} \left|\left|E_{P}\left[ (\bm m(\bm Y_{i},\bm W_{i},\theta,\bm r) -\bm m(\bm Y_{i},\bm W_{i},\tilde{\theta},\tilde{\bm r})) \otimes \bm q^{k_n}(\bm W_{i}) \right] \right| \right|,
\end{align*} 
for all $n$ sufficiently large, and (ii) there is a $(\hat{\theta}_{n},\hat{\bm r}_{n}) \in \mathcal{V}_{n}(P)$ satisfying $Q_{n}(\hat{\theta}_{n},\hat{\bm r}_{n})\leq$ $\inf_{(\theta,\bm r) \in (\Theta \times \mathcal{R}_{n})\cap \mathcal{S}(\vartheta_{n})}$ $ Q_{n}(\theta,\bm r) +$$ o(a_{n}/\sqrt{n})$, uniformly in $P\in \mathcal{P}$. 
\end{assumption}
For the next assumption, define:
\begin{align}
R_{n} := \nu_n J_n B_{n} \sqrt{\frac{k_n\log(1+k_n)}{n}}.\label{eq_capital_Rn}
\end{align}
Here $J_{n}$ and $B_{n}$ are the sequences from Assumption \ref{assumption3.2_DGKR}, and $\nu_n$ is the sequence from Assumption \ref{assumption3.4_DGKR}. The following assumption is similar to Assumption 3.6 in CNS. 
\begin{assumption}\label{assumption3.6_DGKR}
For $\mathcal{S}^*=\mathcal{S}$:$(i)$ $\sqrt{k_{n}\log(1+k_n)} B_n \sup_{P\in \mathcal{P}} J_{[\,\,]}(R_{n}\vee \tau_{n})^{\kappa_{m}},\mathcal{F}_{n},||\,\cdot\,||_{P,2}) = o(a_n)$; $(ii)$	$\sup_{P \in \mathcal{P}} \sup_{(\theta,\bm r) \in \mathcal{I}_{n}^*(P)} \sqrt{n} ||E_{P}[\bm m(\bm Y_{i},\bm W_{i},\theta, \bm r)\otimes \bm q^{k_{n}}(\bm W_{i})|| = o(a_n).$
\end{assumption}
For the next assumption, recall the multiplier bootstrap process $\mathbb{G}_{n}^{b}(\theta,\bm r)$ from \eqref{eq_Gnb} in the main text. The following assumption is similar to Assumption 3.11 in CNS, and is required to hold only for $\mathcal{S}^* = \mathcal{S}(\vartheta_{n})$ along any sequence $\vartheta_{n} \in \Theta_{I}(P)$. 
\begin{assumption}\label{assumption3.11_DGKR}
For any sequence $\vartheta_{n} \in \Theta_{I}(P)$ we have $\sup_{(\theta,\bm r) \in (\Theta \times \mathcal{R})\cap \mathcal{S}(\vartheta_{n})} ||\mathbb{G}_{n}^{b}(\theta,\bm r) - \mathbb{G}_{P}^\star(\theta,\bm r)|| = o_{P}(a_n)$ uniformly in $\Phi\times P$ with $P \in \mathcal{P}$ for $\Phi$ the standard normal distribution, $a_n = o(1)$, and Gaussian $\mathbb{G}_{P}^\star$ (possibly depending on $n$) independent of $\{(\bm Y_{i}, \bm W_{i})\}_{i=1}^{n}$ and having the same distribution as $\mathbb{G}_{P}$. 
\end{assumption}

Define the set:
\begin{align}
\hat{\mathcal{I}}_{n}(\vartheta) := \left\{(\theta,\bm r) \in \Theta \times \mathcal{R}_{n} : Q_{n}(\theta,\bm r) \leq \inf_{(\theta,\bm r) \in (\Theta \times \mathcal{R}_{n}) \cap \mathcal{S}(\vartheta)} Q_{n}(\theta,\bm r) + \tau_{n}\right\}. \label{eq_sample_identified_set}
\end{align}
The following assumption is identical to Assumption CNS Assumption 3.12$(iii)$. CNS Assumption 3.12$(i)$ and 3.12$(ii)$ are not required in our context.
\begin{assumption}\label{assumption3.12_DGKR}
For any sequence $\vartheta_{n} \in \Theta_{I}(P)$ and for the corresponding $\mathcal{V}_{n}(P)$ as in Assumption \ref{assumption3.4_DGKR}, $\text{Pr}_{P}(\hat{\mathcal{I}}_{n}(\vartheta_{n}) \subseteq \mathcal{V}_{n}(P))$ tends to $1$ uniformly in $P \in \mathcal{P}$. 
\end{assumption}

\begin{remark}
Again, our procedure differs from the procedure proposed in CNS; namely, we do not studentize the moment conditions, and we do not make use of the local parameter space. As a result, CNS Assumptions 3.5, 3.7, 3.8, 3.9, 3.10, 3.13, and various components of the other assumptions in CNS are not required in our context. 
\end{remark}

\subsection{Additional Consistency Results}\label{section_additional_consistency}

Recall $Q_{n,P}(\theta,\bm r)$ from \eqref{eq_QP}. Furthermore, let $\Pi_{\mathcal{R}_n}(\mathcal{S}) = \left\{\bm r \in \mathcal{R}_n : \exists \theta \in \Theta \text{ s.t. }(\theta,\bm r) \in \mathcal{S} \right\}$.

\begin{lemma}\label{lemma_consistency}
Suppose Assumptions \ref{assumption3.1_DGKR}, \ref{assumption3.2_DGKR}, and \ref{assumption3.6_DGKR}$(ii)$ hold. Furthermore, suppose that $\exists \delta>0$ and a sequences $0<\nu_n^{-1} =O(1)$ and $b_{n} = o(\nu_n^{-1})$ such that:
\begin{align}
\nu_n^{-1} \min\{\delta, d(\theta,\Theta_{I}(P))\} \leq \inf_{\bm r \in \Pi_{\mathcal{R}_n}(\mathcal{S}) } Q_{n,P}(\theta,\bm r) - \inf_{(\theta',\bm r') \in (\Theta\times \mathcal{R}_n)\cap \mathcal{S}} Q_{n,P}(\theta',\bm r') +b_n,\label{eq_weak_pm}
\end{align}
for every $\theta\in \Theta$ and $P \in \mathcal{P}$. Finally, suppose $\Theta \subset \mathbb{R}^{d_\theta}$ is compact and that $\tau_n$ satisfies:
\begin{align}
\tau_{n} \nu_{n} = o(1), \qquad J_n B_{n} \sqrt{\frac{k_n\log(1+k_n)}{n}} = o(\tau_n), \quad \frac{a_n}{\sqrt{n}} = O(\tau_n).\label{eq_taun}
\end{align}
Then for any $\varepsilon>0$:
\begin{align*}
\limsup_{n\to \infty} \sup_{P \in \mathcal{P}} \text{Pr}_{P}\left( d_{H}(\hat{\Theta}_{I,n}, \Theta_{I}(P)) >\varepsilon\right)=0.
\end{align*}
\end{lemma}
\begin{proof}[Proof of Lemma \ref{lemma_consistency}]
We follow a proof similar to the proof of Lemma S.1.1 in CNS. First note that:
\begin{align}
d_{H}(\hat{\Theta}_{I,n}, \Theta_{I}(P)) = \max \left\{\sup_{\theta \in \hat{\Theta}_{I,n}} d(\theta,\Theta_{I}(P)) , \sup_{\theta \in \Theta_{I}(P)} d(\theta,\hat{\Theta}_{I,n})  \right\}. \label{eq_consistency0}
\end{align}
Now define the set $\mathcal{E}(P,\varepsilon) := \{\theta \in \Theta : d(\theta,\Theta_{I}(P)) \leq \varepsilon \}$. Then: 
\begin{align*}
\text{Pr}_{P} \left(\sup_{\theta \in \hat{\Theta}_{I,n}} d(\theta,\Theta_{I}(P)) > \varepsilon  \right) \leq \text{Pr}_{P} \left( \inf_{\theta \in \Theta\setminus\mathcal{E}(P,\varepsilon) } \inf_{\bm r \in \Pi_{\mathcal{R}_n}(\mathcal{S}) }  Q_{n}(\theta,\bm r)  \leq \inf_{(\theta,\bm r) \in (\Theta \times \mathcal{R}_{n})\cap \mathcal{S} }  Q_{n}(\theta,\bm r) + \tau_{n}\right). 
\end{align*}
Now note that Assumption \ref{assumption3.1_DGKR} implies Assumption 3.1$(i)$ in CNS, and Assumption \ref{assumption3.2_DGKR} implies Assumption 3.2$(i)(iii)$ in CNS. Thus, following a nearly identical proof as the proof of Lemma S.1.2 in CNS, we obtain that there exists random variables $A_{n}$ and $A_{n}'$ satisfying:
\begin{align*}
&\inf_{\theta \in \Theta\setminus\mathcal{E}(P,\varepsilon) } \inf_{\bm r \in \Pi_{\mathcal{R}_n}(\mathcal{S}) }  Q_{n,P}(\theta,\bm r)  - \inf_{\theta \in \Theta\setminus\mathcal{E}(P,\varepsilon) } \inf_{\bm r \in \Pi_{\mathcal{R}_n}(\mathcal{S}) }  Q_{n}(\theta,\bm r) \leq A_{n} = O_{P}(R_{n}'),\\
\text{ and }\quad&\inf_{(\theta,\bm r) \in (\Theta \times \mathcal{R}_{n})\cap \mathcal{S} }  Q_{n}(\theta,\bm r) - \inf_{(\theta,\bm r) \in (\Theta \times \mathcal{R}_{n})\cap \mathcal{S}}  Q_{n,P}(\theta,\bm r) \leq A_{n}' = O_{P}(R_{n}'),
\end{align*}
where $R_{n}':= J_n B_{n} \sqrt{\frac{k_n\log(1+k_n)}{n}}$, and where $A_{n}=O_{P}(R_{n}')$ and $A_{n}'=O_{P}(R_{n}')$ both hold uniformly in $P\in\mathcal{P}$.\footnote{Note our ``$R_n'$'' is the same as CNS's ``$\eta_{n}$.''} Now for any fixed $\gamma \in (0,1)$, let $N_{\gamma1}$ and $M_{\gamma}$ be large enough so that:
\begin{align*}
\sup_{P \in \mathcal{P}}\text{Pr}_{P} \left(|A_{n}/R_{n}'| \leq M_{\gamma}, |A_{n}'/R_{n}'| \leq M_{\gamma} \right) \geq 1-\gamma,
\end{align*}
for all $n\geq N_{\gamma1}$. Furthermore, note that $\nu_n \tau_{n}=o(1)$, $\nu_n R_n' = o(1)$ and $\nu_n b_n = o(1)$ under the conditions in the statement of the theorem. Now let $N_{\gamma2}$ be large enough so that $\nu_n^{-1} \min\{\delta, \varepsilon\} > 2R_{n}' M_{\gamma}+ b_{n} + \tau_{n}$ for all $n\geq N_{\gamma2}$. Then using \eqref{eq_weak_pm}, for $n\geq \max\{N_{\gamma1},N_{\gamma2}\}$:
\begin{align*}
&\sup_{P \in \mathcal{P}}\text{Pr}_{P} \left(\sup_{\theta \in \hat{\Theta}_{I,n}} d(\theta,\Theta_{I}(P)) > \varepsilon  \right) \\
&\leq \sup_{P \in \mathcal{P}}\text{Pr}_{P} \left( \inf_{\theta \in \Theta\setminus\mathcal{E}(P,\varepsilon) } \inf_{\bm r \in \Pi_{\mathcal{R}_n}(\mathcal{S}) }  Q_{n,P}(\theta,\bm r)  \leq \inf_{(\theta,\bm r) \in (\Theta \times \mathcal{R}_{n})\cap \mathcal{S}}  Q_{n,P}(\theta,\bm r) + A_{n} + A_{n}' + \tau_{n}\right) \\
&=\sup_{P \in \mathcal{P}}\text{Pr}_{P} \left(\inf_{\theta \in \Theta\setminus\mathcal{E}(P,\varepsilon) } \inf_{\bm r \in \Pi_{\mathcal{R}_n}(\mathcal{S}) }  Q_{n,P}(\theta,\bm r) - \inf_{(\theta,\bm r) \in (\Theta \times \mathcal{R}_{n})\cap \mathcal{S}}  Q_{n,P}(\theta,\bm r)  + b_{n} \leq  A_{n} + A_{n}' + b_{n} + \tau_{n}\right) \\
&\leq \sup_{P \in \mathcal{P}}\text{Pr}_{P} \left( \nu_n^{-1} \min\{\delta, \varepsilon\} \leq  A_{n} + A_{n}' + b_{n} + \tau_{n}\right) \leq 1\left\{ \nu_n^{-1}\min\{\delta, \varepsilon\} \leq  2 R_{n}' M_{\gamma} + b_{n} + \tau_{n}\right\} + \gamma=\gamma.
\end{align*}
Since $\gamma \in (0,1)$ was arbitrary, conclude that:
\begin{align}
\limsup_{n\to \infty} \sup_{P\in \mathcal{P}}\text{Pr}_{P} \left(\sup_{\theta \in \hat{\Theta}_{I,n}} d(\theta,\Theta_{I}(P)) > \varepsilon  \right) =0. \label{eq_consistency1}
\end{align}
Now note:
\begin{align*}
&\text{Pr}_{P} \left(\sup_{\theta \in \Theta_{I}(P)} d(\theta,\hat{\Theta}_{I,n}) > \varepsilon  \right) 
\leq \text{Pr}_{P}\left(\sup_{\theta \in \Theta_{I}(P)}  \inf_{\bm r \in  \Pi_{\mathcal{R}_n}(\mathcal{S})  }  Q_{n}(\theta, \bm r) > \inf_{(\theta, \bm r) \in (\Theta \times \mathcal{R}_{n})\cap \mathcal{S}} Q_{n}(\theta, \bm r) +\tau_{n} \right).
\end{align*}
Again, following a nearly identical proof as the proof of Lemma S.1.2 in CNS, we obtain that there exists random variables $C_{n}$ and $C_{n}'$ satisfying:
\begin{align*}
&\sup_{\theta \in \Theta_{I}(P)}  \inf_{\bm r \in  \Pi_{\mathcal{R}_n}(\mathcal{S})  }  Q_{n}(\theta, \bm r)    - \sup_{\theta \in \Theta_{I}(P)}  \inf_{\bm r \in  \Pi_{\mathcal{R}_n}(\mathcal{S})  } Q_{n,P}(\theta, \bm r) \leq C_{n} = O_{P}(R_{n}'),\\
\text{ and }\quad& \inf_{(\theta, \bm r) \in (\Theta \times \mathcal{R}_{n})\cap \mathcal{S}} Q_{n,P}(\theta, \bm r)   -  \inf_{(\theta, \bm r) \in (\Theta \times \mathcal{R}_{n})\cap \mathcal{S}} Q_{n}(\theta, \bm r) \leq C_{n}' = O_{P}(R_{n}'),
\end{align*}
where $R_{n}'$ is defined above, and where $C_{n}=O_{P}(R_{n}')$ and $C_{n}'=O_{P}(R_{n}')$ both hold uniformly in $P\in\mathcal{P}$. Furthermore, note that:
\begin{align*}
&\sup_{\theta \in \Theta_{I}(P)}  \inf_{\bm r \in  \Pi_{\mathcal{R}_n}(\mathcal{S})  } Q_{n,P}(\theta, \bm r)\\
 &= \sup_{\theta \in \Theta_{I}(P)}  \inf_{\bm r \in  \Pi_{\mathcal{R}_n}(\mathcal{S})  } || E_{P}[\bm m(\bm Y_{i},\bm W_{i},\theta,\bm r) \otimes \bm q^{k_n}(\bm W_{i})]||\\ 
&\leq \inf_{(\theta, \bm r) \in (\Theta \times \mathcal{R}_{n})\cap\mathcal{S} } || E_{P}[\bm m(\bm Y_{i},\bm W_{i},\theta,\bm r) \otimes \bm q^{k_n}(\bm W_{i})]||+\sup_{\theta \in \Theta_{I}(P)}  \inf_{\bm r \in  \Pi_{\mathcal{R}_n}(\mathcal{S})  } || E_{P}[\bm m(\bm Y_{i},\bm W_{i},\theta,\bm r) \otimes \bm q^{k_n}(\bm W_{i})]||\\
&\leq \inf_{(\theta, \bm r) \in (\Theta \times \mathcal{R}_{n})\cap\mathcal{S} } || E_{P}[\bm m(\bm Y_{i},\bm W_{i},\theta,\bm r) \otimes \bm q^{k_n}(\bm W_{i})]||+E_{n}=\inf_{(\theta, \bm r) \in (\Theta \times \mathcal{R}_{n})\cap \mathcal{S}} Q_{n,P}(\theta, \bm r)+E_{n},
\end{align*}
uniformly in $P\in\mathcal{P}$, where $E_{n} = o(a_{n}/\sqrt{n}) = o(\tau_{n})$ by Assumption \ref{assumption3.6_DGKR}$(ii)$ and \eqref{eq_taun}. Now again, for any fixed $\gamma \in (0,1)$, let $N_{\gamma1}$ and $M_{\gamma}$ be large enough so that:
\begin{align*}
\sup_{P \in \mathcal{P}}\text{Pr}_{P} \left(|C_{n}/R_{n}'| \leq M_{\gamma}, |C_{n}'/R_{n}'| \leq M_{\gamma}\right) \geq 1-\gamma,
\end{align*}
for all $n\geq N_{\gamma1}$. Furthermore, let $N_{\gamma2}$ be large enough so that $2R_{n}' M_{\gamma}/\tau_{n} +E_n/\tau_n < 1$ for all $n\geq N_{\gamma2}$, which is possible since $R_{n}' = o(\tau_n)$ (by \eqref{eq_taun}) and $E_{n} = o(\tau_{n})$. Combining everything, for $n\geq \max\{N_{\gamma1},N_{\gamma2}\}$ we have:
\begin{align*}
&\sup_{P \in \mathcal{P}}\text{Pr}_{P} \left(\sup_{\theta \in \Theta_{I}(P)} d(\theta,\hat{\Theta}_{I,n})> \varepsilon  \right)\\
&\leq \sup_{P \in \mathcal{P}}\text{Pr}_{P}\left(\sup_{\theta \in \Theta_{I}(P)}  \inf_{\bm r \in  \Pi_{\mathcal{R}_n}(\mathcal{S})  } Q_{n}(\theta, \bm r) > \inf_{(\theta, \bm r) \in (\Theta \times \mathcal{R}_{n})\cap \mathcal{S}} Q_{n}(\theta, \bm r) +\tau_{n} \right)\\
&\leq \sup_{P \in \mathcal{P}}\text{Pr}_{P}\left(\sup_{\theta \in \Theta_{I}(P)}  \inf_{\bm r \in  \Pi_{\mathcal{R}_n}(\mathcal{S})  } Q_{n,P}(\theta, \bm r)+C_{n} +C_{n}' > \inf_{(\theta, \bm r) \in (\Theta \times \mathcal{R}_{n})\cap \mathcal{S}} Q_{n,P}(\theta, \bm r) +\tau_{n} \right)\\
&\leq \sup_{P \in \mathcal{P}}\text{Pr}_{P}\left(\inf_{(\theta, \bm r) \in (\Theta \times \mathcal{R}_{n})\cap \mathcal{S}} Q_{n,P}(\theta, \bm r)+C_{n} +C_{n}'+E_{n} > \inf_{(\theta, \bm r) \in (\Theta \times \mathcal{R}_{n})\cap \mathcal{S}} Q_{n,P}(\theta, \bm r) +\tau_{n} \right)\\
& = \sup_{P \in \mathcal{P}}\text{Pr}_{P}\left(C_{n} + C_{n}' + E_{n} >  \tau_{n} \right)\\
& \leq \sup_{P \in \mathcal{P}}\text{Pr}_{P}\left(2R_{n}' M_{\gamma} + E_{n} >  \tau_{n} \right) +\gamma = 1\left\{2R_{n}' M_{\gamma} + E_{n} >  \tau_{n} \right\} +\gamma =\gamma. 
\end{align*}
Since $\gamma \in (0,1)$ was arbitrary, conclude that:
\begin{align}
\limsup_{n\to \infty} \sup_{P\in \mathcal{P}}\text{Pr}_{P} \left(\sup_{\theta \in \Theta_{I}(P)} d(\theta,\hat{\Theta}_{I,n}) > \varepsilon  \right) =0. \label{eq_consistency2}
\end{align}
Combining \eqref{eq_consistency0}, \eqref{eq_consistency1} and \eqref{eq_consistency2}, the result follows.

\end{proof}

\subsection{Additional Inference Results}\label{section_additional_inference}

For any $\vartheta \in \Theta_{I}(P)$, define the quantity:
\begin{align}
U_{n,P}(\vartheta):= \inf_{(\theta,\bm r) \in \mathcal{I}_{n}^*(\vartheta,P)} \left|\left|\mathbb{G}_{P}(\theta, \bm r)\right|\right|,\label{eq_UnP}
\end{align}
where $\mathcal{I}_{n}^*(\vartheta,P)$ is from \eqref{eq_Isets} and $\mathbb{G}_{P}$ is the Gaussian process from Assumption \ref{assumption3.3_DGKR}. Recall the test statistic $T_{n}(\vartheta)$ from \eqref{eq_test_statistic}. The following result is the analog of CNS Theorem 3.1, adjusted for our setting and our assumptions. 
\begin{lemma}\label{lemma_inference_main}
Suppose Assumptions \ref{assumption3.1_DGKR} - \ref{assumption3.6_DGKR} hold. Then for any sequence $\vartheta_{n} \in \Theta_{I}(P)$, we have $T_{n}(\vartheta_{n}) \leq U_{n,P}(\vartheta_{n}) + o_{P}(a_n)$, uniformly in $P\in \mathcal{P}$.
\end{lemma}
\begin{proof}[Proof of Lemma \ref{lemma_inference_main}]
We have:
\begin{align}
&T_{n}(\vartheta_{n}) \leq \inf_{(\theta,\bm r) \in \mathcal{I}_{n}^*(\vartheta_{n},P) } \sqrt{n} Q_{n}(\theta,\bm r) \nonumber\\
&= \inf_{(\theta,\bm r) \in \mathcal{I}_{n}^*(\vartheta_{n},P) } \sqrt{n} Q_{n}(\theta,\bm r)  - \inf_{(\theta,\bm r) \in \mathcal{I}_{n}^*(\vartheta_{n},P)} || \mathbb{G}_{P}(\theta,\bm r) + \sqrt{n} E_{P}[\bm m(\bm Y_{i}, \bm W_{i},\theta,\bm r)\otimes \bm q^{k_{n}}(\bm W_{i})||\nonumber\\
&\qquad\qquad\qquad + \inf_{(\theta,\bm r) \in \mathcal{I}_{n}^*(\vartheta_{n},P)} || \mathbb{G}_{P}(\theta,\bm r) + \sqrt{n} E_{P}[\bm m(\bm Y_{i}, \bm W_{i},\theta,\bm r)\otimes\bm q^{k_{n}}(\bm W_{i})||\nonumber\\
&\leq \inf_{(\theta,\bm r) \in \mathcal{I}_{n}^*(\vartheta_{n},P) } \sqrt{n} Q_{n}(\theta,\bm r)  - \inf_{(\theta,\bm r) \in \mathcal{I}_{n}^*(\vartheta_{n},P)} || \mathbb{G}_{P}(\theta,\bm r) + \sqrt{n} E_{P}[\bm m(\bm Y_{i}, \bm W_{i},\theta,\bm r)\otimes\bm q^{k_{n}}(\bm W_{i})||\nonumber\\
&\qquad\qquad\qquad + \inf_{(\theta,\bm r) \in \mathcal{I}_{n}^*(\vartheta_{n},P)} || \mathbb{G}_{P}(\theta,\bm r) ||+ o_{P}(a_{n}),\label{eq_Tn_result}
\end{align}
uniformly in $P \in \mathcal{P}$, where the last line follows from the triangle inequality and Assumption \ref{assumption3.6_DGKR}$(ii)$. Now by Assumption \ref{assumption3.3_DGKR}$(i)$ and the reverse triangle inequality:
\begin{align*}
&\left| \inf_{(\theta,\bm r) \in \mathcal{I}_{n}^*(\vartheta_{n},P) } \sqrt{n} Q_{n}(\theta,\bm r) - \inf_{(\theta,\bm r) \in \mathcal{I}_{n}^*(\vartheta_{n})  } || \mathbb{G}_{P}(\theta,\bm r) + \sqrt{n} E_{P}[\bm m(\bm Y_{i}, \bm W_{i},\theta,\bm r)\otimes\bm q^{k_{n}}(\bm W_{i})||\right|\\
&\leq \sup_{(\theta,\bm r) \in \mathcal{I}_{n}^*(\vartheta_{n},P) } \left| \sqrt{n} Q_{n}(\theta,\bm r) -  || \mathbb{G}_{P}(\theta,\bm r) + \sqrt{n} E_{P}[\bm m(\bm Y_{i}, \bm W_{i},\theta,\bm r)\otimes\bm q^{k_{n}}(\bm W_{i})||\right|\\
&\leq \sup_{(\theta,\bm r) \in \mathcal{I}_{n}^*(\vartheta_{n},P)  } || \mathbb{G}_n(\theta,\bm r) - \mathbb{G}_{P}(\theta,\bm r)||\leq \sup_{(\theta,\bm r) \in (\Theta \times \mathcal{R}_n)\cap \mathcal{S}(\vartheta_{n}) } || \mathbb{G}_n(\theta,\bm r) - \mathbb{G}_{P}(\theta,\bm r)||=o_{P}(a_n),
\end{align*}
uniformly in $P\in \mathcal{P}$. The result then follows from \eqref{eq_Tn_result}. 
\end{proof}

Recall the multiplier bootstrap process $\mathbb{G}_{n}^{b}(\theta,\bm r)$ from \eqref{eq_Gnb}. Furthermore, recall that the process $\mathbb{G}_{P}^\star$ from Assumption \ref{assumption3.11_DGKR} is independent of $\{(\bm Y_{i}, \bm W_{i})\}_{i=1}^{n}$ and has the same distribution as $\mathbb{G}_{P}$. Now define:
\begin{align}
U_{n,P}^\star(\vartheta):= \inf_{(\theta,\bm r) \in \mathcal{I}_{n}^*(\vartheta,P)} \left|\left|\mathbb{G}_{P}^\star(\theta, \bm r)\right|\right|,&&
\hat{U}_{n}(\vartheta):= \inf_{(\theta,\bm r) \in \hat{\mathcal{I}}_{n}(\vartheta)} \left|\left|\mathbb{G}_{n}^{b}(\theta,\bm r)\right|\right|.\label{eq_UnP_star}
\end{align}
In addition, define the norm:
\begin{align}
|| (\theta, \bm r) ||_{\bm E} = \sup_{P\in \mathcal{P}} (E_{P}[||r(\bm W_{i})||^2])^{1/2} + ||\theta||.\label{eq_E_norm}
\end{align}
The following result is the analog of CNS Theorem 3.2. 
\begin{lemma}\label{lemma_bootstrap_approximation}
Suppose Assumptions \ref{assumption3.1_DGKR} - \ref{assumption3.12_DGKR} hold. Then for any sequence $\vartheta_{n} \in \Theta_{I}(P)$, we have $\hat{U}_{n}(\vartheta_{n}) \geq U_{n,P}^\star(\vartheta_{n}) + o_{P}(a_n),$ uniformly in $\Phi \times P$, with $P\in \mathcal{P}$ and for $\Phi$ the standard normal distribution. 
\end{lemma}
\begin{proof}[Proof of Lemma \ref{lemma_bootstrap_approximation}]
Here we follow the proof of Theorem S.3.1$(i)$ in CNS. First recall $\hat{\mathcal{I}}_{n}(\vartheta)$ from \eqref{eq_sample_identified_set_lambda} in the main text. Since $\hat{\mathcal{I}}_{n}(\vartheta_{n}) \subseteq (\Theta\times \mathcal{R}_n)\cap \mathcal{S}(\vartheta_{n})$, we have by the reverse triangle inequality and Assumption \ref{assumption3.11_DGKR}:
\begin{align*}
&\left|\inf_{(\theta,\bm r) \in \hat{\mathcal{I}}_{n}(\vartheta_{n})} \left|\left|\mathbb{G}_{n}^{b}(\theta,\bm r)\right|\right|- \inf_{(\theta,\bm r) \in \hat{\mathcal{I}}_{n}(\vartheta_{n})} \left|\left|\mathbb{G}_{P}^\star(\theta, \bm r)\right|\right|\right|\\
&\leq \sup_{(\theta,\bm r) \in \hat{\mathcal{I}}_{n}(\vartheta_{n})} \left|\left|\mathbb{G}_{n}^{b}(\theta,\bm r)-\mathbb{G}_{P}^\star(\theta, \bm r)\right|\right|\leq \sup_{(\theta,\bm r) \in (\Theta\times \mathcal{R}_n)\cap \mathcal{S}(\vartheta_{n})} \left|\left|\mathbb{G}_{n}^{b}(\theta,\bm r)-\mathbb{G}_{P}^\star(\theta, \bm r)\right|\right|=o_{P}(a_n),
\end{align*}
uniformly in $\Phi \times P$ with $P\in \mathcal{P}$. This implies:
\begin{align*}
\hat{U}_{n}(\vartheta_{n}) = \inf_{(\theta,\bm r) \in \hat{\mathcal{I}}_{n}(\vartheta_{n})} \left|\left|\mathbb{G}_{P}^\star(\theta, \bm r)\right|\right| +o_{P}(a_n).
\end{align*}
Thus, we can choose $(\hat{\theta}_n,\hat{\bm r}_{n}) \in \hat{\mathcal{I}}_{n}(\vartheta_{n})$ (i.e. a $o_{P}(a_n)$-minimizer, depending on $P$) such that:
\begin{align}
\hat{U}_{n}(\vartheta_{n}) = \left|\left|\mathbb{G}_{P}^\star(\hat{\theta}_n,\hat{\bm r}_{n})\right|\right| +o_{P}(a_n),\label{eq_Unhat_1}
\end{align}
uniformly in $\Phi \times P$ with $P\in \mathcal{P}$. Now note that Assumption \ref{assumption3.1_DGKR}$(i)$ implies CNS Assumption 3.1$(i)$, Assumption \ref{assumption3.2_DGKR} implies CNS Assumption 3.2$(i)(iii)$, Assumption \ref{assumption3.3_DGKR}$(i)$ implies CNS Assumption 3.3$(i)$, Assumption \ref{assumption3.4_DGKR}$(i)$ implies CNS Assumption 3.4$(i)$, Assumption \ref{assumption3.6_DGKR}$(ii)$ implies CNS Assumption 3.6$(ii)$, and Assumption \ref{assumption3.12_DGKR} implies CNS Assumption 3.12$(iii)$. CNS Assumption 3.7 is not needed in our context since we do not studentize the moments. Thus, all of the assumptions required for Corollary S.1.2$(i)$ in CNS hold. 
Recall the sequence $R_{n}$ from \eqref{eq_capital_Rn}. By Assumption \ref{assumption3.6_DGKR}$(i)$ there exists a sequence $\delta_n$ satisfying  $R_{n}\vee \nu_n\tau_n = o(\delta_n)$ and:
\begin{align}
\sqrt{k_{n} \log(1+ k_n)} B_{n} \times \sup_{P\in \mathcal{P}} J_{[\,]}(\delta_n^{\kappa_{m}}, \mathcal{F}_n, ||\,\cdot\,||_{P,2}) = o(a_n).\label{eq_s.62}
\end{align}
Furthermore, by Corollary S.1.2$(i)$ in CNS, there is some $(\theta_{0n}, \bm r_{0n}) \in \mathcal{I}_{n}^*(\vartheta_{n},P)$ such that:
\begin{align}
||(\hat{\theta}_{n},\hat{\bm r}_{n}) - (\theta_{0n}, \bm r_{0n})||_{\mathbf{E}} = o_{P}(\delta_n).\label{eq_s.63}
\end{align}
Now since $||q_{k}||_{\infty} \leq B_{n}$ for all $1\leq k \leq k_{n}$ by Assumption \ref{assumption3.2_DGKR}$(i)$, we obtain from Assumption \ref{assumption3.3_DGKR}$(ii)$ together with \eqref{eq_s.63} that for any instrument function $q_k(\,\cdot\,)$:
\begin{align}
E_{P}[||\bm m(\bm Y_{i},\bm W_{i},\hat{\theta}_n,\hat{\bm r}_{n}) - \bm m(\bm Y_{i},\bm W_{i},\theta_{0n}, \bm r_{0n})||^{2} q_{k}^2(\bm W_{i})]&\leq B_{n}^2 K_{m}^2 ||(\hat{\theta}_n,\hat{\bm r}_{n}) - (\theta_{0n}, \bm r_{0n})||_{\bm E}^{2 \kappa_{m}}\nonumber \\
&\leq B_{n}^2 K_{m}^2 \delta_n^{2 \kappa_{m}},\label{eq_s.65}
\end{align}
with probability approaching 1 uniformly in $P \in \mathcal{P}$, where $||\,\cdot\,||_{\bm E}$ is the norm from Assumption \ref{assumption3.3_DGKR}. Now let $\mathcal{G}_n:=\{ f q_k : f \in \mathcal{F}_n, 1 \leq k\leq k_{n}\}$ and let $\mathbb{G}_P$ be a Gaussian process on $\mathcal{G}_n$ satisfying $E_{P}[\mathbb{G}_{P}(g_1)\mathbb{G}_{P}(g_2)] = E_{P}[g_1(\bm Y_{i},\bm W_{i})g_2(\bm Y_{i},\bm W_{i})]$ and $E_{P}[g_1(\bm Y_{i},\bm W_{i})]=0$ for any $g_1,g_2 \in \mathcal{G}_n$. Since \eqref{eq_s.65} holds with probability tending to 1 uniformly in $P \in \mathcal{P}$, Markov's inequality, result (S.45) in CNS, and the fact that $\delta_n$ satisfies \eqref{eq_s.62} and \eqref{eq_s.63} implies:
\begin{align*}
&\limsup_{n\to \infty} \sup_{P \in \mathcal{P}} \text{Pr}_{P} \left( ||\mathbb{G}_{P}^\star(\hat{\theta}_n,\hat{\bm r}_n) -  \mathbb{G}_{P}^\star(\theta_{0n}, \bm r_{0n})|| > a_n \epsilon \right)\\ 
&\leq \limsup_{n\to \infty} \sup_{P \in \mathcal{P}}\frac{1}{a_n\epsilon} E_{P} \left[||\mathbb{G}_{P}^\star(\hat{\theta}_n,\hat{\bm r}_n) -  \mathbb{G}_{P}^\star(\theta_{0n}, \bm r_{0n})||\right]\\ 
&\leq \limsup_{n\to \infty} \sup_{P \in \mathcal{P}}\frac{1}{a_n\epsilon} E_{P} \left[\sup_{g_1,g_2 \in \mathcal{G}_n : ||g_1 - g_2||_{P,2} \leq B_{n} K_{m} \delta_{n}^{\kappa_m}} |\mathbb{G}_{P}(g_1) - \mathbb{G}_{P}(g_2)| \right]=0. 
\end{align*}
Thus, combine this with \eqref{eq_Unhat_1} to conclude that:
\begin{align*}
 U_{n,P}^\star(\vartheta_{n}) = \inf_{(\theta,\bm r) \in \mathcal{I}_{n}^{*}(\vartheta_{n},P)} \left|\left|\mathbb{G}_{P}^\star(\theta,\bm r)\right|\right| &\leq \left|\left|\mathbb{G}_{P}^\star(\theta_{0n},\bm r_{0n})\right|\right| \leq \left|\left|\mathbb{G}_{P}^\star(\hat{\theta}_{n},\hat{\bm r}_{n})\right|\right| +o_{P}(a_n) \leq \hat{U}_{n}(\vartheta_{n}) +o_{P}(a_n),
\end{align*}
uniformly in $\Phi \times P$ with $P\in \mathcal{P}$. This completes the proof.
\end{proof}

The following Lemma establishes a result similar to Corollary 3.1 in CNS. 
\begin{lemma}\label{lemma_validity}
Suppose Assumptions \ref{assumption3.1_DGKR} - \ref{assumption3.12_DGKR} hold. Furthermore, for any $\delta>0$, let $\hat{q}_{1-\alpha+\delta}(\hat{U}_{n}(\vartheta))$ denote the $1-\alpha+\delta$ quantile of the bootstrap distribution of $\hat{U}_{n}(\vartheta)$. Then:
\begin{align*}
\limsup_{n\to \infty} \sup_{P\in \mathcal{P}} \sup_{\vartheta \in \Theta_{I}(P)} \text{Pr}_{P}(T_{n}(\vartheta) > \hat{q}_{1-\alpha+\delta}(\hat{U}_{n}(\vartheta)) + \delta) \leq \alpha. 
\end{align*}
\end{lemma}
\begin{proof}[Proof of Lemma \ref{lemma_validity}]
From Lemma \ref{lemma_bootstrap_approximation}, for any sequence $\vartheta_{n} \in \Theta_{I}(P)$ we have:
\begin{align}
\hat{U}_{n}(\vartheta_{n}) \geq U_{n,P}^\star(\vartheta_{n}) + o_{P}(a_n),\label{eq_lemma_s.2.3_result}
\end{align}
uniformly in $\Phi\times P$ for $P\in \mathcal{P}$ and for $\Phi$ the standard normal distribution, where $\hat{U}_{n}(\vartheta)$ and $U_{n,P}^\star(\vartheta)$ are defined in \eqref{eq_UnP_star}. Furthermore, from Lemma \ref{lemma_inference_main} we have:
\begin{align}
T_{n}(\vartheta_{n}) \leq U_{n,P}(\vartheta_{n}) + o_{P}(a_n),\label{eq_lemma_s.2.2_result}
\end{align}
uniformly in $P\in \mathcal{P}$, where $U_{n,P}(\vartheta)$ is defined in \eqref{eq_UnP}. Applying Lemma S.3.5 in CNS using \eqref{eq_lemma_s.2.3_result} with $B_n= \hat{U}_{n}(\vartheta_{n})$, $D_{n} = \{(\bm Y_{i}, \bm W_{i})\}_{i=1}^{n}$, and $C_{P,n}^\star = U_{n,P}^\star(\vartheta_{n})$, we have:
\begin{align}
&\liminf_{n\to \infty} \inf_{P\in \mathcal{P}}\inf_{\vartheta \in \Theta_{I}(P)} \text{Pr}_{P} \left(\hat{q}_{1-\alpha+\delta}(\hat{U}_{n}(\vartheta)) + \frac{a_n}{2} > q_{1-\alpha+\delta -\delta_n,P}(U_{n,P}^\star(\vartheta))  \right)\nonumber\\
&=\liminf_{n\to \infty} \inf_{P\in \mathcal{P}} \text{Pr}_{P}\left(\hat{q}_{1-\alpha+\delta}(\hat{U}_{n}(\vartheta_{n})) + \frac{a_n}{2} > q_{1-\alpha+\delta -\delta_n,P}(U_{n,P}^\star(\vartheta_{n}))  \right)=1,\label{eq_Lemma2.3.5}
\end{align}
for some $\delta_n=o(1)$, where $\{\vartheta_{n}\}_{n=1}^{\infty}$ is any infimum sequence (possibly depending on $P$), and $q_{p,P}(U_{n,P}^\star(\vartheta))$ denotes the $p^{th}$ quantile of $U_{n,P}^\star(\vartheta)$. Since $U_{n,P}^\star(\vartheta_{n}) \overset{d}{=} U_{n,P}(\vartheta_{n})$ by Assumption \ref{assumption3.11_DGKR}, we have:
\begin{align*}
&\limsup_{n\to \infty} \sup_{P\in \mathcal{P}}\sup_{\vartheta \in \Theta_{I}(P)}\text{Pr}_{P}(T_{n}(\vartheta) > \hat{q}_{1-\alpha+\delta}(\hat{U}_{n}(\vartheta))+ \delta) \\
&=\limsup_{n\to \infty} \sup_{P\in \mathcal{P}}\text{Pr}_{P}(T_{n}(\vartheta_{n}) > \hat{q}_{1-\alpha+\delta}(\hat{U}_{n}(\vartheta_{n}))+ \delta) \\
&\leq \limsup_{n\to \infty} \sup_{P\in \mathcal{P}} \text{Pr}_{P}\left(\frac{T_{n}(\vartheta_{n}) - U_{n,P}(\vartheta_{n})}{a_n}>\frac{1}{2} \right)\\ 
&\qquad + \limsup_{n\to \infty} \sup_{P\in \mathcal{P}}\text{Pr}_{P}\left(T_{n}(\vartheta_{n}) > \hat{q}_{1-\alpha+\delta}(\hat{U}_{n}(\vartheta_{n}))+ \delta, \frac{T_{n}(\vartheta_{n}) - U_{n,P}(\vartheta_{n})}{a_n}\leq \frac{1}{2} \right)\\
&\leq \limsup_{n\to \infty} \sup_{P\in \mathcal{P}}\text{Pr}_{P}\left(U_{n,P}(\vartheta_{n}) + \frac{a_{n}}{2} > \hat{q}_{1-\alpha+\delta}(\hat{U}_{n}(\vartheta_{n}))+ \delta \right)\\ 
&\leq \limsup_{n\to \infty} \sup_{P\in \mathcal{P}}\text{Pr}_{P}\left(U_{n,P}(\vartheta_{n}) + \frac{a_{n}}{2} >  q_{1-\alpha+\delta -\delta_n,P}(U_{n,P}^\star(\vartheta_{n}))+ \delta - \frac{a_{n}}{2}\right)\\ 
&\leq \limsup_{n\to \infty} \sup_{P\in \mathcal{P}}\text{Pr}_{P}\left(U_{n,P}(\vartheta_{n}) + a_{n} >  q_{1-\alpha+\delta -\delta_n,P}(U_{n,P}(\vartheta_{n}))+ \delta \right)\leq \alpha,
\end{align*}
where the second inequality holds by \eqref{eq_lemma_s.2.2_result}, the third inequality holds by \eqref{eq_Lemma2.3.5}, and the final line holds since for all $n$ sufficiently large we have $q_{1-\alpha+\delta -\delta_n,P}(U_{n,P}(\vartheta_{n})) - a_{n} + \delta\geq q_{1-\alpha+\delta/2,P}(U_{n,P}(\vartheta_{n})) - a_{n} + \delta >   q_{1-\alpha,P}(U_{n,P}(\vartheta_{n}))$, after which the cdf of $U_{n,P}(\vartheta_{n})$ must have a continuity point (possibly depending on $n$) between the values $q_{1-\alpha+\delta/2,P}(U_{n,P}(\vartheta_{n})) - a_{n} + \delta$ and $q_{1-\alpha,P}(U_{n,P}(\vartheta_{n}))$. 
\end{proof}

\subsection{Verification of Main Assumptions}\label{section_verification}

In this section we verify Assumptions \ref{assumption3.1_DGKR} - \ref{assumption3.12_DGKR} from Section \ref{section_inference_appendix_assumptions} using Assumptions \ref{assumption_main}, \ref{assumption_consistency}, and \ref{assumption_inference} from the main text. Assumption \ref{assumption_consistency2} is treated separately, since it is required only for Theorem \ref{theorem_consistency}, and used only to verify the conditions in Lemma \ref{lemma_consistency}. 

\begin{theorem}\label{theorem_assumption_verification}
Suppose Assumptions \ref{assumption_main} and \ref{assumption_consistency} hold, let $0<\frac{3\beta_{k}}{2}<\beta_\tau<\frac{1}{2}-\beta_k$, and let $\tau_{n}\downarrow 0$, $l_n\uparrow \infty$, and $k_n\uparrow \infty$ be sequences satisfying $\tau_n = O(n^{-\beta_\tau})$, $l_n\asymp k_n$, and $l_n\leq k_{n} =O(n^{\beta_k})$. Then Assumptions \ref{assumption3.1_DGKR}, \ref{assumption3.2_DGKR}, \ref{assumption3.3_DGKR}, and \ref{assumption3.6_DGKR}(ii) hold with $J_{n} = O(\sqrt{l_{n}})$ and $B_{n}=O(1)$ and for any nonzero sequence satisfying $a_n = o((\log(n))^{-\beta_a})$ for $\beta_a>0$ from Assumption \ref{assumption_consistency}$(v)$. If Assumption \ref{assumption_inference} also holds, then Assumptions \ref{assumption3.4_DGKR}, \ref{assumption3.6_DGKR}(i), \ref{assumption3.11_DGKR}, and \ref{assumption3.12_DGKR} hold for $\nu_n \asymp k_n^{1/2}$ and any nonzero sequence satisfying $a_n = o((\log(n))^{-\beta_a})$ for $\beta_a>0$ from Assumption \ref{assumption_consistency}$(v)$.
\end{theorem}

\begin{proof}[Proof of Theorem \ref{theorem_assumption_verification}]
Suppose Assumptions \ref{assumption_main} and \ref{assumption_consistency} hold. Assumption \ref{assumption_consistency}$(i)$ is identical to Assumption \ref{assumption3.1_DGKR}$(i)$. Assumption \ref{assumption3.1_DGKR}$(ii)$ holds with $\mathbf{B} = \mathbb{R}^{d_{\theta}} \times \left( \bigtimes_{m=0}^{S} \ell^{\infty}(\mathcal{W}) \right)$ equipped with the norm $||(\theta,\bm r)||_{\bm B} := ||\theta|| + ||\bm r||_{\mathcal{R}}$ where $||\bm r||_{\mathcal{R}} :=||\bm r||_{\infty}$. Assumption \ref{assumption3.1_DGKR}$(iii)$ holds for $\mathcal{S}^* = \mathcal{S}$ by taking $\Upsilon_{\bm F}$ as the zero function and $\Upsilon_{\bm G}$ as the map that takes $(\theta,\bm r)$ and outputs the $d_{\bm G}:=((2^m -1) + (2^{m+1}-1))$-dimensional vector-valued function that consists of the \textit{minus} of the principle minors of the matrices $\bm{H}_{m}^*(\bm{r}(\bm w),\varsigma(\bm w))$ and $\bm{B}_{m}(\bm{r}(\bm w))$ if $S=2m+1$ is odd, or the \textit{minus} of the principle minors of the matrices $\bm{H}_{m}(\bm{r}(\bm w))$ and $\bm{B}_{m}^*(\bm{r}(\bm w),\varsigma(\bm w))$ if $S=2m$ is even. Here, when $S=2m+1$ is odd, $\varsigma(\bm w)$ is a function of $\bm r_{0}(\bm w), \ldots, \bm r_{2m+1}(\bm w)$ that ensures the \textit{minus} of all principle minors of the matrix $\bm{H}_{m}^*(\bm{r}(\bm w),\varsigma(\bm w))$ involving the element $\varsigma(\bm w)$ are less than or equal to zero. When such a choice is not possible, $\varsigma(\bm w)$ can be set to zero. A similar construction can be repeated when $S=2m$ is even. In either case, conclude that $\varsigma(\bm w)$ is either zero, or can be written as the maximum of at most finitely many continuous functions of either $\bm r_{0}(\bm w), \ldots, \bm r_{2m+1}(\bm w)$ (or $\bm r_{1}(\bm w), \ldots, \bm r_{2m}(\bm w)$). Thus, $\Upsilon_{\bm G}$ maps to the Banach space $\bm F = \bigtimes_{g=1}^{d_{\bm G}} \ell^{\infty}(\mathcal{W})$ equipped with the norm $||\,\cdot\,||_{\bm F} := ||\,\cdot\,||_{\infty}$. This verifies Assumption \ref{assumption3.1_DGKR}$(iii)$.

Assumption \ref{assumption3.2_DGKR}$(i)$ is satisfied with $B_{n}=1$ by the choice of instrument functions from \eqref{eq_instrument_functions}, formalized in Assumption \ref{assumption_consistency}$(iv)$. For Assumption \ref{assumption3.2_DGKR}$(ii)$, recall $\mathcal{F}_{n}$ from \eqref{eq_Fn}. Now for fixed $\bm y \in \mathcal{Y}^T$ and $j=1,\ldots,J$, consider the functions of the form:
\begin{align}
f(\bm y, \bm w) = 1\{\bm y = \bm y_{j}\} - \sum_{D \in \mathcal{D}_{l_n}} \left(\sum_{s=0}^{S} g_{s}(\bm y,\bm w,\theta) \delta_{D,s}\right) 1\{\bm w \in  D\}.\label{eq_functions_of_this_form}
\end{align}
Now define:
\begin{align*}
\mathcal{F}_{n,j, \bm y}&:= \bigg\{ f(\bm y, \,\cdot\,):\mathcal{W} \to \mathbb{R} : \text{$f(\bm y,\bm w)$ is of the form \eqref{eq_functions_of_this_form} for some }  \delta_{D,s} \in [0,\overline{\delta}], \, \theta \in \Theta  \bigg\}.
\end{align*}
Note that since $\mathcal{W}$ and $\Theta$ are compact by Assumption \ref{assumption_consistency}$(ii)$ and $(iii)$, and each $g_{s}(\bm y,\bm w,\theta)$ is continuously differentiable in $(\bm w, \theta)$ by Assumption \ref{assumption_main}, we have that $g_{s}(\bm y,\bm w,\theta)$ are Lipschitz continuous and uniformly bounded over $(\bm w, \theta)$ for every $\bm y$ and $s$, and we can take the upper and lower bounds as $\bar{c}$ and $0$. Now let $\overline{C} = \max\{|\overline{c}|,|\overline{\delta}|\}$. Since $\mathcal{F}_n$ is contained in the union of the classes $\mathcal{F}_{n,j,\bm y}$ across $(j,\bm y)$, we have:
\begin{align}
N_{[\,]}(\epsilon, \mathcal{F}_{n},||\,\cdot\,||_{P,2})\leq \sum_{j=1}^{J}\sum_{\bm y \in \mathcal{Y}^T} N_{[\,]}(\epsilon,\mathcal{F}_{n,j,\bm y},||\,\cdot\,||_{P,2}). \label{eq_bracket_sum}
\end{align}
Note for two functions $f^{(1)}, f^{(2)} \in \mathcal{F}_{n,j,\bm y}$:
\begin{align*}
| f^{(1)}(\bm y, \bm w) - f^{(2)}(\bm y, \bm w)|
&\leq  \sum_{s=0}^{S} \max_{D \in \mathcal{D}_{l_n}} \left|g_{s}(\bm y, \bm w,\theta^{(1)}) \delta_{D,s}^{(1)} -g_{s}(\bm y, \bm w,\theta^{(2)}) \delta_{D,s}^{(2)} \right|.
\end{align*}
Now note that:
\begin{align*}
\left|g_{s}(\bm y, \bm w,\theta^{(1)}) \delta_{D,s}^{(1)} -g_{s}(\bm y, \bm w,\theta^{(2)}) \delta_{D,s}^{(2)}  \right|
&\leq \overline{C} \left(\left|\delta_{D,s}^{(1)} - \delta_{D,s}^{(2)} \right| + \left|g_{s}(\bm y, \bm w,\theta^{(1)})  - g_{s}(\bm y, \bm w,\theta^{(2)}) \right|\right)\\
&\leq \overline{C} \left(\left|\delta_{D,s}^{(1)} - \delta_{D,s}^{(2)} \right| + L_{s} ||\theta^{(1)}  - \theta^{(2)}||\right),
\end{align*}
for some constant $L_{s}$ that holds for all $(\bm y,\bm w)$ (by finiteness of $\mathcal{Y}^T$, continuous differentiability of $g_{s}(\bm y,\bm w, \theta)$ in $(\bm w, \theta)$ by Assumption \ref{assumption_main}, and compactness of $\mathcal{W}$ by Assumption \ref{assumption_consistency}$(ii)$). Thus: 
\begin{align*}
| f^{(1)}(\bm y, \bm w) - f^{(2)}(\bm y, \bm w)|
&\leq L \max\{||\bm \delta^{(1)} - \bm \delta^{(2)} ||_{\infty}, ||\theta^{(1)}  - \theta^{(2)}||_{\infty} \},
\end{align*}
for some constant $L$. In other words, $\mathcal{F}_{n,j,\bm y}$ is a parametric class of Lipschitz functions (\textit{in the parameters}) in the $\infty-$norm. From \cite{vaart2023empirical} Theorem 2.7.17:
\begin{align}
N_{[\,]}(2\varepsilon L,\mathcal{F}_{n,j,\bm y},||\,\cdot\,||_{\infty}) \leq N(\varepsilon,[0,\overline{\delta}]^{(S+1) \cdot l_n}\times \Theta,||\,\cdot\,||_{\infty})\leq \left( \frac{\overline{\delta}\cdot C}{\varepsilon}\right)^{(S+1) \cdot l_n + d_{\theta}},\label{eq_lipschtiz_bracket}
\end{align}
where $C\geq 1$ is any value such that $\Theta$ is contained in an $||\,\cdot\,||_{\infty}$-box of side length $\overline{\delta}\cdot C$. 
Finally, note that we can take $||F_{n}||_{P,2} = 1+ (S+1) \cdot\overline{C}^2$ (this envelope works for both $\mathcal{F}_{n}$ and each of the classes $\mathcal{F}_{n,j,\bm y}$). Then we have:
\begin{align*}
&J_{[\,]}(||F_{n}||_{P,2},\mathcal{F}_{n},||\,\cdot\,||_{P,2}) = \int_{0}^{1+ (S+1) \cdot\overline{C}^2}
\sqrt{1+ \log N_{[\,]}(\epsilon,\mathcal{F}_{n},||\,\cdot\,||_{P,2})}\,d\epsilon\\
&\overset{(1)}{\leq} \sqrt{(1+\log(J^2))} \int_{0}^{1+ (S+1) \cdot\overline{C}^2} \max_{j, \bm y}
\sqrt{1+\log N_{[\,]}(\epsilon,\mathcal{F}_{n,j,\bm y},||\,\cdot\,||_{P,2})}\,d\epsilon\\
&\overset{(2)}{\leq} \sqrt{(1+\log(J^2))}(1+ (S+1) \cdot\overline{C}^2) \\ 
&\qquad\qquad+ \sqrt{(1+\log(J^2))} \int_{0}^{1+ (S+1) \cdot\overline{C}^2} \max_{j, \bm y}
\sqrt{\log N_{[\,]}(\epsilon,\mathcal{F}_{n,j,\bm y},||\,\cdot\,||_{P,2})}\,d\epsilon\\
&\overset{(3)}{=} \sqrt{(1+\log(J^2))}(1+ (S+1) \cdot\overline{C}^2) \\ 
&\qquad\qquad+  2L\sqrt{(1+\log(J^2))} \int_{0}^{(1+ (S+1) \cdot\overline{C}^2)/2L} \max_{j, \bm y}
\sqrt{\log N_{[\,]}(2\epsilon L,\mathcal{F}_{n,j,\bm y},||\,\cdot\,||_{P,2})}\,d\epsilon\\
&\overset{(4)}{\leq} \sqrt{(1+\log(J^2))}(1+ (S+1) \cdot\overline{C}^2) \\ 
&\qquad\qquad+  2L\sqrt{(1+\log(J^2))} \int_{0}^{(1+ (S+1) \cdot\overline{C}^2)/2L} \sqrt{\log\left(\left( \frac{\overline{\delta}\cdot C}{\varepsilon}\right)^{(S+1) \cdot l_n + d_{\theta}}\right)}\,d\epsilon\\
&\overset{(5)}{\leq}  \sqrt{(1+\log(J^2))}(1+ (S+1) \cdot\overline{C}^2) \\ 
&\qquad\qquad+  \sqrt{(1+\log(J^2))((S+1) \cdot l_n + d_{\theta})} (1+ (S+1) \cdot\overline{C}^2) \int_{0}^{1} \sqrt{\log\left(\frac{2L\cdot\overline{\delta}\cdot C}{(1+ (S+1) \cdot\overline{C}^2)  \epsilon}\right)}\,d\epsilon\\
&\overset{(6)}{\leq}  \sqrt{(1+\log(J^2))}(1+ (S+1) \cdot\overline{C}^2) \\ 
&\qquad\qquad+  \sqrt{(1+\log(J^2))((S+1) \cdot l_n + d_{\theta})} (1+ (S+1) \cdot\overline{C}^2)  \sqrt{\log\left(\frac{2L\cdot \overline{\delta}\cdot C}{ 1+ (S+1) \cdot\overline{C}^2}\vee 1\right)}\\
&\qquad\qquad\qquad+  \sqrt{(1+\log(J^2))((S+1) \cdot l_n + d_{\theta})} (1+ (S+1) \cdot\overline{C}^2)   \int_{0}^{1} \sqrt{\log\left(\frac{1}{\epsilon}\right)}\,d\epsilon,\\
&=K_{1} + K_{2} \sqrt{l_n},
\end{align*}
for constants $K_{1}$ and $K_{2}$ depending only on $L$, $S$, $\bar{c}$, $C$ and $J$. Here, $(1)$ follows from \eqref{eq_bracket_sum} and the fact $\sqrt{1+\log(xy)}\leq \sqrt{1+\log(x)}\sqrt{1+\log(y)}$ for $x,y\geq 1$, $(2)$ follows from the fact $\sqrt{1+x} \leq 1 + \sqrt{x}$ for $x\geq 0$, $(3)$ follows from a change of variable, $(4)$ follows from \eqref{eq_lipschtiz_bracket}, $(5)$ follows from a change of variable, and $(6)$ uses the fact that $\sqrt{x+y} \leq \sqrt{x} + \sqrt{y}$ for $x,y\geq 0$. The final integral evaluates to $\frac{\sqrt{\pi}}{2}$. Thus Assumption \ref{assumption3.2_DGKR}$(ii)$ is satisfied with $F_{n} = 1+(S+1)\cdot\overline{C}^2$ and $J_n = K_{1} + K_{2} \sqrt{l_n}$. 

For Assumption \ref{assumption3.3_DGKR}$(i)$ we use a slight modification of Lemma S.4.6 in CNS, which in turn employs the coupling result of \cite{zhai2018high}. This assumption is only imposed for $\mathcal{S}^* = \mathcal{S}(\vartheta_{n})$, where $\vartheta_n \in \Theta_I(P)$ is an arbitrary sequence. When $\mathcal{S}^* = \mathcal{S}(\vartheta_{n})$, each function in the class $\mathcal{F}_{n}$ can be written as $f_{n,j}(\bm y, \bm w)  = \vec{\bm f}_{n,j}(\bm y, \bm w,\vartheta_n)^\top \check{\bm \delta}$ where $\check{\bm \delta} \in \{0,1\} \times [0,\overline{\delta}]^{l_n}$ is a vector. Furthermore, note that $||\check{\bm \delta}|| \leq (1+l_n\cdot\overline{\delta}^2)^{1/2}$, $\sup_{\bm w} || \bm q^{k_n}(\bm w) ||\leq 1$, and $\sup_{\bm y, \bm w} ||\vec{\bm f}_{n,j}(\bm y, \bm w,\vartheta_n)|| \leq (1+(S+1)\overline{c}^2)^{1/2}$. Thus, by Lemma S.4.6 in CNS, there exists an isonormal Gaussian process $\mathbb{G}_{P}$ (possibly depending on $n$) such that:\footnote{Inspecting the proof of Lemma S.4.6 in CNS shows that it continues to hold under identical assumptions when the functions $\{r_j\}_{j=1}^{j_n}$ (in their notation) are replaced by functions $\{r_{n,j}\}_{j=1}^{j_n}$, which may now depend on $n$. Redefining their class $\mathcal{G}_{n}$ to accommodate this change, the result is identical with the exception that the corresponding Gaussian process $\mathbb{G}_{P}$ now depends on $n$. }
\begin{align*}
\sup_{f \in \mathcal{F}_{n}} &\left|\left| \frac{1}{n} \sum_{i=1}^{n} \left(f(\bm Y_{i},\bm W_{i})\otimes \bm q^{k_n}(\bm W_{i}) - E_{P}[f(\bm Y_{i},\bm W_{i})\otimes \bm q^{k_n}(\bm W_{i})]\right) - \mathbb{G}_{P}(f\otimes \bm q^{k_n}) \right|\right|= O_{P} \left(\frac{l_n k_n^{1/2} \cdot\log(n)}{\sqrt{n}} \right).
\end{align*}
In particular, in CNS's notation in Lemma S.4.6, we set $C_{n} =  (1+l_n\cdot\overline{\delta}^2)^{1/2}$, $j_n= l_n+1$, $b_{1n}$ as constant, and $b_{2n}$ as constant. Since $l_n\asymp k_n$ by the statement of the theorem, this verifies Assumption \ref{assumption3.3_DGKR}$(i)$ for any $a_n$ satisfying $k_n^{3/2}\log(n)/\sqrt{n} \asymp n^{\frac{3}{2}\beta_k}\log(n)/\sqrt{n}  = o(a_n)$. Since $\beta_k<\frac{1}{5}$ under the constraints in the statement of the theorem, in the worst case we require $a_n$ to satisfy $n^{\frac{3}{10}}\log(n)/\sqrt{n}  = o(a_n)$. Any sequence satisfying $a_n=o((\log(n))^{-\beta_a})$ for some $\beta_a>0$ certainly satisfies this requirement. 

For Assumption \ref{assumption3.3_DGKR}$(ii)$, fix any $(\vartheta_{n},\bm r), (\vartheta_{n},\bm r') \in (\Theta \times \mathcal{R}_{n})\cap \mathcal{S}(\vartheta_{n})$ and $P \in \mathcal{P}$, and recall the norm $||\,\cdot\,||_{\bm E}$ defined in \eqref{eq_E_norm}. Let $||\,\cdot\,||_{2}$ denote the matrix $2-$norm.  Recall from the verification of Assumption \ref{assumption3.2_DGKR}$(ii)$ that $g_{s}(\bm y,\bm w,\theta) \in [0, \overline{c}]$. Conclude that $||\bm G(\bm w, \theta)||_{2} \leq  \sqrt{J} ||\bm G(\bm w, \theta)||_{\infty}\leq \sqrt{J} (S+1) \bar{c}$. Thus:
\begin{align*}
E_{P}[||\bm m(\bm Y_{i},\bm W_{i},\vartheta_{n},\bm r) - \bm m(\bm Y_{i},\bm W_{i},\vartheta_{n},\bm r')||^{2}]&= E_{P}[||\bm G(\bm W_{i},\vartheta_{n}) \bm r(\bm W_{i}) - \bm G(\bm W_{i},\vartheta_{n}) \bm r'(\bm W_{i})||^2]\\
&\leq E_{P}[||\bm G(\bm W_{i},\vartheta_{n})||_{2}^2 \cdot || \bm r(\bm W_{i}) - \bm r'(\bm W_{i})||^{2}]\\
&\leq J (S+1)^2 \bar{c}^2 \sup_{P\in \mathcal{P}} E_{P}[|| \bm r(\bm W_{i}) - \bm r'(\bm W_{i})||^{2}] \\
&= J (S+1)^2 \bar{c}^2 ||(\vartheta_{n},\bm r)- (\vartheta_{n},\bm r')||_{\bm E}^2.
\end{align*}
Thus, Assumption \ref{assumption3.3_DGKR}$(ii)$ is verified with $K_{m}^2=J (S+1)^2 \bar{c}^2$ and with $\kappa_{m}=1$.

Finally, note that for any $(\theta, \bm r) \in \mathcal{I}^*(P)$ we have $||E_{P}[\bm m(\bm Y_{i},\bm W_{i}, \theta, \bm r) \otimes  \bm q^{k_n}( \bm W_{i})]|| = 0$. Thus, by the triangle inequality and the definition of the moment functions from \eqref{eq_mom_definition}, to show that Assumption \ref{assumption3.6_DGKR}$(ii)$ is satisfied, it suffices to show that for every $P_n\in \mathcal{P}$ and $(\theta_n, r_n) \in \mathcal{I}_n^*(P_n)$ there exists a corresponding $(\theta_{n}^*, r_{n}^*) \in \mathcal{I}^*(P_n)$ such that:
\begin{align*}
 \sqrt{n}||E_{P_n}[( \bm G( \bm W_{i}, \theta_n^*) \bm r_n^*(\bm W_{i})- \bm G( \bm W_{i}, \theta_{n}) \bm r_{n}(\bm W_{i}))\otimes  \bm q^{k_n}( \bm W_{i})]|| = o(a_n).
\end{align*} 
Let $(\theta_{n}^*, r_{n}^*)$ be any element satisfying $(\theta_n, r_n) = \Pi_n(\theta_{n}^*, r_{n}^*)$, where $\Pi_{n}:\Theta\times \mathcal{R} \to \Theta\times \mathcal{R}_n$ is the operator that returns the closest (in the norm $||\,\cdot\,||_{\bm E}$ from \eqref{eq_E_norm}) element $(\theta',\bm r')\in\Theta\times \mathcal{R}_n$ in the sieve space to a given pair $(\theta,\bm r) \in \Theta \times \mathcal{R}$. Since there is no sieve on $\Theta$, by definition of $||\,\cdot\,||_{\bm E}$ from \eqref{eq_E_norm}, we must have $\theta_n^* = \theta_n$. Thus:
\begin{align*}
||E_{P_n}[( \bm G(& \bm W_{i}, \theta_n^*) \bm r_n^*(\bm W_{i})- \bm G( \bm W_{i}, \theta_{n}) \bm r_{n}(\bm W_{i}))\otimes  \bm q^{k_n}( \bm W_{i})]||^2\\
&=\sum_{k=1}^{k_n} \sum_{j=1}^{J} (E_{P_n}[\bm g(\bm y_j, \bm W_{i}, \theta_n^*)^\top (\bm r_n^*(\bm W_{i})- \bm r_{n}(\bm W_{i}))q_k( \bm W_{i})])^2\\
&\overset{(1)}{\leq} \sum_{k=1}^{k_n} \sum_{j=1}^{J} E_{P_n}[(\bm g(\bm y_j, \bm W_{i}, \theta_n^*)^\top (\bm r_n^*(\bm W_{i})- \bm r_{n}(\bm W_{i})))^2] E_{P_n}[q_k( \bm W_{i})^2 ]\\
&= \sum_{k=1}^{k_n} E_{P_n}[q_k( \bm W_{i})^2]\sum_{j=1}^{J} E_{P_n}[(\bm g(\bm y_j,\bm W_{i}, \theta_n^*)^\top (\bm r_n^*(\bm W_{i})- \bm r_{n}(\bm W_{i})))^2]\\
&\overset{(2)}{\leq} \sum_{j=1}^{J} E_{P_n}[(\bm g(\bm y_j, \bm W_{i}, \theta_n^*)^\top (\bm r_n^*(\bm W_{i})- \bm r_{n}(\bm W_{i})))^2]\\
&\overset{(3)}{\leq} \bar{c} \cdot J \cdot E_{P_n}[(\bm 1^\top (\bm r_n^*(\bm W_{i})- \bm r_{n}(\bm W_{i})))^2]\\
&= \bar{c} \cdot J \cdot  \sum_{j=1}^{J} \sum_{j'=1}^{J} E_{P_n}\left[(\bm r_{j,n}^*(\bm W_{i})- \bm r_{j,n}(\bm W_{i}))(\bm r_{j',n}^*(\bm W_{i})- \bm r_{j',n}(\bm W_{i}))\right]\\
&\overset{(4)}{\leq} \bar{c} \cdot J \cdot  \sum_{j=1}^{J} \sum_{j'=1}^{J} \left(E_{P_n}\left[(\bm r_{j,n}^*(\bm W_{i})- \bm r_{j,n}(\bm W_{i}))^2\right]\right)^{1/2} \left(E_{P_n}\left[(\bm r_{j',n}^*(\bm W_{i})- \bm r_{j',n}(\bm W_{i}))^2\right]\right)^{1/2}\\
&= \bar{c} \cdot J \cdot  \left( \sum_{j=1}^{J} \left(E_{P_n}\left[(\bm r_{j,n}^*(\bm W_{i})- \bm r_{j,n}(\bm W_{i}))^2\right]\right)^{1/2}\right)^2\\
&\overset{(5)}{\leq} \bar{c} \cdot J^2 \cdot \sum_{j=1}^{J} E_{P_n}\left[(\bm r_{j,n}^*(\bm W_{i})- \bm r_{j,n}(\bm W_{i}))^2\right]\\
&= \bar{c} \cdot J^2 \cdot  E_{P_n}\left[||\bm r_{n}^*(\bm W_{i})- \bm r_{n}(\bm W_{i})||^2\right] = \bar{G} \cdot J^2 \cdot  ||(\theta_n^*,\bm r_n^*) - \Pi_{n}(\theta_n^*,\bm r_n^*) ||_{\bm E}^2. 
\end{align*}
Here $(1)$ and $(4)$ follow from the Cauchy-Schwarz inequality, $(2)$ follows from the fact that the indicators $q_{k}(\bm W_{i}) = 1\{\bm W_{i} \in D_k\}$ are nonnegative and sum to 1 across $1\leq k\leq k_n$, $(3)$ follows from the fact that each element of $\bm g(\bm y_j,\bm W_{i},\theta_n^*)$ is bounded when $\mathcal{W}$ and $\Theta$ are compact (as is the case under Assumption \ref{assumption_consistency}$(ii)$ and $(iii)$), and $(5)$ follows from Jensen's inequality. Thus, Assumption \ref{assumption3.6_DGKR}$(ii)$ follows from Assumption \ref{assumption_consistency}$(v)$ with $a_n= (\log(n))^{-\beta_a}$. 

For the remainder of the proof, we suppose that Assumption \ref{assumption_inference} also holds. 
For Assumption \ref{assumption3.4_DGKR}$(i)$, we use a strategy similar to Lemma S.4.1 in CNS. Note that every $\bm r \in \mathcal{R}_{n}$ is of the form:
\begin{align*}
\bm r(\bm w) =
\begin{bmatrix}
\sum_{D \in \mathcal{D}_{l_n}} \delta_{D,0}\cdot 1\{\bm w \in  D\}\\
\sum_{D \in \mathcal{D}_{l_n}} \delta_{D,1}\cdot 1\{\bm w \in  D\}\\
\vdots\\
\sum_{D \in \mathcal{D}_{l_n}} \delta_{D,S}\cdot 1\{\bm w \in  D\}
\end{bmatrix}
=
\begin{bmatrix}
\bm d^{l_n}(\bm w)^\top \bm \delta_0\\
\bm d^{l_n}(\bm w)^\top \bm \delta_1\\
\vdots\\
\bm d^{l_n}(\bm w)^\top \bm \delta_S
\end{bmatrix} = (\bm I_{S+1} \otimes \bm d^{l_n}(\bm w)^\top) \bm \delta,
\end{align*}
where $\bm I_{S+1}$ is the $(S+1)\times (S+1)$ identity matrix, $\bm d^{l_n}(\bm w)^\top =\left[1\{\bm w \in  D_1\}, \ldots,  1\{\bm w \in  D_{l_n}\} \right]$ and:
\begin{align*}
\bm \delta_{s}^\top =
\begin{bmatrix}
\delta_{D_{1},s}&
\delta_{D_{2},s}&
\ldots &
\delta_{D_{l_{n}},s}
\end{bmatrix},
&&\bm \delta^\top =
\begin{bmatrix}
\bm \delta_{0}^\top&
\bm \delta_{1}^\top&
\ldots&
\bm \delta_{S}^\top
\end{bmatrix}.
\end{align*}
Recall the norm $||\,\cdot\,||_{\bm E}$ from \eqref{eq_E_norm}, let $\mathcal{V}_{n}(P) = (\Theta \times \mathcal{R}_n)\cap \mathcal{S}(\vartheta_{n})$ for $\vartheta_n \in \Theta_I(P)$, and for any $(\vartheta_{n}, \bm r) \in  \mathcal{V}_{n}(P)$ let $\Pi_{n,P}^* (\vartheta_{n},\bm r)$ denote its projection on $\mathcal{I}_{n}^*(\vartheta_{n},P)$ in the norm $||\,\cdot\,||_{\bm E}$ from \eqref{eq_E_norm}. Abusing notation, let $\Pi_{n,P}^* \bm r \in \mathcal{R}_{n}$ be the corresponding element of $\mathcal{R}_{n}$ (which is linked to $\vartheta_n$). Furthermore, suppose $\bm r \in \mathcal{R}_{n}$ and $\Pi_{n,P}^{*} \bm r \in \mathcal{R}_{n}$ have coefficients $\{\bm \delta_s(\bm r)\}_{s=0}^{S}$ and $\{\bm \delta_s(\Pi_{n,P}^* \bm r)\}_{s=0}^{S}$, and recall the matrices $\bm M_{n,P}^{(1)}(\vartheta_n)$, $\bm M_{n,P}^{(2)}$, and $\bm M_{n,P}^{(3)}(\vartheta_n)$ from the main text. 
Now note Assumption \ref{assumption3.4_DGKR}$(i)$ holds trivially if $\bm r = \Pi_{n,P}^{*} \bm r$. For any $(\vartheta_{n},\bm r) \in \mathcal{V}_{n}(P)$ with $\bm r \neq \Pi_{n,P}^{*} \bm r$, for all $n$ sufficiently large we have:
\begin{align*}
&\vec{d}_{H}\left((\vartheta_{n},\bm r),\mathcal{I}_{n}^*(\vartheta_{n},P),||\,\cdot\,||_{\bm E}\right)\\
&=||(\vartheta_{n},\bm r) - \Pi_{n,P}^* (\vartheta_{n},\bm r)||_{\bm E}\\
&=\sup_{Q\in \mathcal{P}} (E_{Q}[||\bm r(\bm W_{i}) - \Pi_{n,P}^* \bm r(\bm W_{i}) ||^2])^{1/2}\\
&=\sup_{Q\in \mathcal{P}} (E_{Q}[||(\bm I_{S+1} \otimes \bm d^{l_n}(\bm W_{i})^\top) (\bm \delta(\bm r) - \bm \delta(\Pi_{n,P}^* \bm r))||^2])^{1/2}\\
&=\sup_{Q\in \mathcal{P}} \left(\sum_{s=0}^{S} (E_{Q}[(\bm d^{l_n}(\bm W_{i})^\top (\bm \delta_s(\bm r) - \bm \delta_s(\Pi_{n,P}^* \bm r)))^2]\right)^{1/2}\\
&=\sup_{Q\in \mathcal{P}} \left(\sum_{s=0}^{S} (\bm \delta_s(\bm r) - \bm \delta_s(\Pi_{n,P}^* \bm r))^\top \bm M_{n,Q}^{(2)} (\bm \delta_s(\bm r) - \bm \delta_s(\Pi_{n,P}^* \bm r))\right)^{1/2}\\
&= \sup_{Q\in \mathcal{P}} \left((\bm \delta(\bm r)  - \bm \delta(\Pi_{n,P}^* \bm r))^\top ( \bm I_{S+1} \otimes \bm M_{n,Q}^{(2)}) (\bm \delta(\bm r)  - \bm \delta(\Pi_{n,P}^* \bm r))\right)^{1/2} \\
&= \sup_{Q\in \mathcal{P}} \left((\bm \delta(\bm r)  - \bm \delta(\Pi_{n,P}^* \bm r))^\top ( \bm I_{S+1} \otimes \bm M_{n,Q}^{(2)} - c_1 k_{n} (\bm M_{n,Q}^{(1)})^\top \bm M_{n,Q}^{(1)}) (\bm \delta(\bm r)  - \bm \delta(\Pi_{n,P}^* \bm r)) \right.\\
&\qquad\qquad\qquad\qquad\qquad\qquad\left.+ (\bm \delta(\bm r)  - \bm \delta(\Pi_{n,P}^* \bm r))^\top ( c_1 k_{n}(\bm M_{n,Q}^{(1)})^\top \bm M_{n,Q}^{(1)}) (\bm \delta(\bm r)  - \bm \delta(\Pi_{n,P}^* \bm r)) \right)^{1/2} \\
&\leq\sup_{Q\in \mathcal{P}}  c_1 k_{n}^{1/2} \left((\bm \delta(\bm r)  - \bm \delta(\Pi_{n,P}^* \bm r))^\top ((\bm M_{n,Q}^{(1)})^\top \bm M_{n,Q}^{(1)}) (\bm \delta(\bm r)  - \bm \delta(\Pi_{n,P}^* \bm r)) \right)^{1/2} \\
&=\sup_{Q\in \mathcal{P}}  c_1 k_{n}^{1/2} \frac{||\bm M_{n,Q}^{(1)}(\vartheta_n)(\bm \delta(\bm r)  - \bm \delta(\Pi_{n,P}^* \bm r))||}{||\bm M_{n,P}^{(1)}(\vartheta_n)(\bm \delta(\bm r)  - \bm \delta(\Pi_{n,P}^* \bm r))||}{||\bm M_{n,P}^{(1)}(\vartheta_n)(\bm \delta(\bm r)  - \bm \delta(\Pi_{n,P}^* \bm r))||} \\
&\leq   c_1 c_2 k_{n}^{1/2} ||\bm M_{n,P}^{(1)}(\vartheta_n)(\bm \delta(\bm r)  - \bm \delta(\Pi_{n,P}^* \bm r))||\\
&= c_1 c_2 k_{n}^{1/2} \left|\left|E_{P}\left[ (\bm q^{k_n}(\bm W_{i}) \otimes  \bm G(\bm W_{i}, \vartheta_{n}))(\bm I_{S+1}\otimes \bm d^{l_n}(\bm W_{i})^\top )(\bm \delta(\bm r)  - \bm \delta(\Pi_{n,P}^* \bm r))\right] \right| \right|\\
&= c_1 c_2 k_{n}^{1/2} \left|\left|E_{P}\left[ (\bm q^{k_n}(\bm W_{i}) \otimes  \bm G(\bm W_{i}, \vartheta_{n}))(\bm r(\bm W_{i})-\Pi_{n,P}^* \bm r(\bm W_{i}) )\right] \right| \right|\\
&\leq \sup_{(\vartheta_{n},\tilde{\bm r})\in \mathcal{I}_{n}^*(\vartheta_{n},P)}c_1 c_2 k_{n}^{1/2} \left|\left|E_{P}\left[ (\bm q^{k_n}(\bm W_{i}) \otimes  \bm G(\bm W_{i}, \vartheta_{n}))(\bm r(\bm W_{i})-\tilde{\bm r}(\bm W_{i}))\right] \right| \right|,
\end{align*}
where the first inequality follows from Assumption \ref{assumption_inference}$(i)$ (which implies that $\bm I_{S+1} \otimes \bm M_{n,Q}^{(2)} - c_1 k_{n} (\bm M_{n,Q}^{(1)})^\top \bm M_{n,Q}^{(1)}$ is negative semidefinite), and the second inequality follows from Assumption \ref{assumption_inference}$(ii)$. This verifies Assumption \ref{assumption3.4_DGKR}$(i)$ with $\nu_n\asymp k_n^{1/2}$. Assumption \ref{assumption3.4_DGKR}$(ii)$ now holds trivially with our choice of $\mathcal{V}_{n}(P)$. 

For Assumption \ref{assumption3.6_DGKR}$(i)$, recall $R_{n}$ from \eqref{eq_capital_Rn}, and note that from the calculation above:
\begin{align*}
J_{[\,]}(R_{n}\vee \nu_n\tau_{n},\mathcal{F}_{n},||\,\cdot\,||_{P,2})
&=O( (R_{n}\vee \nu_n\tau_{n}) \sqrt{l_n}).
\end{align*}
Since $J_n = O(\sqrt{l_n})$ and $l_n\leq k_n$, we have $J_n = O(\sqrt{k_n})$ and $R_n = O(\nu_n k_n \sqrt{\log(1+k_n)/n})$. Also, since $k_n=O(n^{\beta_k})$, we have $R_n = O(\nu_n n^{\beta_k} \sqrt{\log(n)/n})$. By the statement of the theorem, $\tau_n = O(n^{-\beta_{\tau}})$ with $\beta_{\tau}<\frac{1}{2} - \beta_k$. Conclude that $R_n = o(\nu_n\tau_n)$. Thus, if $\nu_n \asymp k_n^{1/2}$, Assumption \ref{assumption3.6_DGKR}$(i)$ is satisfied for any $a_n$ satisfying $k_n \sqrt{\log(1+k_n)} \sqrt{l_n} \tau_{n}  = o(a_n)$. Since $l_n\leq k_n$, $k_n=O(n^{\beta_k})$, and $\tau_n=O(n^{-\beta_\tau})$, this is certainly satisfied for any $a_n$ satisfying $n^{\frac{3\beta_{k}}{2}-\beta_{\tau}} \sqrt{\log(n)} = o(a_n)$. But since $\frac{3\beta_{k}}{2}<\beta_{\tau}$ by the condition in the statement of the theorem, this is certainly satisfied by any sequence satisfying $a_n = o((\log(n))^{-\beta_a})$ for some $\beta_a>0$. 

To verify Assumption \ref{assumption3.11_DGKR}, we rely on Theorem S.7.1 in CNS. Thus, we focus on verifying Assumptions S.7.1 and S.7.2 in CNS, and follow a strategy similar to Lemma S.4.8 in CNS. Consider the array of vector-valued functions $\bm f_{n,P}^{d_{n}}(\bm y, \bm w) = \bm b_{n,j}(\bm y, \bm w,\vartheta_n) - E_{P}[\bm b_{n,j}(\bm y, \bm w,\vartheta_n)]$ for $\vartheta_n \in \Theta_{I}(P)$ with dimension $d_{n} = k_n+\tilde{k}_n$, where $\bm b_{n,j}(\bm y, \bm w, \vartheta_n)$ is from \eqref{eq_bnj}. Since $l_n \asymp k_n$ by the statement of the theorem, and since the partitions that determine the instruments and piecewise constant functions are eventually nested by Assumption \ref{assumption_consistency}$(iv)$, we have $\tilde{k}_{n} \asymp k_n$. In addition, by Assumption \ref{assumption_inference}$(iii)$, the eigenvalues of the covariance matrix $E_{P}[\bm f_{n,P}^{d_{n}}(\bm Y_{i}, \bm W_{i})\bm f_{n,P}^{d_{n}}(\bm Y_{i}, \bm W_{i})^\top ]$ are bounded away from infinity uniformly in $P \in \mathcal{P}$, $\vartheta_n \in \Theta_{I}(P)$, and $n$. This verifies Assumption S.7.1$(i)$ in CNS. Furthermore, note that $\sup_{P \in \mathcal{P}} \max_{1\leq d \leq d_{n}} ||f_{d,n,P}||_{\infty} \leq 2(\bar{c} \vee 1)$, so that Assumption S.7.1$(ii)$ in CNS is also satisfied. Now note that CNS Assumption S.7.2$(i)$ is trivially satisfied by the choice of $\mathcal{F}_{n}$, so we can take $G_{n,P}=0$ and $J_{1n}=1$ (in the notation of CNS). Finally, consider the set:
\begin{align*}
\mathcal{B}_{n}:= \left\{ \beta \in \mathbb{R}_{+}^{d_n} : \beta^\top = (e_k^\top, \gamma^\top) \text{ for some }1\leq k \leq k_n,\,\, \gamma \in \Gamma^{\tilde{k}_n} \right\}, && \Gamma^{\tilde{k}_n}:=\left\{ \gamma \in \mathbb{R}_{+}^{\tilde{k}_n} : ||\gamma||_{\infty}\leq \overline{c}\vee 1 \right\},
\end{align*}
where $e_k$ is a $k_n\times 1$ vector of zeros with a $1$ in the $k^{th}$ position. Let $\mathbb{B}_p^d$ denote the unit $||\,\cdot\,||_{p}-$ball in $d-$dimensions, and without loss of generality assume $\bar{c} \geq 1$ in the following derivation. From \cite{wainwright2019high} Lemma 5.7 we have:
\begin{align*}
N(\epsilon,[0,\bar{c}]^d,||\,\cdot\,||) 
\leq N(\epsilon/\bar{c},[-1,1]^d,||\,\cdot\,||)  \leq \left(\frac{2\bar{c}}{\epsilon}+1\right)^d \frac{1}{\text{vol}(\mathbb{B}_2^d)} = \left(\frac{2\bar{c}}{\epsilon}+1\right)^d  \frac{\Gamma( \frac{d}{2}+1)}{\pi^{d/2}}.
\end{align*}
Then we have:
\begin{align*}
J_{2n} &:= \int_{0}^{\infty} \sqrt{\log (N(\epsilon,\mathcal{B}_{n},||\,\cdot\,||))} \,d\epsilon\\
&\leq \int_{0}^{\infty} \sqrt{\log (k_n N(\epsilon,\Gamma^{\tilde{k}_n},||\,\cdot\,||))} \,d\epsilon\leq \sqrt{\tilde{k}_n\log(k_n)}\bar{c} + \int_{0}^{\bar{c}\sqrt{\tilde{k}_n}} \sqrt{\log (N(\epsilon,[0,\bar{c}]^{\tilde{k}_n},||\,\cdot\,||))} \,d\epsilon\\
&\leq \bar{c}\sqrt{\tilde{k}_n\log(k_n)} + \bar{c}\sqrt{\tilde{k}_n}\sqrt{\log\left(\frac{\Gamma( \frac{\tilde{k}_n}{2}+1)}{\pi^{\tilde{k}_n/2}}\right)} + \sqrt{\tilde{k}_n} \int_{0}^{\bar{c}\sqrt{\tilde{k}_n}} \sqrt{\log\left(\frac{2\bar{c}}{\epsilon} +1 \right)} \,d\epsilon.
\end{align*}
Furthermore:
\begin{align*}
\int_{0}^{\bar{c}\sqrt{\tilde{k}_n}} \sqrt{\log\left(\frac{2\bar{c}}{\epsilon} +1 \right)} \,d\epsilon
\leq 4\bar{c}+  \bar{c}  \sqrt{\tilde{k}_n\log(2)}.
\end{align*}
Also, by \cite{batir2008inequalities} Theorem 1.4, we have $\Gamma(x+1)\leq \beta_o^{-\beta_o} e^{-x} (x+\beta_o)^{x+\beta_o}$ for $\beta_o=e^{-\gamma} = 0.56146...$, where $\gamma$ is Euler's constant. Thus:
\begin{align*}
\bar{c}\sqrt{\tilde{k}_n} \sqrt{\log\left(\frac{\Gamma( \frac{\tilde{k}_n}{2}+1)}{\pi^{\tilde{k}_n/2}}\right)} 
\leq \bar{c}\sqrt{\tilde{k}_n} \sqrt{\left(\frac{\tilde{k}_n}{2}+1\right)\log\left(\frac{\tilde{k}_n}{2}+1\right) }.
\end{align*}
Thus we have:
\begin{align*}
 J_{2n} \leq \bar{c}\sqrt{\tilde{k}_n\log(k_n)} +\bar{c}\sqrt{\tilde{k}_n} \sqrt{\left(\frac{\tilde{k}_n}{2}+1\right)\log\left(\frac{\tilde{k}_n}{2}+1\right) } +4\bar{c}+  \bar{c}  \sqrt{\tilde{k}_n\log(2)} =O(\tilde{k}_n\sqrt{\log(\tilde{k}_n)}).
\end{align*} 
This verifies Assumption S.7.2$(ii)$ in CNS. From Assumption \ref{assumption_inference}$(iii)$, it follows from CNS Theorem 2.7.1$(ii)$ that, since $\sqrt{d_n\log(1+d_n)}/\sqrt{n} = o(1)$, there exists a linear Gaussian $\mathbb{G}_{P}^\star$, possibly depending on $n$, such that:
\begin{align*}
&\sup_{f \in \mathcal{F}_{n}} \left|\left| \frac{1}{n} \sum_{i=1}^{n} \xi_{i}\left(f(\bm Y_{i},\bm W_{i})\bm q^{k_n}(\bm W_{i}) - E_{P}[f(\bm Y_{i},\bm W_{i})\bm q^{k_n}(\bm W_{i})]\right) - \mathbb{G}_{P}^\star (f\bm q^{k_n}) \right|\right|= O_{P} \left(\frac{k_n^2 \log(k_n)}{\sqrt{n}}  \right),
\end{align*}
where here we have used the fact that $\tilde{k}_n\asymp k_n$ as a consequence of Assumption \ref{assumption_consistency}$(iv)$. In CNS's notation, we set $d_{n} = k_n + \tilde{k}_n \asymp  k_{n}$, $J_{1n}=1$, $J_{2n} = O(k_{n} \sqrt{\log(k_n)})$, $\xi_n = k_n$ (by Assumption \ref{assumption_inference}$(iii)$) and $K_{n}$ and $C_{n}$ as constants. This verifies Assumption \ref{assumption3.11_DGKR} for any $a_n$ satisfying  $k_n^{2} \log(k_n)/\sqrt{n} = o(a_n)$. Since $k_n = O(n^{\beta_k})$, this is satisfied for any $a_n$ satisfying  $n^{2\beta_{k}} \log(n)/\sqrt{n} = o(a_n)$. Since $\beta_{k}<\frac{1}{5}$ by the conditions from the statement of the theorem, this is certainly satisfied for any sequence $a_n$ satisfying $a_n = o((\log(n))^{-\beta_a})$ for some $\beta_a>0$. 

Finally, note that Assumption \ref{assumption3.12_DGKR} holds trivially with our choice of $\mathcal{V}_{n}(P)$ above.

\end{proof}

\subsection{An Alternative to Assumption \ref{assumption_consistency2}}\label{section_alternative}

We present an alternative assumption that is sufficient for Assumption \ref{assumption_consistency2} in the main text. The alternative assumption does not depend on the sieve space $\mathcal{R}_n$, and instead emphasizes the role of the instruments. Recall the partition $\mathcal{D}_{k_n}$ of $\mathcal{W}$ introduced in Section \ref{section_consistency} and Assumption \ref{assumption_consistency}(iv). 
\begin{assumption}
\label{ass:cond-sep}
For some constants $\delta_0 > 0$, $\nu_0>0$, and $c_0>0$,  and for every $\varepsilon > 0$, $P \in \mathcal{P}$, and $\theta$ satisfying $d(\theta,\Theta_{I}(P)) \geq \varepsilon$, there exists an integer $N(\varepsilon)\geq 1$ and a set $B_n = B_n(P,\theta,\varepsilon)$ with $B_n \in \sigma(\mathcal{D}_{k_n})$ such that $P(\bm W_i \in B_n) \;\geq\; c_0$ and:
\begin{align*}
\max_{j = 1,\ldots,J} \;
\inf_{r \in \Pi_{\mathcal{R}}(\mathcal{S})}
\left|
E_{P}\!\left[ m_j(\bm Y_i,\bm W_i,\theta,\bm r)
\;\middle|\; \bm W_i \in B_n \right]
\right|
\;\geq\; \nu_{0}^{-1} \min\{\delta_0,\,\varepsilon\},
\end{align*}
for all $n\geq N(\varepsilon)$. 
\end{assumption}
\begin{remark}
The key set $B_n$ in Assumption \ref{ass:cond-sep} must be constructed from the sets in the partition $\mathcal{D}_{k_n}$ used in the instrument functions. Under the condition that $\lim_{n\to \infty}\max_{D \in \mathcal{D}_{k_n}} \text{diam}(D) = 0$, the $\sigma-$algebra $\sigma(\mathcal{D}_{k_n})$ converges to the Borel $\sigma-$algebra on $\mathcal{W}$; that is, the Borel $\sigma-$algebra on $\mathcal{W}$ is the limit of the filtration defined by the $\sigma-$algebras $\sigma(\mathcal{D}_{k_n})$, making this condition quite flexible for many choices of partitions $\mathcal{D}_{k_n}$ and instrument functions.  
\end{remark}

The following Lemma shows that Assumptions \ref{assumption_main}, \ref{assumption_consistency}, and \ref{ass:cond-sep} imply Assumption \ref{assumption_consistency2} when $k_{n}$ satisfies the rate conditions in Theorem \ref{theorem_consistency}. 
\begin{lemma}\label{theorem:label}
Suppose that Assumptions \ref{assumption_main}, \ref{assumption_consistency}, and \ref{ass:cond-sep} hold, and that $k_{n}$ satisfies the rate conditions in Theorem \ref{theorem_consistency}. Then Assumption \ref{assumption_consistency2} holds for $\nu_n \asymp k_n^{1/2}$.
\end{lemma}
\begin{proof}[Proof of Lemma \ref{theorem:label}]
By Assumption \ref{assumption_consistency}(iv), it is without loss of generality to assume that the sequence of partitions is $\mathcal{D}_{k_n}$ nested for all $n$. Assumption \ref{assumption_consistency2} holds trivially if $\theta \in \Theta_{I}(P)$, so consider the case when $d(\theta,\Theta_{I}(P)) \geq \varepsilon$ for some $\varepsilon>0$. By Assumption \ref{ass:cond-sep}, there exist constants $\delta_0 > 0$, $c_0>0$, and $\nu_{0} > 0$ such that, for every $\varepsilon > 0$, $P \in \mathcal{P}$, and $\theta$ satisfying $d(\theta,\Theta_{I}(P)) \geq \varepsilon$, there exists an integer $N(\varepsilon)\geq 1$ and a set $B_n = B_n(P,\theta,\varepsilon)$ with $B_n \in \sigma(\mathcal{D}_{k_n})$ such that $P(\bm W_i \in B_n) \;\geq\; c_0$ and:
\begin{align*}
\max_{j = 1,\ldots,J} \;
\inf_{\bm r \in \Pi_{\mathcal{R}}(\mathcal{S})}
\left|
E_{P}\!\left[ m_j(\bm Y_i,\bm W_i,\theta, \bm r)
\;\middle|\; \bm W_i \in B_n \right]
\right|
\;\geq\; \nu_{0}^{-1} \min\{\delta_0,\,\varepsilon\},
\end{align*}
for all $n\geq N(\varepsilon)$. This implies:
\begin{align*}
 \max_{j = 1,\ldots,J} \; \inf_{\bm r \in \Pi_{\mathcal{R}}(\mathcal{S})} \left|
E_{P}\!\left[ m_j(\bm Y_i,\bm W_i,\theta,\bm r) 1\{\bm W_i \in B_n\} \right]
\right| \geq c_0\nu_{0}^{-1} \min\{\delta_0,\,\varepsilon\}.
\end{align*} 
Now since $B_n \in \sigma(\mathcal{D}_{k_n})$ for $n\geq N(\varepsilon)$, $B_n$ can be written as a disjoint union of sets $\{D_{k,n}\}_{k=1}^{k_n^B}\subset \mathcal{D}_{k_n}$ for $n\geq N(\varepsilon)$. Thus:
\begin{align*}
&\max_{j = 1,\ldots,J} \; \inf_{\bm r \in \Pi_{\mathcal{R}}(\mathcal{S})} \left|
E_{P}\!\left[ m_j(\bm Y_i,\bm W_i,\theta,\bm r) 1\{\bm W_i \in B_n\} \right]
\right|\\ 
&\qquad\leq \inf_{\bm r \in \Pi_{\mathcal{R}}(\mathcal{S})} \max_{j = 1,\ldots,J} \;  \left|
E_{P}\!\left[ m_j(\bm Y_i,\bm W_i,\theta,\bm r) 1\{\bm W_i \in B_n\} \right] \right|\\
&\qquad\leq \inf_{\bm r \in \Pi_{\mathcal{R}}(\mathcal{S})} \left(\sum_{j=1}^{J}  (
E_{P}\!\left[ m_j(\bm Y_i,\bm W_i,\theta,\bm r) 1\{\bm W_i \in B_n\} \right])^2\right)^{1/2}
\\
&\qquad= \inf_{\bm r \in \Pi_{\mathcal{R}}(\mathcal{S})} \left(\sum_{j=1}^{J}  \left(\sum_{k=1}^{k_n^B}
E_{P}\!\left[ m_j(\bm Y_i,\bm W_i,\theta,\bm r) 1\{\bm W_i \in D_{k,n}\} \right]\right)^2\right)^{1/2}
\\
&\qquad\leq (k_{n}^B)^{1/2}\inf_{\bm r \in \Pi_{\mathcal{R}}(\mathcal{S})} \left(\sum_{j=1}^{J}  \sum_{k=1}^{k_n^B}\left(
E_{P}\!\left[ m_j(\bm Y_i,\bm W_i,\theta,\bm r) 1\{\bm W_i \in D_{k,n}\} \right]\right)^2\right)^{1/2}
\\
&\qquad\leq k_{n}^{1/2} \inf_{\bm r \in \Pi_{\mathcal{R}}(\mathcal{S})} Q_{n,P}(\theta,\bm r),
\end{align*}
for all $n$ sufficiently large. Now note:
\begin{align*}
\inf_{\bm r \in \Pi_{\mathcal{R}}(\mathcal{S})} Q_{n,P}(\theta,\bm r)&\leq \inf_{\bm r \in \Pi_{\mathcal{R}_n}(\mathcal{S})} Q_{n,P}(\theta,\bm r) - \inf_{(\theta',\bm r') \in (\Theta\times \mathcal{R}_n)\cap \mathcal{S}}  Q_{n,P}(\theta',\bm r')+b_n,
\end{align*}
where:
$$b_{n}:=\inf_{(\theta',\bm r') \in (\Theta\times \mathcal{R}_n)\cap \mathcal{S}}  Q_{n,P}(\theta',\bm r').$$
Now let $(\theta_n,\bm r_n) \in \mathcal{I}_n^*(P)$, and note that $b_n \leq Q_{n,P}(\theta_n,\bm r_n)$. By Assumption \ref{assumption_consistency}$(v)$, there exists a corresponding $(\theta_{n},\bm r_{n}^*) \in \mathcal{I}^*(P)$ such that $\sqrt{n}(E_{P}\left[||\bm r_{n}^*(\bm W_{i})- \bm r_{n}(\bm W_{i})||^2\right])^{1/2} = o((\log(n))^{-\beta_a})$ for some $\beta_a>0$. For this pair we have:
\begin{align*}
Q_{n,P}(\theta_n,\bm r_n) &= \left|\left| E_{P}[\bm p(\bm W_i) \otimes \bm q^{k_n}(\bm W_i)  ] - E_{P}[\bm G(\bm W_{i},\theta_n)\bm r_{n}(\bm W_{i}) \otimes \bm q^{k_n}(\bm W_i)  ] \right|\right|\\
&= \left|\left| E_{P}[\bm G(\bm W_{i},\theta_n)\bm r_{n}^*(\bm W_{i}) \otimes \bm q^{k_n}(\bm W_i)  ] - E_{P}[\bm G(\bm W_{i},\theta_n)\bm r_{n}(\bm W_{i}) \otimes \bm q^{k_n}(\bm W_i)  ] \right|\right|\\
&\leq  E_{P}[||\bm q^{k_n}(\bm W_i) \otimes \bm G(\bm W_{i},\theta_n)(\bm r_{n}^*(\bm W_{i}) - \bm r_{n}(\bm W_{i}) ) ||] \\
&\leq  E_{P}[||\bm q^{k_n}(\bm W_i) \otimes \bm G(\bm W_{i},\theta_n)||_{2}||\bm r_{n}^*(\bm W_{i}) - \bm r_{n}(\bm W_{i}) || ] \\
&\leq \left(E_{P}[|| \bm q^{k_n}(\bm W_i) \otimes \bm G(\bm W_{i},\theta_n)||_2^2]E_{P}[||\bm r_{n}^*(\bm W_{i}) - \bm r_{n}(\bm W_{i})||^2 ] \right)^{1/2}\\
&= \left(E_{P}[|| \bm G(\bm W_{i},\theta_n)||_2^2]E_{P}[||\bm r_{n}^*(\bm W_{i}) - \bm r_{n}(\bm W_{i})||^2 ] \right)^{1/2}= o\left(\frac{1}{\sqrt{n}\log(n)^{\beta_a}} \right),
\end{align*}
where the first inequality follows from Jensen's inequality, the second follows from a property of the matrix $2-$norm, the third inequality follows the Cauchy-Schwarz inequality, and the final line follows from Assumption \ref{assumption_consistency}$(v)$ and the fact that $\bm G(\bm w,\theta)$ is uniformly bounded under Assumptions \ref{assumption_main} and \ref{assumption_consistency}. Now set $\delta=\delta_0$ and $\nu_{n}  = c_0^{-1} k_{n}^{1/2} \nu_{0}$, and note that $\nu_{n} b_{n} = o(1)$. Combining everything, Assumption \ref{assumption_consistency2} follows. 


\end{proof}

\section{Simulation Exercises}\label{appendix_simulation}

In this section we investigate the power properties of our proposed inference method. We consider four data generating processes (DGPs):
\begin{enumerate}
	\item [] \textbf{DGP1:} AR(1) with $T=2$: $Y_{it} = \mathds{1}\{\alpha_i + \beta Y_{it-1} \geq \epsilon_{it}\}$ for $t=1,2$. 
	\item [] \textbf{DGP2:} AR(1) with $T=3$: $Y_{it} = \mathds{1}\{\alpha_i + \beta Y_{it-1} \geq \epsilon_{it}\}$ for $t=1,2,3$. 
	\item [] \textbf{DGP3:} AR(1), Time Trend, with $T=3$: $Y_{it} = \mathds{1}\{\alpha_i + \beta Y_{it-1} + \gamma t \geq \epsilon_{it}\}$ for $t=1,2,3$. 
	\item [] \textbf{DGP4:} AR(1), Continuous Covariate, with $T=3$: $Y_{it} = \mathds{1}\{\alpha_i + \beta Y_{it-1} + \eta X_{it} \geq \epsilon_{it}\}$ for $t=1,2,3,$ with $X_{it} = \Phi(Z_{it})$ with $\Phi$ the standard normal cdf and $Z_{it} \sim N(\alpha_{i},1)$. 
\end{enumerate}
In each DGP we draw $\epsilon_{it} \overset{i.i.d.}{\sim}\text{Logistic}(0,1)$ and consider two different distributions for $\alpha_{i}$: $(i)$ $\alpha_{i}\overset{i.i.d.}{\sim} N(0,1)$ and $(ii)$ $\alpha_{i} \overset{i.i.d.}{\sim} \text{Uniform}\{-1,-0.8,\ldots,1\}$. We run all simulations for three sample sizes, $n=1,000$, $n=5,000$, and $n=10,000$, using $B=999$ bootstrap iterations. The true values in all DGPs are $(\beta,\gamma,\eta) = (0.5,0.8,-0.8)$, and in all simulations we set $\tilde{\tau}_{n}=0$ (see Remark \ref{remark_tau_zero}) and ``recycle'' all optimal solutions from the test statistic during the bootstrap (see the discussion at the end of Section \ref{section_consistency_and_inference}). For each DGP, we simulate $R=500$ samples, and at each point in a fine grid over the parameter space we run $R=500$ hypothesis tests, one for each sample, to test the null hypothesis that the parameter belongs to the identified set. Below we plot the resulting power curves. 


\begin{figure}[t!]
     \centering
     \begin{subfigure}[b]{0.47\textwidth}
         \centering
         \includegraphics[width=\textwidth]{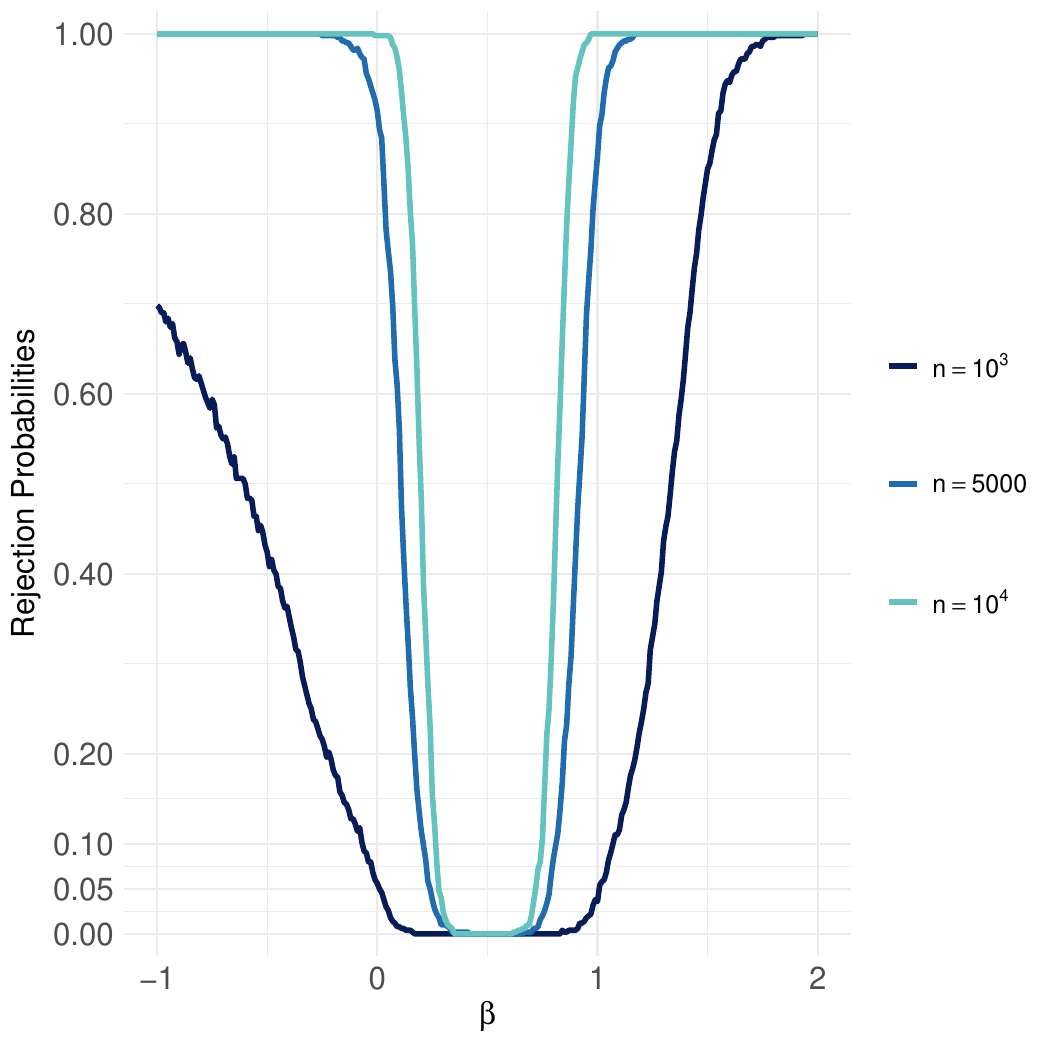}
         \caption{$\alpha_{i}\overset{i.i.d.}{\sim} N(0,1)$.}
         \label{fig_sim_T2_D1}
     \end{subfigure}
     \hfill
     \begin{subfigure}[b]{0.47\textwidth}
         \centering
         \includegraphics[width=\textwidth]{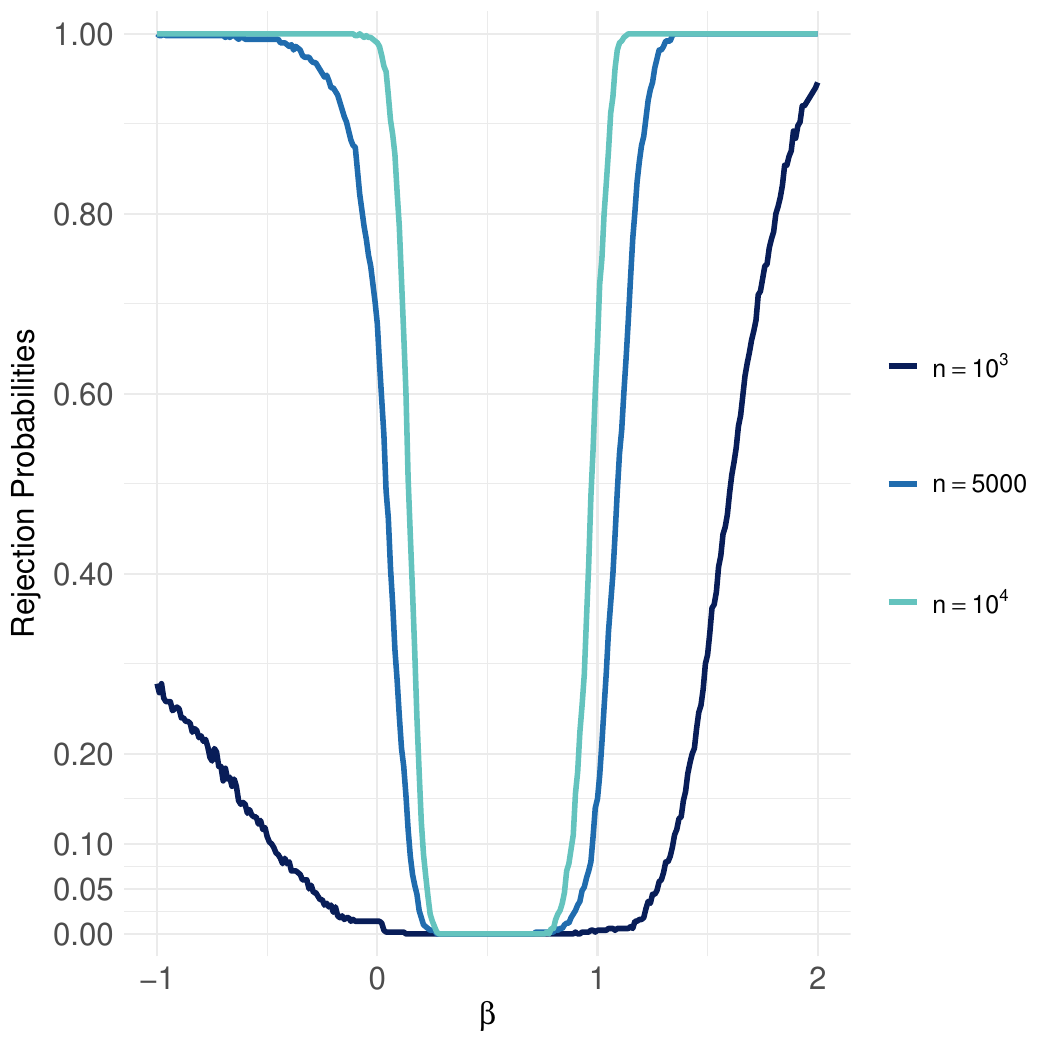}
         \caption{$\alpha_{i}\overset{i.i.d.}{\sim} \text{Uniform}\{-1,-0.8,\ldots,1\}$.}
         \label{fig_sim_T2_D2}
     \end{subfigure}
        \caption{AR(1), $T=2$.}
        \label{fig_sim_T2}
\end{figure}

The power curves for DGP1 are displayed in Figure \ref{fig_sim_T2}. As suggested by the figure, the structural parameter $\beta$ is partially-identified in this DGP. As a result, there are large regions of the parameter space where the rejection probability is exactly zero. The identified set is approximately $[0.43,0.56]$ when $\alpha_{i}\overset{i.i.d.}{\sim} N(0,1)$, and is $[0.41,0.62]$ when $\alpha_{i} \overset{i.i.d.}{\sim} \text{Uniform}\{-1,-0.8,\ldots,1\}$.\footnote{These are computed numerically by simulating from multiple samples of size $n=10^6$, and averaging the resulting end points. } As a result, the confidence sets are wider in Figure \ref{fig_sim_T2}$(b)$ where  $\alpha_{i} \overset{i.i.d.}{\sim} \text{Uniform}\{-1,-0.8,\ldots,1\}$ than in Figure \ref{fig_sim_T2}$(a)$ where $\alpha_{i}\overset{i.i.d.}{\sim} N(0,1)$. However, power increases substantially as the sample size increases. Using our results, we can also calculate the average lower and upper bounds for a $95\%$ confidence interval by computing a $95\%$ confidence interval for each sample, and then averaging the lower bounds and upper bounds across all confidence intervals. For $n=10,000$, the average $95\%$ confidence intervals were $[0.20,0.81]$ for $\alpha_{i}\overset{i.i.d.}{\sim} N(0,1)$ and $[0.14,0.97]$ for $\alpha_{i} \overset{i.i.d.}{\sim} \text{Uniform}\{-1,-0.8,\ldots,1\}$. On average, the computation time required to compute the test statistic (i.e. run a single SDP) in this DGP was $0.0029$ seconds. The average computational cost for each test, including the bootstrap procedure, was $0.040$ seconds.

\begin{figure}[t!]
     \centering
     \begin{subfigure}[b]{0.47\textwidth}
         \centering
         \includegraphics[width=\textwidth]{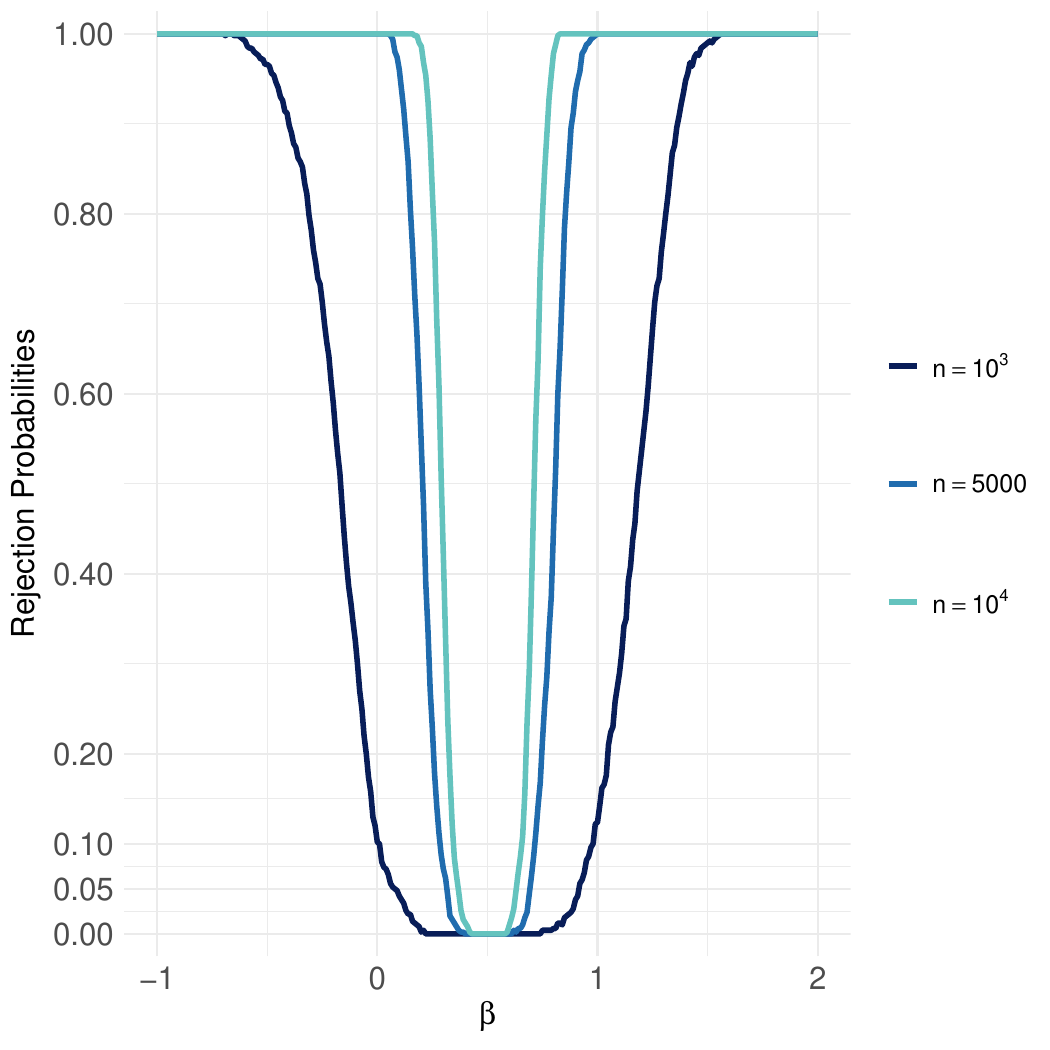}
         \caption{$\alpha_{i}\overset{i.i.d.}{\sim} N(0,1)$.}
         \label{fig_sim_T3_D1}
     \end{subfigure}
     \hfill
     \begin{subfigure}[b]{0.47\textwidth}
         \centering
         \includegraphics[width=\textwidth]{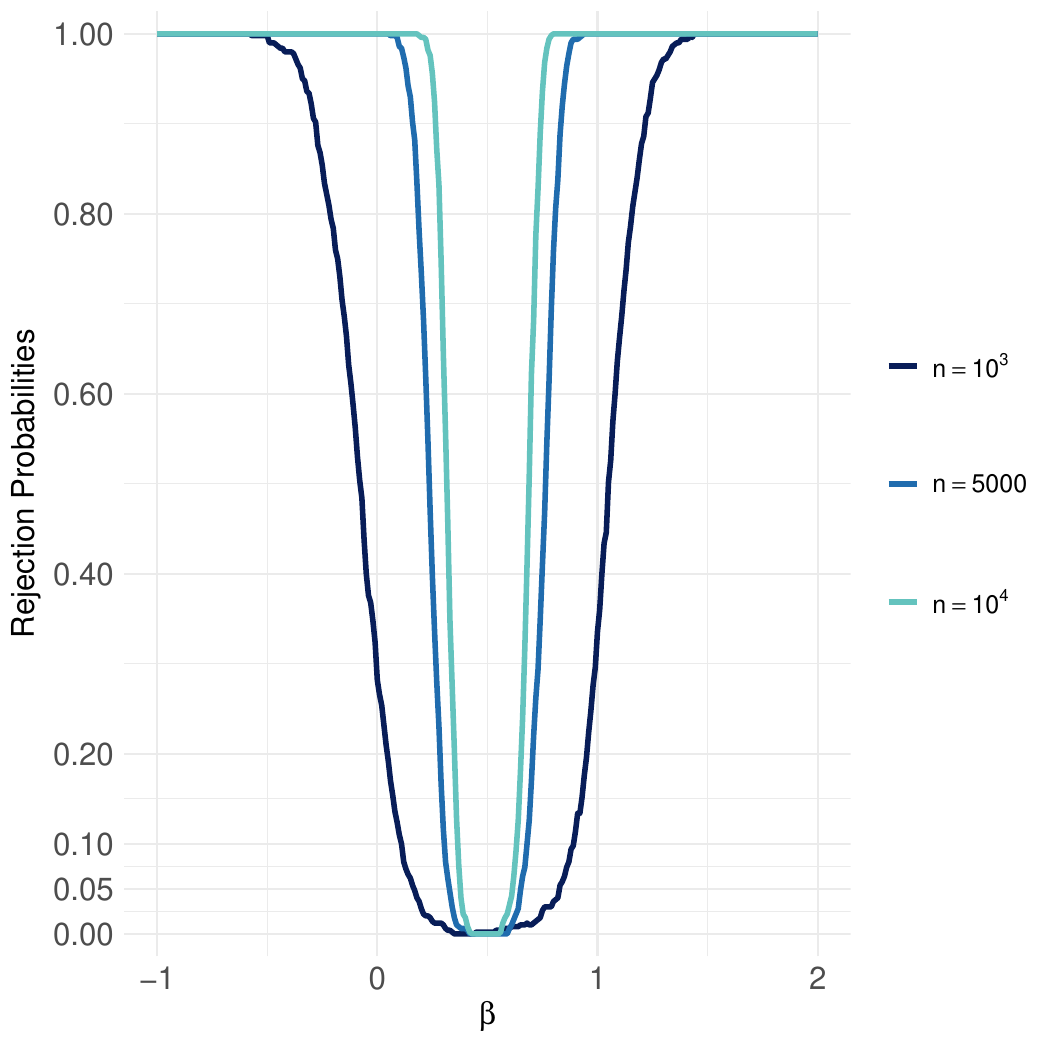}
         \caption{$\alpha_{i}\overset{i.i.d.}{\sim} \text{Uniform}\{-1,-0.8,\ldots,1\}$.}
         \label{fig_sim_T3_D2}
     \end{subfigure}
        \caption{AR(1), $T=3$.}
        \label{fig_sim_T3}
\end{figure}

The results for DGP2 are displayed in Figure \ref{fig_sim_T3}. Unlike DGP1, with $T=3$ the parameter $\beta$ is now point-identified. While the procedure effectively controls size, the fact that there is zero rejection in a small (but vanishing) neighborhood around the true value $\beta=0.5$ suggests that the procedure is conservative. Nevertheless, the rejection probability is still high at values close to $\beta=0.5$, especially at larger sample sizes. For $n=10,000$, the average $95\%$ confidence intervals were $[0.30,0.71]$ for $\alpha_{i}\overset{i.i.d.}{\sim} N(0,1)$ and $[0.32,0.68]$ for $\alpha_{i} \overset{i.i.d.}{\sim} \text{Uniform}\{-1,-0.8,\ldots,1\}$. As expected, these average confidence intervals are tighter than those produced by DGP1. Altogether, we interpret the results as evidence that the procedure is conservative, but still useful and potentially informative.  On average, the computation time required to compute the test statistic in this DGP was $0.0082$ seconds. The average computational cost for each test, including the bootstrap procedure, was $0.049$ seconds. 
\begin{figure}[t!]
     \centering
     \begin{subfigure}[b]{0.47\textwidth}
         \centering
         \includegraphics[width=\textwidth]{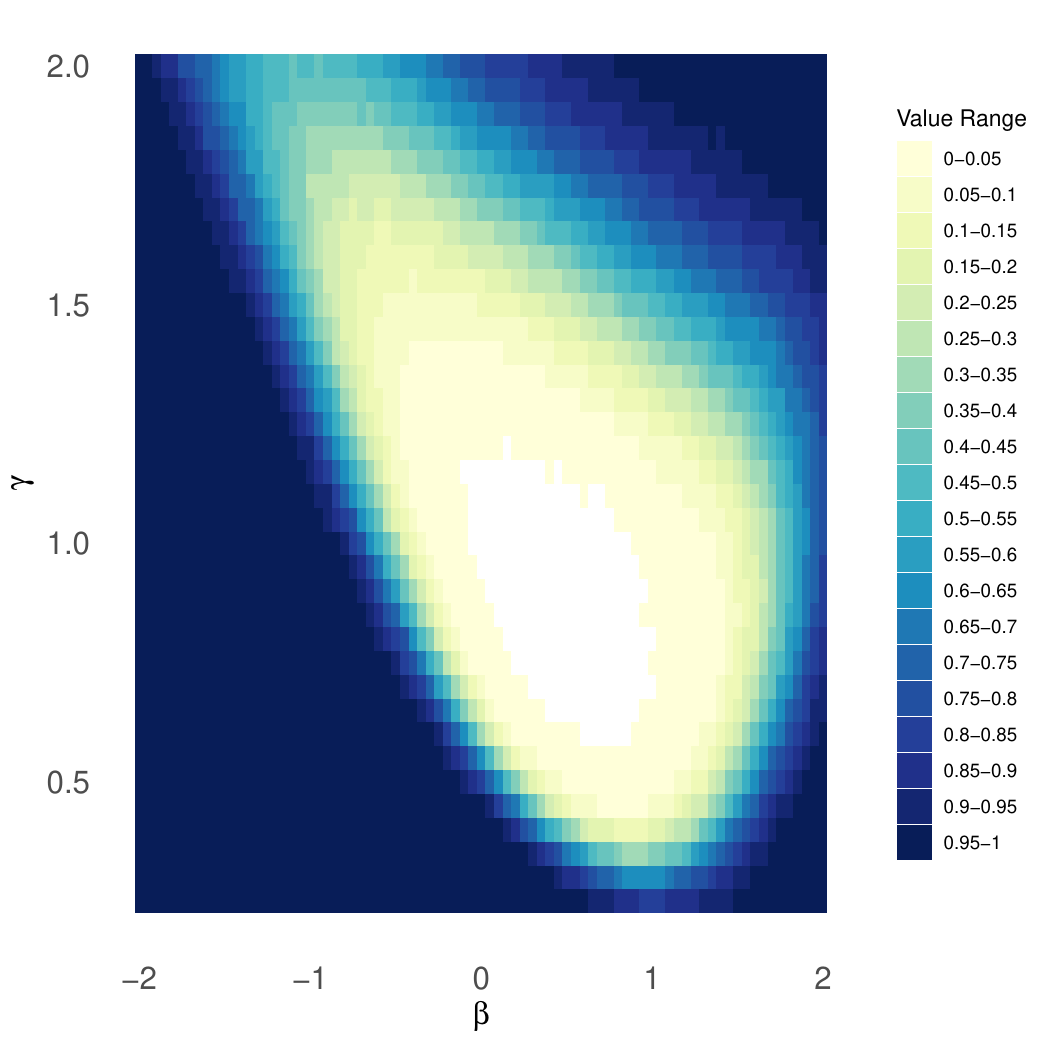}
         \caption{$\alpha_{i}\overset{i.i.d.}{\sim} N(0,1)$.}
         \label{fig_sim_T3_TT_n1e3_D1}
     \end{subfigure}
     \hfill
     \begin{subfigure}[b]{0.47\textwidth}
         \centering
         \includegraphics[width=\textwidth]{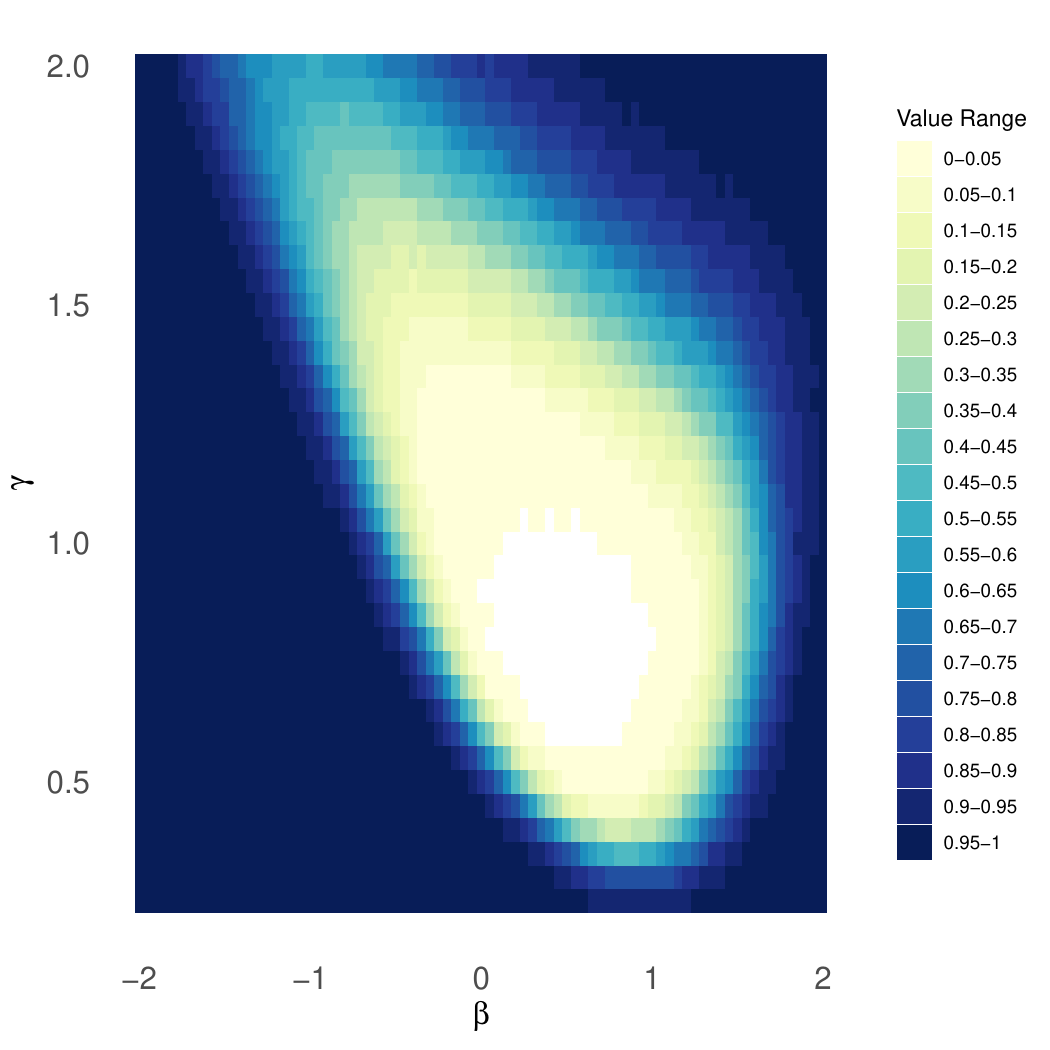}
         \caption{$\alpha_{i}\overset{i.i.d.}{\sim} \text{Uniform}\{-1,-0.8,\ldots,1\}$.}
         \label{fig_sim_T3_TT_n1e3_D2}
     \end{subfigure}
        \caption{AR(1), $T=3$, time trend, $n=1,000$.}
        \label{fig_sim_T3_TT1}
\end{figure}

\begin{figure}[t!]
     \centering
     \begin{subfigure}[b]{0.47\textwidth}
         \centering
         \includegraphics[width=\textwidth]{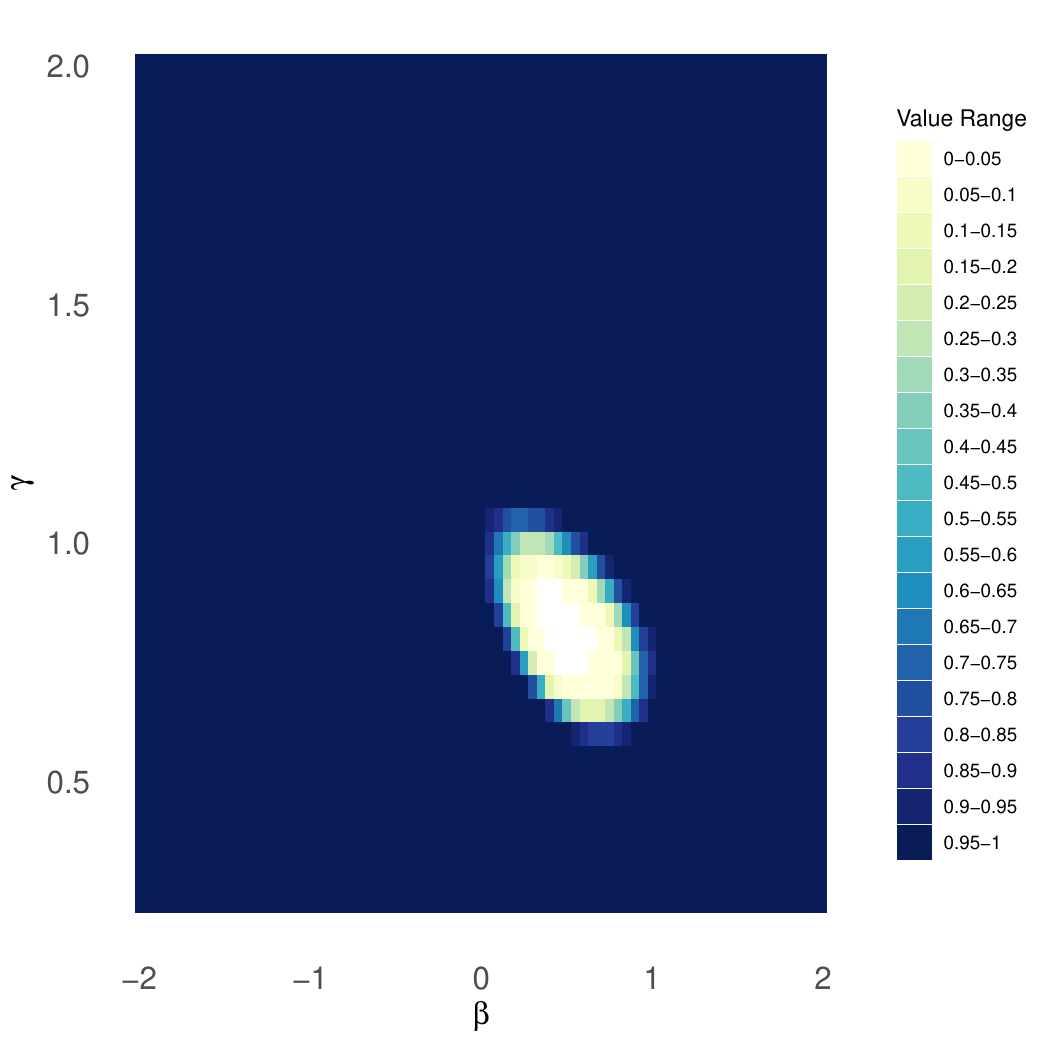}
         \caption{$\alpha_{i}\overset{i.i.d.}{\sim} N(0,1)$.}
         \label{fig_sim_T3_TT_n1e4_D1}
     \end{subfigure}
     \hfill
     \begin{subfigure}[b]{0.47\textwidth}
         \centering
         \includegraphics[width=\textwidth]{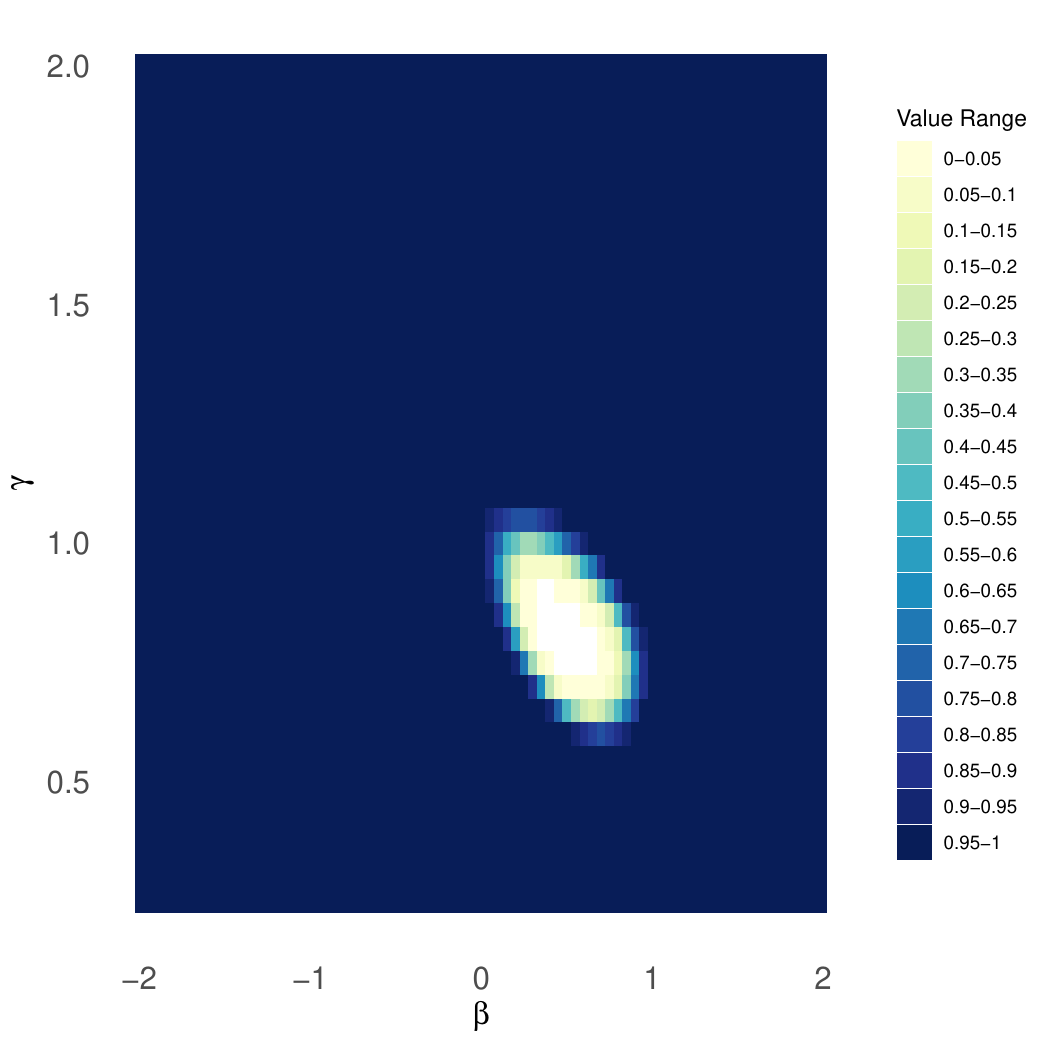}
         \caption{$\alpha_{i}\overset{i.i.d.}{\sim} \text{Uniform}\{-1,-0.8,\ldots,1\}$.}
         \label{fig_sim_T3_TT_n1e4_D2}
     \end{subfigure}
        \caption{AR(1), $T=3$, time trend, $n=10,000$.}
        \label{fig_sim_T3_TT3}
\end{figure}

The results for DGP3 for sample sizes $n=1,000$ and $n=10,000$ are displayed in Figures \ref{fig_sim_T3_TT1} and \ref{fig_sim_T3_TT3}, respectively. The figure for $n=5,000$ is very similar to the figure for $n=10,000$, and so is omitted to save space. In this DGP, there are two potential parameters of interest: the parameter $\beta$ measuring state dependence, and the parameter $\gamma$ measuring the effect of the time trend. Figures \ref{fig_sim_T3_TT1} and \ref{fig_sim_T3_TT3} thus plot the level sets of the power functions for a test of a joint null hypothesis involving these two parameters. Recall that confidence intervals for individual parameters can be constructed via projection of the confidence set \eqref{eq_CI} constructed in Section \ref{section_consistency_and_inference}. The results for $n=1,000$ show that the test has lower power at a number of nearby alternatives. For $\alpha_{i}\overset{i.i.d.}{\sim} N(0,1)$, for instance, this leads to wide average confidence intervals of $[-1.32,1.81]$ for $\beta$ and $[0.34,1.90]$ for $\gamma$.\footnote{For comparison, the average confidence interval for $\beta$ for DGP2 when $n=1,000$ and $\alpha_{i}\overset{i.i.d.}{\sim} N(0,1)$ was $[-0.18,1.18]$.}  
However, the power also appears to improve dramatically with the sample size, as is seen in Figure \ref{fig_sim_T3_TT3} for $n=10,000$. Although there are still small regions with no recorded rejections, when $n=10,000$ and $\alpha_{i}\overset{i.i.d.}{\sim} N(0,1)$ the average length of the (projected) confidence interval is $[0.13,0.89]$ for $\beta$ and $[0.64,1.01]$ for $\gamma$, both less than a quarter of the length of the confidence intervals when $n=1,000$. As is evident in Figure \ref{fig_sim_T3_TT3}, and consistent with the simulation evidence from the previous DGPs, the suggested procedure is likely conservative, but can still be highly informative.  On average, the computation time required to compute the test statistic in this DGP was $0.0071$ seconds. The average computational cost for each test, including the bootstrap procedure, was $0.051$ seconds.


\begin{figure}[t!]
     \centering
     \begin{subfigure}[b]{0.47\textwidth}
         \centering
         \includegraphics[width=\textwidth]{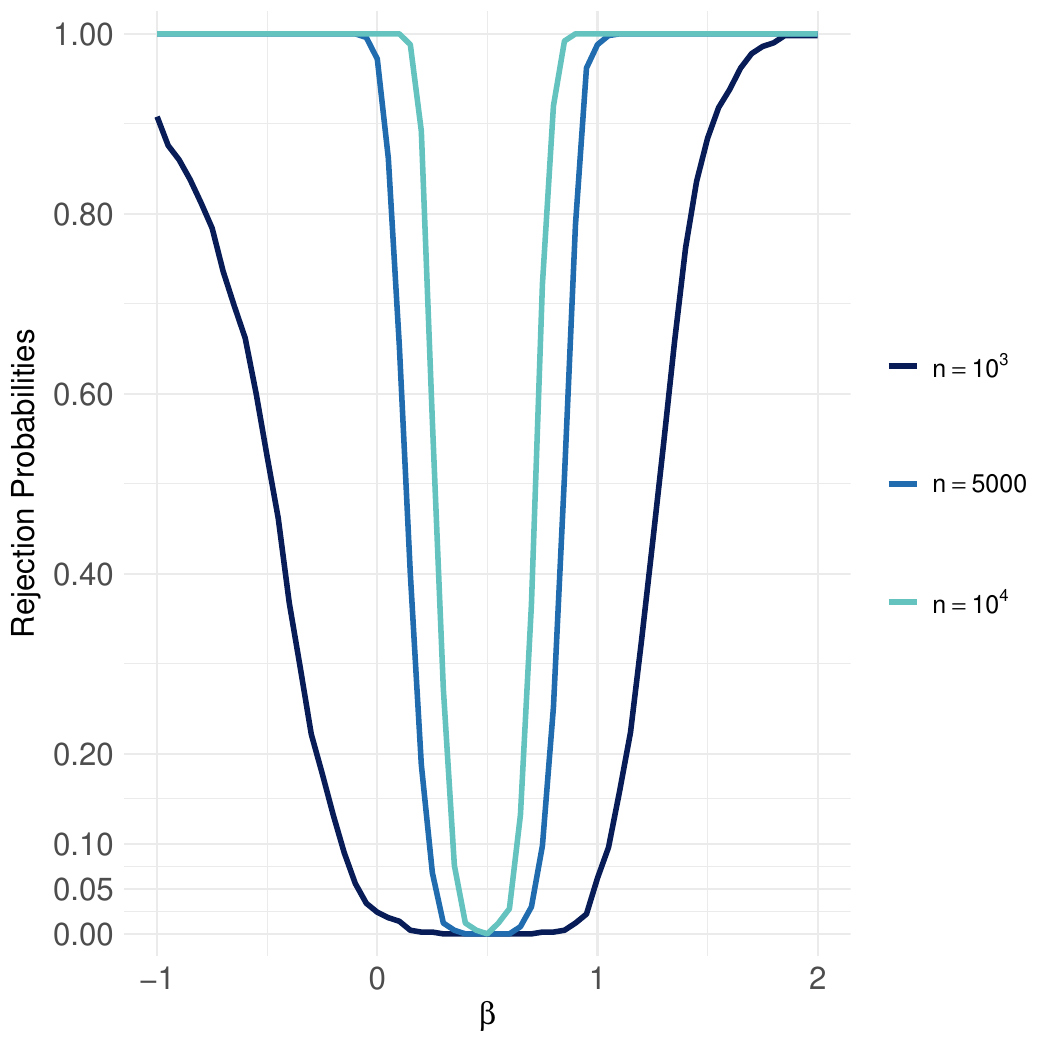}
         \caption{$\alpha_{i}\overset{i.i.d.}{\sim} N(0,1)$.}
         \label{fig_sim_T3_X_D1}
     \end{subfigure}
     \hfill
     \begin{subfigure}[b]{0.47\textwidth}
         \centering
         \includegraphics[width=\textwidth]{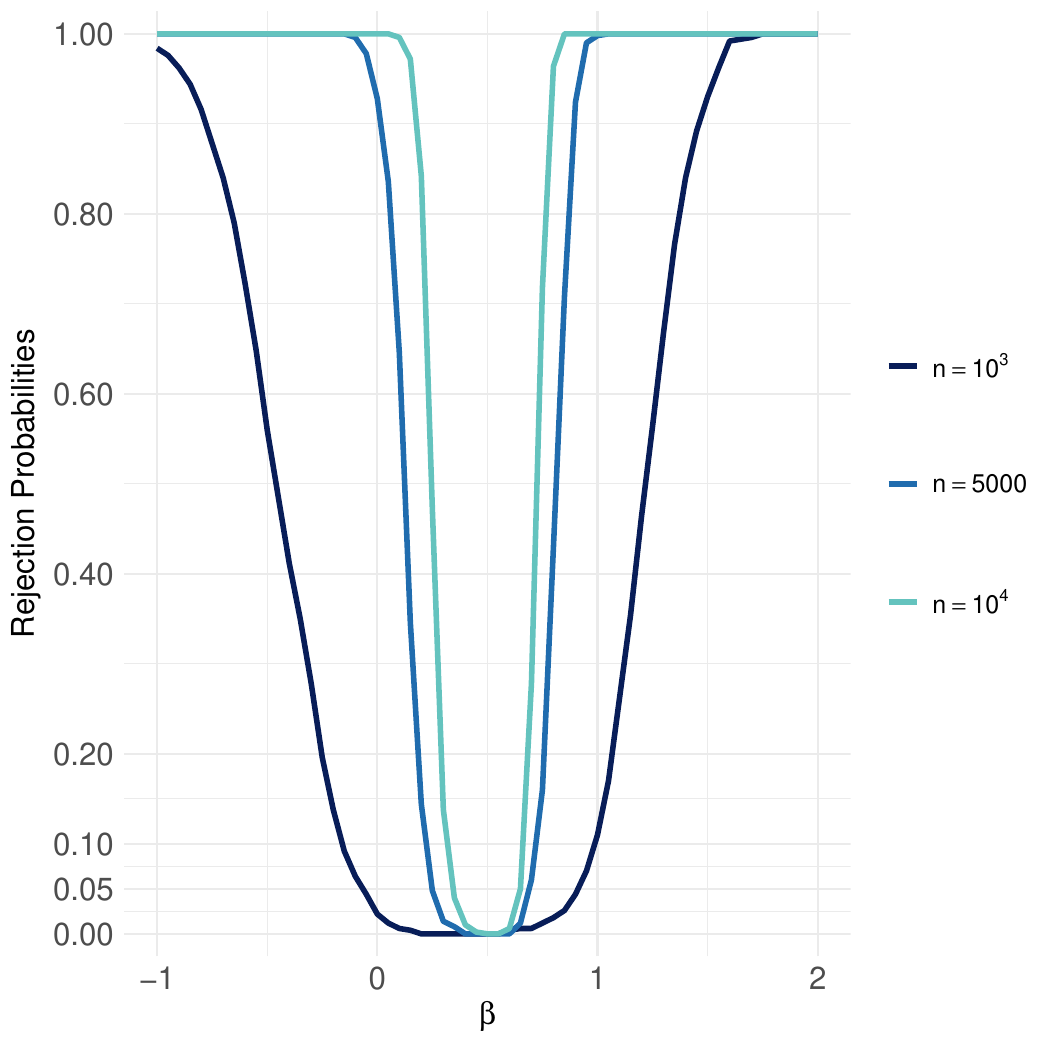}
         \caption{$\alpha_{i}\overset{i.i.d.}{\sim} \text{Uniform}\{-1,-0.8,\ldots,1\}$.}
         \label{fig_sim_T3_X_D2}
     \end{subfigure}
        \caption{AR(1), Continuous Covariate, $T=3$.}
        \label{fig_sim_T3_X}
\end{figure}

Finally, Figure \ref{fig_sim_T3_X} contains the results for DGP4. Since this DGP contains a continuous covariate, consistent with the application in the main text, for the instrument functions we interact indicators $1\{Y_{i0}=0\}$ and $1\{Y_{i0}=1\}$ with indicators of the form $1\{\max\{X_{i1},X_{i2},X_{i3}\} \in D_k\}$ where $D_{k} = \left(\frac{k-1}{k_n}, \frac{k}{k_n}\right]$ for $k=1,\ldots,k_{n} = 1+\lceil n^{1/6} \rceil$. For the piecewise constant approximation to the moment vector, we use a similar partition, but with only $l_n = k_n-1$ subsets. The power curves in Figure \ref{fig_sim_T3_X} are then constructed by counting the number of times, out of $500$ samples, that the null is rejected for the pair $(\beta,\eta)$ for every value of $\eta$.  For $n=10,000$, the average $95\%$ confidence intervals were $[0.29,0.69]$ for $\alpha_{i}\overset{i.i.d.}{\sim} N(0,1)$ and $[0.27,0.70]$ for $\alpha_{i} \overset{i.i.d.}{\sim} \text{Uniform}\{-1,-0.8,\ldots,1\}$.  On average, the computation time required to compute the test statistic in this DGP was $0.044$ seconds. The average computational cost for each test, including the bootstrap procedure, was $0.13$ seconds. 

\clearpage

\renewcommand{\thesection}{T.\arabic{section}}
\setcounter{section}{0}
\supplementtitle{Additional Online Supplementary Material for ``Identification of Dynamic Panel Logit Models with Fixed Effects''}
\addcontentsline{toc}{section}{Additional Online Supplementary Material}
\section{On the Number of Generalized Moments}
Dobronyi, Gu, Kim, Russell (2026) (DGKR hereafter) show that for a class of logit-type models, the likelihood function can be represented by a fraction involving polynomials of $\exp(\alpha)$, where $\alpha$ is the latent fixed effect. The order of the polynomial represents the number of generalized moments, and depends both on $T$ and on the particular model under consideration. We provide details on this relationship in this section. Note that the number of moments does not depend on the support of the covariates $\bm X$, so we omit the covariates in the discussion below.
	
	\subsection{The AR(1) Model} 
	The AR(1) model states $y_t = 1\{\alpha + \beta y_{t-1} \geq \epsilon_t\}$. For notational simplicity, denote $A = \exp(\alpha)$ and $B = \exp(\beta)$. First consider $T = 2$. In this case, the likelihood conditional on $\bm W = Y_0$ takes the form: 
	\[
	f(\bm y \mid \bm w, \alpha; \theta)=
	\begin{cases}
		\frac{1}{1+AB^{y_0}}\frac{1}{1+A}, &\text{ if }\bm y = (0,0),\\
		\frac{AB^{y_0}}{1+AB^{y_0}} \frac{1}{1+AB}, &\text{ if }\bm y = (1,0),\\
		\frac{1}{1+AB^{y_0}} \frac{A}{1+A}, &\text{ if }\bm y = (0,1),\\
		\frac{AB^{y_0}}{1+AB^{y_0}} \frac{AB}{1+AB}, &\text{ if }\bm y = (1,1).
	\end{cases}
	\]
	Extracting the common denominator of all terms, we have $\kappa(\bm w, \alpha, \theta) = \{(1+A)(1+AB)(1+AB^{y_0})\}^{-1}$. Multiplying the likelihood function by $\kappa(\bm w, \alpha, \theta)^{-1}$, we obtain the corresponding polynomials in $A$:
	\[
	\kappa(\bm w, \alpha, \theta)^{-1} f(\bm y \mid \bm w, \alpha; \theta)=
	\begin{cases}
		(1+AB), &\text{ if }\bm y = (0,0),\\
		AB^{y_0}(1+A), &\text{ if }\bm y = (1,0),\\
		A(1+AB), &\text{ if }\bm y = (0,1),\\
		AB^{y_0} AB (1+A), &\text{ if }\bm y = (1,1).
	\end{cases}
	\]
	Hence the number of moments for $T=2$, which corresponds to the highest order of $A$ in the polynomials above, is $S_2 = 3$.\footnote{DKGR uses the order $S$ in Assumption 2.1, but here we make it explicitly depend on $T$ in the notation.} 

	We now add one more period and derive a recursive relationship. When we add one more period, we must append 0 and 1 to all previous choice paths. Doing so leads to additional terms in the likelihood, highlighted by the square boxes below:
	\[
	f(\bm y \mid \bm w, \alpha; \theta)=
	\begin{cases}
		 \frac{1}{1+AB^{y_0}}\frac{1}{1+A}\boxed{\frac{1}{1+A}}, &\text{ if }\bm y = (0,0, \boxed{0}),\\[10pt]
		  \frac{AB^{y_0}}{1+AB^{y_0}} \frac{1}{1+AB} \boxed{\frac{1}{1+A}}, &\text{ if }\bm y = (1,0,\boxed{0}),\\[10pt]
		  \frac{1}{1+AB^{y_0}} \frac{A}{1+A}\boxed{\frac{1}{1+AB}}, &\text{ if }\bm y = (0,1,\boxed{0}),\\[10pt]
		  \frac{AB^{y_0}}{1+AB^{y_0}} \frac{AB}{1+AB}\boxed{\frac{1}{1+AB}}, &\text{ if }\bm y = (1,1,\boxed{0}),\\[10pt]
		  \frac{1}{1+AB^{y_0}}\frac{1}{1+A}\boxed{\frac{A}{1+A}} , &\text{ if }\bm y = (0,0, \boxed{1}),\\[10pt]
		  \frac{AB^{y_0}}{1+AB^{y_0}} \frac{1}{1+AB} \boxed{\frac{A}{1+A}}, &\text{ if }\bm y = (1,0,\boxed{1}),\\[10pt]
		  \frac{1}{1+AB^{y_0}} \frac{A}{1+A}\boxed{\frac{AB}{1+AB}}, &\text{ if }\bm y = (0,1,\boxed{1}),\\[10pt]
		  \frac{AB^{y_0}}{1+AB^{y_0}} \frac{AB}{1+AB}\boxed{\frac{AB}{1+AB}}, &\text{ if }\bm y = (1,1,\boxed{1}).
	\end{cases}
	\]
	This adds two extra terms to $\kappa(\bm w, \alpha, \theta)$: $\{(1+A)(1+AB)\}^{-1}$. Multiplying the likelihood function by the new $\kappa(\bm w, \alpha, \theta)^{-1}$, we have:
	\[
	\kappa(\bm w, \alpha, \theta)^{-1} f(\bm y \mid \bm w, \alpha; \theta)=
	\begin{cases}
		 (1+AB)\boxed{(1+AB)}, &\text{ if }\bm y = (0,0, \boxed{0}),\\
		  AB^{y_0}(1+A)\boxed{(1+AB)}, &\text{ if }\bm y = (1,0,\boxed{0}),\\
		  A(1+AB) \boxed{AB(1+A)}, &\text{ if }\bm y = (0,1,\boxed{0}),\\
		  AB^{y_0}  AB (1+A) \boxed{AB(1+A)}, &\text{ if }\bm y = (1,1,\boxed{0}),\\
		  (1+AB)\boxed{A(1+AB)}, &\text{ if }\bm y = (0,0, \boxed{1}),\\
		  AB^{y_0}(1+A)\boxed{A(1+AB)}, &\text{ if }\bm y = (1,0,\boxed{1}),\\
		  A(1+AB\boxed{AB(1+A)}, &\text{ if }\bm y = (0,1,\boxed{1}),\\
		  AB^{y_0} AB (1+A)\boxed{AB(1+A)}, &\text{ if }\bm y = (1,1,\boxed{1}).
	\end{cases}
	\]
	Thus, we see there are $S_3 = S_2 + 2$ moments. Continuing in this way, the recursive relationship becomes clear: with $T$ periods we will $S_T = 3 + 2(T-2) = 2T-1$ moments. 
	
	\subsection{The AR(2) Model} 
	
	The AR(1) model states $y_t = 1\{\alpha + \beta_1 y_{t-1} + \beta_2 y_{t-2}\geq \epsilon_t\}$. Let $B_1 = \exp(\beta_1)$ and $B_2 = \exp(\beta_2)$, and let $\bm W = Y^{(0)}$, the initial condition, with $Y^{(0)} = (Y_{-1}, Y_0)$. Fix $Y^{(0)} = (y_0, y_{-1})$, and consider $T = 2$. The likelihood function takes the form:
	\[
	f(\bm y \mid \bm w, \alpha; \theta)=
	\begin{cases} 
		\frac{1}{1+AB_1^{y_0}B_2^{y_{-1}}} \frac{1}{1+AB_2^{y_0}}, &\text { if }\bm y=(0,0),\\[10pt]
		\frac{AB_1^{y_0}B_2^{y_{-1}}}{1+AB_1^{y_0}B_2^{y_{-1}}} \frac{1}{1+AB_1B_2^{y_0}}, &\text { if }\bm y=(1,0),\\[10pt]
		\frac{1}{1+AB_1^{y_0}B_2^{y_{-1}}} \frac{AB_2^{y_0}}{1+AB_2^{y_0}}, &\text { if }\bm y=(0,1),\\[10pt]
		\frac{AB_1^{y_0}B_2^{y_{-1}}}{1+AB_1^{y_0}B_2^{y_{-1}}} \frac{AB_1B_2^{y_0}}{1+AB_1B_2^{y_0}}, &\text { if }\bm y=(1,1).
	\end{cases} 
	\]
	Extracting the common denominator of all terms, we have $\kappa(\bm w, \alpha, \theta) = \{(1+AB_1^{y_0}B_2^{y_{-1}})(1+AB_2^{y_0})(1+AB_1B_2^{y_0})\}^{-1}$. Multiplying the likelihood function by $\kappa(\bm w, \alpha, \theta)^{-1}$, we have the number of moments of $A$ for $T=2$ is $S_2 = 3$. We now add one more period and again derive a recursive relationship. When we add one more period, using the same practice as in the AR(1) model, we must append 0 and 1 to all existing choice paths. Doing so leads to additional terms in the likelihood, highlighted in the square boxes below: 
	\[
	f(\bm y \mid \bm w, \alpha; \theta)=
	\begin{cases}
		\frac{1}{1+AB_1^{y_0}B_2^{y_{-1}}} \frac{1}{1+AB_2^{y_0}}\boxed{\frac{1}{1+A}}, &\text{ if }\bm y =(0,0, \boxed{0}),\\[10pt]
		\frac{AB_1^{y_0}B_2^{y_{-1}}}{1+AB_1^{y_0}B_2^{y_{-1}}} \frac{1}{1+AB_1B_2^{y_0}} \boxed{\frac{1}{1+AB_2}}&\text{ if }\bm y =(1,0,\boxed{0}),\\[10pt]
		\frac{1}{1+AB_1^{y_0}B_2^{y_{-1}}} \frac{AB_2^{y_0}}{1+AB_2^{y_0}}\boxed{\frac{1}{1+AB_1}}&\text{ if }\bm y =(0,1,\boxed{0}),\\[10pt]
		\frac{AB_1^{y_0}B_2^{y_{-1}}}{1+AB_1^{y_0}B_2^{y_{-1}}} \frac{AB_1B_2^{y_0}}{1+AB_1B_2^{y_0}}\boxed{\frac{1}{1+AB_1B_2}}&\text{ if }\bm y =(1,1,\boxed{0}), \\[10pt]
		\frac{1}{1+AB_1^{y_0}B_2^{y_{-1}}} \frac{1}{1+AB_2^{y_0}}\boxed{\frac{A}{1+A}}&\text{ if } \bm y =(0,0, \boxed{1}), \\[10pt]
		\frac{AB_1^{y_0}B_2^{y_{-1}}}{1+AB_1^{y_0}B_2^{y_{-1}}} \frac{1}{1+AB_1B_2^{y_0}} \boxed{\frac{AB_2}{1+AB_2}}&\text{ if }\bm y =(1,0,\boxed{1}), \\[10pt]
		\frac{1}{1+AB_1^{y_0}B_2^{y_{-1}}} \frac{AB_2^{y_0}}{1+AB_2^{y_0}}\boxed{\frac{AB_1}{1+AB_1}}&\text{ if }\bm y =(0,1,\boxed{1}), \\[10pt]
		\frac{AB_1^{y_0}B_2^{y_{-1}}}{1+AB_1^{y_0}B_2^{y_{-1}}} \frac{AB_1B_2^{y_0}}{1+AB_1B_2^{y_0}}\boxed{\frac{AB_1B_2}{1+AB_1B_2}}&\text{ if }\bm y =(1,1,\boxed{1}).
	\end{cases}
	\]
	The new terms added to $\kappa(\bm w, \alpha, \theta)$ are $\{(1+A)(1+AB_1)(1+AB_2)(1+AB_1B_2)\}^{-1}$. Multiplying this new $\kappa(\bm w, \alpha, \theta)$ with the likelihood yields $S_3 = S_2 + 4$ moments. Continuing in this way, we will always add the same new terms to $\kappa(\bm w, \alpha, \theta)$.  We conclude that $S_T = 4T-5$ for $T \geq 2$.  
	
	\subsection{The AR(p) Model}
	For AR(p) model with any $p \geq 2$, the model states $y_t = 1\{\alpha +\beta_1 y_0 + \beta_2 y_{-1} + \dots + \beta_p y_{1-p} \geq \epsilon_t\}$. Denote $\bm W = Y^{(0)}$, the initial conditions, with $Y^{(0)} = (Y_{1-p}, Y_{2-p}, \dots, Y_{-1}, Y_0)$, and let $B_k = \exp(\beta_k)$ for $k = 1, \dots, p$. Fix any value of $y^{(0)} = (y_{1-p}, y_{2-p}, \dots, y_0)$. Starting with $T = 2$, we have: 
	\[
	f(\bm y \mid \bm w, \alpha; \theta)=
	\begin{cases}
		\frac{1}{1+AB_1^{y_0}B_2^{y_{-1}}\dots B_p^{y_{1-p}}}\frac{1}{1+AB_2^{y_0}B_3^{y_{-1}}\dots B_p^{y_{2-p}}},&\text{ if }\bm y =(0,0), \\[10pt]
		\frac{AB_1^{y_0}B_2^{y_{-1}}\dots B_p^{y_{1-p}}}{1+AB_1^{y_0}B_2^{y_{-1}}\dots B_p^{y_{1-p}}}\frac{1}{1+AB_1B_2^{y_0}B_3^{y_{-1}}\dots B_p^{y_{2-p}}},&\text{ if }\bm y =(1,0), \\[10pt]
		\frac{1}{1+AB_1^{y_0}B_2^{y_{-1}}\dots B_p^{y_{1-p}}}\frac{AB_2^{y_0}B_3^{y_{-1}}\dots B_p^{y_{2-p}}}{1+AB_2^{y_0}B_3^{y_{-1}}\dots B_p^{y_{2-p}}},&\text{ if }\bm y =(0,1), \\[10pt]
		\frac{AB_1^{y_0}B_2^{y_{-1}}\dots B_p^{y_{1-p}}}{1+AB_1^{y_0}B_2^{y_{-1}}\dots B_p^{y_{1-p}}}\frac{AB_1B_2^{y_0}B_3^{y_{-1}}\dots B_p^{y_{2-p}}}{1+AB_1B_2^{y_0}B_3^{y_{-1}}\dots B_p^{y_{2-p}}},&\text{ if }\bm y =(1,1). 
	\end{cases}
	\]
	The common denominator of all terms is $\kappa(\bm w, \alpha, \theta)= \{(1+AB_1^{y_0}B_2^{y_{-1}}\dots B_p^{y_{1-p}})(1+AB_2^{y_0}B_3^{y_{-1}}\dots B_p^{y_{2-p}})(1+AB_1B_2^{y_0}B_3^{y_{-1}}\dots B_p^{y_{2-p}})\}^{-1}$. Multiplying the likelihood by $\kappa(\bm w, \alpha, \theta)^{-1}$ gives:
	\begin{align*}
	&\kappa(\bm w, \alpha, \theta)^{-1}f(\bm y \mid \bm w, \alpha; \theta)\\[10pt]
	&=
	\begin{cases}
		(1+AB_1B_2^{y_0}B_3^{y_{-1}}\dots B_p^{y_{2-p}}),&\text{ if }\bm y =(0,0),\\
		AB_1^{y_0}B_2^{y_{-1}}\dots B_p^{y_{1-p}} (1+AB_2^{y_0}B_3^{y_{-1}}\dots B_p^{y_{2-p}}),&\text{ if }\bm y =(1,0),\\
		AB_2^{y_0}B_3^{y_{-1}}\dots B_p^{y_{2-p}} (1+AB_1B_2^{y_0}B_3^{y_{-1}}\dots B_p^{y_{2-p}}), &\text{ if }\bm y =(0,1),\\
		AB_1^{y_0}B_2^{y_{-1}}\dots B_p^{y_{1-p}} AB_1B_2^{y_0}B_3^{y_{-1}}\dots B_p^{y_{2-p}} (1+AB_2^{y_0}B_3^{y_{-1}}\dots B_p^{y_{2-p}}),&\text{ if }\bm y =(1,1).
	\end{cases}
	\end{align*}
	Thus, $S_2 = 3$. Adding one more period, we obtain the following extra terms in the likelihood:
	\[
	f(\bm y \mid \bm w, \alpha; \theta)=
	\begin{cases}
		\frac{1}{1+AB_1^{y_0}B_2^{y_{-1}}\dots B_p^{y_{1-p}}}\frac{1}{1+AB_2^{y_0}B_3^{y_{-1}}\dots B_p^{y_{2-p}}} \boxed{\frac{1}{1+AB_3^{y_0}B_4^{y_{-1}}\dots B_p^{y_{3-p}}}}, &\text{ if }\bm y=(0,0,\boxed{0}),\\[10pt]
		\frac{AB_1^{y_0}B_2^{y_{-1}}\dots B_p^{y_{1-p}}}{1+AB_1^{y_0}B_2^{y_{-1}}\dots B_p^{y_{1-p}}}\frac{1}{1+AB_1B_2^{y_0}B_3^{y_{-1}}\dots B_p^{y_{2-p}}}\boxed{\frac{1}{1+AB_2B_3^{y_0}B_4^{y_{-1}}\dots B_p^{y_{3-p}}}}, &\text{ if }\bm y=(1,0,\boxed{0}),\\[10pt]
		\frac{1}{1+AB_1^{y_0}B_2^{y_{-1}}\dots B_p^{y_{1-p}}}\frac{AB_2^{y_0}B_3^{y_{-1}}\dots B_p^{y_{2-p}}}{1+AB_2^{y_0}B_3^{y_{-1}}\dots B_p^{y_{2-p}}}\boxed{\frac{1}{1+AB_1B_3^{y_0}B_4^{y_{-1}}\dots B_p^{y_{3-p}}}}, &\text{ if }\bm y=(0,1,\boxed{0}),\\[10pt]
		\frac{AB_1^{y_0}B_2^{y_{-1}}\dots B_p^{y_{1-p}}}{1+AB_1^{y_0}B_2^{y_{-1}}\dots B_p^{y_{1-p}}}\frac{AB_1B_2^{y_0}B_3^{y_{-1}}\dots B_p^{y_{2-p}}}{1+AB_1B_2^{y_0}B_3^{y_{-1}}\dots B_p^{y_{2-p}}} \boxed{\frac{1}{1+AB_1B_2B_3^{y_0}B_4^{y_{-1}}\dots B_p^{y_{3-p}}}}, &\text{ if }\bm y=(1,1,\boxed{0}),\\[10pt]
		\frac{1}{1+AB_1^{y_0}B_2^{y_{-1}}\dots B_p^{y_{1-p}}}\frac{1}{1+AB_2^{y_0}B_3^{y_{-1}}\dots B_p^{y_{2-p}}} \boxed{\frac{AB_3^{y_0}B_4^{y_{-1}}\dots B_p^{y_{3-p}}}{1+AB_3^{y_0}B_4^{y_{-1}}\dots B_p^{y_{3-p}}}}, &\text{ if }\bm y=(0,0,\boxed{1}),\\[10pt]
		\frac{AB_1^{y_0}B_2^{y_{-1}}\dots B_p^{y_{1-p}}}{1+AB_1^{y_0}B_2^{y_{-1}}\dots B_p^{y_{1-p}}}\frac{1}{1+AB_1B_2^{y_0}B_3^{y_{-1}}\dots B_p^{y_{2-p}}} \boxed{\frac{AB_2B_3^{y_0}B_4^{y_{-1}}\dots B_p^{y_{3-p}}}{1+AB_2B_3^{y_0}B_4^{y_{-1}}\dots B_p^{y_{3-p}}}}, &\text{ if }\bm y=(1,0,\boxed{1}),\\[10pt]
		\frac{1}{1+AB_1^{y_0}B_2^{y_{-1}}\dots B_p^{y_{1-p}}}\frac{AB_2^{y_0}B_3^{y_{-1}}\dots B_p^{y_{2-p}}}{1+AB_2^{y_0}B_3^{y_{-1}}\dots B_p^{y_{2-p}}} \boxed{\frac{AB_1B_3^{y_0}B_4^{y_{-1}}\dots B_p^{y_{3-p}}}{1+AB_1B_3^{y_0}B_4^{y_{-1}}\dots B_p^{y_{3-p}}}}, &\text{ if }\bm y=(0,1,\boxed{1}),\\[10pt]
		\frac{AB_1^{y_0}B_2^{y_{-1}}\dots B_p^{y_{1-p}}}{1+AB_1^{y_0}B_2^{y_{-1}}\dots B_p^{y_{1-p}}}\frac{AB_1B_2^{y_0}B_3^{y_{-1}}\dots B_p^{y_{2-p}}}{1+AB_1B_2^{y_0}B_3^{y_{-1}}\dots B_p^{y_{2-p}}} \boxed{\frac{AB_1B_2B_3^{y_0}B_4^{y_{-1}}\dots B_p^{y_{3-p}}}{1+AB_1B_2B_3^{y_0}B_4^{y_{-1}}\dots B_p^{y_{3-p}}}}, &\text{ if }\bm y=(1,1,\boxed{1}).
	\end{cases}
	\]
	This adds four extra terms in the $\kappa(\bm w, \alpha, \theta)$: $\{(1+AB_3^{y_0}\dots B_p^{y_{3-p}})(1+AB_1B_3^{y_0}\dots B_p^{y_{3-p}}) \allowbreak(1+AB_2 B_3^{y_0}\dots B_p^{y_{3-p}})(1+AB_1B_2B_3^{y_0}\dots B_p^{y_{3-p}})\}^{-1}$. Thus, $S_3 = S_2 + 4 = 7.$ Adding one more period, the new terms contributing to $\kappa(\bm w, \alpha, \theta)$ will be $\{(1+A)(1+AB_1 B_4^{y_0}\dots B_p^{y_{4-p}})(1+AB_2 B_4^{y_0}\dots B_p^{y_{4-p}})(1+AB_3 B_4^{y_0}\dots B_p^{y_{4-p}})(1+AB_1B_3 B_4^{y_0}\dots B_p^{y_{4-p}})(1+AB_2B_3 B_4^{y_0}\dots B_p^{y_{4-p}}) + (1+AB_1B_2B_3 B_4^{y_0}\dots B_p^{y_{4-p}}\}^{-1}$. Thus, $S_4 = S_3 + (1+3+3+1) = 7 + 8 = 15$. Continuing, the number of moments is summarized in the following table: 
	\begin{table}[H]
		\centering
		\begin{tabular}{c c}
			\hline
			$t$ & $S_t$ \\
			\hline
			1 & 1 \\
			2 & $S_{1}+ \sum_{k=0}^1\binom{2}{k}$ \\
			3 &  $S_{2} + \sum_{k=0}^2\binom{2}{k}$\\
			4 & $S_{3} + \sum_{k=0}^3\binom{3}{k}$ \\
			5  & $S_{4} + \sum_{k=0}^4\binom{4}{k}$ \\
			$\vdots$ & $\vdots$ \\
			$p+1$ & $S_{p} + \sum_{k=0}^p\binom{p}{k}$\\
			$p+2$ & $S_{p+1} + \sum_{k=0}^p\binom{p}{k}$\\
			$\vdots $&$\vdots$\\
			t & $S_{t-1} + \sum_{k=0}^p\binom{2p}{k}$\\
			\hline
		\end{tabular}
	\end{table}

	\noindent Summarizing: 
	\[
	S_T = \begin{cases}
		1+2+2^2 + \dots 2^{T-1} = 2^T-1 &\text{ if } T \leq p+1,\\
		(2^p-1) + 2^p(T-p) &\text{ if } T > p+1.
	\end{cases}
	\]
	\subsection{The AR(1) Ordered Dynamic Logit Model}
	The model states 
	\[
	y_t = \begin{cases} \,\,1, &\text{ if } \alpha + \sum_{j=1}^M \beta_j 1\{y_{t-1} = j\} + \epsilon_t \in (-\infty, r_1],\\
		\,\,2, &\text{ if } \alpha + \sum_{j=1}^M \beta_j 1\{y_{t-1} = j\} + \epsilon_t \in (r_1,r_2],\\
		\,\,\vdots & \qquad\qquad\qquad\qquad\quad\vdots \\
		M, &\text{ if } \alpha + \sum_{j=1}^M \beta_j 1\{y_{t-1} = j\} + \epsilon_t \in (r_{M-1}, +\infty).
	\end{cases} 
	\]
	Denote $A = \exp(\alpha), B_j = \exp(\beta_j)$ for $j = 1, 2, \dots, M$, and $R_s = \exp(-\gamma_s)$ for $s = 1, 2, \dots, M-1$. Let $\bm W = Y_0$ and consider $T = 1$. The likelihood function takes the form: 
	\[
	f(y \mid \bm w, \alpha; \theta)=
	\begin{cases} 
		\frac{1}{1+AB_{y_0}R_1}, &\text{ if }y=1,\\
		\frac{AB_{y_0}(R_1-R_2)}{(1+AB_{y_0}R_1)(1+AB_{y_0}R_2)}, &\text{ if }y=2,\\
		\qquad\vdots & \qquad\vdots \\
		\frac{AB_{y_0}R_{M-1}}{1+AB_{y_0}R_{M-1}}, &\text{ if }y=M.
	\end{cases} 
	\]
	The common denominator is $\kappa(\bm w, \alpha, \theta)=\{\prod_{s = 1}^{M-1} (1+AB_{y_0}B_s)\}^{-1}$. Multiplying the likelihood by $\kappa(\bm w, \alpha, \theta)^{-1}$ produces a polynomial in $A$ of order $M-1$. Adding one more period, we must 1 to all previous choices, and then add 2, and so on until we add $M$. This produces $(M-1)M$ extra terms in $\kappa(\bm w, \alpha, \theta)$, with the following pattern:
	\[
	\kappa_{T=2}(w,\alpha,\theta) = \frac{\kappa_{T=1}(w,\alpha,\theta)}{\prod_{s=1}^{M-1} \prod_{j = 1}^M (1+AB_j R_s)}
	\]
	Here we index $\kappa$ by $T$ to reflect the recursive pattern explicitly. 
	Multiplying the likelihood by this term adds $M(M-1)$ moments of $A$, i.e. $S_2 = M(M-1) + (M-1)$. More generally, we have $S_T = M -1 + (T-1) M(M-1)$.
	
	\section{Automated Construction of the Matrix $\mathbf G(\mathbf{w}, \theta)$}
	
	We now consider how to automate the construction of the matrix $\bm G(w,\theta)$ in various models considered in DKGR. In the following, we index $\kappa$ and $\bm G$ by $T$.

	 \subsection{The AR(1) Model Without Covariates} \label{automation_G_AR1}
	

Fix $W = w = y_0$. For $T = 1$, we have: 
 \[
 f(0\mid y_0, \alpha;\beta) = 1/(1+AB^{y_0}), \quad f(1\mid y_0, \alpha; \beta) = AB^{y_0}/(1+AB^{y_0}).
 \]
 Thus we can choose $1/\kappa_{T=1}(w,\alpha,\beta) = (1+AB^{y_0})$, which is just the common denominator between the two likelihoods. The corresponding matrix $\bm G_{T=1}(w,\beta)$ is given by: 
 \[
 \bm G_{T=1}(w,\beta) = \begin{bmatrix} 1 & 0 \\ 0 & B^{y_0}\end{bmatrix}.
 \]
 For $T = 2$, we have: 
	 \begin{align*}
		 	f((0,0)\mid y_0, \alpha; \beta) = f(0\mid y_0, \alpha; \beta) \frac{1}{1+A} , & \quad f((1,0)\mid y_0, \alpha; \beta) = f(1\mid y_0, \alpha; \beta) \frac{1}{1+AB},\\
		 	f((0,1)\mid y_0, \alpha; \beta) = f(0\mid y_0, \alpha; \beta) \frac{A}{1+A}, & \quad f((1,1)\mid y_0, \alpha; \beta) = f(1\mid y_0, \alpha; \beta) \frac{AB}{1+AB}.
		 \end{align*}
	Thus we update the choice of $\kappa$ by: 
	 \[
	 \kappa_{T=2}(w,\alpha,\beta) = \kappa_{T=1}(w,\alpha,\beta)/ (1+A)(1+AB).
	 \]
	 Again, this choice is just the common denominator between all likelihood terms. This choice implies:
	 \begin{align*}
		 	\frac{f((0,0)\mid y_0, \alpha, \beta)}{\kappa_{T=2}(y_0,\alpha,\beta)} &= 1 \times (1+AB) , && \quad \frac{f((1,0)\mid y_0, \alpha, \beta)}{\kappa_{T=2}(y_0,\alpha,\beta)} =AB^{y_0}\times  (1+A),\\[10pt]
		 	\frac{f((0,1)\mid y_0, \alpha, \beta)}{\kappa_{T=2}(y_0,\alpha,\beta)}  &=1 \times A(1+AB), && \quad \frac{f((1,1)\mid y_0, \alpha, \beta)}{\kappa_{T=2}(y_0,\alpha,\beta)}  = AB^{y_0}\times AB(1+A),
		 \end{align*}
	 and hence we can update the matrix $\bm G(w,\theta)$ by:
	 \[
	 \bm G_{T=2}(w,\beta) = \begin{bmatrix} \bm G_{T=1}(w,\beta) \star \begin{bmatrix} 1 & B & 0\\
			 		1 & 1 & 0\end{bmatrix} \\[10pt]
			\bm G_{T=1}(w,\beta) \star \begin{bmatrix} 0 & 1 & B \\
			 		0 & B & B \end{bmatrix} 
		 	\end{bmatrix} = \begin{bmatrix} 1 & B & 0 & 0 \\ 0 & B^{y_0} & B^{y_0} & 0 \\ 0 & 1 & B & 0 \\ 0 & 0 & B^{y_0+1} & B^{y_0+1}\end{bmatrix}, 
	 \]
	 where $\star$ represents vector convolutions.\footnote{Let $\bm u = (u_1, \dots, u_j)^\top \in \R^j$ and $\bm v = (v_1, \dots, v_s)^\top \in \R^s$. Then $u \star v =(c_1, \dots, c_{j+s-1})$ with $c_x = \sum_j u_j v_{x-j+1}$. For example, the convolution of the $2\times 1$ vectors $\bm u = (u_1,u_2)^\top$ and $\bm v= (v_1,v_2)^\top$ gives the $3\times 1$ vector $\bm u \star \bm v = (u_1v_2, u_1v_2+u_2v_1, u_2v_2)^\top$. Convolution of two matrices is done by performing the convolution between each of their corresponding row vectors.} The same pattern persists as $T$ increases, and we can update iteratively by setting: 
	 \[
	 \kappa_{T=t}(w,\alpha,\beta) = \kappa_{T=t-1}(w,\alpha,\beta)/ (1+A)(1+AB),
	 \]
	 and: 
	 \[
	 \bm G_{T=t}(w,\beta) = 
	 \begin{bmatrix} 
	 \bm G_{T=t-1}(w,\beta) \star 
		\begin{bmatrix} \bm 1_{2^{t-2}} \otimes 
			\begin{bmatrix} 
			1 & B & 0
			\end{bmatrix} \\[10pt]
		\bm 1_{2^{t-2}} \otimes 
			\begin{bmatrix}	
			1 & 1 & 0
			\end{bmatrix}  
		\end{bmatrix} \\\\
	\bm G_{T=t-1}(w,\beta) \star 
		\begin{bmatrix} 
		\bm 1_{2^{t-2}} \otimes 
			\begin{bmatrix} 
			0 & 1 & B 
			\end{bmatrix}  \\[10pt]
	 	\bm 1_{2^{t-2}} \otimes 
		 	\begin{bmatrix}	
		 	0 & B & B 
		 	\end{bmatrix}  
	 	\end{bmatrix} 
		 \end{bmatrix}, 
	 \]
	 where $\bm 1_m$ denotes an $m\times 1$ vector of ones. Since the length of the convolution of two vectors of length $j$ and $s$ produces a vector of length $j+s-1$, the number of columns of $\bm G_{T=t}(y_0,\beta)$ grows according to $2+(3-1)\times (t-1) = 2t$, which confirms that the matrix $\bm G(y_0,\beta)$ for general $T$ is of dimension $2^T \times 2T$. 

	\subsection{The AR(1) Model With Covariates}
	For simplicity, consider the case of a scalar covariate, and define $A := \exp(\alpha)$, $B := \exp(\beta)$, and $C := \exp(\gamma)$. Consider $T = 1$ and fix $\bm W = \bm w = (y_0, \bm x)$. Then we have: 
	\[
	f(0 \mid \bm w, \alpha; \theta) = \frac{1}{(1+AB^{y_0}C^{x_1})}, \quad f(1 \mid \bm w, \alpha; \theta) = \frac{AB^{y_0}C^{x_1}}{(1+AB^{y_0}C^{x_1})}.
	\]
	Now pick $\kappa_{T=1}(\bm w, \alpha,\theta) = 1/(1+AB^{y_0}C^{x_1})$ which implies:
	\[
	\bm G_{T=1}(\bm w,\theta) = \begin{bmatrix} 1 & 0 \\ 0 & B^{y_0}C^{x_1}\end{bmatrix}. 
	\]
	For $T = 2$, we pick $\kappa_{T=2}(\bm w,\alpha,\theta) = \kappa_{T=1}(\bm w,\alpha,\theta) /(1+AC^{x_2})(1+ABC^{x_2})$, which implies:
	\begin{align*}
		\frac{f( (0,0) \mid A, \bm w; \theta)}{\kappa_{T=2}(\bm w,\alpha,\theta)} &= 1 \times (1+ABC^{x_2}) , &&\frac{f((1,0)|A, \bm w; \theta)}{\kappa_{T=2}(\bm w,\alpha,\theta) } = AB^{y_0}C^{x_1} \times (1+AC^{x_2}),\\\\
		\frac{f((0,1)|A, \bm w; \theta)}{\kappa_{T=2}(\bm w,\alpha,\theta) } &= 1\times AC^{x_2} (1+ABC^{x_2}) , &&\frac{f((1,1)|A, \bm w; \theta)}{\kappa_{T=2}(\bm w,\alpha,\theta) } = AB^{y_0}C^{x_1} \times ABC^{x_2} (1+AC^{x_2}).
	\end{align*}
	Thus:
	\begin{align*}
		\bm G_{T=2}(\bm w,\theta) = 
		\begin{bmatrix} 
			1 & BC^{x_2} & 0 & 0\\ 
			0 & B^{y_0}C^{x_1} &B^{y_0}C^{x_1}C^{x_2} & 0\\
			0 & C^{x_2} & BC^{2x_2} & 0\\
			0 & 0 & B^{y_0+1}C^{x_1}C^{x_2}  & B^{y_0+1}C^{x_1}C^{2x_2}
		\end{bmatrix}. 
	\end{align*}
	In the general case, as we increase from $t-1$ periods to $t$ periods we have the updating rule:
	\[
	\kappa_{T=t}(\bm w,\alpha,\theta) = \frac{\kappa_{T=t-1}(\bm w,\alpha,\theta)}{(1+AC^{x_t})(1+ABC^{x_t})}.
	\]
	This is due to the fact that each time we add a new period, the previous period's outcome can either be $0$ or $1$, hence the additional common denominator for the added period is $\tfrac{1}{(1+AC^{x_t})(1+ABC^{x_t})}$. 
	Furthermore, we always append $0$ to all existing histories in period $t-1$, and then append 1. This implies that when a new period is added, we will have $2^{t-2}$ choice histories with $y_{t-1} = 0, y_t = 0$, $2^{t-2}$ choice histories with $y_{t-1} = 1, y_{t} = 0$, $2^{t-2}$ choice histories with $y_{t-1}=0, y_t = 1$, and $2^{t-2}$ choice histories with $y_{t-1} = 1, y_t = 1$. This implies that the updating of $\bm G$ is given by:
	\[
	\bm G_{T=t}(\bm w,\theta) = \begin{bmatrix} 
		\bm G_{T=t-1}(\bm w,\theta) \star 
		\begin{bmatrix} 
			\bm{1}_{2^{t-2}} \otimes 
			\begin{bmatrix} 1 & BC^{x_t}& 0 
			\end{bmatrix} \\[10pt]
			\bm{1}_{2^{t-2}} \otimes 
			\begin{bmatrix} 1 & C^{x_t} & 0 
			\end{bmatrix} 
		\end{bmatrix} \\
		\\
		\bm G_{T=t-1}(\bm w,\theta) \star  
		\begin{bmatrix} \bm{1}_{2^{t-2}} \otimes 
			\begin{bmatrix} 0 & C^{x_t} & BC^{2x_t}
			\end{bmatrix} \\[10pt] 
			\bm{1}_{2^{t-2}} \otimes 
			\begin{bmatrix} 0 & BC^{x_t} & BC^{2x_t}
			\end{bmatrix} 
		\end{bmatrix} 
	\end{bmatrix}.
	\]

	
	\subsection{The AR(p) Model}
	
	We first show how to construct the matrix $\bm G(\bm w ,\theta)$ for $p = 2$, and then we comment on how to adapt the procedure to the case of general $p$. For $p = 2$, the model is $y_{t} = 1\{ \alpha + \beta_1 y_{t-1} + \beta_2 y_{t-2} + x_t\gamma \geq \epsilon_{t}\}$. First consider $T = 1$, fix $\bm W = \bm w =(y_{-1}, y_0)$, and define $A := \exp(\alpha)$, $B_1 := \exp(\beta_1)$, $B_2 := \exp(\beta_2)$ and $C = \exp(\gamma)$. Then: 
	\begin{align*}
		f(0 \mid \bm w, \alpha;\theta) = \frac{1}{1+AB_1^{y_0}B_2^{y_{-1}}C^{x_1}}, && f(1 \mid \bm w, \alpha;\theta) = \frac{AB_1^{y_0}B_2^{y_{-1}}}{1+AB_1^{y_0}B_2^{y_{-1}} C^{x_1}}.
	\end{align*}
Taking the common denominator, we set $\kappa_{T=1}(\bm w, \alpha,\theta) = 1/(1+AB_1^{y_0}B_2^{y_{-1}}C^{x_1})$, which implies:
	\begin{align*}
		\bm G_{T=1}(\bm w, \theta) = 
		\begin{bmatrix} 1&0\\0 & B_1^{y_0}B_2^{y_{-1}}C^{x_1}\end{bmatrix}.
	\end{align*}
	For $T=2$ we have:
	\begin{align*}
		&f((0,0)\mid \bm w, \alpha;\theta) = f(0\mid \bm w, \alpha;\theta) \frac{1}{1+AB_2^{y_0}C^{x_2}}, &&f(1,0)\mid \bm w, \alpha;\theta) = f(1\mid \bm w, \alpha;\theta) \frac{1}{1+AB_1B_2^{y_0}C^{x_2}},\\\\
		&f((0,1)\mid \bm w, \alpha;\theta) = f(0\mid \bm w, \alpha;\theta) \frac{AB_2^{y_0}C^{x_2}}{1+AB_2^{y_0}C^{x_2}}, &&f((1,1)\mid \bm w, \alpha;\theta) = f(1\mid \bm w, \alpha;\theta) \frac{AB_1B_2^{y_0}C^{x_2}}{1+AB_1B_2^{y_0}C^{x_2}}.
	\end{align*}
	This suggests the new common denominator due to adding one more period is $(1+AB_1B_2^{y_0}C^{x_2})(1+B_2^{y_0}C^{x_2})$. Thus:
	\begin{align*}
		\kappa_{T=2}( \bm w, \alpha, \theta) = \frac{\kappa_{T=1}(\bm w, \alpha, \theta)}{(1+AB_1B_2^{y_0}C^{x_2})(1+AB_2^{y_0}C^{x_2})},
	\end{align*}
	and we can construct the matrix $\bm G_{T=2}(\bm w, \theta)$ as: 
	\begin{align*}
		\bm G_{T=2}(\bm w,\theta) &=  
		\begin{bmatrix} 
			\bm G_{T=1}(\bm w,\theta) \star 
			\begin{bmatrix} 
				(1,0)\star (1, B_1B_2^{y_0}C^{x_2})\\
				(1,0)\star (1, B_2^{y_0}C^{x_2}) 
			\end{bmatrix} \\
			\\
			\bm G_{T=1}(\bm w,\theta) \star 
			\begin{bmatrix} (0, B_2^{y_0}C^{x_2}) \star (1, B_1B_2^{y_0}C^{x_2})\\ 
				(0, B_1B_2^{y_0}C^{x_2}) \star (1, B_2^{y_0}C^{x_2})
		\end{bmatrix} \end{bmatrix}\\
		\\
		&=\begin{bmatrix} 
			\bm G_{T=1}(\bm w,\theta) \star 
			\begin{bmatrix} 1 & B_1B_2^{y_0}C^{x_2} & 0 \\
				1 & B_2^{y_0}C^{x_2} & 0 
			\end{bmatrix} \\
			\\
			\bm G_{T=1}(\bm w,\theta) \star 
			\begin{bmatrix} 0 & B_2^{y_0}C^{x_2} & B_1 B_2^{2y_0}C^{2x_2} \\ 0& B_1B_2^{y_0}C^{x_2} & B_1B_2^{2y_0}C^{2x_2}\end{bmatrix} 
		\end{bmatrix}.
	\end{align*}
	Adding one more period, for $T = 3$ we can update the likelihood function as:
	\begin{align*}
		f((0,0,0)\mid \bm w, \alpha;\theta) &= f((0,0)\mid \bm w, \alpha;\theta) \left(\frac{1}{1+AC^{x_3}}\right),\\ 
		f((1,0,0)\mid \bm w, \alpha;\theta) &= f((1,0)\mid \bm w, \alpha;\theta)\left(\frac{1}{1+AB_2C^{x_3}}\right),\\
		f((0,1,0)\mid \bm w, \alpha;\theta) &= f((0,1)\mid \bm w, \alpha;\theta) \left(\frac{1}{1+AB_1C^{x_3}}\right),\\
		f((1,1,0)\mid \bm w, \alpha;\theta) &= f((1,1)\mid \bm w, \alpha;\theta) \left(\frac{1}{1+AB_1B_2C^{x_3}}\right),\\
		f((0,0,1)\mid \bm w, \alpha;\theta) &= f((0,0)\mid \bm w, \alpha;\theta) \left(\frac{AC^{x_3}}{1+AC^{x_3}}\right),\\ 
		f((1,0,1)\mid \bm w, \alpha;\theta) &= f((1,0)\mid \bm w, \alpha;\theta) \left(\frac{AB_2C^{x_3}}{1+AB_2C^{x_3}}\right),\\
		f((0,1,1)\mid \bm w, \alpha;\theta) &= f((0,1)\mid \bm w, \alpha;\theta) \left(\frac{AB_1C^{x_3}}{1+AB_1C^{x_3}}\right),\\ 
		f((1,1,1)\mid \bm w, \alpha;\theta) &= f((1,1)\mid \bm w, \alpha;\theta) \left(\frac{AB_1B_2C^{x_3}}{1+AB_1B_2C^{x_3}}\right).
	\end{align*} 
	Taking the common deminator for the added period, we can set $\kappa_{T=3}(\bm w, \alpha, \theta)$ as:
	\begin{align*}
		\kappa_{T=3}(\bm w, \alpha, \theta)= \frac{\kappa_{T=2}(\bm w, \alpha, \theta)}{(1+AC^{x_3})(1+AB_1C^{x_3})(1+AB_2C^{x_3})(1+AB_1B_2C^{x_3})},
	\end{align*}
	in which case we have: 
	\begin{align*}
		\bm G_{T=3}(\bm w,\theta) = 
		\begin{bmatrix} 
			\bm G_{T=2}(\bm w,\theta) \star 
			\begin{bmatrix} (1,0) \star (1, B_1C^{x_3}) \star (1, B_2C^{x_3}) \star (1, B_1B_2C^{x_3})\\
				(1,0) \star (1, C^{x_3}) \star (1, B_1C^{x_3}) \star (1, B_1B_2C^{x_3})\\
				(1,0) \star (1, C^{x_3}) \star (1, B_2 C^{x_3}) \star (1, B_1B_2C^{x_3})\\
				(1,0) \star (1, C^{x_3}) \star (1, B_1C^{x_3}) \star (1, B_2C^{x_3})\end{bmatrix} \\
			\\
			\bm G_{T=2}(\bm w,\theta) \star 
			\begin{bmatrix} (0, C^{x_3})\star (1, B_1C^{x_3}) \star (1, B_2C^{x_3}) \star (1, B_1B_2C^{x_3})\\
				(0, B_2C^{x_3}) \star (1, C^{x_3}) \star (1, B_1C^{x_3}) \star (1, B_1B_2C^{x_3}) \\
				(0, B_1C^{x_3}) \star (1, C^{x_3}) \star (1, B_2C^{x_3}) \star (1, B_1B_2C^{x_3})\\
				(0, B_1B_2C^{x_3}) \star (1, C^{x_3}) \star (1, B_1C^{x_3}) \star (1, B_2C^{x_3})
			\end{bmatrix} 
		\end{bmatrix}. 
	\end{align*}
	We can see that for $T \in \{1, 2\}$, the initial conditions $(y_{-1},y_0)$ have an impact on the updating of $\kappa(\bm w, \alpha, \theta)$ and $\bm G(\bm w,\theta)$, but for any $T \geq 3$ we have a generic updating rule given by: 
	\begin{align*}
		\kappa_{T=t}(\bm w, \alpha, \theta) = \frac{\kappa_{T=t-1}(\bm w, \alpha, \theta)}{(1+AC^{x_t})(1+AB_1C^{x_t})(1+AB_2C^{x_t})(1+AB_1B_2C^{x_t})},
	\end{align*}
	and:
	\begin{align*}
		\bm G_{T=t}(\bm w,\theta) = \begin{bmatrix} \bm G_{T=t-1}(\bm w,\theta) \star \begin{bmatrix} \bm{1}_{2^{t-3}} \otimes \left((1,0) \star (1, B_1C^{x_t}) \star (1, B_2C^{x_t}) \star (1, B_1B_2C^{x_t})\right)\\
				\bm{1}_{2^{t-3}}\otimes\left(	(1,0) \star (1, C^{x_t}) \star (1, B_1C^{x_t}) \star (1, B_1B_2C^{x_t})\right)\\
				\bm{1}_{2^{t-3}}\otimes\left(	(1,0) \star (1, C^{x_t}) \star (1, B_2 C^{x_t}) \star (1, B_1B_2C^{x_t})\right)\\
				\bm{1}_{2^{t-3}}\otimes\left(	(1,0) \star (1, C^{x_t}) \star (1, B_1C^{x_t}) \star (1, B_2C^{x_t})\right)\end{bmatrix} \\
			\\
			\bm G_{T=t-1}(\bm w,\theta) \star \begin{bmatrix} \bm{1}_{2^{t-3}}\otimes\left((0, C^{x_t})\star (1, B_1C^{x_t}) \star (1, B_2C^{x_t}) \star (1, B_1B_2C^{x_t})\right)\\
				\bm{1}_{2^{t-3}} \otimes\left(	(0, B_2C^{x_t}) \star (1, C^{x_t}) \star (1, B_1C^{x_t}) \star (1, B_1B_2C^{x_t})\right) \\
				\bm{1}_{2^{t-3}} \otimes\left(	(0, B_1C^{x_t}) \star (1, C^{x_t}) \star (1, B_2C^{x_t}) \star (1, B_1B_2C^{x_t})\right)\\
				\bm{1}_{2^{t-3}} \otimes \left((0, B_1B_2C^{x_t}) \star (1, C^{x_t}) \star (1, B_1C^{x_t}) \star (1, B_2C^{x_t})\right)
			\end{bmatrix} 
		\end{bmatrix}. 
	\end{align*}
	The updating rule on $\bm G$ reflects that, as we increase from $t-1$ to $t$ for any $t \geq 3$, the first $2^{t-3}$ histories satisfy $y_t = 0, y_{t-1} = 0, y_{t-2} = 0$, the next $2^{t-3}$ histories satisfy $y_t = 0, y_{t-1} = 0, y_{t-2} = 1$, the next $2^{t-3}$ histories satisfy $y_t = 0, y_{t-1} = 1, y_{t-2} = 0$ and the next $2^{t-3}$ histories satisfy $y_t = 0, y_{t-1} = 1, y_{t-2} = 1$. Afterwards, the pattern repeats with $y_t = 1$; that is, for the next $2^{t-3}$ histories we have $y_t = 1, y_{t-1} = 0, y_{t-2} = 0$, followed with $2^{t-3}$ histories satisfying $y_t = 1, y_{t-1} = 0, y_{t-2} = 1$, followed with $2^{t-3}$ histories satisfying $y_t = 1, y_{t-1} = 1, y_{t-2} = 0$, followed with $2^{t-3}$ histories satisfying $y_t = 1, y_{t-1} = 1, y_{t-2} = 1$. For general $p$, the updating rule can be constructed similarly: the initial conditions $(y_{1-p}, \dots, y_0)$ have an impact on the updating of $\kappa(\bm w, \alpha, \theta)$ and $\bm G(\bm w,\theta)$ for $1\leq T \leq p$, and afterwards there is a generic updating rule starting at $T \geq p+1$. We omit these details.
	
	
	\subsection{The Dynamic Ordered Logit Model with Covariates} 
	
	The model states: 
	\[
	y_t = \begin{cases} 1 &\text{ if } \alpha + \sum_{j = 1}^M\beta_j 1\{y_{t-1}=j\} + x_t \eta + \epsilon_t \in (-\infty, r_1], \\
		2 &\text{ if } \alpha + \sum_{j = 1}^M\beta_j 1\{y_{t-1}=j\} + x_t \eta + \epsilon_t  \in (r_1, r_2],\\
		\vdots & \qquad\qquad\qquad\qquad\vdots \\
		M &\text{ if } \alpha + \sum_{j = 1}^M\beta_j 1\{y_{t-1}=j\} + x_t \eta + \epsilon_t \in (r_{M-1}, +\infty).
	\end{cases} 
	\]
	For $i,j \in \{1,2,\dots, M\}$, let $\pi_{ij}(x)$ denote the probability of choosing $j$ in the current period given the choice was $i$ in the previous period, and given the current covariates are $X_t = x$. Denote $A := \exp(\alpha)$, $B_j := \exp(\beta_j)$ for $j = 1, 2, \dots, M$, $R_s:= \exp(-\gamma_s)$ for $s =1,2,\dots M-1$, and $C := \exp(\eta)$. Then for $j = 1,2,\dots,M$ and $s = 2,3,\dots, M-1$:
	\begin{align*}
		\pi_{i1}(x) &= P(Y_t = 1\mid Y_{t-1}=i, X_t = x, \alpha) = \Lambda(\gamma_1 - \alpha - \beta_i - x \eta) =  \frac{1}{1+AB_iR_1C^{x}},\\
		\pi_{is}(x) & = P(Y_t = s\mid Y_{t-1} = i, X_t = x, \alpha) = \Lambda(\gamma_s - \alpha - \beta_i - x \eta)-  \Lambda(\gamma_{s-1} - \alpha - \beta_i - x \eta) = \\
		& = \frac{AB_i(R_{s-1}-R_{s})C^{x}}{(1+AB_iR_{s-1}C^x)(1+AB_iR_{s}C^x)}, s\in \{2,3,\dots, M-1\}\\
		\pi_{iM}(x) & = P(Y_{t} = M\mid Y_{t-1}=i, X_t = x, \alpha) =1 - \Lambda(\gamma_{M-1} - \alpha - \beta_i - x \eta) =  \frac{AB_iR_{M-1}C^x}{1+AB_iR_{M-1}C^x}.
	\end{align*}
	Let $[M]:=\{1, 2, \dots, M\}$ and define the set $\C= \{(s,j): s \in [M-1], j \in [M] \} = [M-1] \times [M]$, and let the symbol $\bigstar_{m \in \mathcal{M}} v_m$ represent vector convolutions for all vectors $v_{m}$ indexed by the set $\mathcal{M}$. For example, if $\mathcal{M} = \{1,2,3\}$, then $\bigstar_{m \in \mathcal{M}} v_m= v_1 \star v_2 \star v_3$.
	Now consider $T = 1$, and fix $\bm W = \bm w = (y_0, x_1)$. Then we have: 
	\begin{align*}
		f(1 | \bm w, \alpha, \theta) & = \frac{1}{1+AB_{y_0}R_1C^{x_1}},\\[10pt]
		f(2|\bm w, \alpha, \theta) & = \frac{AB_{y_0}(R_1-R_2)C^{x_1}}{(1+AB_{y_0}R_1C^{x_1})(1+AB_{y_0}R_2C^{x_1})},\\[10pt]
		\vdots\\[10pt]
		f(M|\bm w, \alpha, \theta) & = \frac{AB_{y_0}R_{M-1}C^{x_1}}{1+AB_{y_0}R_{M-1}C^{x_1}}.
	\end{align*}
	Now choose:
	\begin{align*}
		\kappa_{T=1}(\bm w, \alpha, \theta) = \frac{1}{(1+AB_{y_0}R_1C^{x_1})\times \dots \times (1+AB_{y_0}R_{M-1}C^{x_1})}, 
	\end{align*}
which implies:
	\begin{align*}
		\bm G_{T=1}(\bm w,\theta) = 
		\begin{bmatrix} 
			(1,0) \star \left(\underset{s \in [M-1]\setminus\{1\}}{\bigstar} (1, B_{y_0}R_sC^{x_1})\right)\\[10pt]
			(0, B_{y_0}(R_1-R_2)C^{x_1}) \star \left(\underset{s \in [M-1]\setminus\{1,2\}}{\bigstar}(1, B_{y_0}R_sC^{x_1})\right)\\[10pt]
			(0, B_{y_0}(R_1-R_2)C^{x_1}) \star \left(\underset{s \in [M-1]\setminus\{2,3\}}{\bigstar}(1, B_{y_0}R_sC^{x_1})\right)\\[10pt]
			\vdots \\[5 pt]
			(0, B_{y_0}R_{M-1}C^{x_1}) \star \left(\underset{s \in [M-1]\setminus\{M-3,M-2\}}{\bigstar}(1, B_{y_0}R_sC^{x_1})\right)\\[10pt]
			(0, B_{y_0}R_{M-1}C^{x_1}) \star \left(\underset{s \in [M-1]\setminus\{M-1\}}{\bigstar}(1, B_{y_0}R_sC^{x_1})\right)
		\end{bmatrix}. 
	\end{align*}
	Then the highest power for $A$ is $M-1$. Thus, $\bm G_{T=1}(\bm w,\theta)$ is of dimension $M \times M$. For $T = 2$, we follow the same routine of first adding $1$ to all existing choice sequences, and then $2$ and so on. We can update the likelihood as: 
	\[
	f((i,j)\mid y_0, x_1,x_2, \alpha;\theta) = f(i \mid y_0,x_1, \alpha;\theta) \pi_{ij}(x_2).
	\]
	By taking the common denominator for all pairs of $(i,j)$ in $\pi_{i,j}(\cdot)$, we update: 
	\begin{align*}
		\kappa_{T=2}(\bm w, \alpha, \theta) = \frac{\kappa_{T=1}(\bm w, \alpha, \theta)}{\prod_{s=1}^{M-1} \prod_{j=1}^M (1+AB_{j}R_sC^{x_2})},
	\end{align*}
	and:
	
	\resizebox{0.8\textwidth}{!}{$
		\bm G_{T=2}(\bm w,\theta) = \begin{bmatrix} 
			\bm G_{T=1}(\bm w,\theta) \star 
			\begin{bmatrix} (1,0) \star \left( \underset{(s,j) \in \C\setminus\{(1,1)\}}{\bigstar} (1, B_jR_sC^{x_2})\right)\\[15pt]
				(1,0) \star \left( \underset{(s,j) \in \C\setminus\{(1,2)\}}{\bigstar}(1, B_jR_sC^{x_2})\right)\\[15pt]
				\vdots\\[15pt]
				(1,0) \star \left(\underset{(s,j) \in \C\setminus\{(1,M)\}}{\bigstar}(1, B_jR_sC^{x_2})\right)
			\end{bmatrix} \\
			\\
			\bm G_{T=1}(\bm w,\theta) \star 
			\begin{bmatrix} 
				(0, B_1(R_1-R_2)C^{x_2}) \star \left(\underset{(s,j) \in \C\setminus\{(1,1),(2,1)\}}{\bigstar} (1, B_jR_sC^{x_2})\right)\\[15pt]
				(0, B_2(R_1-R_2) C^{x_2}) \star \left(\underset{(s,j) \in \C\setminus\{(1,2),(2,2)\}}{\bigstar} (1, B_jR_sC^{x_2})\right)\\[15pt]
				\vdots\\
				(0, B_M(R_1-R_2)C^{x_2}) \star \left(\underset{(s,j) \in \C\setminus\{(1,M),(2,M)\}}{\bigstar} (1, B_jR_sC^{x_2})\right)
			\end{bmatrix} \\
			\\
			\vdots \\
			\\
			\bm G_{T=1}(\bm w,\theta) \star	\begin{bmatrix} 
				(0, B_1(R_{M-2}-R_{M-1})C^{x_2}) \star \left(\underset{(s,j) \in \C\setminus\{(M-2,1),(M-1,1)\}}{\bigstar} (1, B_jR_sC^{x_2})\right)\\[15pt]
				(0, B_2(R_{M-2}-R_{M-1})C^{x_2}) \star \left(\underset{(s,j) \in \C\setminus\{(M-2,2),(M-1,2)\}}{\bigstar} (1, B_jR_sC^{x_2})\right)\\[15pt]
				\vdots\\
				(0, B_M(R_{M-2}-R_{M-1})C^{x_2}) \star \left(\underset{(s,j) \in \C\setminus\{(M-2,M),(M-1,M)\}}{\bigstar} (1, B_jR_sC^{x_2})\right)
			\end{bmatrix} \\
			\\
			\bm G_{T=1}(\bm w,\theta) \star \begin{bmatrix} 
				(0, B_1R_{M-1}C^{x_2}) \star \left(\underset{(s,j) \in \C\setminus\{(M-1,1)\}}{\bigstar} (1, B_jR_sC^{x_2})\right)\\[15pt]
				(0, B_2R_{M-1}C^{x_2}) \star \left(\underset{(s,j) \in \C\setminus\{(M-1,2)\}}{\bigstar} (1, B_jR_sC^{x_2})\right)\\[15pt]
				\vdots\\
				(0, B_MR_{M-1}C^{x_2}) \star \left(\underset{(s,j) \in \C\setminus\{(M-1,M)\}}{\bigstar} (1, B_jR_sC^{x_2})\right)
			\end{bmatrix}
		\end{bmatrix} 
		$}
	~\\~\\
	For general $t$, use the updates:
	\begin{align*}
		\kappa_{T=t}(\bm w,\alpha,\theta)=\frac{\kappa_{T=t-1}(\bm w,\alpha,\theta)}{\prod_{s=1}^{M-1} \prod_{j=1}^M (1+AB_{j}R_sC^{x_T})}, 
	\end{align*}
	and
	\resizebox{0.9\textwidth}{!}{$
		\bm G_{T=t}(\bm w,\theta) = 
		\begin{bmatrix} \bm G_{T=t-1}(\bm w,\theta) \star 
			\begin{bmatrix} 
				\bm{1}_{Q^{t-2}} \otimes \left((1,0) \star \left( \underset{(s,j) \in \C\setminus\{(1,1)\}}{\bigstar} (1, B_jR_sC^{x_t})\right)\right)\\[15pt]
				\bm{1}_{Q^{t-2}} \otimes \left((1,0) \star \left( \underset{(s,j) \in \C\setminus\{(1,2)\}}{\bigstar} (1, B_jR_sC^{x_t})\right)\right)\\[15pt]
				\vdots\\[15pt]
				\bm{1}_{Q^{t-2}} \otimes \left((1,0) \star \left( \underset{(s,j) \in \C\setminus\{(1,M)\}}{\bigstar} (1, B_jR_sC^{x_t})\right)\right)
			\end{bmatrix} \\
			\\
			\bm G_{T=t-1}(\bm w,\theta) \star 
			\begin{bmatrix} 
				\bm{1}_{M^{t-2}} \otimes \left((0, B_1(R_1-R_2)C^{x_t}) \star \left(\underset{(s,j) \in \C\setminus\{(1,1),(2,1)\}}{\bigstar} (1, B_jR_sC^{x_t})\right)\right)\\[15pt]
				\bm{1}_{M^{t-2}} \otimes \left((0, B_2(R_1-R_2) C^{x_t}) \star \left(\underset{(s,j) \in \C\setminus\{(1,2),(2,2)\}}{\bigstar} (1, B_jR_sC^{x_t})\right)\right)\\[15pt]
				\vdots\\[15pt]
				\bm{1}_{M^{t-2}} \otimes \left((0, B_M(R_1-R_2)C^{x_t}) \star \left(\underset{(s,j) \in \C\setminus\{(1,M),(2,M)\}}{\bigstar} (1, B_jR_sC^{x_t})\right)\right)
			\end{bmatrix} \\
			\\
			\vdots \\
			\\
			\bm G_{T=t-1}(\bm w,\theta) \star	
			\begin{bmatrix} 
				\bm{1}_{M^{t-2}} \otimes \left((0, B_1(R_{M-2}-R_{M-1})C^{x_t}) \star \left(\underset{(s,j) \in \C\setminus\{(M-2,1),(M-1,1)\}}{\bigstar} (1, B_jR_sC^{x_t})\right)\right)\\[15pt]
				\bm{1}_{M^{t-2}} \otimes \left((0, B_2(R_{M-2}-R_{M-1})C^{x_t}) \star \left(\underset{(s,j) \in \C\setminus\{(M-2,2),(M-1,2)\}}{\bigstar} (1, B_jR_sC^{x_t})\right)\right)\\[15pt]
				\vdots\\[15pt]
				\bm{1}_{M^{t-2}} \otimes \left((0, B_M(R_{M-2}-R_{M-1})C^{x_t}) \star \left(\underset{(s,j) \in \C\setminus\{(M-2,M),(M-1,M)\}}{\bigstar} (1, B_jR_sC^{x_t})\right)\right)
			\end{bmatrix} \\
			\\
			\bm G_{T=t-1}(\bm w,\theta) \star 
			\begin{bmatrix}
				\bm{1}_{M^{t-2}} \otimes \left((0, B_1R_{M-1}C^{x_t}) \star \left(\underset{(s,j) \in \C\setminus\{(M-1,1)\}}{\bigstar} (1, B_jR_sC^{x_t})\right)\right)\\[15pt]
				\bm{1}_{M^{t-2}} \otimes \left((0, B_2R_{M-1}C^{x_t}) \star \left(\underset{(s,j) \in \C\setminus\{(M-1,2)\}}{\bigstar} (1, B_jR_sC^{x_t})\right)\right)\\[15pt]
				\vdots\\[15pt]
				\bm{1}_{M^{t-2}} \otimes \left((0, B_MR_{M-1}C^{x_t}) \star \left(\underset{(s,j) \in \C\setminus\{(M-1,M)\}}{\bigstar} (1, B_jR_sC^{x_t})\right)\right)
			\end{bmatrix}
		\end{bmatrix}. 
		$}
	~\\~\\
	The highest power for $A$ when $T = 1$ is $M-1$, denoted by $q_1 = M-1$. Furthermore, the number of columns in $\bm G_{T=1}$ is $q_1 + 1 = M$. Each time we add one more period, with the chosen $\kappa(\bm w ,\alpha,\theta)$ we add an additional $(M-1)M$ moments of $A$, denoted by $q_2 = (M-1)+(M-1)M$, so that $\text{ncol}(\bm G_{T=2}) = M^2$. For general $t\geq 2$, we update by $q_t = q_{t-1} + (M-1)M$ with $q_1 = M-1$; thus, $q_t = q_1 + (t-1)(M-1)M$, and $\text{ncol}(\bm G_t) = q_t + 1$. Putting everything together, the dimension of $\bm G_{T=t}(\bm w,\theta)$ is $M^t \times (M+(t-1)(M-1)M) = M^t \times ((t-1)M^2 - (t-2)M)$ for any $t \geq 2$.
	
	\section{Verification of Assumption 4.2} 
	In this section, we show that Assumption 4.2 in DKGR holds for the AR(1) model with $T = 2$ and $T = 3$. For simplicity, we consider the model without covariates. 
	
	\subsection{The AR(1) Model with $T = 3$}
	Here we first prove: 
	\begin{align*}
		\inf_{\bm r \in \mathcal{R}_{n}} Q_{n,P}(\theta) \geq  \sqrt{\bm p^\top \bm L (\bm L^\top \bm L)^{-1} \bm L^\top \bm p},
	\end{align*}
	where $\bm L_\theta$ is any basis for the left null space of $\bm G(\theta)$, where $\bm G(\theta)$ is the block-diagonal matrix with $\bm G(0,\theta)$ and $\bm G(1, \theta)$ as its blocks. To see this, note that $\inf_{\bm r\in \mathcal{R}_n} Q_{n,P}(\theta,\bm r)$ is nothing but the norm of the (constrained) least squares residual in a regression of $\bm p$ on the columns of $\bm G(\theta)$:
	\begin{align*}
		\inf_{\bm r \in \Pi_{\mathcal{R}_{n}}(\mathcal{S})} Q_{n,P}(\theta) = \inf_{\bm r \in \Pi_{\mathcal{R}_{n}}(\mathcal{S})} ||\bm p - \bm G(\theta) \bm r|| \geq \inf_{\bm r \in \mathbb{R}^{S+1}} ||\bm p - \bm G(\theta) \bm r||.
	\end{align*}
	Now let $\mathcal{G}$ denote the column space of $\bm G(\theta)$, let $\bm p_{\mathcal{G}}$ denote the projection of $\bm p$ onto the space spanned by the columns of $\bm G(\theta)$, and let $\bm p_{\mathcal{G}^\perp}$ denote the projection of $\bm p$ onto the space orthogonal to the columns of $\bm G(\theta)$. By Pythagoras' Theorem:
	\begin{align*}
		||\bm p - \bm G(\theta) \bm r||^2 = || \bm p_{\mathcal{G}} - \bm G(\theta) \bm r||^2 + ||\bm p_{\mathcal{G}^\perp}||^2.
	\end{align*}
	Thus, we have:
	\begin{align*}
		\inf_{\bm r \in \Pi_{\mathcal{R}_{n}}(\mathcal{S})}Q_{n,P}(\theta) \geq \inf_{\bm r \in \mathbb{R}^{S+1}} ||\bm p - \bm G(\theta) \bm r|| = ||\bm p_{\mathcal{G}^\perp}||.
	\end{align*}
	By the Fundamental Theorem of Linear Algebra, the space $\mathcal{G}^\perp$ is exactly the left null space of $\bm G(\theta)$. Thus, if $\bm L_\theta$ is a matrix whose columns form the basis of the left null space of $\bm G$, then $\bm p_{\mathcal{G}^\perp} = \bm L_\theta (\bm L\theta^\top \bm L_\theta)^{-1} \bm L_\theta^\top \bm p$, so that:
	\begin{align*}
		||\bm p_{\mathcal{G}^\perp}|| =  \sqrt{\bm p^\top \bm L_\theta (\bm L_\theta^\top \bm L_\theta)^{-1} \bm L_\theta^\top \bm p}.
	\end{align*}
For the AR(1) model with $T=3$ and $\theta=\beta$, we have:
	\begin{align*}
		\bm p^\top \bm L_{\beta} (\bm L_{\beta}^\top \bm L_{\beta})^{-1} \bm L_{\beta}^\top \bm p
		&=\frac{(P((0,1,0) \mid Y_{i0}=0)-P((1,0,0) \mid Y_{i0}=0))^2}{2}\\
		&\qquad+
		\frac{(P((0,1,1) \mid Y_{i0}=0)-\exp(\beta)\,P((1,0,1) \mid Y_{i0}=0))^2}{\exp(2\beta)+1}\\
		&\qquad\qquad+
		\frac{(\exp(\beta)\,P((0,1,0) \mid Y_{i0}=1)-P((1,0,0) \mid Y_{i0}=1))^2}{\exp(2\beta)+1}\\
		&\qquad\qquad\qquad+
		\frac{(P((0,1,1) \mid Y_{i0}=1)-P((1,0,1) \mid Y_{i0}=1))^2}{2}.
	\end{align*}
	Only the first and third term depend on the model parameters, and these terms are zero precisely when:
	\begin{align*}
		\beta = \log \left( \frac{P((0,1,1) \mid Y_{i0}=0)}{P((1,0,1) \mid Y_{i0}=0)} \right)  = \log \left( \frac{P((1,0,0) \mid Y_{i0}=1)}{P((0,1,0) \mid Y_{i0}=1)} \right).
	\end{align*}
	Note this is exactly Chamberlain's formula for the AR(1) model. If we consider some alternative $\beta' = \beta + \Delta$ for some $\Delta \in \mathbb{R}\setminus\{0\}$, then after some simplification we have:
	\begin{align*}
		&\bm p^\top \bm L_{\beta'} (\bm L_{\beta'}^\top \bm L_{\beta'})^{-1} \bm L_{\beta'}^\top \bm p\\ 
		&\geq \frac{(P((0,1,1) \mid Y_{i0}=0)-\exp(\beta+\Delta)\,P((1,0,1) \mid Y_{i0}=0))^2}{\exp(2\beta+2\Delta)+1}\\
		&\qquad +
		\frac{(\exp(\beta+\Delta)\,P((0,1,0) \mid Y_{i0}=1)-P((1,0,0) \mid Y_{i0}=1))^2}{\exp(2\beta+2\Delta)+1}\\
		& = \frac{P((0,1,0) \mid Y_{i0}=1)^2 (1-e^\Delta)^2(P((0,1,1) \mid Y_{i0}=0)^2+P((1,0,0) \mid Y_{i0}=1)^2)}{e^{2\Delta}P((1,0,0) \mid Y_{i0}=1)^2+P((0,1,0) \mid Y_{i0}=1)^2}.
	\end{align*}
	Now suppose that all elements of $\bm p$ are strictly larger than some $\varepsilon_p>0$. Then when $\Delta >0$, on the range $[0,\delta]$ we have:
	\begin{align*}
		&\frac{P((0,1,0) \mid Y_{i0}=1)^2 (1-e^\Delta)^2(P((0,1,1) \mid Y_{i0}=0)^2+P((1,0,0) \mid Y_{i0}=1)^2)}{e^{2\Delta}P((1,0,0) \mid Y_{i0}=1)^2+P((0,1,0) \mid Y_{i0}=1)^2}\\
		&\geq \left(\frac{P((0,1,0) \mid Y_{i0}=1)^2 (P((0,1,1) \mid Y_{i0}=0)^2+P((1,0,0) \mid Y_{i0}=1)^2)}{e^{2\delta}P((1,0,0) \mid Y_{i0}=1)^2+P((0,1,0) \mid Y_{i0}=1)^2} \right) \Delta^2\\
		&\geq \left(\frac{2\varepsilon_p^4}{(e^{2\delta}+1)(1-\varepsilon_p)^2} \right) \Delta^2.
	\end{align*}
	Here we used the fact that $\exp(\Delta)-1 \geq \Delta$ when $\Delta > 0$. When $\Delta <0$, on the range $[-\delta,0]$ for $\delta<1$ we have:
	\begin{align*}
		&\frac{P((0,1,0) \mid Y_{i0}=1)^2 (1-e^\Delta)^2(P((0,1,1) \mid Y_{i0}=0)^2+P((1,0,0) \mid Y_{i0}=1)^2)}{e^{2\Delta}P((1,0,0) \mid Y_{i0}=1)^2+P((0,1,0) \mid Y_{i0}=1)^2}\\
		&\geq \frac{P((0,1,0) \mid Y_{i0}=1)^2 (P((0,1,1) \mid Y_{i0}=0)^2+P((1,0,0) \mid Y_{i0}=1)^2)}{P((1,0,0) \mid Y_{i0}=1)^2+P((0,1,0) \mid Y_{i0}=1)^2} \left(\frac{\Delta^2}{4}\right)\\
		&\geq \frac{\varepsilon_{p}^4 }{(1-\varepsilon_p)^2} \left(\frac{\Delta^2}{4}\right).
	\end{align*}
	Here we used the fact that $\exp(\Delta)-1 \geq \frac{\Delta}{4}$ when $\Delta \in [-1,0]$. Thus, taking square roots and simplifying, Assumption 4.2 is satisfied for the AR(1) model with $T=3$ and no covariates for $\delta=1$ and (after noticing $\sqrt{e^{2}+1} \leq 3$):
	\begin{align*}
		\nu^{-1} = \frac{\varepsilon_p^2}{2(1-\varepsilon_p)}.
	\end{align*}
	However, other choices of $\delta$ and $\nu$ are also possible. 
	
	\subsection{The AR(1) Model with $T = 2$}
	The argument used for the AR(1) model with $T=3$ does not work for the AR(1) model with $T=2$, because the left null space of the matrix $\bm G(\theta)$ is the zero vector for almost all values of the structural parameters. Here we show that Assumption 4.2 still holds for the AR(1) model with $T=2$ (with no covariates). To see this, again note that $\inf_{\bm r\in \mathcal{R}_n} Q_{n,P}(\theta,\bm r)$ is nothing but the norm of the (constrained) least squares residual in a regression of $\bm p$ on the columns of $\bm G(\theta)$:
	\begin{align*}
		\inf_{\bm r \in \Pi_{\mathcal{R}_{n}}(\mathcal{S})} Q_{n,P}(\theta) = \inf_{r \in \Pi_{\mathcal{R}_{n}}(\mathcal{S})} ||\bm p - \bm G(\theta) \bm r||.
	\end{align*}
	Assume that $\theta=\beta$ is bounded away from zero. Then $\bm G(\theta)$ is invertible in this model. Now define $\bm r_{\theta}^* := \bm G(\theta)^{-1}\bm p$. Also, define $\bm r_{\theta}^*(y_0) := \bm G(y_0,\theta)^{-1}\bm p(y_0)$. Then:
	\begin{align*}
		\inf_{r \in \Pi_{\mathcal{R}_{n}}(\mathcal{S})} ||\bm p - \bm G(\theta) \bm r||&=\inf_{\bm r \in \Pi_{\mathcal{R}_{n}}(\mathcal{S})}  || \bm G(\theta)\bm r_{\theta}^* - \bm G(\theta) \bm r|| \\
		&= \inf_{\bm r \in \Pi_{\mathcal{R}_{n}}(\mathcal{S})}  || \bm G(\theta)(\bm r_{\theta}^* - \bm r) ||\\
		&= \inf_{r \in \Pi_{\mathcal{R}_{n}}(\mathcal{S})}  \frac{|| \bm G(\theta)(\bm r_{\theta}^* - \bm r) ||}{||\bm r_{\theta}^* - \bm r||} ||\bm r_{\theta}^* - \bm r||\\
		&\geq \sigma_{min}(\bm G(\theta))\cdot\inf_{\bm r \in \Pi_{\mathcal{R}_{n}}(\mathcal{S})}||\bm r_{\theta}^* - \bm r||. 
	\end{align*}
	Note  that $||\bm r_{\theta}^* - \bm r|| \geq \max\{||\bm r_{\theta}^*(0) - \bm r(0)||,||\bm r_{\theta}^*(1) - \bm r(1)||\}$. For the remainder, we focus on the case with $y_0=0$ (the condition can be checked in a similar way for $y_0 = 1$). The model restrictions on the moment vector---coming from the restrictions on the Hankel matrices---are:
	\begin{align*}
		C_1 & :=\{ \bm r: g_1(\bm r) = r_0 \geq 0\},\\
		C_2 & :=\{ \bm r: g_2(\bm r) = r_1 \geq 0\},\\
		C_3 & :=\{ \bm r: g_3(\bm r) = r_2 \geq 0\},\\
		C_4 & :=\{ \bm r: g_4(\bm r) = r_3 \geq 0\},\\
		C_5& :=\{ \bm r: g_5(\bm r) = r_0 r_1 - r_2^2 \geq 0\},\\
		C_6 & :=\{ \bm r: g_6(\bm r) =  r_1 r_3 - r_2^2 \geq 0\}.
	\end{align*}
	In particular, the set $S:= \cap_{k=1}^6 C_k$ represents the \textit{closure} of the moment space. We then have: 
	\begin{align*}
		\inf_{\bm r \in \Pi_{\mathcal{R}_{n}}(\mathcal{S})}||\bm r_{\theta}^* - \bm r|| & \geq  \underset{\bm r \in S}{\min}\;  \| \bm r_{\theta}^*(0) - \bm r\| \geq \underset{k \in [6]}{\min} \; \underset{\bm r \in C_k}{\min}\; \|\bm r_{\theta}^*(0) - \bm r\|.
	\end{align*} 
	For a given constraint $k$ from above, define the constraint function $h_k(\theta) = g_k(\bm r(\theta))$, and consider any $\theta_0$ that lies on the boundary of the $k^{th}$ constraint set; that is, $h_k(\theta_0) = 0$. Now consider a point $\theta_k :=\theta_0 + \Delta_k$ that violates the $k^{th}$ constraint. 
	
	\begin{lemma}\label{lemma:label}
		Suppose that $g_k$ is Lipschitz continuous on the line segment between $\bm r(\theta_0 + \Delta_k)$ and $\bm r(\theta_0)$ with Lipschitz constant $M_k$, and suppose that $h_k$ is continuously differentiable with $h'_k(\theta_0)\neq 0$ (both conditions will be verified below for each $k$). Then there exists $\bar \delta_k >0$ such that for all $|\Delta_k| \leq \bar \delta_k$: 
		\[
		\underset{\bm r \in C_k}{\min}\; \|\bm r_{\theta_k}^*(0) - \bm r\|  \geq \frac{|h_k'(\theta_0)|}{2M_k}|\Delta_k|.
		\]
		
	\end{lemma} 
	
	\begin{proof}[Proof of Lemma \ref{lemma:label}]
		Note that:
		\begin{align}
			\underset{\bm r \in C_k}{\min}\; \|\bm r_{\theta_0+\Delta_k}^*(0) - \bm r\| &\geq \frac{|g_k(\bm r_{\theta_0}(0)) -g_k(\bm r_{\theta_0+\Delta_k}(0))|}{M_k}\\
			&=\frac{|-g_k(\bm r_{\theta_0+\Delta_k}(0))|}{M_k}\\
			& = \frac{|-h_k(\theta_0+\Delta_k)|}{M_k}.\label{bound}
		\end{align}
		Now, provided $h_k'(\theta_0) \neq 0$, there exists $\bar \delta>0$ such that $|\theta-\theta_0| \leq \bar \delta$ implies that $|h_k'(\theta) - h_k'(\theta_0)| \leq \frac{1}{2} |h_k'(\theta_0)|$. Then for any $|\Delta_k|<\bar \delta$, the mean value theorem, the reverse triangle inequality, and the bound $|h_k'(\theta) - h_k'(\theta_0)| \leq \frac{1}{2} |h_k'(\theta_0)|$ together imply:
		\begin{align*}
			|h_k(\theta_0+\Delta_k)| = |h_k'(\zeta)|\cdot|\Delta_k| &\geq |h_k'(\theta_0) - (h_k'(\zeta)-h_k'(\theta)) |\cdot|\Delta_k| \\
			&\geq | \,\,|h_k'(\theta_0)| - |(h_k'(\zeta)-h_k'(\theta)) |\,\,| \cdot |\Delta_k| \\
			&\geq (|h_k'(\theta_0)| - |(h_k'(\zeta)-h_k'(\theta)) | )\cdot |\Delta_k| \\
			&\geq \left(|h_k'(\theta_0)| - \frac{1}{2} |h_k'(\theta_0)| \right)\cdot |\Delta_k| \\
			&=\frac{1}{2} |h_k'(\theta_0)| |\Delta_k|,
		\end{align*} 
		for $\zeta$ between $\theta_0$ and $\theta_0+\Delta_k$. Combined with \eqref{bound}, this completes the proof. 
	\end{proof} 
	
	Note we can always take $M_k = \sup_{\bm r} \| \nabla g_k(\bm r)\|$ provided $\bm r$ is bounded, which is guaranteed in the paper by the use of piecewise constant functions with bounded coefficients. We thus focus on showing $h_k'(\theta_0) \neq 0$ for all $k$. Let $\exp(\theta) = B$, and $\bm p = (p_1, p_2, p_3,p_4) = (p(0,0), p(1,0), p(0,1), p(1,1)) > 0$. We consider six cases:
	\begin{enumerate}
		\item For $k = 1$: here we have $h_1(\theta) = \frac{(B-1)p_1-B^2p_2+Bp_3}{B-1}$, and we have that $B_0 = \exp(\theta_0)$ is determined by the roots of $p_2 B_0^2 = (p_1+p_3)B_0-p_1$. Thus: 
		\begin{align*}
			h_1'(\theta_0) = \frac{B_0(B_0(2p_2-p_1-2p_3)+(p_1+p_3))}{(B_0-1)^2}\neq 0,
		\end{align*} 
		provided $(p_1+p_3)^2 - 4p_1p_2 \neq 0$.
		\item For $k = 2$: here we have $h_2(\theta) = \frac{Bp_2-p_3}{B-1}$, and $B_0= \exp(\theta_0)$ is determined by $B_0 p_2 =p_3$. Thus: 
		\begin{align*}
			h_2'(\theta_0) = \frac{B_0(p_3-p_2)}{(B_0-1)^2} \neq 0,
		\end{align*}  
		provided $p_2 \neq p_3$. 
		\item For $k = 3$: here we have $h_3(\theta) = \frac{p_3-p_2}{B-1}=\frac{A^2}{(1+A)^2(1+AB)}$ where $A=\exp(\alpha)$. For $\theta=\beta$ bounded away from zero, we will never have $h_3(\theta)=0$, so we can ignore this case. 
		\item For $k = 4$: here we have $h_4(\theta) = \frac{p_2-p_3}{B-1}+\frac{p_4}{B}$, and $B_0 = \exp(\theta_0)$ is determined by $B_0 = \frac{p_4}{p_2-p_3+p_4}$ (provided $p_2-p_3+p_4>0$ to ensure $B_0>0$). Thus: 
		\begin{align*}
			h_4'(\theta_0) = \frac{p_3-p_2}{(B_0-1)^2} \neq 0,
		\end{align*}
		provided $p_2\neq p_3$.
		\item For $k = 5$: here we have $h_5(\theta) = \frac{-B p_2p_3-p_1p_2+p_1p_3+p_3^2}{B-1}$ and $B_0 = \frac{p_1p_3-p_1p_2+p_3^2}{p_2p_3}$ (provided $p_1(p_3-p_2) + p_3^2 > 0$.). Thus:
		\begin{align*}
			h_5'(\theta_0) = \frac{B_0 (p_1+p_3)(p_2-p_3)}{(B_0-1)^2} \neq 0,
		\end{align*}
		provided $p_2 \neq p_3$. 
		\item For $k = 6$: here we have $h_6(\theta) = \frac{Bp_2(p_2 - p_3+p_4)-p_3p_4}{B(B-1)}$ and $B_0 =\exp(\theta_0)= \frac{p_3p_4}{p_2(p_2-p_3+p_4)}$ (provided $p_2-p_3+p_4>0$). Thus: 
		\[
		h_6'(\theta_0) = \frac{p_2(p_2-p_3+p_4)}{B_0-1} \neq 0,
		\]
		provided $p_2-p_3+p_4\neq 0$.
	\end{enumerate}
	For AR(1) model with $T=2$, $p_2/p_3 = \frac{1+AB}{1+A} \neq 1$ provided $B \neq 1$. If we can also rule out DGPs with $p_2-p_3+p_4 =0$ and $(p_1+p_3)^2 - 4p_1p_2=0$, each corresponding to the distribution of $A$ being degenerate, then Assumption 4.2 is satisfied. 

\end{document}